\documentclass{article}
\usepackage[a4paper, total={6in, 8in}]{geometry}
\usepackage[dvipsnames]{xcolor}
\usepackage{comment}
\usepackage{todonotes}
\usepackage[export]{adjustbox}
\usepackage[english]{babel}
\usepackage[utf8]{inputenc}
\usepackage{capt-of}
\usepackage{amsmath}
\usepackage{tikz}
\usepackage{mathtools}
\usepackage{amsmath}
\usepackage{mathtools}
\usepackage{natbib}
\usepackage{bibentry}
\usepackage{adjustbox}
\usepackage{array}
\usepackage{booktabs}
\usepackage{dsfont}
\usepackage{multirow}
\usepackage{hhline}
\usepackage[utf8]{inputenc} % allow utf-8 input
\usepackage[T1]{fontenc}    % use 8-bit T1 fonts      % hyperlinks
\usepackage[hidelinks]{hyperref}
\usepackage{url}            % simple URL typesetting
\usepackage{doi}
\usepackage{booktabs}       % professional-quality tables
\usepackage{amsfonts}       % blackboard math symbols
\usepackage{nicefrac}       % compact symbols for 1/2, etc.
\usepackage{microtype}      % microtypography
\usepackage{amsthm}
\theoremstyle{definition}
\newtheorem{definition}{Definition}[section]
\theoremstyle{theorem}

\usepackage{float}
\usepackage[font=small,labelfont=bf]{caption}
\usepackage[caption = false]{subfig}
\usepackage{graphicx}

\newtheorem{lemma}{Lemma}

\usepackage[toc,page]{appendix}
\usepackage{algorithm}
\usepackage{algpseudocode}
\usepackage{amsthm,mathtools}
\theoremstyle{definition}

\definecolor{bleu}{RGB}{0,101,189}
\definecolor{vert}{HTML}{004D40}
\definecolor{rose}{HTML}{D81B60}
\definecolor{bleuTOL}{HTML}{332288}

\hypersetup{colorlinks,linkcolor={rose},citecolor={bleu},urlcolor={bleuTOL}}

\title{Optimal Transport for Counterfactual Estimation:\\A Method for Causal Inference}

\author{Arthur Charpentier$^{a*}$, Emmanuel Flachaire$^{b}$ \& Ewen Gallic$^{b}$ \\
        \small $^{a}$ Université du Québec à Montréal (UQAM), Montréal (Québec), Canada \\
        \small $^{b}$ Aix Marseille Univ, CNRS, AMSE, Marseille, France \\\\
        \small $^{*}$Corresponding author: \tt{charpentier.arthur@uqam.ca}}

\date{} %leave blank

\begin{document}

\maketitle
\begin{abstract} 
\noindent Many problems ask a question that can be formulated as a causal question: {\em what would have happened if\ldots}? For example, {\em would the person have had surgery if he or she had been Black}? To address this kind of questions, calculating an average treatment effect (ATE) is often uninformative, because one would like to know how much impact a variable (such as skin color) has on a specific individual, characterized by certain covariates. Trying to calculate a conditional ATE (CATE) seems more appropriate. In causal inference, the propensity score approach assumes that the treatment is influenced by $\boldsymbol{x}$, a collection of covariates. Here, we will have the dual view: doing an intervention, or changing the treatment (even just hypothetically, in a thought experiment, for example by asking what would have happened if a person had been Black) can have an impact on the values of $\boldsymbol{x}$. We will see here that optimal transport allows us to change certain characteristics that are influenced by the variable we are trying to quantify the effect of. We propose here a {\em mutatis mutandis} version of the CATE, which will be done simply in dimension one by saying that the CATE must be computed relative to a level of probability, associated to the proportion of ${x}$ (a single covariate) in the control population, and by looking for the equivalent quantile in the test population. In higher dimension, it will be necessary to go through transport, and an application will be proposed on the impact of some variables on the probability of having an unnatural birth (the fact that the mother smokes, or that the mother is Black). 
\end{abstract}

\ 


\section{Introduction}

\subsection{From intervention to counterfactuals}

In \cite{pearl2018book}, a ``ladder of causation'' is introduced, to describe the three levels of causal reasoning. The first level, named ``{\em association}'', discusses associations (not to use the word ``{\em correlation}'') between variables. Questions such as ``{\em is variable $X$ associated with variable $Y$}?'' can be answered at this level. Econometric models are usually simply based on such associations. 
The second level is labelled ``{\em intervention}''. Reasoning on this level answers questions of the form ``{\em if I make the intervention $T$, how will this affect the level of the outcome $Y$?}'' For example, the question ``{\em would a patient heal faster at home or at the hospital, after some surgery?}'' is a standard question on this second level of the ladder of causation. This kind of reasoning invokes causality and can be used to investigate more questions than the reasoning of the first level. 
The third level of the ``ladder of causation'' is labelled ``{\em counterfactuals}'' and involves answering questions which ask what might have been, had circumstances been different. %An example of counterfactual questions given in the book is ``{\em would Kennedy be alive if Oswald had not killed him?}''.
Counterfactual modeling implies that, to each individual in the control space, described through variables $\boldsymbol{x}$ and $y$, we will associate a counterfactual version of that individual in the hypothetical space. More formally, we will use notations of causal inference to answer counterfactual questions, such as ``{\em would that person have had surgery if she had been Afro-American?}'' 

\subsection{Causal inference framework}

Consider, as in \cite{rubin1974estimating} or
\cite{hernan2010causal}, the following framework: let $t$ denote some binary treatment, $t\in\{0,1\}$, with respectively, the control and the treatment. Let $\boldsymbol{x}$  be some covariates,  $y$ the observed outcome, with $y_{T\leftarrow 1}^\star$ and $y_{T\leftarrow 0}^\star$ the potential outcomes (also denoted $y(1)$ and $y(0)$ in \cite{imbens2015causal} or \cite{imai2018quantitative}, or $y^1$ and $y^0$ in \cite{morgan2014counterfactuals} or \cite{cunningham2021causal}, even $y_{t=1}$ and $y_{t=0}$ in \cite{pearl2018book}), realized either under treatment condition ($t=1$) or under control condition ($t=0$). Note that the observed outcome is $y=y_{T\leftarrow t}^\star$, or $y=t\cdot y_{T\leftarrow 1}^\star+(1-t)\cdot y_{T\leftarrow 0}^\star$. An illustration is reported in Table~\ref{Tab:potential:outcome}.

\begin{table}[!ht]\centering
    \begin{tabular}{lcccccccc}\hline\hline
    & Treatment & \multicolumn{3}{c}{Outcome} & Age & Gender & Height & Weight \\
    \cmidrule(lr){3-5}
    & $t_i$ & $y_i$ & $y_{i,T\leftarrow 1}^\star$ & $y_{i,T\leftarrow o}^\star$ & $x_{1,i}$ &  $x_{2,i}$ &  $x_{3,i}$ &  $x_{4,i}$ \\ \midrule
    1 & 1 & 121 & 121 & {\bf ?} & 37 & F & 160 & 56 \\
    2 & 0 & 109 & {\bf ?} & 109 & 28 & F & 156 & 54 \\
    3 & 1 & 162 & 162 & {\bf ?} & 53 & M & 190 & 87 \\ \hline\hline
    \end{tabular}
    \caption{Potential outcome framework of causal inference, with one binary treatment $t_i$, the observed outcome variable $y_i$ and the two potential outcomes $y_{i,T\leftarrow 1}^\star$ and $y_{i,T\leftarrow 0}^\star$, as well as some covariates $\boldsymbol{x}_i$. One of the two potential outcomes is observed, and the other is missing, indicated by the question mark in the table.}\label{Tab:potential:outcome}
\end{table}

We will use the term ``treatment'' (and letter $t$) even if interventions are not possible, so it is no {\em per se} a ``treatment''. In this article, we try to answer a hypothetical question, like most questions asked at the third level of the ``ladder of causality''. For instance, in a context of quantifying discrimination, the ``treatment'' will denote the sensitive attribute, as in \cite{charpentier2023fairness}, such as the race of an individual, \textit{e.g.}, ``{\em what would have been the outcome if that person had been Afro-American?}'' Since our approach proposes an improvement on the metrics used in causal inference literature, we will use similar notations.

There will be a significant impact of treatment $t$ on $y$ if $y^\star_{T\leftarrow0}\neq y^\star_{T\leftarrow1}$. More specifically, the causal effect for individual $i$ is ${\tau_i=y^\star_{i,T\leftarrow1} - y^\star_{i,T\leftarrow0}}$. The {average treatment effect} (ATE) can the be defined as follows:
$$
\tau = \text{ATE} = \mathbb{E}\big[ Y^\star_{i,T\leftarrow1} - Y^\star_{i,T\leftarrow0}\big].
$$
Its empirical counterpart, the {sample average treatment effect} (SATE) writes:
$$
\widehat{\tau} = 
\text{SATE} = \frac{1}{n}\sum_{i=1}^n y^\star_{i,T\leftarrow1} - y^\star_{i,T\leftarrow0}.
$$
Unfortunately, the latter is not directly observable, since one of the two is always missing, but some techniques can be used to provide some robust estimate of that quantity (we will present some of them in the next section).

Lastly, in the context of possibly heterogeneous effects, captured through covariates $\boldsymbol{x}$ (that can be a subset of the entire set of covariates), the {conditional average treatment effect} (CATE) is defined as the functional
$$
\tau(\boldsymbol{x}) = \text{CATE}(\boldsymbol{x}) = \mathbb{E}\big[ Y^\star_{T\leftarrow1} - Y^\star_{T\leftarrow0}\big\vert \boldsymbol{X}=\boldsymbol{x}\big]
$$
that can be written
$$
\tau(\boldsymbol{x}) = \text{CATE}(\boldsymbol{x}) = \mathbb{E}\big[ Y^\star_{T\leftarrow1} \big\vert \boldsymbol{X}=\boldsymbol{x}\big]- \mathbb{E}\big[ Y^\star_{T\leftarrow0} \big\vert \boldsymbol{X}=\boldsymbol{x}\big],
$$
as introduced in \cite{hahn1998role} and \cite{heckman1998matching}. More recently, \cite{hitsch2018heterogeneous} used that measure to quantify heterogeneous treatment effects to evaluate optimal targeting policies, as well as \cite{powers2018some} and \cite{fan2022estimation}. \cite{wager2018estimation},  \cite{athey2019estimating} and \cite{athey2019generalized}  suggested to use random forests to estimate this quantity, inspired by \cite{davis2017using}. See also \cite{kunzel2019metalearners} or \cite{hsu2022counterfactual} for additional discussion on that quantity.

% Among extensions,  \cite{abadie2002instrumental} and \cite{chernozhukov2005iv} suggested some conditional quantile treatment effects (CQTE), defined as $Q_{\tau}\big[ Y^\star_{i,T\leftarrow1} \big\vert \boldsymbol{X}=\boldsymbol{x}\big]-Q_{\tau}\big[ Y^\star_{i,T\leftarrow0} \big\vert \boldsymbol{X}=\boldsymbol{x}\big]$, defined as the difference between two quantile regressions.

\ 

% The $k$-nearest neighbor estimate of $\tau(\boldsymbol{x})$ is
% $$
% \widehat{\tau}_k(\boldsymbol{x}) =\frac{1}{k}\sum_{i\in V^k_1(\boldsymbol{x})} y_i - \frac{1}{k}\sum_{i\in V^k_0(\boldsymbol{x})} y_i,
% $$
% where $V_j^k(\boldsymbol{x})$ is the set of indices of the $k$ closest neighbors of $\boldsymbol{x}$ in the group of individuals with treatment $t=j$.

A classical assumption is that $(t_i,y_i,\boldsymbol{x}_i)$ is a random sample of size $n$ from some joint random vector $(T,Y,\boldsymbol{X})$. \cite{rosenbaum1983central} suggested a strong ``ignorable treatment assignment'' assumption defined as a conditional independence between $(Y^\star_{T\leftarrow0},Y^\star_{T\leftarrow1})$ and $T$, conditional on the covariates $\boldsymbol{X}$.

% The $k$-nearest neighbor estimate of $\tau(\boldsymbol{x})$, when $\boldsymbol{x}$ is considered in the control group (0), is
% $$
% \widehat{\tau}_k(\boldsymbol{x}) =\frac{1}{k}\sum_{i\in V^k_1(\widehat{T}(\boldsymbol{x}))} y_i - \frac{1}{k}\sum_{i\in V^k_0(\boldsymbol{x})} y_i
% $$
% where $V_0^k(\boldsymbol{x})$ is the set of indices of the $k$ closest neighbors of $\boldsymbol{x}$ in the control group, and $V_1^k(T(\boldsymbol{x}))$ is the set of indices of the $k$ closest neighbors of $T(\boldsymbol{x})$ in the treated group.

\subsection{Agenda}

In Section~\ref{sec:exogeneous}, and more specifically in Section~\ref{subsec:1}, we will discuss further the (possible) connection between covariates $\boldsymbol{x}$,  treatment $t$ and the outcome $y$. Following our example on discrimination, the treatment variable $t$ (such as skin color) is an ``exogenous variable'', in the sense that it cannot be influenced either by covariates $\boldsymbol{x}$ or by the outcome $y$. Using the terminology from directed acyclic graphs (DAGs), $t$ will have no parent, so in a sense, it will be easier to pretend that an hypothetical intervention on $t$ is possible. In most applications, $t$ will have an impact on the outcome $y$, but not only. More precisely, it is possible that $t$ might influence some covariates $\boldsymbol{x}$, and those covariates can, in turn, impact the outcome $y$. In Section~\ref{subsec:3}, we suggest an extension from the standard {\em ceteris paribus} $\text{CATE}(\boldsymbol{x})$ defined as the difference $
\mathbb{E}\big[Y^*_{T\leftarrow 1}\big|\boldsymbol{x}\big] - 
\mathbb{E}\big[Y^*_{T\leftarrow 0}\big|\boldsymbol{x}\big]$, to some {\em mutatis mutandis} $\text{CATE}(\boldsymbol{x})$ defined as the difference $
\mathbb{E}\big[Y^*_{T\leftarrow 1}\big|\boldsymbol{x}_{T\leftarrow 1}\big] - 
\mathbb{E}\big[Y^*_{T\leftarrow 0}\big|\boldsymbol{x}\big]$, where, if $\boldsymbol{x}$ is considered with respect to the control group, the counterfactual in the treated population should be based on a different version of $\boldsymbol{x}$, in the treated space.
As discussed in Section~\ref{subsec:5}, the classical tool used in econometrics is the propensity score, based on $\mathbb{P}[T=1|\boldsymbol{X}=\boldsymbol{x}]$, that is usually considered to take into account the association that exists between the treatment and the covariates. At the second stage of the ``ladder of causation'' --the intervention-- we consider the fact that $\boldsymbol{x}$ might influence $t$. When answering the question ``{\em would a patient heal faster at home or at the hospital, after some surgery?}'', it might be relevant to assume that the propensity score can be used to correct for the bias we have in the data, since some patient have been healing at the hospital, not by choice, but because of some $\boldsymbol{x}$. At the third stage of the ladder --the counterfactuals-- some sort of dual version should be considered, since $t$ is not influenced by $\boldsymbol{x}$, quite the opposite: some $\boldsymbol{x}$ might be influenced by $t$. A simple toy example, based on a Gaussian structural equation model (SEM), is presented in Section~\ref{sec:gaussian:toy}, while in Section~\ref{subsec:7}, we briefly present real data that we will use in the next sections to illustrate various algorithms, based on births in the United States. The variable of interest $y$ is a binary variable, indicating whether a birth was natural, or not. The covariates $\boldsymbol{x}$ considered here will be the weight of the newborn, and the weight gain of the mother. And various ``treatments'' are considered: whether the mother is Afro-American, or not; whether the mother is a smoker, or not; whether the baby is a girl, or not (results for the last two are reported in Appendix~\ref{app:3}).

In Section~\ref{sec:opt:transp:univarie}, we will focus on the case where only one covariate $x$ is considered. We will start with classical matching techniques in  Section~\ref{sub:sec:univ:classical:coupling}, used to match each point in $(y_i,x_i,t_i=0)$ --in the control group-- with another one in $(y_j,x_j,t_j=1)$ --in the treated group-- when the two groups have the same size. In Section~\ref{sub:sec:univ:optimal:coupling}, we will suggest on ``optimal'' matching algorithm, to associate individual $i$ (in the control group) to $j$ (in the treated group), that we will denote $j_i^\star$. Then, in Section~\ref{sub:sec:univ:optimal:matching}, we will discuss the case where the two groups have different sizes, that will be called optimal ``coupling''. In Section~\ref{sub:sec:univ:cate}, we will define an estimator, the {\em mutatis mutandis} CATE, $\widehat{m}_1\big(\widehat{\mathcal{T}}(x)\big) - \widehat{m}_0\big(x\big)$, where $\widehat{\mathcal{T}}(x)= \widehat{F}_1^{-1}\circ \widehat{F}_0(x)$, with $ \widehat{F}_0$ and $ \widehat{F}_1$ denoting the empirical distribution functions of $x$ conditional on $t=0$ and $t=1$, respectively. We will use quantiles to optimally ``transport'' $\boldsymbol{x}$'s from the control group to the treated group, formally through the $\mathcal{T}$ mapping.  Finally, in Section~\ref{sub:ex}, we will illustrate this on probability to have a non-natural baby delivery, on our dataset.

In Section~\ref{sec:opt:transp:multivarie}, we will extend our previous approach to the case where several covariates $\boldsymbol{x}$ are considered. Formally, we will use optimal transport techniques to get a proper counterfactual of $\boldsymbol{x}$, not in the control group, but in the treated group. In Section~\ref{sect:multi:ot}, we will define the optimal transport problem for any number of dimensions and then, in Section~\ref{sect:multi:matching}, we will explain how to optimally associate each observation $\boldsymbol{x}_i$ in the control group (when $t=0$) with a single counterfactual observation $\boldsymbol{x}_j$ in the treated group (when $t=1$) when the two groups have the same size. This can be related to the Gaussian SEM discussed in Section~\ref{sec:gaussian:toy}. In Section~\ref{sect:multi:coupling}, we will see the extension when the two groups have different sizes. 
Unfortunately, those approach do not provide an explicit mapping $\mathcal{T}$, but simply a matching of a single individual $\boldsymbol{x}_i$ (in the control group) to a weighted sum of multiple $\boldsymbol{x}_j$ (in the treated group). As we will see in Section~\ref{sec:gaussian}, it will be possible to get explicit formulation for the mapping $\mathcal{T}$ (from the space of covariates in the control group to the space of covariates in the treated group) when we assume that $\boldsymbol{X}$ conditional on $T$ has Gaussian distributions. In Section~\ref{sub:ex:multi}, those techniques will be further discussed in the context of the application to non-natural birth\footnote{See \href{https://github.com/3wen/counterfactual-estimation-optimal-transport}{\sffamily https://github.com/3wen/counterfactual-estimation-optimal-transport} for more details}..

\section{Ceteris Paribus vs. Mutatis Mutandis}\label{sec:exogeneous}

Before introducing another concept of CATE, we will formalize a little bit more the connections between the ``treatment'' $t$, the outcome $y$ and the covariates $\boldsymbol{x}$.

\subsection{Exogeneity, endogeneity and causal graphs}\label{subsec:1}

As discussed earlier, when presenting the second stage of the ``ladder of causation'', $t$ is a treatment. For example, in epidemiology, $t$ may be a treatment given to patients, possibly resulting from an intervention. At the third level, the treatment would be more a thought experiment (the ``\textit{gedankenexperiment}'' in \cite{mach1893science}), to answer a question such as ``{\em what if $t$ had taken another value?}'', without being able to make an experiment. \cite{chisholm1946contrary} introduced the idea of ``{\em contrary-to-fact conditional}'', coined as ``{\em counterfactual}'' in \cite{goodman1947problem}. A classical example would be when $t\in\{\text{smoker},\text{non-smoker}\}$, since it is not ethically possible to force someone to smoke, but it can also be used on inherent variables, such as the gender or the race of a person, that cannot be changed in a real experiment, to quantify possible discrimination.

Covariates $\boldsymbol{x}$ are available variables that have an impact on the outcome $y$. It is necessary here to distinguish two kinds of covariates, with variables that are influenced by the value of $t$, that might be seen as ''endogenous'', and those that are not influenced by the value of $t$, that might be seen as ``exogenous''. For example, the weight of the baby $x$ is an endogenous variable with respect to the variable indicating whether the mother is a smoker or not. Using a terminology used on causal graphs, ``endogeneous'' covariates $x$ are mediator variables (between $t$ and $y$), while ``exogeneous'' ones are variables colliding with $t$ on $y$, sometimes called collider variables (see Figure~\ref{Fig:DAG:3}).

The Markov assumption, on causal networks, states that each variable is conditionally independent of its non-descendants, given its parents. In Figure~\ref{Fig:DAG:3}, in the `cofounder' case (with the fork $t\to x$ and $t\to y$), and in the `mediator' case (with the chain $t\to x\to y$), $y$ is independent of $t$, conditional on $x$. But in the ``colider'' case (with $x\to y$ and $t\to y$), while $x$ and $t$ are independent, they become conditionally dependent, conditional on $y$. We will not discuss here the construction of the causal graphs, that is supposed to be given (see, e.g., \cite{vowels2021d} for a survey on techniques used to discover causal structures).

% \begin{figure}[!ht]
% \centering
% \include{DAG.tex}
% \vspace{-.8cm}\caption{Three basic graphs, with three variables, $t$, $x$ and $y$. }\label{Fig:DAG:3ex}
% \end{figure}

\begin{figure}[!ht]
\centering
\tikzset{every picture/.style={line width=0.75pt}} %set default line width to 0.75pt        

\begin{tikzpicture}[x=0.75pt,y=0.75pt,yscale=-1,xscale=1]
%uncomment if require: \path (0,454); %set diagram left start at 0, and has height of 454

%Shape: Circle [id:dp647219415878747] 
\draw  [fill={rgb, 255:red, 80; green, 227; blue, 194 }  ,fill opacity=1 ] (198,91) .. controls (198,82.16) and (205.16,75) .. (214,75) .. controls (222.84,75) and (230,82.16) .. (230,91) .. controls (230,99.84) and (222.84,107) .. (214,107) .. controls (205.16,107) and (198,99.84) .. (198,91) -- cycle ;
%Shape: Circle [id:dp4623343393101351] 
\draw  [fill={rgb, 255:red, 184; green, 233; blue, 134 }  ,fill opacity=1 ] (117,91) .. controls (117,82.16) and (124.16,75) .. (133,75) .. controls (141.84,75) and (149,82.16) .. (149,91) .. controls (149,99.84) and (141.84,107) .. (133,107) .. controls (124.16,107) and (117,99.84) .. (117,91) -- cycle ;
%Straight Lines [id:da489138465932506] 
\draw    (133,41) -- (133,72) ;
\draw [shift={(133,75)}, rotate = 270] [fill={rgb, 255:red, 0; green, 0; blue, 0 }  ][line width=0.08]  [draw opacity=0] (8.93,-4.29) -- (0,0) -- (8.93,4.29) -- cycle    ;
%Shape: Circle [id:dp1555112104444264] 
\draw  [fill={rgb, 255:red, 248; green, 231; blue, 28 }  ,fill opacity=1 ] (117,25) .. controls (117,16.16) and (124.16,9) .. (133,9) .. controls (141.84,9) and (149,16.16) .. (149,25) .. controls (149,33.84) and (141.84,41) .. (133,41) .. controls (124.16,41) and (117,33.84) .. (117,25) -- cycle ;
%Straight Lines [id:da8328501702889709] 
\draw    (146,33) -- (202.6,75.21) ;
\draw [shift={(205,77)}, rotate = 216.71] [fill={rgb, 255:red, 0; green, 0; blue, 0 }  ][line width=0.08]  [draw opacity=0] (8.93,-4.29) -- (0,0) -- (8.93,4.29) -- cycle    ;
%Straight Lines [id:da549031465320101] 
\draw    (151,90) -- (195,90.94) ;
\draw [shift={(198,91)}, rotate = 181.22] [fill={rgb, 255:red, 0; green, 0; blue, 0 }  ][line width=0.08]  [draw opacity=0] (8.93,-4.29) -- (0,0) -- (8.93,4.29) -- cycle    ;
%Straight Lines [id:da787546021358317] 
\draw    (148.1,149.3) -- (199.75,103.98) ;
\draw [shift={(202,102)}, rotate = 138.73] [fill={rgb, 255:red, 0; green, 0; blue, 0 }  ][line width=0.08]  [draw opacity=0] (8.93,-4.29) -- (0,0) -- (8.93,4.29) -- cycle    ;
%Shape: Circle [id:dp1986918865995252] 
\draw  [fill={rgb, 255:red, 184; green, 233; blue, 134 }  ,fill opacity=1 ] (116,153) .. controls (116,144.16) and (123.16,137) .. (132,137) .. controls (140.84,137) and (148,144.16) .. (148,153) .. controls (148,161.84) and (140.84,169) .. (132,169) .. controls (123.16,169) and (116,161.84) .. (116,153) -- cycle ;
%Straight Lines [id:da22431902649218183] 
\draw    (149,25) -- (200.1,25.28) ;
\draw [shift={(203.1,25.3)}, rotate = 180.32] [fill={rgb, 255:red, 0; green, 0; blue, 0 }  ][line width=0.08]  [draw opacity=0] (8.93,-4.29) -- (0,0) -- (8.93,4.29) -- cycle    ;
%Shape: Circle [id:dp39745456397102774] 
\draw  [fill={rgb, 255:red, 184; green, 233; blue, 134 }  ,fill opacity=1 ] (202,26) .. controls (202,17.16) and (209.16,10) .. (218,10) .. controls (226.84,10) and (234,17.16) .. (234,26) .. controls (234,34.84) and (226.84,42) .. (218,42) .. controls (209.16,42) and (202,34.84) .. (202,26) -- cycle ;

% Text Node
\draw (208,85.4) node [anchor=north west][inner sep=0.75pt]  [xscale=0.9,yscale=0.9]  {$y$};
% Text Node
\draw (124,85.4) node [anchor=north west][inner sep=0.75pt]  [xscale=0.9,yscale=0.9]  {$\boldsymbol{x}^{m}$};
% Text Node
\draw (129,18.4) node [anchor=north west][inner sep=0.75pt]  [xscale=0.9,yscale=0.9]  {$t$};
% Text Node
\draw (123,147.4) node [anchor=north west][inner sep=0.75pt]  [xscale=0.9,yscale=0.9]  {$\boldsymbol{x}^{c}$};
% Text Node
\draw (47,17) node [anchor=north west][inner sep=0.75pt]  [xscale=0.9,yscale=0.9] [align=left] {treatment};
% Text Node
\draw (234,82) node [anchor=north west][inner sep=0.75pt]  [xscale=0.9,yscale=0.9] [align=left] {outcome};
% Text Node
\draw (9,74) node [anchor=north west][inner sep=0.75pt]  [xscale=0.9,yscale=0.9] [align=left] {\begin{minipage}[lt]{71.4pt}\setlength\topsep{0pt}
\begin{flushright}
mediator\\(endogeneous)
\end{flushright}
\end{minipage}};
% Text Node
\draw (18,138) node [anchor=north west][inner sep=0.75pt]  [xscale=0.9,yscale=0.9] [align=left] {\begin{minipage}[lt]{65.14pt}\setlength\topsep{0pt}
\begin{flushright}
colider\\(exogeneous)
\end{flushright}

\end{minipage}};
% Text Node
\draw (209,20.4) node [anchor=north west][inner sep=0.75pt]  [xscale=0.9,yscale=0.9]  {$\boldsymbol{x}^{p}$};
% Text Node
\draw (240,8) node [anchor=north west][inner sep=0.75pt]  [xscale=0.9,yscale=0.9] [align=left] {confounding\\(noise, proxy)};

\end{tikzpicture}
\vspace{-.8cm}\caption{Distinction of covariates, with confounding variables that will not influence $y$ on the top right, and two sets of explanatory variables that will influence $y$, that are influenced, or not, by ``treatment'' $t$, with mediators and colliders, at the bottom left.}\label{Fig:DAG:3}
\end{figure}

% FAIRE EXEMPLE AVEC FUMER $t$ ET POIDS $x$, EN SUPPOSANT QUE FUMER FASSE GROSSIR...

\subsection{Impact of a treatment $t$ on $y$ and $\boldsymbol{x}$, and CATE}\label{subsec:3}

\begin{figure}[!ht]
\centering
\tikzset{every picture/.style={line width=0.75pt}} %set default line width to 0.75pt        

\begin{tikzpicture}[x=0.75pt,y=0.75pt,yscale=-1,xscale=1]
%uncomment if require: \path (0,454); %set diagram left start at 0, and has height of 454

%Shape: Circle [id:dp13658452349878047] 
\draw  [fill={rgb, 255:red, 80; green, 227; blue, 194 }  ,fill opacity=1 ] (92,89) .. controls (92,80.16) and (99.16,73) .. (108,73) .. controls (116.84,73) and (124,80.16) .. (124,89) .. controls (124,97.84) and (116.84,105) .. (108,105) .. controls (99.16,105) and (92,97.84) .. (92,89) -- cycle ;
%Shape: Circle [id:dp25163412737494506] 
\draw  [fill={rgb, 255:red, 184; green, 233; blue, 134 }  ,fill opacity=1 ] (11,89) .. controls (11,80.16) and (18.16,73) .. (27,73) .. controls (35.84,73) and (43,80.16) .. (43,89) .. controls (43,97.84) and (35.84,105) .. (27,105) .. controls (18.16,105) and (11,97.84) .. (11,89) -- cycle ;
%Straight Lines [id:da08399469689704131] 
\draw [line width=0.75]    (27,39) -- (27,70) ;
\draw [shift={(27,73)}, rotate = 270] [fill={rgb, 255:red, 0; green, 0; blue, 0 }  ][line width=0.08]  [draw opacity=0] (8.93,-4.29) -- (0,0) -- (8.93,4.29) -- cycle    ;
%Straight Lines [id:da9103590392048093] 
\draw [line width=0.75]    (40,31) -- (96.6,73.21) ;
\draw [shift={(99,75)}, rotate = 216.71] [fill={rgb, 255:red, 0; green, 0; blue, 0 }  ][line width=0.08]  [draw opacity=0] (8.93,-4.29) -- (0,0) -- (8.93,4.29) -- cycle    ;
%Straight Lines [id:da2857949781798549] 
\draw [line width=0.75]    (45,88) -- (89,88.94) ;
\draw [shift={(92,89)}, rotate = 181.22] [fill={rgb, 255:red, 0; green, 0; blue, 0 }  ][line width=0.08]  [draw opacity=0] (8.93,-4.29) -- (0,0) -- (8.93,4.29) -- cycle    ;
%Straight Lines [id:da24214997121229342] 
\draw    (42.1,147.3) -- (93.75,101.98) ;
\draw [shift={(96,100)}, rotate = 138.73] [fill={rgb, 255:red, 0; green, 0; blue, 0 }  ][line width=0.08]  [draw opacity=0] (8.93,-4.29) -- (0,0) -- (8.93,4.29) -- cycle    ;
%Shape: Circle [id:dp9158670635817864] 
\draw  [fill={rgb, 255:red, 184; green, 233; blue, 134 }  ,fill opacity=1 ] (10,151) .. controls (10,142.16) and (17.16,135) .. (26,135) .. controls (34.84,135) and (42,142.16) .. (42,151) .. controls (42,159.84) and (34.84,167) .. (26,167) .. controls (17.16,167) and (10,159.84) .. (10,151) -- cycle ;
%Shape: Circle [id:dp4099844373128174] 
\draw  [fill={rgb, 255:red, 80; green, 227; blue, 194 }  ,fill opacity=1 ] (253,89) .. controls (253,80.16) and (260.16,73) .. (269,73) .. controls (277.84,73) and (285,80.16) .. (285,89) .. controls (285,97.84) and (277.84,105) .. (269,105) .. controls (260.16,105) and (253,97.84) .. (253,89) -- cycle ;
%Shape: Circle [id:dp10943440435106111] 
\draw  [fill={rgb, 255:red, 184; green, 233; blue, 134 }  ,fill opacity=1 ] (172,89) .. controls (172,80.16) and (179.16,73) .. (188,73) .. controls (196.84,73) and (204,80.16) .. (204,89) .. controls (204,97.84) and (196.84,105) .. (188,105) .. controls (179.16,105) and (172,97.84) .. (172,89) -- cycle ;
%Straight Lines [id:da4908633877217007] 
\draw [line width=1.5]    (188,39) -- (188,69) ;
\draw [shift={(188,73)}, rotate = 270] [fill={rgb, 255:red, 0; green, 0; blue, 0 }  ][line width=0.08]  [draw opacity=0] (11.61,-5.58) -- (0,0) -- (11.61,5.58) -- cycle    ;
%Straight Lines [id:da05843908704245282] 
\draw [line width=1.5]    (201,31) -- (256.79,72.61) ;
\draw [shift={(260,75)}, rotate = 216.71] [fill={rgb, 255:red, 0; green, 0; blue, 0 }  ][line width=0.08]  [draw opacity=0] (11.61,-5.58) -- (0,0) -- (11.61,5.58) -- cycle    ;
%Straight Lines [id:da018368302396613467] 
\draw [line width=1.5]    (206,88) -- (249,88.91) ;
\draw [shift={(253,89)}, rotate = 181.22] [fill={rgb, 255:red, 0; green, 0; blue, 0 }  ][line width=0.08]  [draw opacity=0] (11.61,-5.58) -- (0,0) -- (11.61,5.58) -- cycle    ;
%Straight Lines [id:da5576703758545092] 
\draw  [dash pattern={on 0.84pt off 2.51pt}]  (203.1,147.3) -- (254.75,101.98) ;
\draw [shift={(257,100)}, rotate = 138.73] [fill={rgb, 255:red, 0; green, 0; blue, 0 }  ][line width=0.08]  [draw opacity=0] (8.93,-4.29) -- (0,0) -- (8.93,4.29) -- cycle    ;
%Shape: Circle [id:dp02714753080436394] 
\draw  [fill={rgb, 255:red, 184; green, 233; blue, 134 }  ,fill opacity=1 ] (171,151) .. controls (171,142.16) and (178.16,135) .. (187,135) .. controls (195.84,135) and (203,142.16) .. (203,151) .. controls (203,159.84) and (195.84,167) .. (187,167) .. controls (178.16,167) and (171,159.84) .. (171,151) -- cycle ;
%Rounded Rect [id:dp2750710829395441] 
\draw  [fill={rgb, 255:red, 248; green, 231; blue, 28 }  ,fill opacity=1 ] (173.1,15.88) .. controls (173.1,12.8) and (175.6,10.3) .. (178.68,10.3) -- (195.42,10.3) .. controls (198.5,10.3) and (201,12.8) .. (201,15.88) -- (201,33.42) .. controls (201,36.5) and (198.5,39) .. (195.42,39) -- (178.68,39) .. controls (175.6,39) and (173.1,36.5) .. (173.1,33.42) -- cycle ;
%Shape: Circle [id:dp5505222531288511] 
\draw  [fill={rgb, 255:red, 248; green, 231; blue, 28 }  ,fill opacity=1 ] (11,23) .. controls (11,14.16) and (18.16,7) .. (27,7) .. controls (35.84,7) and (43,14.16) .. (43,23) .. controls (43,31.84) and (35.84,39) .. (27,39) .. controls (18.16,39) and (11,31.84) .. (11,23) -- cycle ;

% Text Node
\draw (102,83.4) node [anchor=north west][inner sep=0.75pt]  [xscale=0.9,yscale=0.9]  {$y$};
% Text Node
\draw (18,83.4) node [anchor=north west][inner sep=0.75pt]  [xscale=0.9,yscale=0.9]  {$\boldsymbol{x}^{m}$};
% Text Node
\draw (23,16.4) node [anchor=north west][inner sep=0.75pt]  [xscale=0.9,yscale=0.9]  {$t$};
% Text Node
\draw (17,145.4) node [anchor=north west][inner sep=0.75pt]  [xscale=0.9,yscale=0.9]  {$\boldsymbol{x}^{c}$};
% Text Node
\draw (263,83.4) node [anchor=north west][inner sep=0.75pt]  [xscale=0.9,yscale=0.9]  {$y'$};
% Text Node
\draw (179,83.4) node [anchor=north west][inner sep=0.75pt]  [xscale=0.9,yscale=0.9]  {$\boldsymbol{x} '^{m}$};
% Text Node
\draw (184,16.4) node [anchor=north west][inner sep=0.75pt]  [xscale=0.9,yscale=0.9]  {$t'$};
% Text Node
\draw (178,145.4) node [anchor=north west][inner sep=0.75pt]  [xscale=0.9,yscale=0.9]  {$\boldsymbol{x}^{c}$};
% Text Node
\draw (293,84.4) node [anchor=north west][inner sep=0.75pt]  [xscale=0.9,yscale=0.9]  {$y'=h( t',\boldsymbol{x} '^{m} ,\boldsymbol{x}^{c})$};
% Text Node
\draw (57,180) node [anchor=north west][inner sep=0.75pt]  [xscale=0.9,yscale=0.9] [align=left] {DAG};
% Text Node
\draw (202,179) node [anchor=north west][inner sep=0.75pt]  [xscale=0.9,yscale=0.9] [align=left] {intervention on the treatment};

\end{tikzpicture}
\vspace{-.8cm}\caption{A causal graph on the left, and the impact of an intervention on the treatment $t$ on the right.}\label{Fig:DAG:intervention}
\end{figure}

Consider some treatment $t$. Let $\boldsymbol{x}^m$ denote the set of mediator variables and $\boldsymbol{x}^c$ denote the set of collider variables, as in Figure~\ref{Fig:DAG:intervention}.
Following the SEM terminology used in causal inference, consider data generated according to the equations on the left below (real world), prior to intervention on $t$. The right hand equations describe the data generating process with an intervention on $t$ (denoted $do(t)$ in \cite{pearl2018book}):
\begin{center}
  \begin{tabular}{cc}
  real world ~~~~~~~~~ & with intervention ($do(t)$)\\
$\displaystyle{\begin{cases}
T = h_t(U_t) \\
\boldsymbol{X}^m = h_m(T,\boldsymbol{U}_m) \\
\boldsymbol{X}^c = h_c(\boldsymbol{U}_c) \\
Y = h_y(T,\boldsymbol{X}^m,\boldsymbol{X}^c,U_y) \\
\end{cases}}$& $\displaystyle{\begin{cases}
T = t \\
\boldsymbol{X}^m_{T\leftarrow t} = h_m(t,\boldsymbol{U}_m) \\
\boldsymbol{X}^c = h_c(\boldsymbol{U}_c) \\
Y_{T\leftarrow t} = h_y(t,\boldsymbol{X}^m_{T\leftarrow t},\boldsymbol{X} ^c,U_y) \\
\end{cases}}$
  \end{tabular}
\end{center}

Consider some independent noise variables $\{U_t,\boldsymbol{U}_m,\boldsymbol{U}_c,U_y\}$ (that can be assumed to be centered Gaussian to be close to the econometric literature). In the ``real world'', $T$ is a function of $U_t$, and $U_t$ only, through some $h_t:\mathbb{R}\to\{0,1\}$ function, $h_t(u) = \boldsymbol{1}(u>\text{threshold})$. Then we have two possible explanatory variables: mediator (endogenous) and collider (exogenous). If $\boldsymbol{X}^c$ are functions of the noise $\boldsymbol{U}_c$ only (through function $h_c$), $\boldsymbol{X}^m$ are functions of the noise $\boldsymbol{U}_m$ and the treatment $T$ (through function $h_c$). And finally, the outcome $Y$ is function of $\boldsymbol{X}^c$ and $\boldsymbol{X}^m$, also possibly $T$, and some idiosyncratic noise $U_y$.  

%\subsection{Conditional Average Treatment Effect }\label{subsec:4}

In a {\em ceteris paribus} approach, $\text{CATE}(x)$ is equal to $\mathbb{E}\big[Y^*_{T\leftarrow 1}\big|{x}\big] - \mathbb{E}\big[Y^*_{T\leftarrow 0}\big|{x}\big]$. In a {\em mutatis mutandis} version, we should not consider $x$, but a version of $x$ that should be influenced by the treatment $t$, denoted ${x}_{T\leftarrow 1}$.
In a general setting, we have the following definition: 

\begin{definition}
The {\em mutatis mutandis} CATE is
$$
\text{CATE}(\boldsymbol{x})=
\mathbb{E}\big[Y^*_{T\leftarrow 1}\big|\boldsymbol{x}_{T\leftarrow 1}\big] - 
\mathbb{E}\big[Y^*_{T\leftarrow 0}\big|\boldsymbol{x}\big]
$$
(we might denote $\boldsymbol{x}_{T\leftarrow 0}$ instead of $x$ to avoid confusion for the second term).
\end{definition}

More specifically, when we ask the question ``{\em what would have been the probability to have a non-natural delivery for a baby with weight $x$ if the mother had been smoking?}'', we have to take into account the fact that if the mother had been smoking, the weight of the baby would have been impacted. The original weight $x$, associated with a non-Black mother, would become $\boldsymbol{x}_{T\leftarrow 1}$ (instead of $x$) if we seek a counterfactual version of $x$ in the treated population.

\subsection{Propensity score weighting}\label{subsec:5}

The classical approach in causal inference is based on the idea that $T$ is not really exogenous, and can be influenced by $\boldsymbol{x}$. Therefore, 
the average treatment effect $\text{ATE} = \mathbb{E}[Y^\star_{T\leftarrow 1}-Y^\star_{T\leftarrow 0}]$, that can be written
$$\text{ATE} =
\mathbb{E}\left[\frac{TY}{p(\boldsymbol{X})}-\frac{(1-T)Y}{1-p(\boldsymbol{X})}\right]
$$
would be estimated by
$$
\text{SATE}=
\frac{1}{n}\sum_{i=1}^n \frac{t_iy_i}{\widehat{p}(\boldsymbol{x}_i)}-\frac{(1-t_i)y_i}{1-\widehat{p}(\boldsymbol{x}_i)},
$$
where $p(\boldsymbol{x})$ is a ``propensity score'' defined as $p(\boldsymbol{x})=\mathbb{P}[T=1|\boldsymbol{X}=\boldsymbol{x}]$, that can be estimated using, for instance, a logistic regression
    $$
    \widehat{p}(\boldsymbol{x})=\frac{\exp[
   \boldsymbol{x}^\top\widehat{\boldsymbol{\beta}}]}{1+\exp[
   \boldsymbol{x}^\top\widehat{\boldsymbol{\beta}}]}.
    $$
Thus, the SATE can be seen as the difference between two weighted averages of $y_i$'s.
As discussed in \cite{abrevaya2015estimating}, it can be used to estimate $\text{CATE}(x)$, on a subset of features, with a local estimate of the average
$$
\text{CATE}(x) = \frac{1}{\sum K_{h}(x_{i}-x)}\sum \left(\frac{t_iy_i}{\widehat{p}(\boldsymbol{x}_i)}-\frac{(1-t_i)y_i}{1-\widehat{p}(\boldsymbol{x}_i)}\right) K_{h}(x_{i}-x),
$$
using some kernel function $K_h$. A $k$-nearest neighbors estimate can also be considered:
$$
\text{CATE}(x) = \frac{1}{k}\sum_{i\in\mathcal{V}_{k}(x)} \left(\frac{t_iy_i}{\widehat{p}(\boldsymbol{x}_i)}-\frac{(1-t_i)y_i}{1-\widehat{p}(\boldsymbol{x}_i)}\right),
$$
where $i\in\mathcal{V}_{k}(x)$ when $x_i$ is among the $k$-nearest neighbors of $x$.
If $y$ is binary (as the example we will use later on), the ATE is a difference between two probabilities, and logtistic regressions can be used to properly estimate $\mathbb{E}[Y^\star_{T\leftarrow t}|\boldsymbol{X}=\boldsymbol{x}]$, with weights in the regressions, that would be either the inverse of $1-\widehat{p}(\boldsymbol{x}_i)$ if $t_i=0$ or the inverse of $\widehat{p}(\boldsymbol{x}_i)$ if $t_i=1$, as in \cite{li2018balancing}.

% https://www.math.umd.edu/~slud/s818M-MissingData/PropensityScoreWeightingR.pdf

\subsection{A toy (Gaussian) example}\label{sec:gaussian:toy}

\begin{figure}[!ht]
\centering
\includegraphics[width=.48\textwidth]{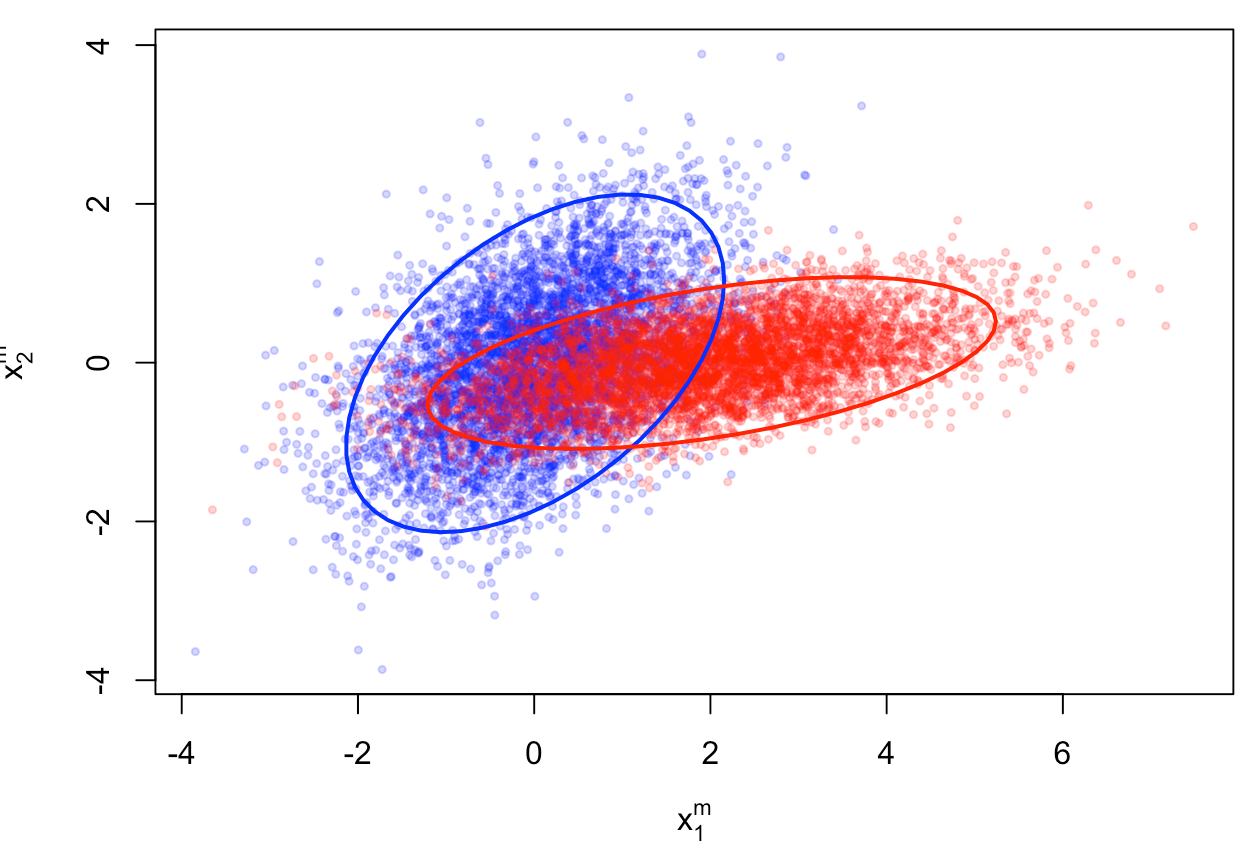}\includegraphics[width=.48\textwidth]{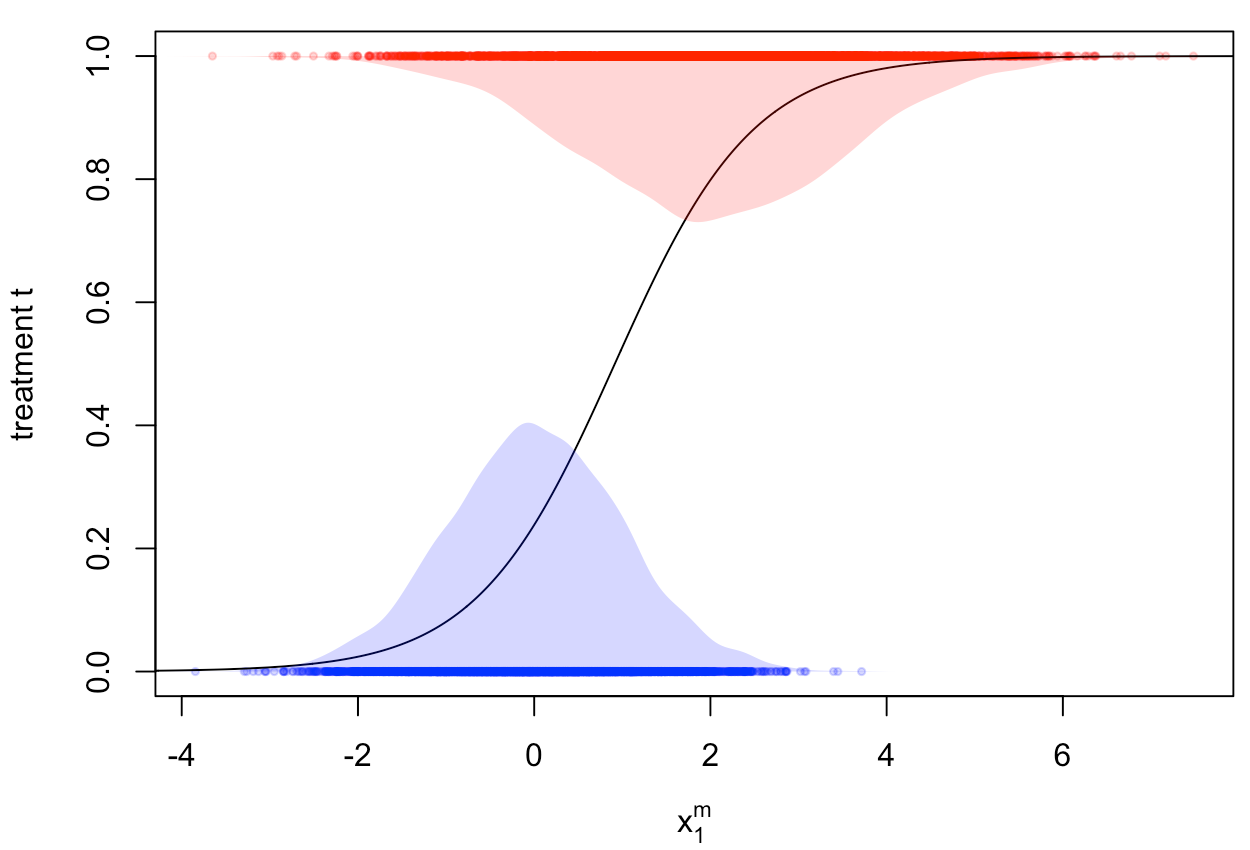}
\caption{Scatter plot of $\boldsymbol{x}^m=(x^m_1,x^m_2)$ with blue points when $t=0$, and red points when $t=1$, on the left, and the logistic regression of $t$ against $x_1^m$ on the right. Toy dataset generated from Equation~(\ref{eq:SEM}).}\label{Fig:ex:1}
\end{figure}

To illustrate our approach, as an alternative to the use of a propensity score, consider the following toy example, with three explanatory variables, two endogenous (and correlated) ones, and an exogenous one, with some linear model (a Gaussian structural equation model, SEM):

\begin{equation}\label{eq:SEM}
\displaystyle{\begin{cases}
T = \boldsymbol{1}(U_t<0),~U_t\sim\mathcal{N}(0,1) \\
\boldsymbol{X}^m =\boldsymbol{\mu}_T+\boldsymbol{\Sigma}_T^{1/2}\boldsymbol{U}_{m},~\boldsymbol{U}_m\sim\mathcal{N}(\boldsymbol{0},\mathbb{I}) \\
{X}^c = \mu+\sigma {U}_c,~U_c\sim\mathcal{N}(0,1)  \\
    Y = \alpha+(\boldsymbol{\beta}_m,\beta_c)(\boldsymbol{X}^m , X^c)^\top +\gamma T+U_y,~U_y\sim\mathcal{N}(0,1) \\
\end{cases}}\end{equation}
where all the noises $(U_t,\boldsymbol{U}_m,U_c,U_y)$ are assumed to be centered, and independent. Here $\boldsymbol{\Sigma}_0^{1/2}$ is Cholesky decomposition of $\boldsymbol{\Sigma}_0$, so that $\boldsymbol{X}^m$  conditional on $T=t$ has distribution $\mathcal{N}(\boldsymbol{\mu}_t,\boldsymbol{\Sigma}_t)$.
Treatment $T$ is a binary variable, well-balanced since $\mathbb{P}(T=0)=\mathbb{P}(T=1)$. Conditional on $T=t$, the mediator (endogenous) variables $\boldsymbol{X}^m$ have a Gaussian distribution, with mean $\boldsymbol{\mu}_t$ and variance matrix $\boldsymbol{\Sigma}_t$. A collider variable $X^c$ is supposed to be independent of the other ones. And finally, $Y$ is a Gaussian variable where the average is a linear combination of $\boldsymbol{X}^m$ and $X^c$, plus $\gamma$ when $T=1$.
In Figure~\ref{Fig:ex:1}, the left-hand panel shows a scatter plot of $\boldsymbol{x}^m=(x^m_1,x^m_2)$ with blue points when $t=0$ and red points when $t=1$. The right-hand panel shows $(x^m_1,t)$ on a scatter plot, with the two conditional densities, as well as the logistic regression of $t$ against $x_1^m$ (that could be seen as the propensity score).

The two interventions yield
\begin{center}
  \begin{tabular}{cc}
  $do(T=0)$~~~~~~~~~~ & $do(T=1)$~~~~~~~~~~\\
$\displaystyle{\begin{cases}
T \leftarrow 0 \\
\boldsymbol{X}^m =\boldsymbol{\mu}_0+\boldsymbol{\Sigma}_0^{1/2}\boldsymbol{U}_{m}\\
{X}^c = \mu+\sigma {U}_c  \\
Y = \alpha+(\boldsymbol{\beta}_m,\beta_c)(\boldsymbol{X}^m , X^c)^\top +U_y \\
\end{cases}}$& $\displaystyle{\begin{cases}
T\leftarrow 1\\
\boldsymbol{X}^m =\boldsymbol{\mu}_1+\boldsymbol{\Sigma}_1^{1/2}\boldsymbol{U}_{m}' \\
{X}^c = \mu+\sigma {U}_c' \\
Y = \alpha+(\boldsymbol{\beta}_m,\beta_c)(\boldsymbol{X}^m , X^c)^\top +\gamma +U_y'\\
\end{cases}}$
  \end{tabular}
\end{center}
 more precisely, in that model with three covariates, $\boldsymbol{X}^m=(X_1^m,X_2^m)$, and since
 $$
 \boldsymbol{\Sigma}_t=\begin{pmatrix}
 \sigma_{t1}^2 & r_t\sigma_{t1}\sigma_{t2} \\
 r_t\sigma_{t1}\sigma_{t2} & \sigma_{t2}^2  \\
 \end{pmatrix}\text{ and }
 \boldsymbol{\Sigma}_t^{1/2}=\begin{pmatrix}
 \sigma_{t1} & 0 \\
 \sigma_{t2}r_t & \sigma_{t2}\sqrt{1-r_t^2}  \\
 \end{pmatrix}
 $$
we can write
\begin{center}
  \begin{tabular}{cc}
  $do(T=0)$~~~~~~~~~~ & $do(T=1)$~~~~~~~~~~\\
$\displaystyle{\begin{cases}
T \leftarrow 0 \\
X_1^m = \mu_{01} + \sigma_{01} U_{1}^m\\
X_2^m = \mu_{02} + \sigma_{02}(r_0 U_{1}^m+\sqrt{1-r_0^2}U_{2}^m)\\
{X}^c = \mu+\sigma {U}_c  \\
Y = \alpha+\beta_1^m X_1^m+\beta_2^m X_2^m+\beta^cX^c +U_y \\
\end{cases}}$& $\displaystyle{\begin{cases}
T \leftarrow 1 \\
X_1^m = \mu_{11} + \sigma_{11} U_{1}^{m'}\\
X_2^m = \mu_{12} + \sigma_{12}(r_1 U_{1}^{m'}+\sqrt{1-r_1^2}U_{2}^{m'})\\
{X}^c = \mu+\sigma {U}_c'  \\
Y = \alpha+\beta_1^m X_1^m+\beta_2^m X_2^m+\beta^cX^c+\gamma +U_y' \\
\end{cases}}$
  \end{tabular}
\end{center}
% In example~\ref{ex:sme}, if we consider $do(T=0)$, when $X_1^m=x_1$, we have
% $$\displaystyle{\begin{cases}
% T \leftarrow 0 \\
% X_1^m \leftarrow x_1\\
% \displaystyle{X_2^m = \mu_{02} + \sigma_{02}(r_0 \sigma_{01}^{-1}[x_1-\mu_{01}]+\sqrt{1-r_0^2}U_{2}^m)}\\
% {X}^c = \mu+\sigma {U}_c  \\
% Y_{T\leftarrow 0} = \alpha+\beta_1^m X_1^m+\beta_2^m X_2^m+\beta^cX^c +U_y \\
% \end{cases}}$$
% while if we consider $do(T=1)$, when $X_1^m=x'_1$, we have
% $$\displaystyle{\begin{cases}
% T \leftarrow 1 \\
% X_1^m \leftarrow x_1'\\
% \displaystyle{X_2^m = \mu_{12} + \sigma_{12}(r_1 \sigma_{11}^{-1}[x_1'-\mu_{11}]+\sqrt{1-r_1^2}U_{2}^{m'})}\\
% {X}^c = \mu+\sigma {U}_c  \\
% Y_{T\leftarrow 0} = \alpha+\beta_1^m X_1^m+\beta_2^m X_2^m+\beta^cX^c+\gamma +U_y \\
% \end{cases}}$$
and therefore
$$\displaystyle{\begin{cases}
Y_{T\leftarrow 0} =  \alpha+\beta_1^m x_1+\beta_2^m \big(\mu_{02} + \sigma_{02}(r_0 \sigma_{01}^{-1}[x_1-\mu_{01}]+\sqrt{1-r_0^2}U_{2}^m)\big)+\beta^c(\mu+\sigma {U}_c) +U_y\\
Y_{T\leftarrow 1} =  \alpha+\beta_1^m x_1'+\beta_2^m \big(\mu_{12} + \sigma_{12}(r_1 \sigma_{11}^{-1}[x_1'-\mu_{11}]+\sqrt{1-r_1^2}U_{2}^{m'})\big)+\beta^c(\mu+\sigma {U}_c')+\gamma +U_y'\\
\end{cases}}.$$
Hence,
$$
\text{ATE} = \mathbb{E}[Y_{T\leftarrow 1}-Y_{T\leftarrow 0}]=\gamma.
$$
For conditional average treatment effects,
$$\displaystyle{\begin{cases}
\mathbb{E}[Y_{T\leftarrow 0}|X_1^m = x_1] =  \alpha+\beta_1^m x_1+\beta_2^m \big(\mu_{02} + \sigma_{02}r_0 \sigma_{01}^{-1}[x_1-\mu_{01}]\big)+\beta^c\mu\\
\mathbb{E}[Y_{T\leftarrow 1}|X_1^m = x_1']  =  \alpha+\beta_1^m x_1'+\beta_2^m \big(\mu_{12} + \sigma_{12}r_1 \sigma_{11}^{-1}[x_1'-\mu_{11}]\big)+\beta^c\mu +\gamma \\
\end{cases}}.$$
{\em Ceteris paribus}, we suppose that $x_1'=x_1$, then
$$
\text{CATE}_{cp}(x_1)=\mathbb{E}[Y_{T\leftarrow 1}|X_1^m = x_1] -\mathbb{E}[Y_{T\leftarrow 0}|X_1^m = x_1] =\text{ATE} +\delta x_1 +\kappa,
$$
where
$$
\begin{cases}
\kappa = \beta_2^m \big(\mu_{12} + \sigma_{02}r_0 \sigma_{01}^{-1}\mu_{01} - \sigma_{12}r_1 \sigma_{11}^{-1}\mu_{11} - \mu_{02} \big)\\
\delta = \beta_2^m \big( \sigma_{12}r_1 \sigma_{11}^{-1}- \sigma_{02}r_0 \sigma_{01}^{-1}\big)
\end{cases}.
$$
{\em Mutatis mutandis}, since $X_1^m = \mu_{01} + \sigma_{01} U_{1}^m$ when $T=0$ while $X_1^m = \mu_{11} + \sigma_{11} U_{1}^m$ when $t=1$, it is legitimate to consider that $x_1'=x_{1:T\leftarrow 1}=\mu_{11} + \sigma_{11}(\sigma_{01}^{-1}[x_1-\mu_{01}])$. Therefore, {\em mutatis mutandis},$$
\text{CATE}_{mm}(x_1) =\text{ATE}+ \delta' x_1 +\kappa',
$$
where
$$
\begin{cases}
\kappa' = \kappa +  \beta_1^m [\mu_{11} - \sigma_{11}\sigma_{01}^{-1}\mu_{01}] \kappa+k\\
\delta' = \delta + \beta_1^m (\sigma_{11}\sigma_{01}^{-1}-1)=\delta+d
\end{cases},
$$
so that we can also write
$$
\text{CATE}_{mm}(x_1) = \text{CATE}_{cp}(x_1)+\big(dx_1+k\big).
$$

\begin{figure}[!ht]
\centering
\includegraphics[width=.48\textwidth]{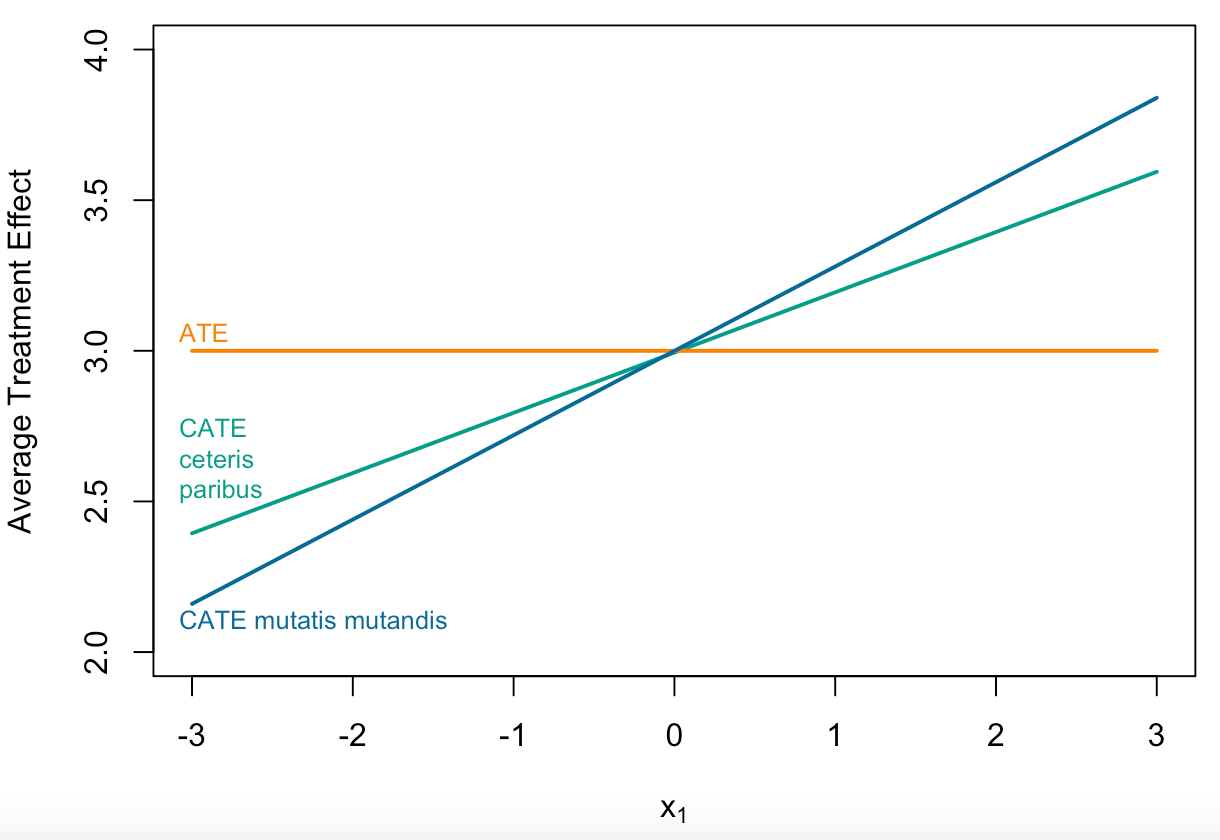}
\includegraphics[width=.48\textwidth]{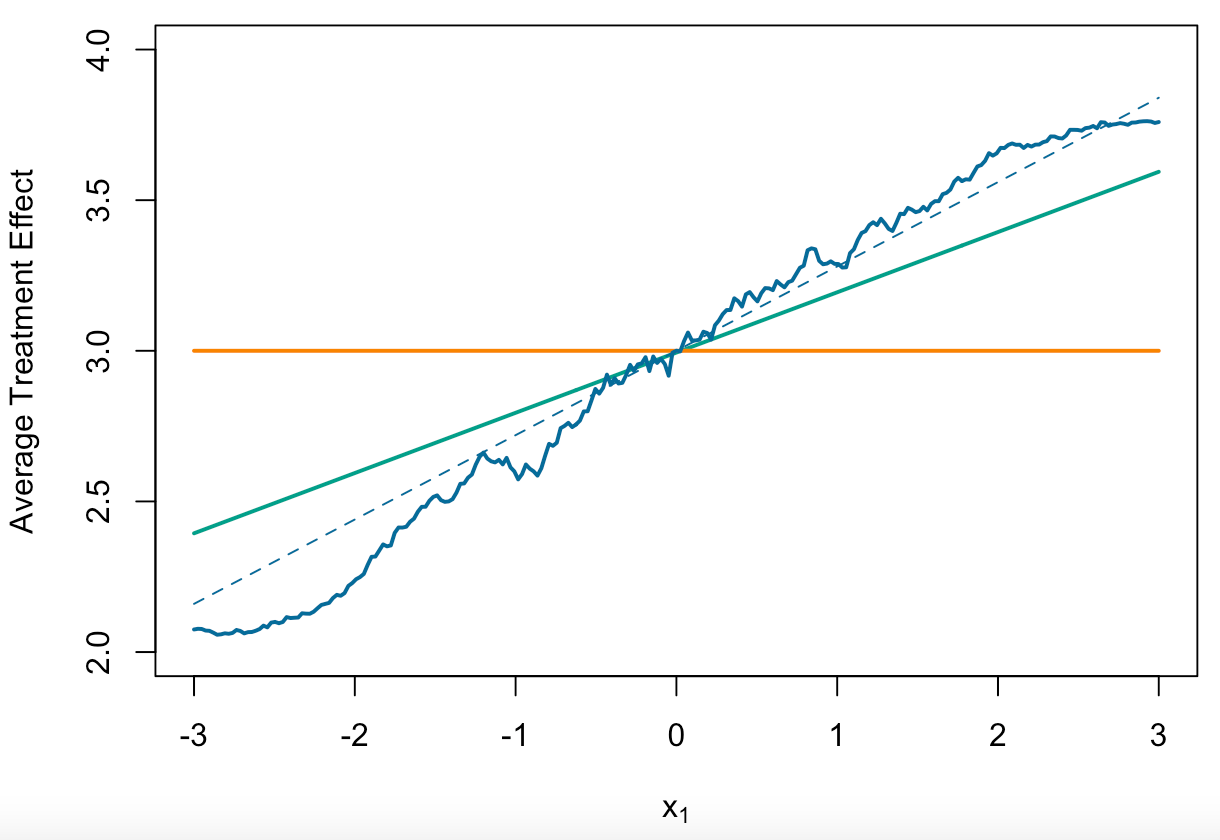}
\caption{ATE, {\em ceteribs paris} $\text{CATE}_{cp}(x_1)$ and {\em mutatis mutandis} $\text{CATE}_{mm}(x_1)$  on the left, with an estimate of {\em mutatis mutandis} $\text{CATE}_{mm}(x_1)$ on the right, from the toy dataset from example~\ref{eq:SEM}. Numerical details are given in Appendix~\ref{app:2}.}\label{Fig:ex:3}
\end{figure}

In Figure~\ref{Fig:ex:3}, the horizontal orange line is the true average treatment effect (ATE). The green line is the true {\em ceteris paribus} CATE, while the blue line is the true {\em mutatis mutandis} CATE, both function of $x_1^m$. The dashed and erratic lines on the right-hand graph are estimations of the CATE function using two techniques, described in the next section.

% A BOUGER 

% \begin{figure}[!ht]
%     \centering
%     \include{dessin/match- dessin-1.tex}
%     \caption{Optimal coupling, with $n_0=n_1=n$.}
%     \label{fig:match:1}
% \end{figure}

% \begin{figure}[!ht]
%     \centering
%     \include{dessin/match-dessin-cdf-1.tex}
%     \caption{Optimal coupling, with $n_0=n_1=n$, with the cumulative distribution functions.}
%     \label{fig:match:1:cdf}
% \end{figure}

% \begin{example}
% On the same framework as Example~\ref{ex:sme}
% \end{example}

% \begin{figure}[!ht]
% \centering
% \includegraphics[width=.48\textwidth]{dessin/simulation-points-SEM-5.png}\includegraphics[width=.48\textwidth]{dessin/simulation-points-SEM-tau-1.png}
% \caption{xxxxxx The dashed horizontal line is the ATE, the straight line is the theoretical $CATE(x_1)$ (mutatis mutandis), and the plain curve is the estimate. Toy dataset from example~\ref{ex:sme}.}\label{Fig:ex:2}
% \end{figure}

\subsection{Application on birth data}\label{subsec:7}

Let us now consider the dataset of all deliveries in the U.S. in 2013.\footnote{\href{https://www.cdc.gov/nchs/data_access/Vitalstatsonline.htm}{https://www.cdc.gov/nchs/data\_access/Vitalstatsonline.htm}}
Those data have been intensively used to discuss the ``low birth weight paradox''.
As explained in \cite{wilcox1993birth,wilcox2001importance}, low birth weight of babies $x$ is strongly associated with increased neonatal mortality $y$. However, low birth weight infants born to mothers who smoke $t=1$ usually have lower mortality rates than low birth weight infants born to nonsmoking mothers $t=0$. \cite{hernandez2006birth} discussed the birth weight paradox based on causal directed acyclic graphs as a conceptual framework. Multiple causal models have been considered. Figure~\ref{Fig:BWP:3ex} illustrates four situations, using directed acyclic graphs. In the first case (Figure~\ref{Fig:BWP:3ex}a), birth weight $x$ has a direct effect on mortality $y$, while smoking $t$ has not. It is also possible to consider a second case where birth weight $x$, and possibly smoking $t$, have a direct effect on mortality $y$ (Figure~\ref{Fig:BWP:3ex}b). To increase the plausibility of this scenario, some known common causes of lower birth weight and mortality, denoted $z$, can be added (Figure~\ref{Fig:BWP:3ex}c). In this third case, \cite{hernandez2006birth} claims that the variables $z$ might induce an association between smoking and mortality, conditional on birth weight $x$. Lastly, a fourth situation that combines the second and the third can be considered (Figure~\ref{Fig:BWP:3ex}d). %In Section~\ref{sec:4}, we will distinguish covariates $x$ and $z$, since $x$ is influenced by $t$, while $z$ is not.

\begin{figure}[!ht]
\include{BWP.tex}
\vspace{-1cm}
\caption{Directed acyclic graphs for the birth weight paradox, when $y$ is the mortality indicator, $x$ the birth weight and $t$ a smoking indicator. $z$ denotes some possible common causes of infant death, from \cite{hernandez2006birth}.}\label{Fig:BWP:3ex}
\end{figure}

Here, instead of focusing on newborn mortality (which is an unbalanced variable, with less than $0.5\%$ mortality rate), we consider $y=\boldsymbol{1}(\text{non-natural delivery})$. As can be seen in Table~\ref{Tab:Stat:Des:XY-T}, about a third of all deliveries can be considered as ``un-natural'' (or ``complicated'', involving a least a C-section). Among possible explanatory variables, we consider the weight of the newborn infant $x_1$ and the weight gain of the mother $x_2$. Conditional densities, of $\boldsymbol{x}=\begin{bmatrix} x_1 &  x_2\end{bmatrix}$ given $y$ can be visualized in Figure~\ref{fig:densite-1x3-conditional-k}. To illustrate various techniques based on optimal transport, we will consider $\text{CATE}(\boldsymbol{x})$,
$$\tau(\boldsymbol{x}) = \text{CATE}(\boldsymbol{x}) = \mathbb{P}\big[ Y^\star_{T\leftarrow1} =1\big\vert \boldsymbol{X}=\boldsymbol{x}\big]- \mathbb{P}\big[ Y^\star_{T\leftarrow0}=1 \big\vert \boldsymbol{X}=\boldsymbol{x}\big],
$$
for several possible ``treatment'' $t$, that can be visualized in Figure~\ref{Fig:NN:1ex}, with either a smoker indicator (for the mother) or a variable indicating whether the newborn is a boy or not. However, emphasis will be placed on a variable indicating whether the mother is Black (Afro-American) or not. Conditional densities of $\boldsymbol{x}$ given $t$ can be visualized in Figure~\ref{fig:densite-2x3-conditional-t}.
In a nutshell, we want to address the following questions ``{\em what would have been the probability of a non-natural delivery for a baby of weight $x_1$ whose mother gained weight $x_2$ during pregnancy, if the mother had been Afro-American?}'' or ``{\em if the mother had been smoking?}''

\begin{table}[!ht]\centering
\begin{tabular}{lrrrrrr}\hline\hline
& \multicolumn{6}{c}{Variable of interest}\\
\cmidrule(lr){2-7}
& \multicolumn{3}{c}{$y=0$ (natural)}  & \multicolumn{3}{c}{$y=1$ (non-natural)}  \\
\cmidrule(lr){2-4}\cmidrule(lr){5-7}
$n$ number of observations & \multicolumn{3}{c}{2,221,522 (65.70\%)} & \multicolumn{3}{c}{1,159,776  (34.30\%)}  \\ 
$x_1$ weight of newborn & \multicolumn{3}{c}{average 3,299 g.} & \multicolumn{3}{c}{average 3,231 g.} \\ 
$x_2$ weight gain of mother & \multicolumn{3}{c}{average 30.02 lbs.} & \multicolumn{3}{c}{average  31.16 lbs.} \\
\hline
& \multicolumn{6}{c}{``Treatment''}\\
\cmidrule(lr){2-7}
& \multicolumn{3}{c}{$t=0$}  & \multicolumn{3}{c}{$t=1$}  \\
\cmidrule(lr){2-4}\cmidrule(lr){5-7}
Afro-American  variable & non-Black & 2,980,387 & (88.14\%) & Black & 400,911& (11.86\%) \\ 
smoker variable & non-smoker & 2,959,847 & (91.54\%)& smoker & 273,685 & (8.46\%)  \\ 
sex variable & baby boy & 1,730,837 & (51.18\%) & baby girl & 1,650,461 & (48.82\%) \\
\hline\hline
\end{tabular}
\caption{Statistics about the variable of interest $y$, indicating a non-natural delivery, and two explanatory variables, the weight of the newborn child ($x_1$) and the weight gain of the mother $(x_2)$, on top; and statistics about the ``treatment'' considered at the bottom.}\label{Tab:Stat:Des:XY-T}
\end{table}

%\begin{table}[!ht]\centering
%\begin{tabular}{lrr}\hline\hline
 %&{$y=0$ (natural)}  &$y=1$ (non-natural)  \\\hline
%$n$ number of observations & 2,221,522  (65.70\%) & 1,159,776    (34.30\%)  \\ 
%$x_1$ weight of newborn & average 3,299 g. & average 3,231 g. \\ 
%$x_2$ weight gain of mother & average 30.02 lbs. & average  31.16 lbs. \\  \hline\hline
%\end{tabular}
%\caption{Statistics about the variable of interest $y$, indicating a non-natural delivery, and two explanatory variables, the weight of the newborn child ($x_1$) and the weight gain of the mother $(x_2)$.}\label{Tab:Stat:Des:XY}
%\end{table}
%
%\begin{table}[!ht]\centering
%\begin{tabular}{lcccccc}\hline\hline
% & \multicolumn{3}{c}{$t=0$}  & \multicolumn{3}{c}{$t=1$}  \\\hline
%Afro-American  variable & non-Black & 2,980,387 & (88.14\%) & Black & 400,911& (11.86\%) \\ 
%smoker variable & non-smoker & 2,959,847 & (91.54\%)& smoker & 273,685 & (8.46\%)  \\ 
%gender variable & baby boy & 1,730,837 & (51.18\%) & baby girl & 1,650,461 & (48.82\%) \\ \hline\hline
%\end{tabular}
%\caption{Statistics about the various ``treatment'' considered here.}\label{Tab:Stat:Des:T}
%\end{table}

\begin{figure}[!ht]
\centering
\tikzset{every picture/.style={line width=0.75pt}} %set default line width to 0.75pt        

\begin{tikzpicture}[x=0.75pt,y=0.75pt,yscale=-1,xscale=1]
%uncomment if require: \path (0,367); %set diagram left start at 0, and has height of 367

%Shape: Circle [id:dp7073551278848534] 
\draw  [fill={rgb, 255:red, 248; green, 231; blue, 28 }  ,fill opacity=1 ] (47,82) .. controls (47,73.16) and (54.16,66) .. (63,66) .. controls (71.84,66) and (79,73.16) .. (79,82) .. controls (79,90.84) and (71.84,98) .. (63,98) .. controls (54.16,98) and (47,90.84) .. (47,82) -- cycle ;
%Shape: Circle [id:dp8817370144003989] 
\draw  [fill={rgb, 255:red, 184; green, 233; blue, 134 }  ,fill opacity=1 ] (126,65) .. controls (126,56.16) and (133.16,49) .. (142,49) .. controls (150.84,49) and (158,56.16) .. (158,65) .. controls (158,73.84) and (150.84,81) .. (142,81) .. controls (133.16,81) and (126,73.84) .. (126,65) -- cycle ;
%Straight Lines [id:da0538598597186436] 
\draw    (79,82) -- (120.57,72.66) ;
\draw [shift={(123.5,72)}, rotate = 167.33] [fill={rgb, 255:red, 0; green, 0; blue, 0 }  ][line width=0.08]  [draw opacity=0] (8.93,-4.29) -- (0,0) -- (8.93,4.29) -- cycle    ;
%Straight Lines [id:da3430937630106825] 
\draw    (163,73) -- (202.01,99.32) ;
\draw [shift={(204.5,101)}, rotate = 214.01] [fill={rgb, 255:red, 0; green, 0; blue, 0 }  ][line width=0.08]  [draw opacity=0] (8.93,-4.29) -- (0,0) -- (8.93,4.29) -- cycle    ;
%Shape: Circle [id:dp5410995502172932] 
\draw  [fill={rgb, 255:red, 80; green, 227; blue, 194 }  ,fill opacity=1 ] (208,108) .. controls (208,99.16) and (215.16,92) .. (224,92) .. controls (232.84,92) and (240,99.16) .. (240,108) .. controls (240,116.84) and (232.84,124) .. (224,124) .. controls (215.16,124) and (208,116.84) .. (208,108) -- cycle ;
%Shape: Circle [id:dp47352325503811743] 
\draw  [fill={rgb, 255:red, 184; green, 233; blue, 134 }  ,fill opacity=1 ] (127,137) .. controls (127,128.16) and (134.16,121) .. (143,121) .. controls (151.84,121) and (159,128.16) .. (159,137) .. controls (159,145.84) and (151.84,153) .. (143,153) .. controls (134.16,153) and (127,145.84) .. (127,137) -- cycle ;
%Straight Lines [id:da6706009011606401] 
\draw    (159,137) -- (199.9,113.49) ;
\draw [shift={(202.5,112)}, rotate = 150.11] [fill={rgb, 255:red, 0; green, 0; blue, 0 }  ][line width=0.08]  [draw opacity=0] (8.93,-4.29) -- (0,0) -- (8.93,4.29) -- cycle    ;
%Straight Lines [id:da40816585260325744] 
\draw    (79,82) -- (124.3,123.96) ;
\draw [shift={(126.5,126)}, rotate = 222.81] [fill={rgb, 255:red, 0; green, 0; blue, 0 }  ][line width=0.08]  [draw opacity=0] (8.93,-4.29) -- (0,0) -- (8.93,4.29) -- cycle    ;
%Straight Lines [id:da5171044386097509] 
\draw    (79,82) -- (198.06,106.4) ;
\draw [shift={(201,107)}, rotate = 191.58] [fill={rgb, 255:red, 0; green, 0; blue, 0 }  ][line width=0.08]  [draw opacity=0] (8.93,-4.29) -- (0,0) -- (8.93,4.29) -- cycle    ;

% Text Node
\draw (58,75.4) node [anchor=north west][inner sep=0.75pt]  [xscale=0.9,yscale=0.9]  {$t$};
% Text Node
\draw (134,56.4) node [anchor=north west][inner sep=0.75pt]  [xscale=0.9,yscale=0.9]  {$x_{1}$};
% Text Node
\draw (120,160) node [anchor=north west][inner sep=0.75pt]  [xscale=0.9,yscale=0.9] [align=left] {\begin{minipage}[lt]{32.22pt}\setlength\topsep{0pt}
\begin{center}
weight\\gain
\end{center}

\end{minipage}};
% Text Node
\draw (198,50) node [anchor=north west][inner sep=0.75pt]  [xscale=0.9,yscale=0.9] [align=left] {\begin{minipage}[lt]{54.37pt}\setlength\topsep{0pt}
non-natural
\begin{center}
delivery
\end{center}

\end{minipage}};
% Text Node
\draw (117,6) node [anchor=north west][inner sep=0.75pt]  [xscale=0.9,yscale=0.9] [align=left] {\begin{minipage}[lt]{41.88pt}\setlength\topsep{0pt}
\begin{center}
newborn\\weight
\end{center}

\end{minipage}};
% Text Node
\draw (217,100.4) node [anchor=north west][inner sep=0.75pt]  [xscale=0.9,yscale=0.9]  {$y$};
% Text Node
\draw (135,128.4) node [anchor=north west][inner sep=0.75pt]  [xscale=0.9,yscale=0.9]  {$x_{2}$};
% Text Node
\draw (1,104) node [anchor=north west][inner sep=0.75pt]  [xscale=0.9,yscale=0.9] [align=left] {\begin{minipage}[lt]{73.66pt}\setlength\topsep{0pt}
\begin{center}
smoker\\\textcolor{BrickRed}{black mother}\\gender of infant
\end{center}

\end{minipage}};

\end{tikzpicture}
\vspace{-.8cm}\caption{Directed acyclic graphs to explain non-natural deliveries, when $y=\boldsymbol{1}(\text{non-natural delivery})$, $x$ is either the birth weight of the infant ($x_1$), or the weight gain of the pregnant mother ($x_2$), and $t$ is either a smoker indicator (for the mother), or an indicator that the mother is Black (Afro-American), or that the baby is a boy.}\label{Fig:NN:1ex}
\end{figure}

\section{Quantile based matching}\label{sec:opt:transp:univarie}

In this section, we consider the simple case where $x$ is univariate. This allows us to introduce properties that will be extended more formally in higher dimension in the next section. Following the example of Section~\ref{sec:gaussian:toy}, we will propose some techniques to generate a counterfactual version of $(x,y,t=0)$, or $(x,y_{T\leftarrow 0}^\star)$,  that will be $(x_{T\leftarrow 1},y_{T\leftarrow 1}^\star)$. In Section~\ref{sub:sec:univ:classical:coupling}, we will discuss classical matching techniques, used to match each point in $(y_i,x_i,t_i=0)$ --in the control group-- with another one in $(y_j,x_j,t_j=1)$ --in the treated group-- when the two groups have the same size. In Section~\ref{sub:sec:univ:optimal:coupling}, we will suggest on optimal matching algorithm, to associate individual $i$ (in the control group) to $j$ (in the treated group), or $j_i^\star$. Then, in Section~\ref{sub:sec:univ:optimal:matching}, we will discuss the case where the two groups have different sizes, that will be called optimal ``coupling''. In Section~\ref{sub:sec:univ:cate}, we will define an estimator, the {\em mutatis mutandis} CATE, $\widehat{m}_1\big(\widehat{\mathcal{T}}(x)\big) - \widehat{m}_0\big(x\big)$, where $\widehat{\mathcal{T}}(x)= \widehat{F}_1^{-1}\circ \widehat{F}_0(x)$, with $ \widehat{F}_0$ and $ \widehat{F}_1$ denoting the empirical distribution functions of $x$ conditional on $t=0$ and $t=1$, respectively. Thus, we will use quantiles to optimal ``transport'' $x$'s from the control group to the treated group, formally through the $\mathcal{T}$ mapping.  Finally, in Section~\ref{sub:ex}, we will illustrate this on the probability that a non-natural baby delivery occurs.

\subsection{Classical matching techniques}\label{sub:sec:univ:classical:coupling}

To estimate the average treatment effect $\displaystyle{\tau =  \mathbb{E}\big[ Y^\star_{T\leftarrow1} - Y^\star_{T\leftarrow0}\big]}$, a standard technique is to consider matching techniques to match each point in $(y_i,x_i,t_i=0)$ or $(y_i^{(0)},x_i^{(0)})$ with  another one in $(y_j,x_j,t_j=1)$, or $(y_j^{(1)},x_j^{(1)})$. In this coupling approach, we assume that there are $n$ treated and $n$ non-treated individuals. A treated individual $i$ ($t_i=1$) is matched to someone in the non-treated group ($t_j=0$) that is close enough for some distance on the set of covariates $\mathcal{X}$, $j^\star_i=\displaystyle{\underset{j:t_j=0}{\text{argmin}}\{d(x_i^{(0)},x_j^{(1)})\}}$, so that
$$
\widehat{\tau} = \frac{1}{n}\sum_{i=1}^{n} \big(y^{(1)}_{j^\star_i}-y_{i}^{(0)}\big)= \frac{1}{n}\sum_{i=1}^{n}y_{j^\star_i}^{(1)}-\frac{1}{n}\sum_{i=1}^{n} y^{(0)}_i=\overline{y}^{(1)}-\overline{y}^{(0)},
$$
since we simply consider a re-ordering of the treated population. But interestingly, that approach provides a counterfactual version of $(x_i,y_i)$ in the treated population, $(x_{j^\star_i},y_{j^\star_i})$. An algorithm performing such a matching would be Algorithm~\ref{alg:ate:0}. 

%REPRENDRE ALGO de matching, output = rangs + algo soit global soit local

\begin{algorithm}[!ht]
\caption{Counterfactual matching -- ``1:1 nearest neighbor matching'' (classical)}\label{alg:ate:0}
\begin{algorithmic}
 \State$\mathcal{D} \gets \{(y_i,\boldsymbol{x}_i,t_i)\}$
\Function{Counterfactual1}{$\mathcal{D}$}
\State $\mathcal{D}_0 \gets $ subset of $\mathcal{D}$ when $t=0$ (size $n$) shuffled, with indices $i$
\State $\mathcal{D}_1 \gets $ subset of $\mathcal{D}$ when $t=1$ (size $n$), with indices $j$
\For{$i=1,2,\cdots,n$}
\State $j^\star_i=\displaystyle{\underset{j:t_j=1}{\text{argmin}}\{d(\boldsymbol{x}_i,\boldsymbol{x}_j)\}}$ in $\mathcal{D}_1$,
\State $L_i \gets (i,j^\star_i,y_{j^\star_i}^{(1)}-y^{(0)}_i)$
\State remove observation $j^\star_i$ from $\mathcal{D}_1$
\EndFor
\State \Return matrix $L$ ~~ ($n\times 3$, with $L=(L_i)$) 
\EndFunction
\end{algorithmic}
\end{algorithm}

This algorithm, introduced by \cite{rubin1973matching}, is described in \cite{stuart2010matching} under the name ``1:1 nearest neighbor matching'', and properties are discussed in \cite{ho2007matching} or \cite{dehejia1999causal} that focuses on the problem of not removing selected observations (also called ``Greedy Matching'').

% \begin{enumerate}
%     \item consider a permutation of all treated individuals ($t_i=1$),
%     \item for treated individual $i$ ($t_i=1$), match that individual to someone in the non-treated group ($t_j=0$), 
%     \item then, remove the untreated observation from the database, and iterate (so as to match all treated individuals with an untreated person.
% \end{enumerate}
Quite naturally, it is possible to define some local version of the previous quantity using weights or some $k$ nearest neighbors approach, to derive an estimate of the CATE $\widehat{\tau}(x)$, as in Algorithm~\ref{alg:cate:0}
$$
\widehat{\tau}(x) \propto \sum_{i=1}^{n} \omega_{i}(x)\big(y_{j^\star_i}^{(1)}-y^{(0)}_i\big),
$$
where weight $\omega_i(x)$ are all the higher that $x_i$ is close to $x$, either based on a $k$-nearest neighbors approach ($\omega_i(x) = \boldsymbol{1}(i\in V_{{x}}^k)$, as in Algorithm~\ref{alg:cate:0}) or based on a kernel approach ($\omega_i(x) =K(|{x}-{x}_i|)$ for some kernel $K$).

\begin{algorithm}[!ht]
\caption{Estimate SCATE (classical, with $k$-NN)}\label{alg:cate:0}
\begin{algorithmic}
\State dataset $\mathcal{D}\gets\{(y_i,\boldsymbol{x}_i,t_i)\}$, 
\Function{scate1}{$\mathcal{D},k,\boldsymbol{x}$}
\State $L \gets $ {\sc Counterfactual1}$(\mathcal{D})$
\State $V_{\boldsymbol{x}}^k\gets$ list of $k$ nearest neighbors of  $\boldsymbol{x}_i$'s in $\mathcal{D}_0$ close to $\boldsymbol{x}$
\For{$i\in V_{\boldsymbol{x}}^k$}
\State $d_i\gets L\$d(i)$
\EndFor
\State\Return $\displaystyle{\frac{1}{k}\sum_{i\in V_{\boldsymbol{x}}^k} d_i}$
\EndFunction
\end{algorithmic}
\end{algorithm}

Unfortunately, that matching mechanism can be very sensitive to the initial permutation: individuals picked first will have a counterfactual in the treated group close to them, but it might not be the case for the individuals picked last. In the next section, we will consider some optimal matching among individuals in the two populations.

\subsection{Optimal matching}\label{sub:sec:univ:optimal:coupling}

The matching procedure described previously is characterized by some $n\times n$ permutation matrix, $P$, with entries in $\{0,1\}$, satisfying $\mathbb{P}\boldsymbol{1}_n=\boldsymbol{1}_n$ and  $\mathbb{P}^{\star\top}\boldsymbol{1}_n=\boldsymbol{1}_n$, see \cite{brualdi2006combinatorial}. Hence, there is a permutation $\sigma$ of $\{1,\cdots,n\}$ such that $j_i^\star = \sigma(i)$, and $P$ is the matrix associated with $\sigma$ (that satisfies $\boldsymbol{e}_i P=\boldsymbol{e}_{\sigma(i)}$, where $\boldsymbol{e}_i$'s denote the standard basis vector, i.e., a row vector of length $n$ with $1$ in the $i$-th position and $0$ in every other position). It is possible to seek an ``optimal'' permutation: if $C$ is the $n\times n$ matrix that quantifies the distance between individuals in the two groups, $C_{i,j}=d(x_i^{(0)},x_j^{(1)})=\delta(x_i^{(0)}-x_j^{(1)})$, the optimal matching is solution of 
$$
\min_{P\in\mathcal{P}} \langle P,C\rangle = 
\min_{P\in\mathcal{P}} \sum_{i,j} P_{i,j}C_{i,j},
$$
where $\mathcal{P}$ is the set of permutation matrices, and $\langle \cdot,\cdot\rangle$ is the Frobenius dot-product. This is also called Kantorovich’s optimal transport problem, from \cite{kantorovich1942translocation}. If $\delta$ is (strictly) convex --as is the standard Euclidean distance-- it can be proven that this optimal transport problem has a simple solution.
Instead of using $(y_i^{(0)},x_i^{(0)})$, let $r_i^{(0)}$ denote the rank of $x_i^{(0)}$ in $\{x_1^{(0)},\cdots,x_n^{(0)}\}$. Similarly, let $r_i^{(1)}$ denote the rank of $x_i^{(1)}$ in the treated dataset $\{x_1^{(1)},\cdots,x_n^{(1)}\}$. The procedure then becomes simply a matching based on ranks, in the sense that $j_i^\star$ satisfies $r_{j_i^\star}^{(1)}=r_i^{(0)}$, as discussed in Chapter 2 of \cite{santambrogio2015optimal}. Since ranks are defined on $\{1,2,\cdots,n\}$, vectors $\boldsymbol{r}^{(0)}$ and $\boldsymbol{r}^{(1)}$ correspond to two permutations of $\{1,2,\cdots,n\}$, that we can denote $\sigma_0$ and $\sigma_1$, respectively. The optimal coupling is based on permutation $\sigma=\sigma_1\circ\sigma_0^{-1}$ in the sense that $x_i^{(0)}$ is associated to $x_{\sigma(i)}^{(1)}$. %This can be visualized in Figure~\ref{fig:match:1}: 
If the $\boldsymbol{x}^{(0)}$'s and the $\boldsymbol{x}^{(1)}$'s are sorted, then $P=\mathbb{I}_n$, i.e., $x_i^{(0)}$ is coupled with $x_i^{(1)}$. Or, if $\widehat{F}_0$ and $\widehat{F}_1$ are the cumulative distribution functions associated with sample $\boldsymbol{x}^{(0)}$ and $\boldsymbol{x}^{(1)}$, we can see that if $u\in(0,1)$ is such that $\widehat{F}_0^{-1}(u)=x_i^{(0)}$, then $\widehat{F}_1^{-1}(u)=x_i^{(1)}$, with the exact same $i$.

\subsection{Optimal coupling}\label{sub:sec:univ:optimal:matching}

The previous procedure can be extended in the case where the two groups do not necessarily have the same size. If the two groups $({x}_i,t_i=0)$ and $({x}_j,t_j=1)$ have different sizes, namely $n_0$ and $n_1$, respectively, it is possible to define some matching using weights, and weighted mean of individuals in the two groups.

 In a very general setting, if $\boldsymbol{a}_0\in\mathbb{R}_+^{n_0}$ and $\boldsymbol{a}_1\in\mathbb{R}_+^{n_1}$ satisfy $\boldsymbol{a}_0  ^\top\boldsymbol{}{1}_{n_0}=\boldsymbol{a}_1  ^\top\boldsymbol{}{1}_{n_1}$ (identical sums), define
 $$
 U(\boldsymbol{a}_0,\boldsymbol{a}_1)=\big\lbrace
 M\in\mathbb{R}_+^{n_0\times n_1}:M\boldsymbol{1}_{n_1}=\boldsymbol{a}_{0}\text{ and }{M}^\top\boldsymbol{1}_{n_0}=\boldsymbol{a}_{1}
 \big\rbrace.
 $$
 This set of matrices is a convex polytope (see \cite{brualdi2006combinatorial}). The optimal coupling is matrix $P^\star$ solution of
 $$
 \min_{P\in  U(\boldsymbol{a}_0,\boldsymbol{a}_1)} \left\lbrace \langle C,P\rangle \right\rbrace,
 $$
 which is solved using linear programming, by casting matrix $P\in\mathbb{R}_+^{n_0\times n_1}$ as a vector $\boldsymbol{p}\in\mathbb{R}_+^{n_0n_1}$ such that $\boldsymbol{p}_{i+n(j-1)}=P_{i,j}$, and similarly for the cost matrix $C$. The constraint $P\in  U(\boldsymbol{a}_0,\boldsymbol{a}_1)$ becomes equivalently 
 $$\begin{pmatrix}
 \boldsymbol{1}_{n_0}^\top\otimes\mathbb{I}_{n_1}\\
\mathbb{I}_{n_0}\otimes\boldsymbol{1}_{n_1}^\top
 \end{pmatrix}\boldsymbol{p} =
A\boldsymbol{p} =(\boldsymbol{a}_0,\boldsymbol{a}_1)^\top= \begin{pmatrix}
\boldsymbol{a}_0\\
\boldsymbol{a}_1
 \end{pmatrix},
 $$
 where $A$ is some $(n_0+n_1)\times(n_0n_1)$ matrix. The optimal matching problem is then simply
 $$
\min\left\lbrace \boldsymbol{c}^\top\boldsymbol{p} \right\rbrace\text{ subject to }A\boldsymbol{p} =(\boldsymbol{a}_0,\boldsymbol{a}_1)^\top.
 $$

In our case, let $U_{n_0,n_1}$ denote $U(\boldsymbol{1}_0,\frac{n_0}{n_1}\boldsymbol{1}_1)$
\begin{equation}\label{prog:match}
P^* \in \underset{P\in  U_{n_0,n_1}}{\text{argmin}}  \langle P,C\rangle \text{ ou }
\underset{P\in  U_{n_0,n_1}}{\text{argmin}}\sum_{i=1}^{n_0} \sum_{j=1}^{n_1} P_{i,j}C_{i,j}.
\end{equation}
One can notice that this matrix optimisation problem does not depend on the dimension of space, so it will easily be extended to the case where $x$ is multivariate. Nevertheless, in the univariate setting, this approach can be related to quantile functions.

\subsection{From optimal matching to CATE}\label{sub:sec:univ:cate}

Let $F_0$ and $F_1$ denote the two conditional distributions of $X$, an absolutely continuous variable, in the control group ($t=0$) and in the treatment group ($t=1$), respectively. Then the optimal matching between the two groups is based on transformation $\mathcal{T}:x_0\mapsto x_1 = F_1^{-1}\circ F_0(x_0)$. From the probability integral transform property: if $X_0\sim F_0$, then $F_0(X_0)$ is uniform on the unit interval $[0,1]$, and then $X_1 = \mathcal{T}(X_0)\sim F_1$.

\begin{lemma}
If $X_0\sim F_0$, then $X_1 = \mathcal{T}(X_0)\sim F_1$, where $\mathcal{T}:x_0\mapsto x_1 = F_1^{-1}\circ F_0(x_0)$.
\end{lemma}

\begin{definition}
The {\em mutatis mutandis} quantile-based CATE is
\begin{equation}\label{eq:CATE:quantile}
\text{QCATE}(u) = 
\mathbb{E}\big[Y^*_{T\leftarrow 1}\big|X=F_{1}^{-1}(u)\big] - 
\mathbb{E}\big[Y^*_{T\leftarrow 0}\big|X=F_{0}^{-1}(u)\big],
\end{equation}
where $F_t$ is the cumulative distribution function of $X$, conditional on $T=t$, or
\begin{equation}\label{eq:CATE:quantile:2}
\text{CATE}(x) = 
\mathbb{E}\big[Y^*_{T\leftarrow 1}\big|X=\mathcal{T}(x)\big] - 
\mathbb{E}\big[Y^*_{T\leftarrow 0}\big|X=x\big],~ \mathcal{T} = F_1^{-1}\circ F_0
\end{equation}
where $x$ is considered with respect to the control group.
\end{definition}

Thus, $\text{CATE}(x) =\text{QCATE}(F_0(x))$.

\begin{definition}
Consider two models, $\widehat{m}_0(x)$ and $\widehat{m}_1(x)$, that estimate, respectively, $\mathbb{E}[Y|X=x,T=0]$ and  $\mathbb{E}[Y|X=x,T=1]$.
A natural estimator of the {\em mutatis mutandis} CATE is
$$
\text{SCATE}(x)=\widehat{m}_1\big(\widehat{\mathcal{T}}(x)\big) - \widehat{m}_0\big(x\big)
$$
where $\widehat{\mathcal{T}}(x)= \widehat{F}_1^{-1}\circ \widehat{F}_0(x)$, $ \widehat{F}_0$ with $ \widehat{F}_1$ denoting the empirical distribution functions of $x$ conditional on $t=0$ and $t=1$, respectively.
\end{definition}

Note that a simple parametric transformation can be obtained, based on the assumption that $X$ conditional on $T$ is Gaussian. More precisely, if $X_1\overset{\mathcal{L}}{=}{X}|t=1\sim\mathcal{N}({\mu}_1,{\sigma}_1^2)$ and $X_0\overset{\mathcal{L}}{=}{X}|t=0\sim\mathcal{N}({\mu}_0,{\Sigma}_0)$, 
$$
\mu_1+\sigma_1\cdot \frac{X_0-\mu_0}{\sigma_0} \overset{\mathcal{L}}{=} X_1
$$

\begin{definition}
Consider two models, $\widehat{m}_0(x)$ and $\widehat{m}_1(x)$, that estimate respectively $\mathbb{E}[Y|X=x,T=0]$ and  $\mathbb{E}[Y|X=x,T=0]$.
A Gaussian estimator of the {\em mutatis mutandis} CATE is
$$
\text{SCATE}_{\mathcal{N}}(x)=\widehat{m}_1\big(\widehat{\mathcal{T}}_{\mathcal{N}}(x)\big) - \widehat{m}_0\big(x\big)
$$
where $\widehat{\mathcal{T}}_{\mathcal{N}}(x)= \overline{x}_1+s_1s_0^{-1} (x-\overline{x}_0)$,
$\overline{x}_0$ and $\overline{x}_0$ being respectively the averages of $x$ in the two sub-populations, and $s_0$ and $s_1$ the sample standard deviations.
\end{definition}

An algorithm to compute that estimator is Algorithm~\ref{alg:1:gauss:1} (in higher dimension).

\subsection{Application to non-natural deliveries}\label{sub:ex}

\begin{figure}[!ht]
    \centering

    \centering    
    \includegraphics[width=.49\textwidth]{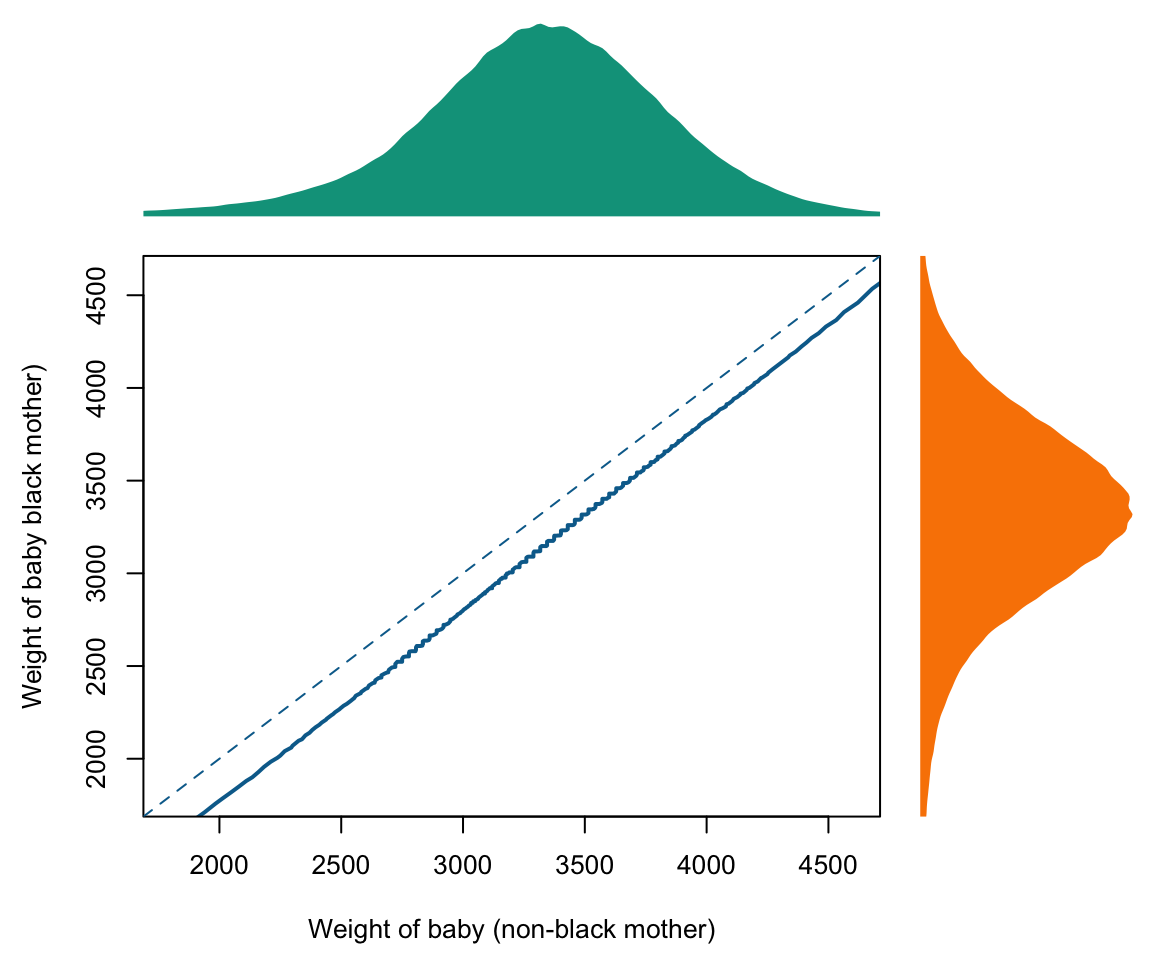}
    \includegraphics[width=.49\textwidth]{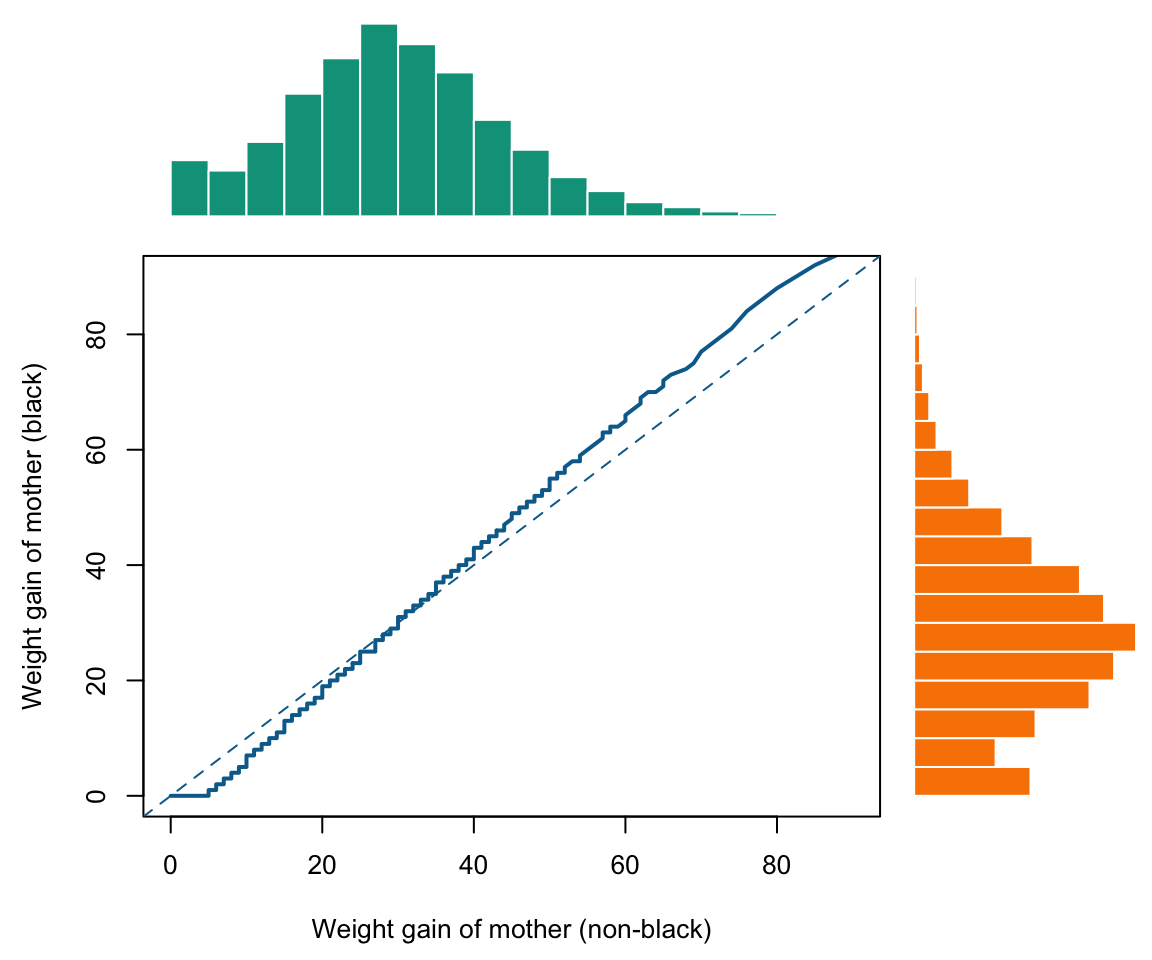}

    % \centering
    % \includegraphics[width=.49\textwidth]{figures/joint-transport-gender-baby-weight-1}
    % \includegraphics[width=.49\textwidth]{figures/joint-transport-gender-baby-weight-gain-1}
    
    \caption{Optimal transport (quantile based) when $X$ is the weight of the newborn infant on the left, and the weight gain of the mother on the right, when $T$ indicates whether the mother is Black or not in the middle. The solid line depicts the transported values while the dashed line is the identity line. See Figure~\ref{fig:opt-transport-quantiles-2x3:appendix} in Appendix~\ref{app:3} for similar graphs when $T$ indicates whether the mother is a smoker or not, or indicates the sex of the newborn.}
    \label{fig:opt-transport-quantiles-2x3}
\end{figure}

In Figure~\ref{fig:opt-transport-quantiles-2x3}, we can visualize $x\mapsto\widehat{\mathcal{T}}(x)$ when $x$ is either the weight of the newborn infant on the left, or the weight gain of the mother on the right, when $t$ indicates whether the mother is Black or not. The $x$-axis is the value of $x$ in the control group ($t=0$) and the $y$-axis is the value of $x$ in the treated group ($t=1$). On the left, observe that  $x\mapsto\widehat{\mathcal{T}}(x)$ is almost linear, parallel to the first diagonal, below. This corresponds to the fact that the distribution of $X$ conditional on $T=0$ and $T=1$ are similar, up to a translation (same standard deviation but different mean if a Gaussian transport $\widehat{\mathcal{T}}_{\mathcal{N}}$ was considered).  On the right, $x\mapsto\widehat{\mathcal{T}}(x)$ is single-crossing the first diagonal. This corresponds to the fact that the distribution of $X$ conditional on $T=0$ and $T=1$ have different variances.

\begin{figure}[!ht]
    \centering
    \includegraphics[width=.49\textwidth]{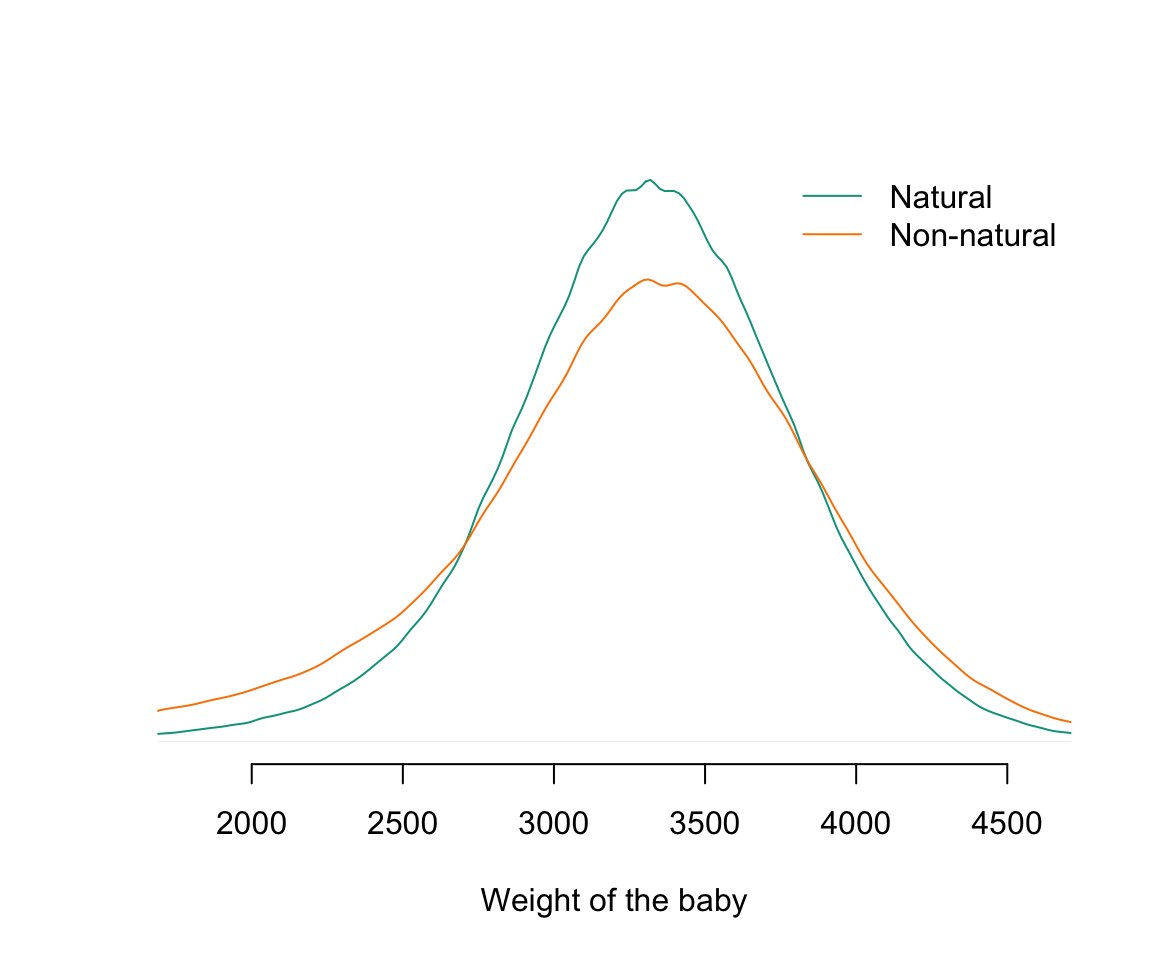}
     \includegraphics[width=.49\textwidth]{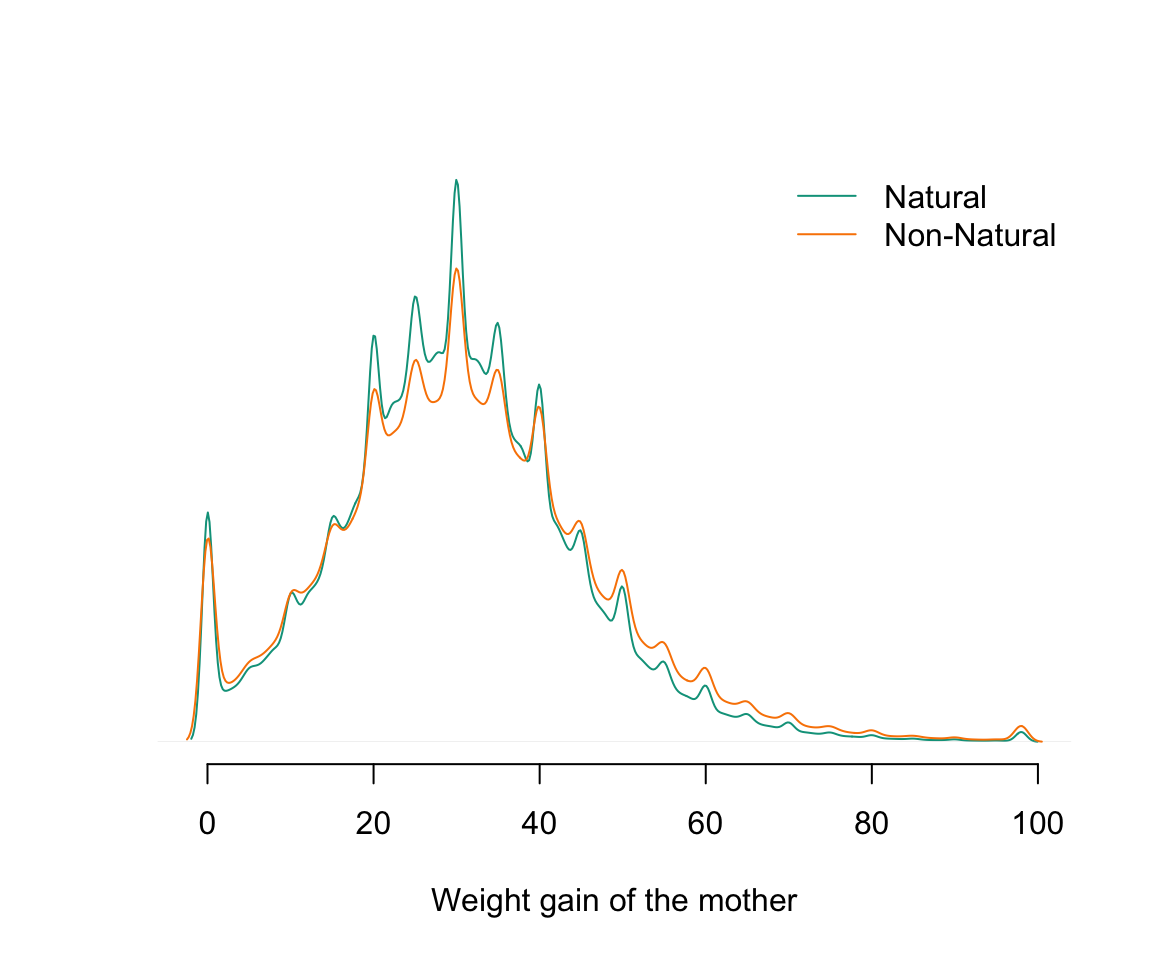}
    
    \caption{Distribution of the weight of the newborn infant (in grams) on the left and distribution of the weight gain of the mother on the right, conditional on the delivery mode, $Y=\boldsymbol{1}(\text{non-natural delivery})$.}
    \label{fig:densite-1x3-conditional-k}
\end{figure}

\begin{figure}[!ht]
    \centering

    \centering
    \includegraphics[width=.49\textwidth]{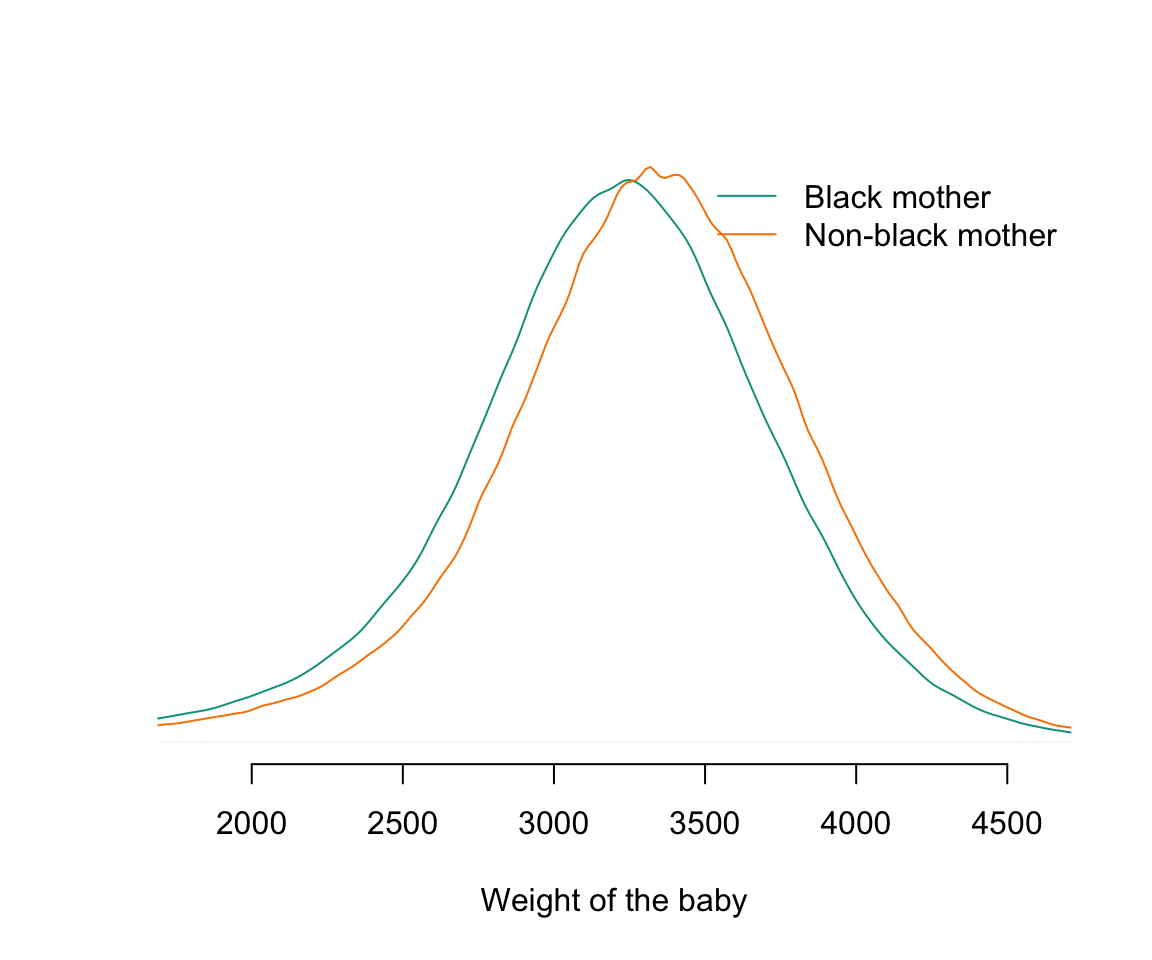}
    \includegraphics[width=.49\textwidth]{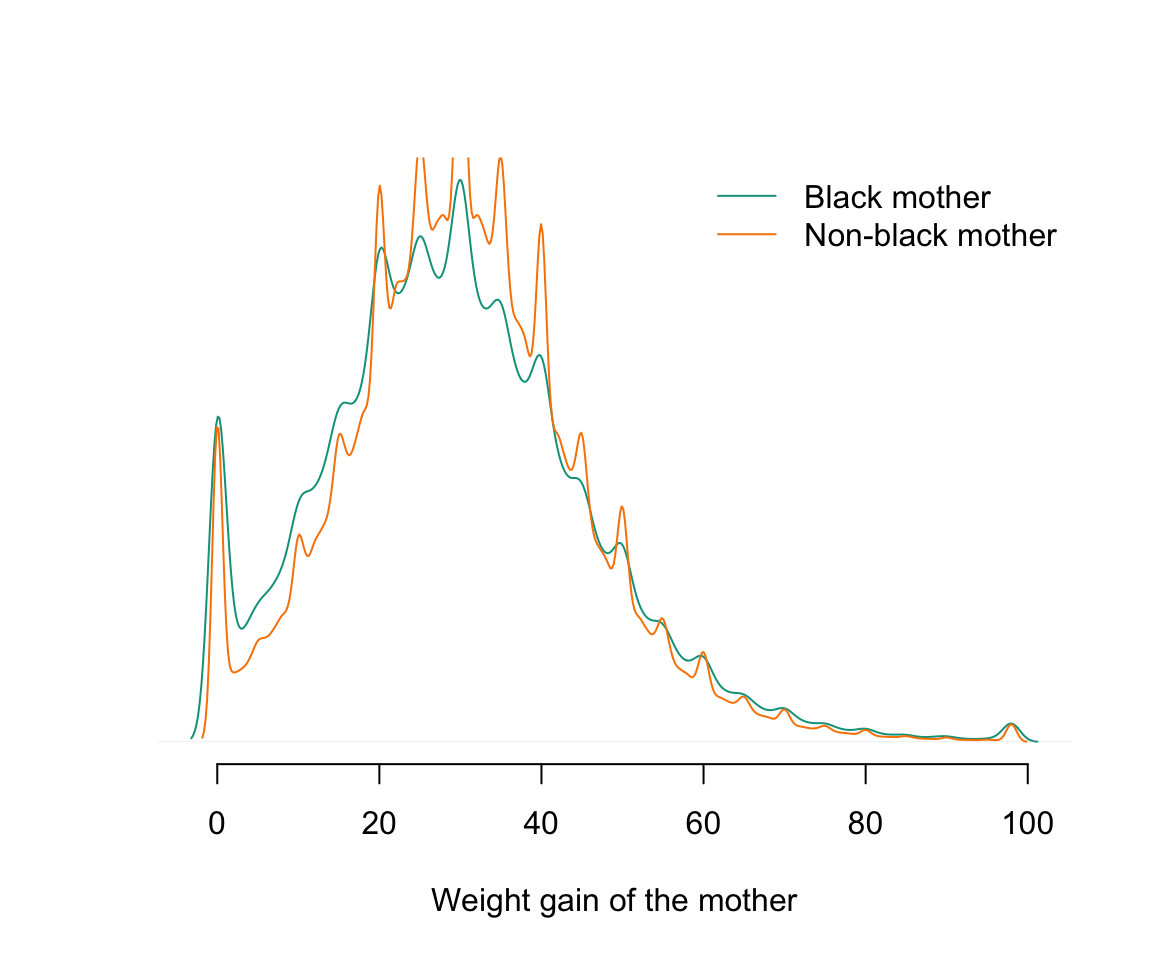}

    % \centering
    % \includegraphics[width=.49\textwidth]{figures/density-weight-gender-baby-1.png}
    % \includegraphics[width=.49\textwidth]{figures/density-weight-gain-gender-baby-1.png}
    
    \caption{Distribution of the weight of the newborn infant (in grams) on the left and distribution of the weight gain of the mother on the right, whether the mother is Black or not. See Figure~\ref{fig:densite-2x3-conditional-t:appendix} in Appendix~\ref{app:3} for similar graphs when $T$ indicates whether the mother is a smoker or not, or indicates the sex of the newborn.}
    \label{fig:densite-2x3-conditional-t}
\end{figure}

In Figure~\ref{fig:densite-1x3-conditional-k}, we can visualize the conditional distributions of $x$, when $y=0$ and $y=1$ (natural and non-natural deliveries, respectively), when $x$ is the weight of the baby (on the left) and the weight gain of the mother (on the right). In Figure~\ref{fig:densite-2x3-conditional-t}, we can visualize the conditional distributions of $x$, when $y=0$ and $y=1$, when $t=0$ and $t=1$, where $t$ denotes whether the mother is Afro-American or not.

In Figure~\ref{fig:opt-transport-quantiles-2x3}, we can visualize the empirical optimal coupling function $\widehat{\mathcal{T}}:x_0\mapsto x_1 = \widehat{F}_1^{-1}\circ \widehat{F}_0(x_0)$, where $ \widehat{F}_0$ and $ \widehat{F}_1$ denote the empirical distribution functions of $x$ conditional on $t=0$ and $t=1$, respectively.

\begin{figure}[!ht]
    % \centering
    %  \includegraphics[width=.49\textwidth]{figures/gam-univariate-smoker-weight-compare-transport-pb-min.png}
    %  \includegraphics[width=.49\textwidth]{figures/gam-univariate-diff-CATE-smoker-weight-compare-transport-1}

    \centering        
    \includegraphics[width=.49\textwidth]{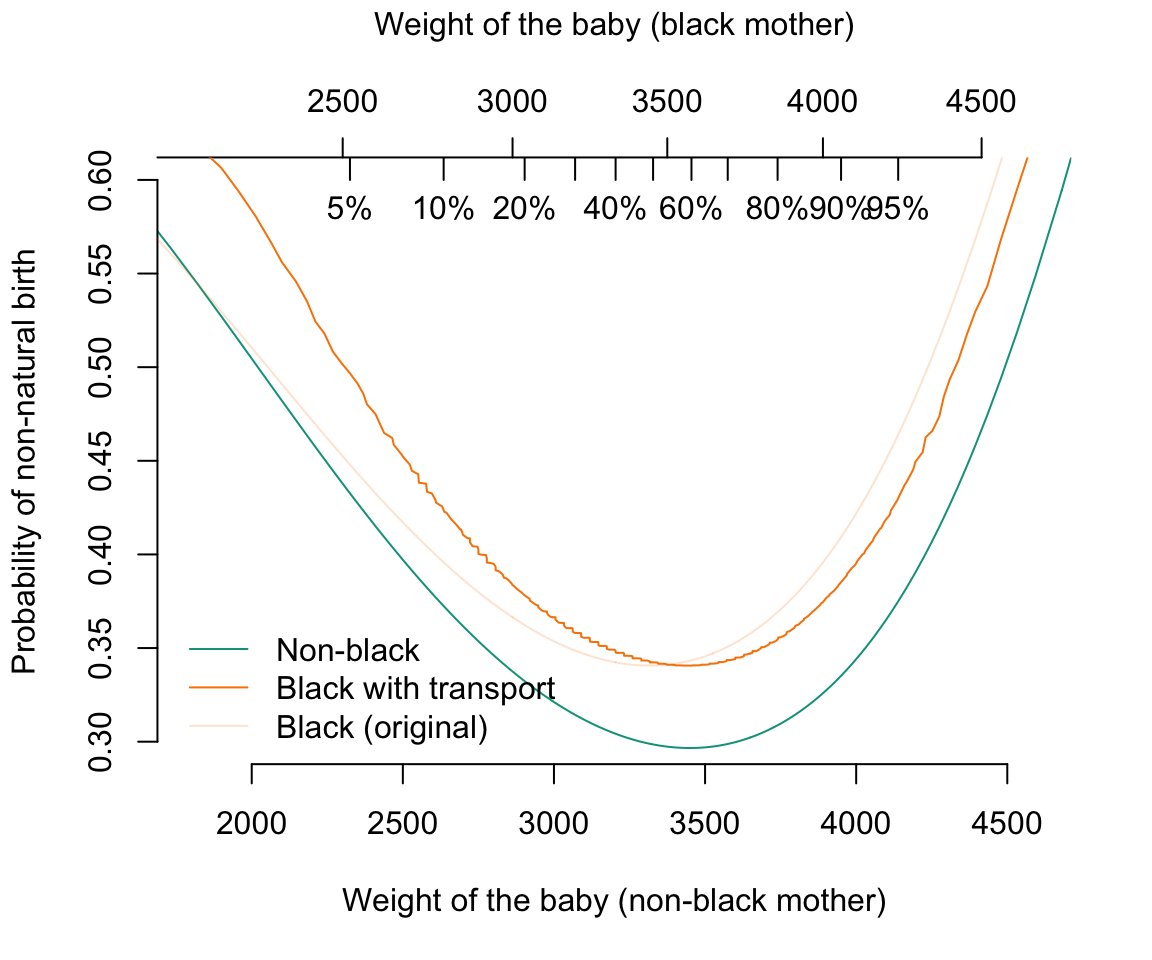}
     \includegraphics[width=.49\textwidth]{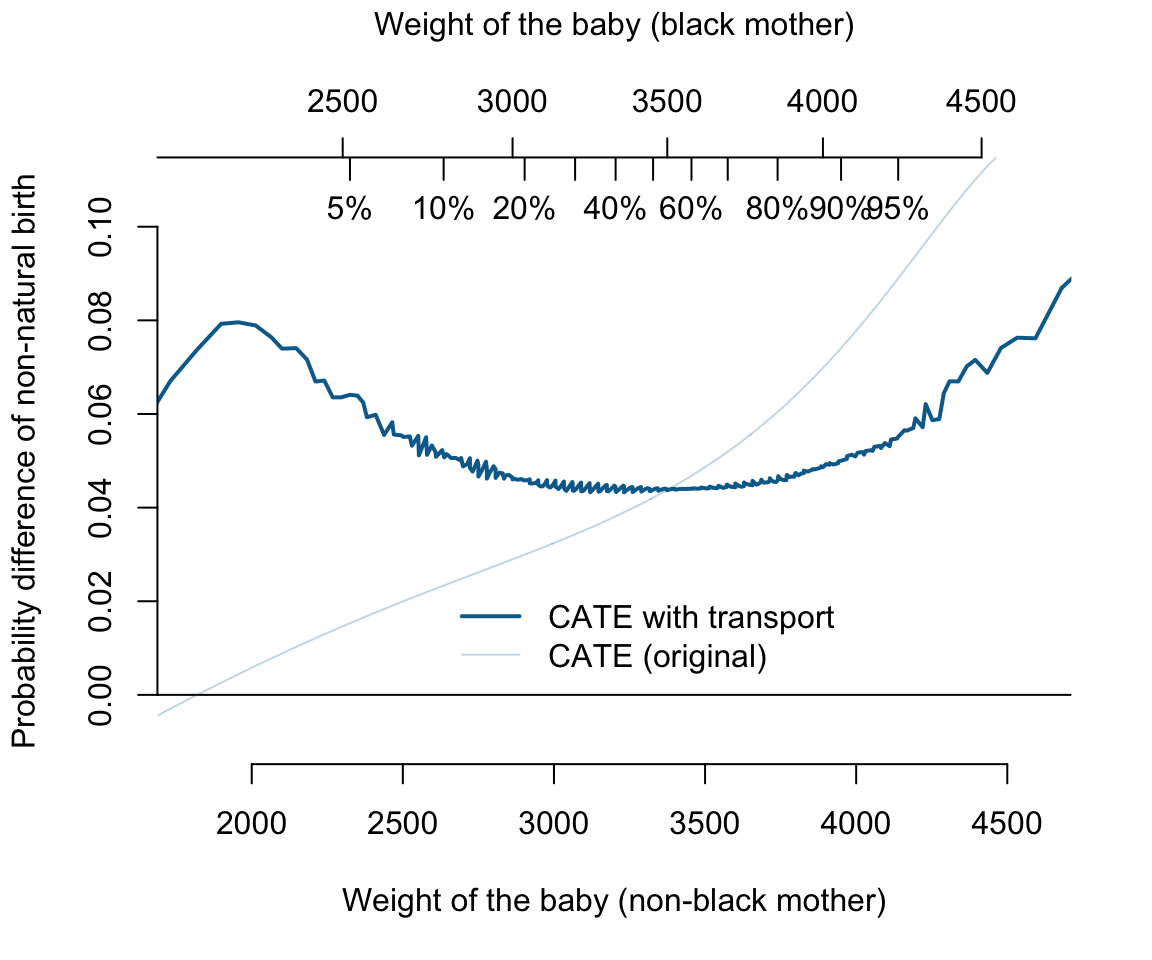}

    % \centering
    %  \includegraphics[width=.49\textwidth]{figures/gam-univariate-gender-baby-weight-compare-transport-1}
    %  \includegraphics[width=.49\textwidth]{figures/gam-univariate-diff-CATE-gender-baby-weight-compare-transport-1-1.png}

    \caption{On the left, evolution of $x\mapsto\mathbb{E}[Y|X_{T\leftarrow t}=x,T=t]$, estimated using a logistic GAM model, when $Y=\boldsymbol{1}(\text{non-natural delivery})$, and $X$ is the weight of the newborn infant, respectively when $T$ indicates whether the mother is Black or not. On the right, evolution of  $x\mapsto\text{SCATE}[Y|X=x]$. See Figure~\ref{fig:CATE-compare-quantiles-2x3-weight:appendix} in Appendix~\ref{app:3} for similar graphs when $T$ indicates whether the mother is a smoker or not, or indicates the sex of the newborn.}
    \label{fig:CATE-compare-quantiles-2x3-weight}
\end{figure}

\begin{figure}[!ht]
    % \centering
    %  \includegraphics[width=.49\textwidth]{figures/gam-univariate-smoker-weight-gain-compare-transport-1}
    %  \includegraphics[width=.49\textwidth]{figures/gam-univariate-diff-CATE-smoker-weight-gain-compare-transport-1}

    \centering        
    \includegraphics[width=.49\textwidth]{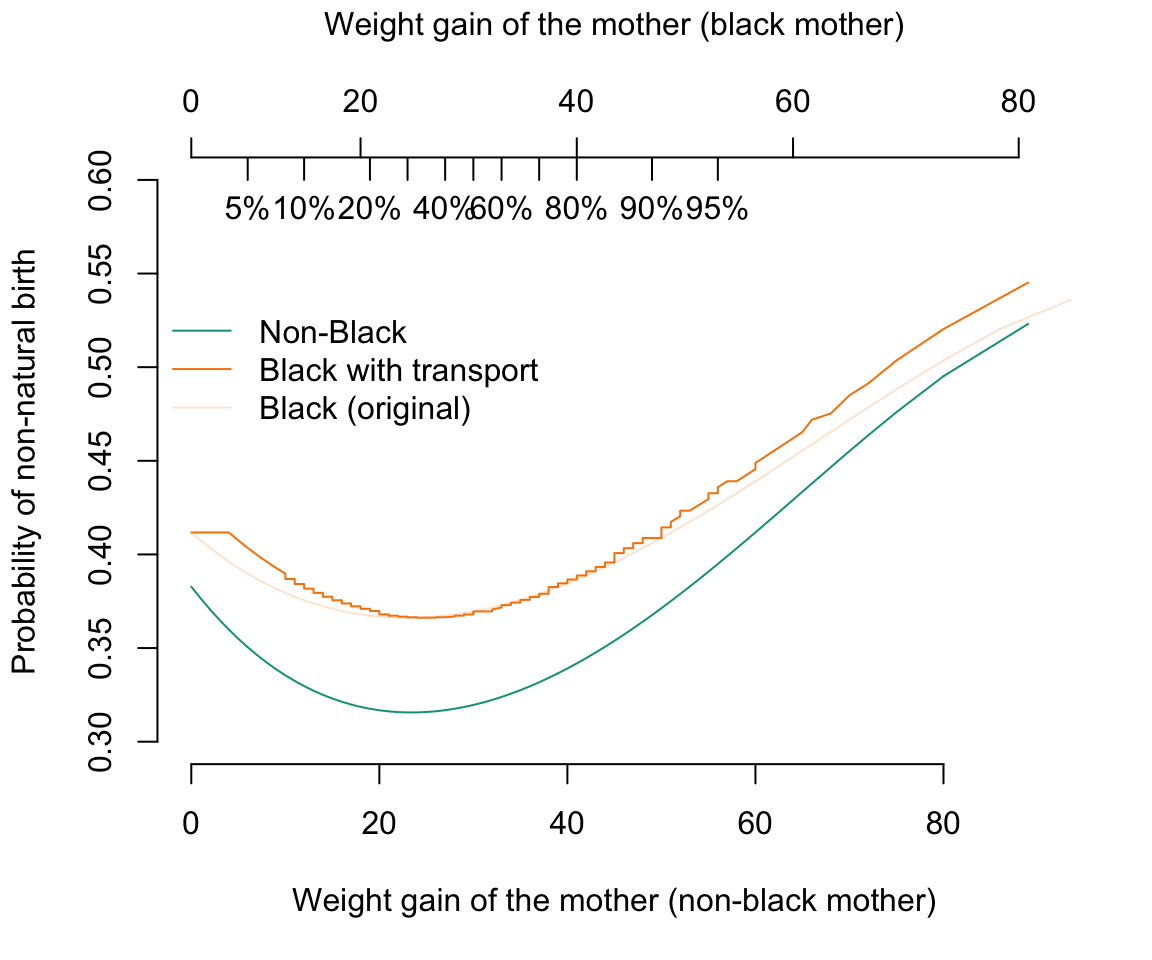}
     \includegraphics[width=.49\textwidth]{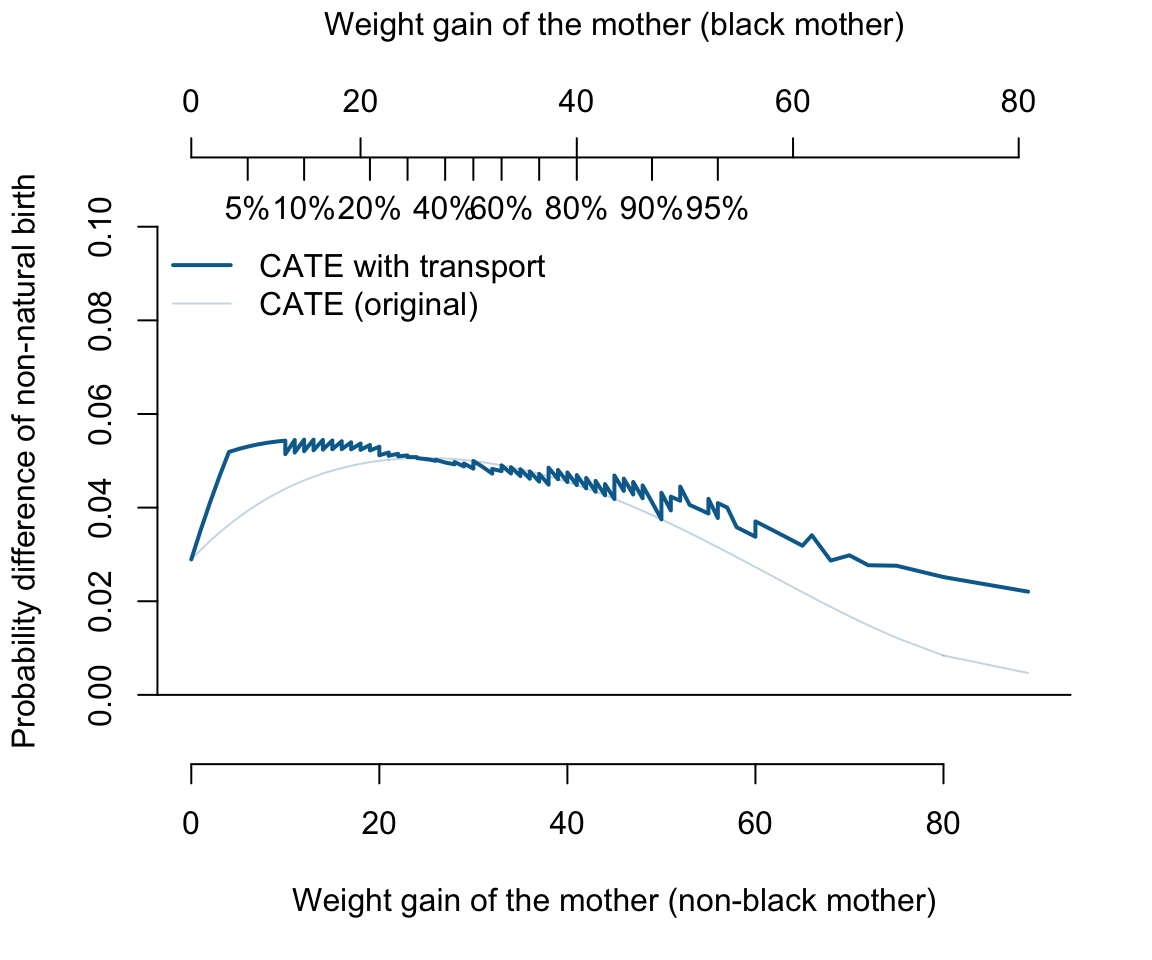}

    % \centering
    %  \includegraphics[width=.49\textwidth]{figures/gam-univariate-gender-baby-weight-gain-compare-transport-1-1.png}
    %  \includegraphics[width=.49\textwidth]{figures/gam-univariate-diff-CATE-gender-baby-weight-gain-compare-transport-1-1.png}

    \caption{On the left, evolution of $x\mapsto\mathbb{E}[Y|X_{T\leftarrow t}=x,T=t]$, estimated using a logistic GAM model, when $Y=\boldsymbol{1}(\text{non-natural delivery})$, and $X$ is the weight gain of the mother, respectively when $T$ indicates whether the mother is Black or not. On the right, evolution of  $x\mapsto\text{SCATE}[Y|X=x]$. See Figure~\ref{fig:CATE-compare-quantiles-2x3-weight-gain:appendix} in Appendix~\ref{app:3} for similar graphs when $T$ indicates whether the mother is a smoker or not, or indicates the sex of the newborn.}
    \label{fig:CATE-compare-quantiles-2x3-weight-gain}
\end{figure}

In Figures~\ref{fig:CATE-compare-quantiles-2x3-weight} and~\ref{fig:CATE-compare-quantiles-2x3-weight-gain}, we can visualize $\widehat{m}_0(x)$ and $\widehat{m}_1\big(\widehat{\mathcal{T}}(x)\big)$ on the left, when $t$ indicates whether the mother is Afro-American or not, when $x$ the weight of the newborn infant in Figure~\ref{fig:CATE-compare-quantiles-2x3-weight} and when $x$ is the weight gain of the mother in Figure~\ref{fig:CATE-compare-quantiles-2x3-weight-gain}. On the right, we can visualize $x\mapsto \text{CATE}(x)=\widehat{m}_1\big(\widehat{\mathcal{T}}(x)\big) - \widehat{m}_0\big(x\big)$ as a function of $x$. The light curve in the back is $\widehat{m}_1\big(x\big) - \widehat{m}_0\big(x\big)$. Numerical values are given in Table~\ref{tab:num:CATE:w} when $x$ is the weight of the newborn, and Table~\ref{tab:num:CATE:g} when $x$ is the weight gain of the pregnant mother. For instance, a baby weighting $2500$g (7.46\% quantile in the non-Black population) corresponds to a baby weighting $2301$g if the mother had been Black. The probability to have a non-natural delivery has then an additional $5.5\%$ compared with non-Black mother, using the GAM-SCATE approach. Using a Gaussian transport, the counterfactual in the Black population is a $2297$g baby, and the probability to have a non-natural delivery has then an additional $5.60\%$ compared with a non-Black mother, using the GAM-$\text{SCATE}_{\mathcal{N}}$ approach. Similarly, a baby weighting $3500$g (64.13\% quantile in the non-Black population) corresponds to a baby weighting $3375$g had the mother been Black (about $3.6\%$ less). The probability to have a non-natural delivery has then an additional $4.42\%$ compared with non-Black mother, using the GAM-SCATE approach. Using a Gaussian transport, estimates are similar.

\begin{table}[!ht]
    \centering
    \begin{tabular}{c|cccccc}\hline\hline
    \multicolumn{7}{c}{$t$: mother is Afro-American}\\
    \hline
  $x$ (newborn's weight) & 2000 & 2500 & 3000 & 3500 & 4000 & 4500 \\\hline
  
 % smoker & $u$ & 2.75\% &  7.44\% & 24.99\% & 64.14\% & 91.75\% & 98.86\% \\
 % & $\text{CATE}_0(x)$ (GAM) & -4.41\% & -2.55\% & -0.69\% &  0.79\% &  1.97\% &  2.50\%  \\
 % & $\text{CATE}(x)$ (propensity) & -5.69\% &  -2.33\% &  0.35\% &   1.94\% &   2.43\% &   0.93\%   \\
 %  &  $\widehat{\mathcal{T}}(x)$ & 1775   &   2280   &   2802   &   3317 &     3830  &    4337\\
 % & $\text{SCATE}(x)$ (GAM) & -0.08\% &      1.15\% &      1.15\% &      0.50\% &     -1.07\% &     -4.39\%  \\
 %  &  $\widehat{\mathcal{T}}_{\mathcal{N}}(x)$ &  1786& 2295& 2805& 3314& 3824& 4333\\
 % & $\text{SCATE}_{\mathcal{N}}(x)$ (GAM) &  -0.28\% &  0.88\% &  1.12\% &  0.50\% & -1.15\% & -4.53\%  \\
 % & $\text{SCATE}_{\mathcal{N}}(x)$ (kernel) &   -0.80\% &  0.24\% &  1.72\% &  0.15\% & -1.75\% & -2.78\%     \\\hline
   $u$ & 2.67\% &  7.46\% & 25.13\% & 64.13\% & 91.73\% & 98.87\% \\
  $\text{CATE}_0(x)$ (GAM) & 0.58\% &  1.99\% &  3.24\% &  4.86\% &  7.78\% & 11.70\% \\
 % & $\text{CATE}(x)$ (propensity) & -6.64\% &  -3.67\% &  -0.56\% &   1.85\% &   3.39\% &   2.98\%     \\
    $\widehat{\mathcal{T}}(x)$ &1595   &   2301  &    2863   &   3375  &    3890  &    4415\\
  $\text{SCATE}(x)$ (GAM) &   7.94\% &    5.53\% &     4.53\% &      4.42\% &      5.16\% &      7.46\%  \\
    $\widehat{\mathcal{T}}_{\mathcal{N}}(x)$  &1758 &2297 &2836& 3376& 3915& 4455\\
  $\text{SCATE}_{\mathcal{N}}(x)$ (GAM) &  5.15\%& 5.60\%& 4.82\%& 4.42\%& 5.71\%& 9.41\% \\
 $\text{SCATE}_{\mathcal{N}}(x)$ (kernel) & 6.98\%& 6.64\%& 4.34\%& 4.53\%& 5.34\%& 7.13\%   \\\hline\hline
 %  gender& $u$ & 2.79\% &  7.26\% & 23.00\% & 60.32\% & 90.04\% & 98.55\% \\
 % & $\text{CATE}_0(x)$ (GAM)  &-0.24\% & -1.66\% & -2.24\% & -2.00\% & -0.77\% &  2.38\%   \\
 % & $\text{CATE}(x)$ (propensity) &  -4.22\% &  -2.51\% &  -0.12\% &   1.24\% &   0.28\% &  -5.39\%    \\
 %  &  $\widehat{\mathcal{T}}(x)$& 1960 &     2438   &   2892   &   3374    &  3856   &   4338  \\
 % & $\text{SCATE}(x)$ (GAM) &  0.71\% &      -0.38\% &      -0.96\% &      -2.04\% &      -3.38\% &      -5.14\%  \\
 %  &  $\widehat{\mathcal{T}}_{\mathcal{N}}(x)$  &1947& 2424& 2901& 3377 &3854& 4331\\
 % & $\text{SCATE}_{\mathcal{N}}(x)$ (GAM) &  1.02\% & -0.08\% & -1.07\% & -2.05\% & -3.41\% & -5.43\% \\
 % & $\text{SCATE}_{\mathcal{N}}(x)$ (kernel) &  2.06\%& -0.27\%& -1.02\%& -2.30\%& -3.53\% &-5.10\%    \\\hline
    \end{tabular}
    \caption{Estimation of the conditional average treatment (CATE), on the probability to have a non-natural birth ($y$), as a function of the weight of the baby ($x$, in g.), when the mother is Afro-American. Several weights $x$ are considered, from $2$ to $4.5$kg. $u$ is the probability associated with $x$, in the baseline population ($t=0$). $\text{CATE}_0$ is simply the difference  $\widehat{m}_1(x)-\widehat{m}_0(x)$, where both $\widehat{m}_0$ and $\widehat{m}_1$ are GAMs. $\widehat{\mathcal{T}}(x)$ is the quantile based transport function ($\widehat{\mathcal{T}}(x)= \widehat{F}_1^{-1}\circ \widehat{F}_0(x)$), while $\widehat{\mathcal{T}}_{\mathcal{N}}(x)$ is the Gaussian one. Thus, $\text{SCATE}(x)$ is the {\em mutatis mutandis CATE} $
\text{SCATE}(x)=\widehat{m}_1\big(\widehat{\mathcal{T}}(x)\big) - \widehat{m}_0\big(x\big)
$, while $
\text{SCATE}_{\mathcal{N}}(x)=\widehat{m}_1\big(\widehat{\mathcal{T}_{\mathcal{N}}}(x)\big) - \widehat{m}_0\big(x\big)
$, where both $\widehat{m}_0$ and $\widehat{m}_1$ are GAMs. Finally, the last estimate is obtained when $\widehat{m}_0$ and $\widehat{m}_1$ are simple local averages, using kernels. See Table~\ref{tab:num:CATE:w:appendix} in Appendix~\ref{app:3} for similar table when $T$ indicates whether the mother is a smoker or not, or indicates the sex of the newborn.}
    \label{tab:num:CATE:w}
\end{table}

% & $\text{CATE}_0(x)$ (GAM) & 0.06\% &  0.84\% &  0.69\% & -0.08\% & -1.26\% & -2.59\%   \\
 
\begin{table}[!ht]
    \centering
    \begin{tabular}{c|cccccc}\hline\hline
    \multicolumn{7}{c}{$t$: mother is Afro-American}\\
    \hline
  $x$ (weight gain of the mother) &  5& 15& 25& 35& 45& 55 \\\hline\hline
 % smoker & $u$ & 4.61\% & 14.50\% & 37.49\% & 67.12\% & 86.54\% & 95.05\% \\
 % & $\text{CATE}_0(x)$ (GAM) & 0.06\% &  0.84\% &  0.69\% & -0.08\% & -1.26\% & -2.59\%   \\
 % & $\text{CATE}(x)$ (propensity) & 1.36\% &   2.02\% &   1.72\% &   0.81\% &  -0.45\% &  -1.77\%      \\
 % &  $\widehat{\mathcal{T}}(x)$ &1     &   13 &       25  &      37      &  49   &     60 \\
 % & $\text{CATE}(x)$ (GAM) &  1.60\% &      1.21\% &      0.69\% &      0.14\% &     -0.38\% &    -1.08\%   \\
 % &  $\widehat{\mathcal{T}}_{\mathcal{N}}(x)$ &1 &13 &24 &36 &48 &59 \\
 % & $\text{CATE}_{\mathcal{N}}(x)$ (GAM) &  1.63\% &  1.29\% &   0.71\% &   0.01\% &  -0.71\% &  -1.33\% \\
 % & $\text{CATE}_{\mathcal{N}}(x)$ (kernel) &   0.31\% &  1.03\% &  0.98\% &  0.29\% & -1.30\% & -1.08\%     \\\hline
  $u$   & 4.57\% & 14.34\% & 37.15\% & 66.81\% & 86.34\% & 94.94 \\
  $\text{CATE}_0(x)$  (GAM) & 3.79\% & 4.79\% & 5.06\% & 4.82\% & 4.18\% & 3.26\%  \\
 % & $\text{CATE}(x)$ (propensity) &  -0.10\% &   0.95\% &   1.19\% &   0.86\% &   0.11\% &  -0.92\%     \\
   $\widehat{\mathcal{T}}(x)$  & 1  &      12  &      24  &      35    &    47   &     58 \\
  $\text{CATE}(x)$  (GAM) &  5.25\% &      5.25\% &      5.04\% &      4.82\% &      4.69\% &      4.19\%    \\
   $\widehat{\mathcal{T}}_{\mathcal{N}}(x)$  &1& 12& 23& 34& 46& 57\\
  $\text{CATE}_{\mathcal{N}}(x)$ (GAM) & 5.22\% &  5.21\% &  5.03\% &  4.74\% &  4.33\% &  3.78\%    \\
  $\text{CATE}_{\mathcal{N}}(x)$ (kernel) &  3.78\% & 5.49\% & 5.31\% & 4.49\% & 4.12\% & 3.61\%   \\\hline
 %   gender & $u$   & 4.68\% & 14.40\% & 36.73\% & 65.81\% & 85.58\% & 94.55\%  \\
 % & $\text{CATE}_0(x)$ (GAM) &  -1.79\% &     -1.60\% &     -1.60\% &     -1.73\% &     -1.90\% &     -2.02\%   \\
 % & $\text{CATE}(x)$ (propensity) & 0.19\% &   0.24\% &   0.14\% &  -0.04\% &  -0.22\% &  -0.32\%     \\
 %  & $\widehat{\mathcal{T}}(x)$ & 5& 15& 25& 35& 45& 55\\
 % & $\text{CATE}(x)$  (GAM)&  -1.79\% &     -1.60\% &     -1.60\% &     -1.73\% &     -1.90\% &     -2.02\%   \\
 % &  $\widehat{\mathcal{T}}_{\mathcal{N}}(x)$ & 4 &14 &24 &34 &44 &54\\
 % & $\text{CATE}_{\mathcal{N}}(x)$ (GAM) &   -1.52\% &  -1.47\% &  -1.61\% &  -1.87\% &  -2.17\% &  -2.41\%  \\
 % & $\text{CATE}_{\mathcal{N}}(x)$ (kernel) &  -1.58\% & -1.37\% & -1.66\% & -1.79\% & -2.22\% & -2.88\%    \\\hline
    \end{tabular}
\caption{Estimation of the conditional average treatment (CATE), on the probability to have a non-natural birth ($y$), as a function of the weight gain of the mother ($x$, in lbs), when the mother is Afro-American. Several weight gains $x$ are considered, from $5$ to $55$lbs. $u$ is the probability associated with $x$, in the baseline population ($t=0$). $\text{CATE}_0$ is simply the difference  $\widehat{m}_1(x)-\widehat{m}_0(x)$, where both $\widehat{m}_0$ and $\widehat{m}_1$ are GAMs. $\widehat{\mathcal{T}}(x)$ is the quantile based transport function ($\widehat{\mathcal{T}}(x)= \widehat{F}_1^{-1}\circ \widehat{F}_0(x)$), while $\widehat{\mathcal{T}}_{\mathcal{N}}(x)$ is the Gaussian one. Thus, $\text{SCATE}(x)$ is the {\em mutatis mutandis} CATE $
\text{SCATE}(x)=\widehat{m}_1\big(\widehat{\mathcal{T}}(x)\big) - \widehat{m}_0\big(x\big)
$, while $
\text{SCATE}_{\mathcal{N}}(x)=\widehat{m}_1\big(\widehat{\mathcal{T}_{\mathcal{N}}}(x)\big) - \widehat{m}_0\big(x\big)
$, where both $\widehat{m}_0$ and $\widehat{m}_1$ are GAMs. Finally, the last estimate is obtained when $\widehat{m}_0$ and $\widehat{m}_1$ are simple local averages, using kernels. See Table~\ref{tab:num:CATE:g:appendix} in Appendix~\ref{app:3} for similar table when $T$ indicates whether the mother is a smoker or not, or indicates the sex of the newborn.}
    \label{tab:num:CATE:g}
\end{table}

\begin{figure}[!ht]
    % \centering
    %  \includegraphics[width=.49\textwidth]{figures/gam-univariate-smoker-weight-gain-compare-gaussian-transport-1}
    %  \includegraphics[width=.49\textwidth]{figures/gam-univariate-CATE-smoker-weight-gain-compare-gaussian-transport-1}

    \centering        
    \includegraphics[width=.49\textwidth]{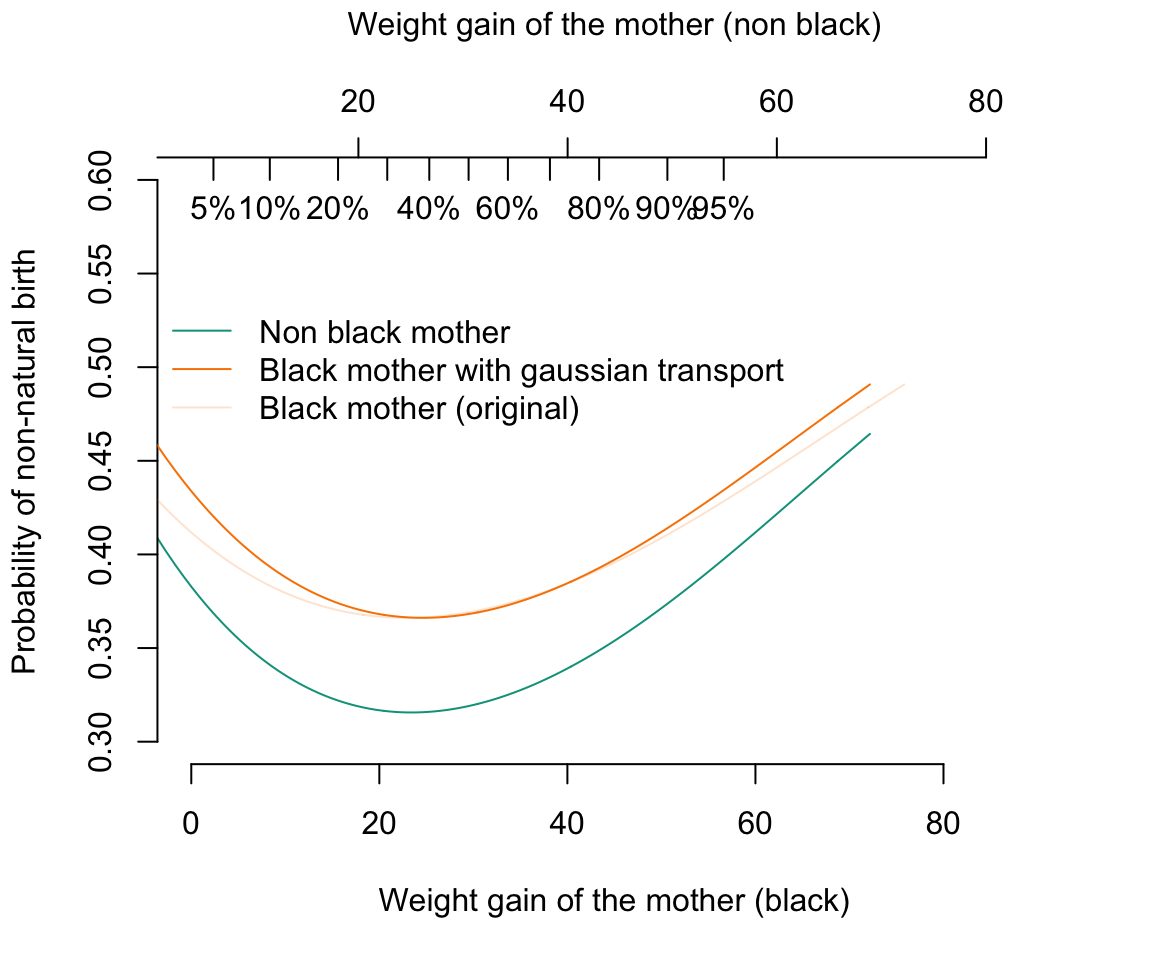}
     \includegraphics[width=.49\textwidth]{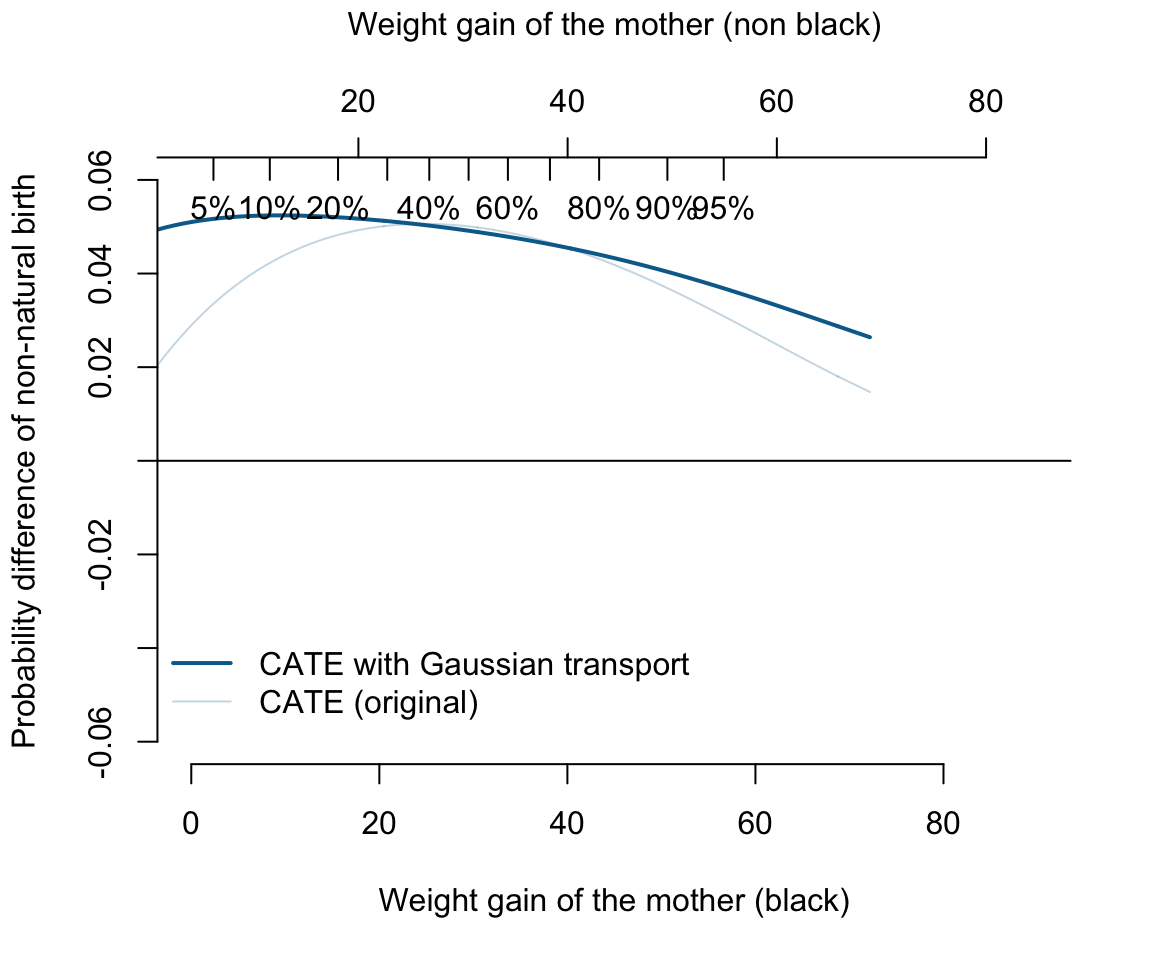}

    \caption{On the left, evolution of $x\mapsto\mathbb{E}[Y|X_{T\leftarrow t}=x,T=t]$, estimated using a logistic GAM model, when $Y=\boldsymbol{1}(\text{non-natural delivery})$, and $X$ is the weight gain of the mother, respectively when $T$ indicates whether the mother is Black or not. On the right, evolution of  $x\mapsto\text{SCATE}_{\mathcal{N}}[Y|X=x]$. See Figure~\ref{fig:CATE-compare-gaussian-2x2-weight-gain:appendix} in Appendix~\ref{app:3} for similar graphs when $T$ indicates whether the mother is a smoker or not.}
    \label{fig:CATE-compare-gaussian-2x2-weight-gain}
\end{figure}

In Figure~\ref{fig:CATE-compare-gaussian-2x2-weight-gain}, as previously, $\widehat{m}_0(x)$ and $\widehat{m}_1\big(\widehat{\mathcal{T}}_{\mathcal{N}}(x)\big)$ can be visualized on the left, when $t$ indicates whether the mother is Afro-American or not, and when $x$ is the gain weight of the mother. On the right, we can visualize $x\mapsto \text{CATE}(x)=\widehat{m}_1\big(\widehat{\mathcal{T}}(x)\big) - \widehat{m}_0\big(x\big)$ as a function of $x$. Numerical values are given in Table~\ref{tab:num:CATE:w} when $x$ is the weight of the newborn, and Table~\ref{tab:num:CATE:g} when $x$ is the weight gain of the pregnant mother.

\begin{figure}[!ht]
    % \centering
    %  \includegraphics[width=.49\textwidth]{figures/kernel-univariate-smoker-weight-compare-gaussian-transport-1.png}
    %  \includegraphics[width=.49\textwidth]{figures/kernel-univariate-CATE-smoker-weight-compare-gaussian-transport-1.png}

    \centering        
    \includegraphics[width=.49\textwidth]{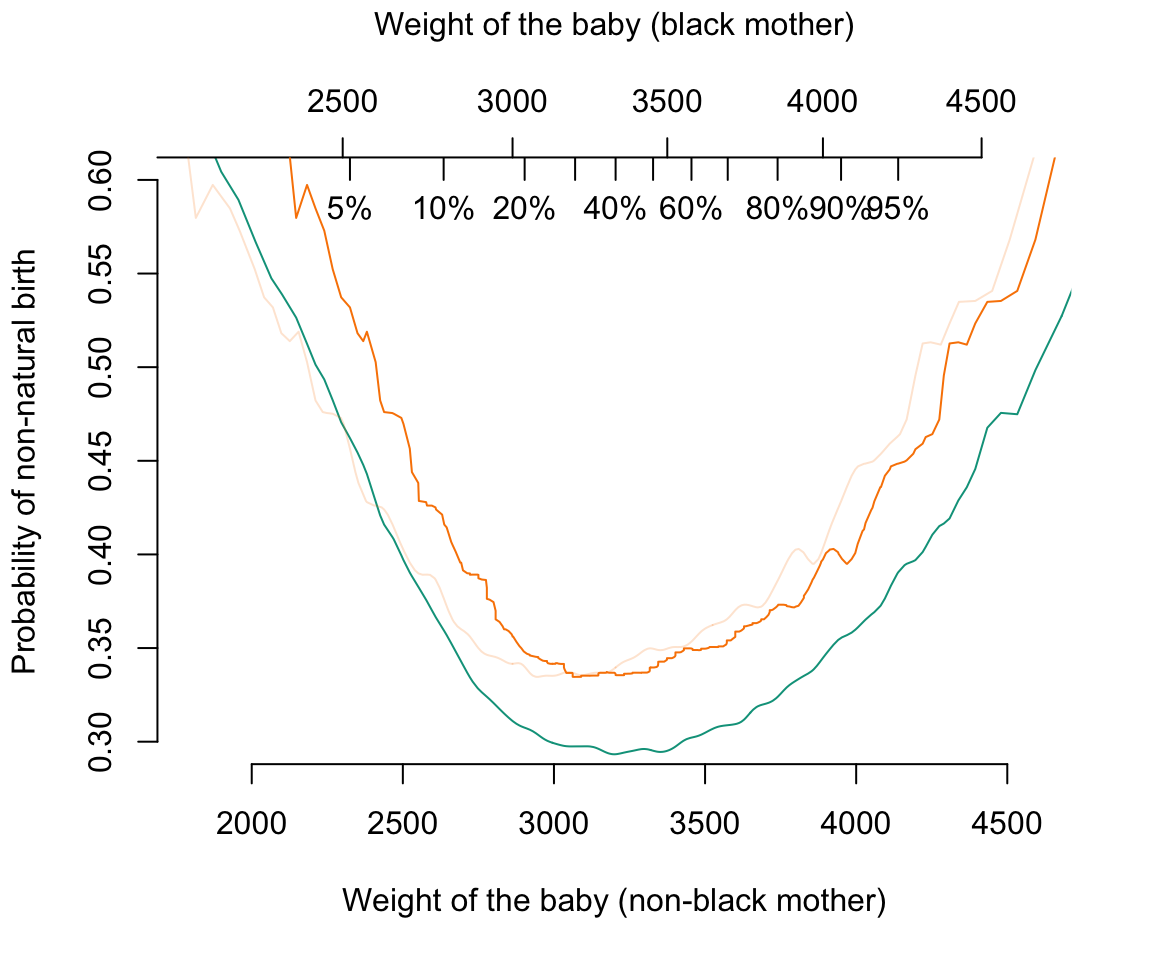}
     \includegraphics[width=.49\textwidth]{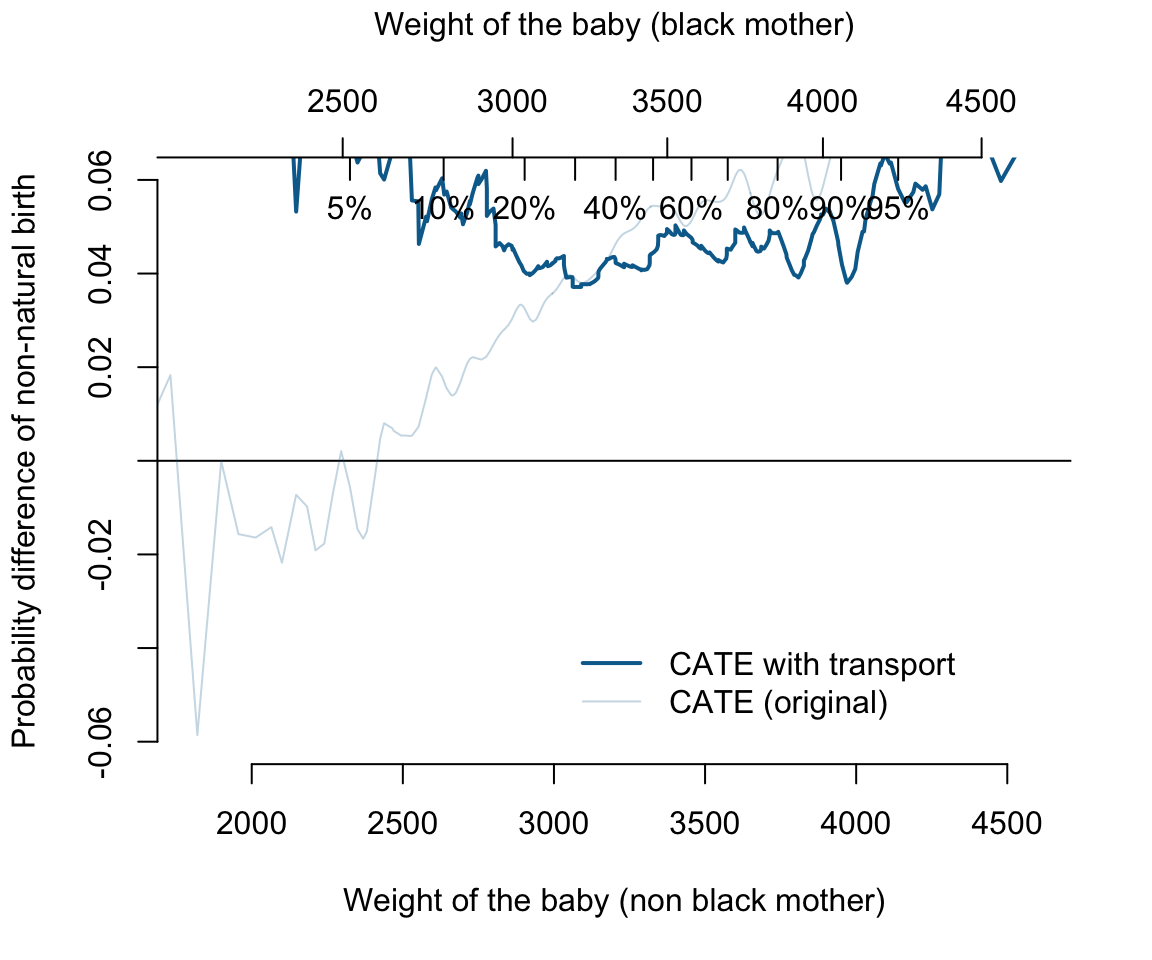}

    % \centering
    %  \includegraphics[width=.49\textwidth]{figures/kernel-univariate-gender-baby-weight-compare-gaussian-transport-1.png}
    %  \includegraphics[width=.49\textwidth]{figures/kernel-univariate-CATE-gender-baby-weight-compare-gaussian-transport-1.png}

    \caption{On the left, evolution of $x\mapsto\mathbb{E}[Y|X_{T\leftarrow t}=x,T=t]$, estimated using a kernel based local average, when $Y=\boldsymbol{1}(\text{non-natural delivery})$, and $X$ is the weight of the newborn infant, respectively when $T$ indicates whether the mother is a smoker or not (on top), when the mother is Black or not in the middle, and the sex of the infant below. On the right, evolution of  $x\mapsto\text{SCATE}_{\mathcal{N}}[Y|X=x]$ with an without transport, based on a Gaussian transport. See Figure~\ref{fig:CATE-compare-kernel-2x3-weight:appendix} in Appendix~\ref{app:3} for similar graphs when $T$ indicates whether the mother is a smoker or not, or indicates the sex of the newborn.}
    \label{fig:CATE-compare-kernel-2x3-weight}
\end{figure}

%In Figure~\ref{fig:CATE-compare-knn-2x1-weight-gain} we can visualize $\widehat{m}_0(x)$ and $\widehat{m}_1\big(\widehat{\mathcal{T}}_{\mathcal{N}}(x)\big)$ on the left, obtained using $k$-nearest neighbors, with $x\mapsto \text{CATE}(x)=\widehat{m}_1\big(\widehat{\mathcal{T}}(x)\big) - \widehat{m}_0\big(x\big)$ as a function of $x$ on the right. Those estimates are too volatile to be considered here. 
In Figure~\ref{fig:CATE-compare-kernel-2x3-weight}, some local kernels are used to estimate $\widehat{m}_0(x)$ and $\widehat{m}_1\big(\widehat{\mathcal{T}}_{\mathcal{N}}(x)\big)$ on the left. Numerical values are given in Table~\ref{tab:num:CATE:w} when $x$ is the weight of the newborn, and Table~\ref{tab:num:CATE:g} when $x$ is the weight gain of the pregnant mother.

% In Figure~\ref{fig:CATE-sample-size-ranks-weight-black}, subsamples of size $n=4,000$, $20,000$ and $100,000$ are generated,  On the left, we can visualize the evolution of $\widehat{\mathcal{T}}(x)$ for several values of $x$, and on the right, the evolution of $\text{CATE}(x)$ on those subsamples.

% \subsection{Extension to CQTE}

% Given $\tau\in(0,1)$, one can define a CQTE estimate that quantile-based,  expressed as
% \begin{equation}\label{eq:CQTE:quantile}
% \text{CQTE}(u) = 
% Q_{\tau}\big[Y^*_{T\leftarrow 1}\big|X=F_{1}^{-1}(u)\big] - 
% Q_{\tau}\big[Y^*_{T\leftarrow 0}\big|X=F_{0}^{-1}(u)\big],
% \end{equation}
% where $F_t$ is the cumulative distribution function of $X$, conditional on $T=t$.

% \begin{figure}[!ht]
%     \centering
%      \includegraphics[width=.49\textwidth]{figures/knn-univariate-smoker-weight-gain-compare-gaussian-transport-1.png}
%      \includegraphics[width=.49\textwidth]{figures/knn-univariate-CATE-smoker-weight-gain-compare-gaussian-transport-1.png}

%     \caption{On the left, evolution of $x\mapsto\mathbb{E}[Y|X_{T\leftarrow t}=x,T=t]$, estimated using $k$-nearest neighbors, when $Y=\boldsymbol{1}(\text{non-natural delivery})$, and $X$ is the weight gain of the mother, when $T$ is the indicator that the mother is a smoker or not . On the right, evolution of  $x\mapsto\text{SCATE}_{\mathcal{N}}[Y|X=x]$ with and without transport.}
%     \label{fig:CATE-compare-knn-2x1-weight-gain}
% \end{figure}

\section{Optimal transport based matching}\label{sec:opt:transp:multivarie}

In this section, we will extend what was derived in the previous section. Heuristically, optimal matching of margins components of $\boldsymbol{x}$ will probably not work, and the mapping should be multivariate. % in the coupling problem, let $\sigma_{0:1}$ and $\sigma_{1:1}$ denote the permutations of $x_1$ in the two groups, control and treated, respectively. Similarly, consider $\sigma_{0:2}$ and $\sigma_{1:2}$, the permutations of $x_2$ in the two groups. According to the results of the previous section, the optimal coupling will associate $(x^{(0)}_{i:1},x^{(0)}_{i:2})$ to $(x^{(1)}_{\sigma_{1:1}\circ\sigma_{0:1}^{-1}(i):1},x^{(1)}_{\sigma_{1:2}\circ\sigma_{0:2}^{-1}(i):2})$. Unfortunately, if $\sigma_{1:1}\circ\sigma_{0:1}^{-1}\neq \sigma_{1:2}\circ\sigma_{0:2}^{-1}$, it is possible that this point does not exist in the sample.
We will therefore use optimal transport techniques to get a proper counterfactual of $\boldsymbol{x}$, not in the control group, but in the treated group. In Section~\ref{sect:multi:ot}, we will define properly the optimal transport problem (in any dimension). Then, in Section~\ref{sect:multi:matching}, we will describe how to optimally associate each observation $\boldsymbol{x}_i$ in the control group (when $t=0$) with a single counterfactual observation $\boldsymbol{x}_j$ in the treated group (when $t=1$), when two groups have the same size. This can be related to the Gaussian SEM discussed in Section~\ref{sec:gaussian:toy}. In Section~\ref{sect:multi:coupling}, we will present the extension when the two groups have different sizes. In Section~\ref{sec:gaussian}, we will give an explicit formulation for $\mathcal{T}$ when we the distribution of $\boldsymbol{X}$ conditional on $T$ is assumed to be Gaussian. The application to non-natural deliveries will finally be discussed in Section~\ref{sub:ex:multi}.

\subsection{Optimal transport}\label{sect:multi:ot}

In the mathematical formulation of \cite{monge1781memoire}'s problem, we want to push a distribution from $\mathbb{P}_0$ to $\mathbb{P}_1$ (distributions on  $\mathbb{R}^k$, not necessarily in $\mathbb{R}$ as considered in the previous section). 
Given $\mathcal{T}:\mathbb{R}^k\rightarrow\mathbb{R}^k$, define the ``{\em push-forward}'' measure, 
$$
\mathbb{P}_1(A)=  \mathcal{T}_{\#}\mathbb{P}_0(A)= \mathbb{P}_0\big(\mathcal{T}^{-1}(A)\big),~\forall A\subset \mathbb{R}^k.
$$
For instance, when $k=1$, if $F$ is the cumulative distribution of a univariate random variable $X$ under $\mathbb{P}$ (i.e., $F(x)=\mathbb{P}[X\leq x]$) then $\mathbb{Q}=F_{\#}\mathbb{P}$ is the uniform distribution on the unit interval $[0,1]$ as well as $\mathbb{Q}'=\overline F_{\#}\mathbb{P}$, where $\overline F$ is the survival function  associated with $F$ (i.e., $\overline F(x)=\mathbb{P}[X> x]$). Similarly, or conversely, if $Q$ is the quantile function associated with $F$ --$Q(u)=F^{-1}(u)$ for any $u\in(0,1)$-- then if $\mathbb{P}$ is the uniform distribution on the unit interval $[0,1]$, $\mathbb{Q}=Q_{\#}\mathbb{P}$ satisfies $\mathbb{Q}[X\leq x]=Q^{-1}(x)=F(x)$, and similarly for $\overline Q$ where $\overline Q(u)=F^{-1}(1-u)$. 

Observe that if $\mathbb{P}_0$ and $\mathbb{P}_1$ have densities $f_0$ and $f_1$, respectively, and if $T$ is continuously differentiable, $\mathbb{P}_1=  \mathcal{T}_{\#}\mathbb{P}_0$ is any only if $f_0(\boldsymbol{x})=f_1(\mathcal{T}(\boldsymbol{x}))\cdot |\det \nabla \mathcal{T}(\boldsymbol{x})|$, for all $\boldsymbol{x}$. This non-linear function is a special case of the so-called Monge-Ampère partial differential equations.

An optimal transport $\mathcal{T}^\star$ (in Brenier's sense,  from \cite{brenier1991polar}, see \cite{villani2009optimal} or \cite{galichon2016optimal}) from $\mathbb{P}_0$ towards $\mathbb{P}_1$ will be solution of 
$$
\mathcal{T}^\star\in \underset{\mathcal{T}:\mathcal{T}_{\#}\mathbb{P}_0=\mathbb{P}_1}{\text{arginf}}\left\lbrace\int_{\mathbb{R}^k} \|\boldsymbol{x}-\mathcal{T}(\boldsymbol{x})\|^2d\mathbb{P}_0(\boldsymbol{x})\right\rbrace,
$$
for a quadratic cost, or more generally,
$$
\mathcal{T}^\star\in \underset{\mathcal T:\mathcal T_{\#}\mathbb{P}_0=\mathbb{P}_1}{\text{arginf}}\left\lbrace\int_{\mathbb{R}^k} \gamma(\boldsymbol{x},\mathcal{T}(\boldsymbol{x}))d\mathbb{P}_0(\boldsymbol{x})\right\rbrace,
$$
for some cost function $\gamma:\mathbb{R}^k\times\mathbb{R}^k\to\mathbb{R}_+$.

If $k=1$, and if the cost function $\gamma$ can be written $\gamma(x,y)=h(|x-y|)$ for some strictly convex and positive function $h$, then $T^\star$ is an increasing function, and more precisely, if $F_0(x)=\mathbb{P}_0[X\leq x]$ and $F_1(x)=\mathbb{P}_1[X\leq x]$, with $F_0$ absolutely continuous, then {$\mathcal{T}^\star (x) = F_1^{-1}\circ F_0(x)$} satisfies $\mathcal{T}^\star_{\#}\mathbb{P}_0=\mathbb{P}_1$ (since $F_1(x)=F_0(T^{\star-1} (x))$ and $\mathcal{T}^\star$ is optimal. the quadratic cost function (when $h(x)=x^2$) is a particular case. The case where $h$ is concave was discussed in \cite{mccann1999exact}.
%  In the univariate case, if $\mathbb{P}_0$ to $\mathbb{P}_1$ have densities $f_0$ and $f_1$, respectively, since $\mathcal{T}^\star (x) = F_1^{-1}\circ F_0(x)$, observe that
%  $$
% \frac{d\mathcal{T}^\star (x)}{dx} = \frac{f_0(x)}{f_1(\mathcal{T}^\star (x))} =\frac{}$$

In higher dimension, for a quadratic cost, one can prove (see \cite{villani2003optimal,villani2009optimal} or \cite{galichon2016optimal}) that $\mathcal{T}^\star=\nabla \psi$ where $\psi$ is a convex function.

% Note that we can write similarly
% $$
% {\min_{\text{copula }C}\int_{\mathbb{R}^2} \gamma(x,y)dF(x,y)},\text{ where }F(x,y)=C(F_0(x),F_1(y)),
% $$
% but we no longer see the push-forward mapping $T$.

% \begin{figure}[!ht]
%     \centering
%     \includegraphics[width=\textwidth]{dessin/smoker-weight-age.png}
%     \caption{Distribution of the weight of the newborn infant (in grams) when the mother is a smoker or not, on the left; distribution of age of the mother at delivery when the mother is a smoker or not, in the middle; joint scatterplot of the  weight of the newborn infant ($x$) and the age of the mother ($y$).}
%     \label{fig:weight:distribution}
% \end{figure}

\subsection{Empirical version of optimal matching}\label{sect:multi:matching}

This transport can be seen as a matching between individuals in the two groups, both of size $n$, $(\boldsymbol{x}_i,t_i=0)$ and $(\boldsymbol{x}_j,t_j=1)$, instead of two distributions $\mathbb{P}_0$ and $\mathbb{P}_1$. If $C$ is a $n\times n$ matrix that quantifies the distance between individuals in the two groups, $C_{i,j}=d(\boldsymbol{x}_i,\boldsymbol{x}_j)$, the optimal matching is solution of 
$$
\min_{P\in\mathcal{P}} \langle P,C\rangle = 
\min_{P\in\mathcal{P}} \sum_{i=1}^n\sum_{j=1}^n P_{i,j}C_{i,j},
$$
where $\mathcal{P}$ is the set of permutation matrices, and $\langle \cdot,\cdot\rangle$ is the Frobenius dot-product. This is also called Kantorovich’s optimal transport problem, from \cite{kantorovich1942translocation}. Interestingly, there are some algorithms that can be used to find that optimal coupling, or matching, which can, in turn, be used to get a counterfactual for all individuals in each group.

% \begin{figure}[!ht]
% \centering
% \includegraphics[width=.48\textwidth]{dessin/simulation-points-SEM-1.png}
% \includegraphics[width=.48\textwidth]{dessin/simulation-points-SEM-2.png}
% \caption{Scatter plot of $\boldsymbol{x}^m=(x^m_1,x^m_2)$ with blue points when $t=0$, red when $t=1$, on the left, an optimal matching of points in the two groupes. Toy dataset generated from SEM (\ref{eq:SEM}).}\label{Fig:ex:2}
% \end{figure}

\subsection{Empirical version of optimal coupling}\label{sect:multi:coupling}

If the two groups $(\boldsymbol{x}_i,t_i=0)$ and $(\boldsymbol{x}_j,t_j=1)$ have different sizes, namely $n_0$ and $n_1$, respectively, it is possible to define some matching using weights. In the coupling case, described previously, $P$ was some $n\times n$ permutation matrix. But here, as in Section~\ref{sub:sec:univ:optimal:matching} some $n_0\times n_1$ matrices will be involved, and similar problems are considered
\begin{equation}\label{prog:match:2}
\min_{P\in  U(\boldsymbol{a}_0,\boldsymbol{a}_1)}  \langle P,C\rangle = 
\min_{P\in  U(\boldsymbol{a}_0,\boldsymbol{a}_1)}  \sum_{i=1}^{n_0} \sum_{j=1}^{n_1} P_{i,j}C_{i,j}.
\end{equation}

And again, assuming Gaussian distributions for $\boldsymbol{X}$ conditional on $T$ will provide an explicit simple transport formula that can be used to get an estimation of the {\em mutatis mutandis} CATE. This algorithm is given by Algorithm~\ref{alg:1:empirical}, used to compute the Average Treatment Effect. %Algorithm~\ref{alg:2} presents a bootstrap procedure that can be used to compute a confidence interval for the {\em mutatis mutandis} CATE.

\begin{algorithm}[!ht]
\caption{Counterfactual matching, with optimal matching}\label{alg:ATE:1}
\begin{algorithmic}
 \State$\mathcal{D} \gets \{(y_i,\boldsymbol{x}_i,t_i)\}$
\Function{Counterfactual2}{$\mathcal{D}$}
\State $\mathcal{D}_0 \gets $ subset of $\mathcal{D}$ when $t=0$ (size $n_0$), with indices $i$
\State $\mathcal{D}_1 \gets $ subset of $\mathcal{D}$ when $t=1$ (size $n_1$), with indices $j$
\State  $C \gets $ matrix $n_0\times n_1$, $C_{i,j}=d(\boldsymbol{x}_i,\boldsymbol{x}_j)$ between points in $\mathcal{D}_0$ and $\mathcal{D}_1$
    \State  $P^* \gets$ solution of Problem (\ref{prog:match}) 
\State \Return matrix $P^*$ ~~ ($n_0\times n_1$)
\EndFunction
\end{algorithmic}
\end{algorithm}

\begin{algorithm}[!ht]
\caption{Estimate SCATE (optimal matching based)}\label{alg:1:empirical}
\begin{algorithmic}
\State dataset $\mathcal{D}\gets\{(y_i,\boldsymbol{x}_i,t_i)\}$, 
\Function{scate2}{$\mathcal{D},k,\boldsymbol{x}$}
\State $P \gets $ {\sc Counterfactual2}$(\mathcal{D})$
\State $V_{\boldsymbol{x}}^k\gets$ list of $k$ nearest neighbors of  $\boldsymbol{x}_i$'s in $\mathcal{D}_0$ close to $\boldsymbol{x}$
\State\Return $\displaystyle{\frac{1}{k}\sum_{i\in V_{\boldsymbol{x}}^k} y_i^1-{P}_i^{\top}\boldsymbol{y}^0}$
\EndFunction
\end{algorithmic}
\end{algorithm}

% \begin{itemize}
%     \item resample from dataset $\mathcal{D}$, get a dataset $\mathcal{D}^{(b)}$ and use the previous algorithm to compute $\text{CATE}^{(b)}_{\mathcal{N}}(\boldsymbol{x}^m,\boldsymbol{x}^c)$ and $\text{CATE}^{(b)}(\boldsymbol{x})$
% \end{itemize}

% \begin{algorithm}[!ht]
% \caption{Confidence interval for SCATE}\label{alg:2}
% \begin{algorithmic}
% \Require dataset $\mathcal{D}=\{(y_i,\boldsymbol{x}_i,t_i)\}$ and $\boldsymbol{x}$
% \For{$b=1,2,\cdots,B$}
% \State $\mathcal{D}^{(b)}\gets$ resampling of $\mathcal{D}$
% \State $\text{SCATE}^{(b)}(\boldsymbol{x})\gets$ output of Algorithm~\ref{alg:1:empirical} (with dataset $\mathcal{D}^{(b)}$) 
% \EndFor
% \State $m_c(\boldsymbol{x})\gets \displaystyle{\frac{1}{B}\sum_{b=1}^B}\text{SCATE}^{(b)}(\boldsymbol{x})$
% and $s^2_c(\boldsymbol{x})\gets \displaystyle{\frac{1}{B-1}\sum_{b=1}^B}\left(\text{SCATE}^{(b)}(\boldsymbol{x})-m_c(\boldsymbol{x})\right)^2$
% \State $\text{IC}\gets \displaystyle{\big[m_c(\boldsymbol{x})\pm 1.96 \sqrt{s^2_c(\boldsymbol{x})}\big]}$
% \end{algorithmic}
% \end{algorithm}

\subsection{Counterfactuals for Gaussian covariates}\label{sec:gaussian}

In the general case, there are no simple construction and interpretation of the optimal mapping $\mathcal{T}^*$, as the one we had in the univariate case, based on quantiles. If it is possible, following \cite{hallin2021distribution}, to define multivariate quantiles (and therefore to extend concepts defined in Section \ref{sub:sec:univ:cate}). But here, we will simply consider the multivariate Gaussian case.
Suppose that $\boldsymbol{X}|t=1\sim\mathcal{N}(\boldsymbol{\mu}_1,\boldsymbol{\Sigma}_1)$ and $\boldsymbol{X}|t=0\sim\mathcal{N}(\boldsymbol{\mu}_0,\boldsymbol{\Sigma}_0)$. There is an explicit expression for the optimal transport, which is simply an affine map (see \cite{villani2003optimal} for more details). In the univariate case, $x_1 = \mathcal{T}^*_{\mathcal{N}}(x_0) = \mu_1+ \displaystyle{\frac{\sigma_1}{\sigma_0}(x_0-\mu_0)}$, while in the multivariate case, an analogous expression can be derived:
$$
\boldsymbol{x}_1 = \mathcal{T}^*_{\mathcal{N}}(\boldsymbol{x}_0)=\boldsymbol{\mu}_1 + \boldsymbol{A}(\boldsymbol{x}_0-\boldsymbol{\mu}_0),
$$
where $\boldsymbol{A}$ is a symmetric positive matrix that satisfies $\boldsymbol{A}\boldsymbol{\Sigma}_0\boldsymbol{A}=\boldsymbol{\Sigma}_1$, which has a unique solution given by $\boldsymbol{A}=\boldsymbol{\Sigma}_0^{-1/2}\big(\boldsymbol{\Sigma}_0^{1/2}\boldsymbol{\Sigma}_1\boldsymbol{\Sigma}_0^{1/2}\big)^{1/2}\boldsymbol{\Sigma}_0^{-1/2}$, where $\boldsymbol{M}^{1/2}$ is the square root of the square (symmetric) positive matrix $\boldsymbol{M}$ based on the Schur decomposition ($\boldsymbol{M}^{1/2}$ is a positive symmetric matrix), as described in \cite{higham2008functions}.

\begin{definition}
Consider two models, $\widehat{m}_0(\boldsymbol{x})$ and $\widehat{m}_1(\boldsymbol{x})$, that estimate, respectively, $\mathbb{E}[Y|\boldsymbol{X}=\boldsymbol{x},T=0]$ and  $\mathbb{E}[Y|\boldsymbol{X}=\boldsymbol{x},T=1]$.
A Gaussian estimator of the {\em mutatis mutandis} CATE is
$$
\text{SCATE}_{\mathcal{N}}(\boldsymbol{x})=\widehat{m}_1\big(\widehat{\mathcal{T}}_{\mathcal{N}}(\boldsymbol{x})\big) - \widehat{m}_0\big(\boldsymbol{x}\big),
$$
where $\widehat{\mathcal{T}}_{\mathcal{N}}(\boldsymbol{x})= \overline{\boldsymbol{x}}_1 +\widehat{\boldsymbol{A}}(\boldsymbol{x}-\overline{\boldsymbol{x}}_0)$, with
$\overline{\boldsymbol{x}}_0$ and $\overline{\boldsymbol{x}}_1$ being, respectively, the averages of $x$ in the two sub-populations, and $\widehat{\boldsymbol{A}}=\widehat{\boldsymbol{\Sigma}}_0^{-1/2}\big(\widehat{\boldsymbol{\Sigma}}_0^{1/2}\widehat{\boldsymbol{\Sigma}}_1\widehat{\boldsymbol{\Sigma}}_0^{1/2}\big)^{1/2}\widehat{\boldsymbol{\Sigma}}_0^{-1/2}$ where $\widehat{\boldsymbol{\Sigma}}_0$ and $\widehat{\boldsymbol{\Sigma}}_1$ denote the sample variance.
\end{definition}

The algorithm to compute that estimate is Algorithm~\ref{alg:1:gauss:1}.

% For example (see Section~\ref{sec:application} for more details on the data used), if $x$ denotes the weight at birth of newborn infants, and $t$ the smoking status, $x_1=x_0-145$ (in grams), since variances can be considered as equal, while the differences between the means is 145 grams (95\% confidence interval $[104~;~187]$). A VIRER 

\begin{algorithm}[!ht]
\caption{Optimal Gaussian Transport}\label{alg:gauss:-}
\begin{algorithmic}
\State dataset $\mathcal{D}\gets\{(y_i,\boldsymbol{x}_i,t_i)\}$, 
\Function{Tgaussian}{$\mathcal{D}$}
\State $\mathcal{D}_0 \gets $ subset of $\mathcal{D}$ when $t=0$
\State $\mathcal{D}_1 \gets $ subset of $\mathcal{D}$ when $t=1$
    \State  estimate moments of $\boldsymbol{x}$'s $\hat{\boldsymbol{\mu}}_0$, $\hat{\boldsymbol{\mu}}_1$, $\hat{ \boldsymbol {\Sigma}}_0$ and $\hat{\boldsymbol{\Sigma}}_1$, in $\mathcal{D}_0$ and $\mathcal{D}_1$
    \State $\hat{\boldsymbol{A}} \gets \hat{\boldsymbol{\Sigma}}_0^{-1/2}\big(\hat{\boldsymbol{\Sigma}}_0^{1/2}\hat{\boldsymbol{\Sigma}}_1\hat{\boldsymbol{\Sigma}}_0^{1/2}\big)^{1/2}\hat{\boldsymbol{\Sigma}}_0^{-1/2}$  
\Function{T}{$\boldsymbol{x}$} 
\State\Return $\hat{\boldsymbol{\mu}}_1 + \hat{\boldsymbol{A}}(\boldsymbol{x}-\hat{\boldsymbol{\mu}}_0)$,
\EndFunction
\State \Return function {\sc T}
\EndFunction
\end{algorithmic}
\end{algorithm}

\begin{algorithm}[!ht]
\caption{Parametric Estimate $\text{SCATE}_{\mathcal{N}}$ (Gaussian transport)}\label{alg:1:gauss:1}
\begin{algorithmic}
\State dataset $\mathcal{D}\gets\{(y_i,\boldsymbol{x}_i,t_i)\}$,
\State $\mathcal{D}_0 \gets $ subset of $\mathcal{D}$ when $t=0$
\State $\mathcal{D}_1 \gets $ subset of $\mathcal{D}$ when $t=1$
\State $\widehat{m}_0 \gets $ model to predict $y$ based on $\boldsymbol{x}$, trained on $\mathcal{D}_0$
\State $\widehat{m}_1 \gets $ model to predict $y$ based on $\boldsymbol{x}$, trained on $\mathcal{D}_1$
\State $T\gets $ {\sc Tgaussian}($\mathcal{D}$)
\Function{Scate3}{$\widehat{m}_0,\widehat{m}_1,T,\boldsymbol{x}$}
\State\Return $\displaystyle{\widehat{m}_1(T(\boldsymbol{x})) - \widehat{m}_0({\boldsymbol{x}})}$
\EndFunction
\end{algorithmic}
\end{algorithm}

We should probably stress here that, in the very general case, we should transport {\em only} endogenous variables $\boldsymbol{x}^m$ (or mediators) and not exogenous ones $\boldsymbol{x}^c$ (or coliders), as discussed in Section \ref{subsec:1} (and Figure \ref{Fig:DAG:intervention}).

\subsection{Application to non-natural deliveries}\label{sub:ex:multi}

\begin{figure}[!ht]
    \centering
     \includegraphics[width=.49\textwidth]{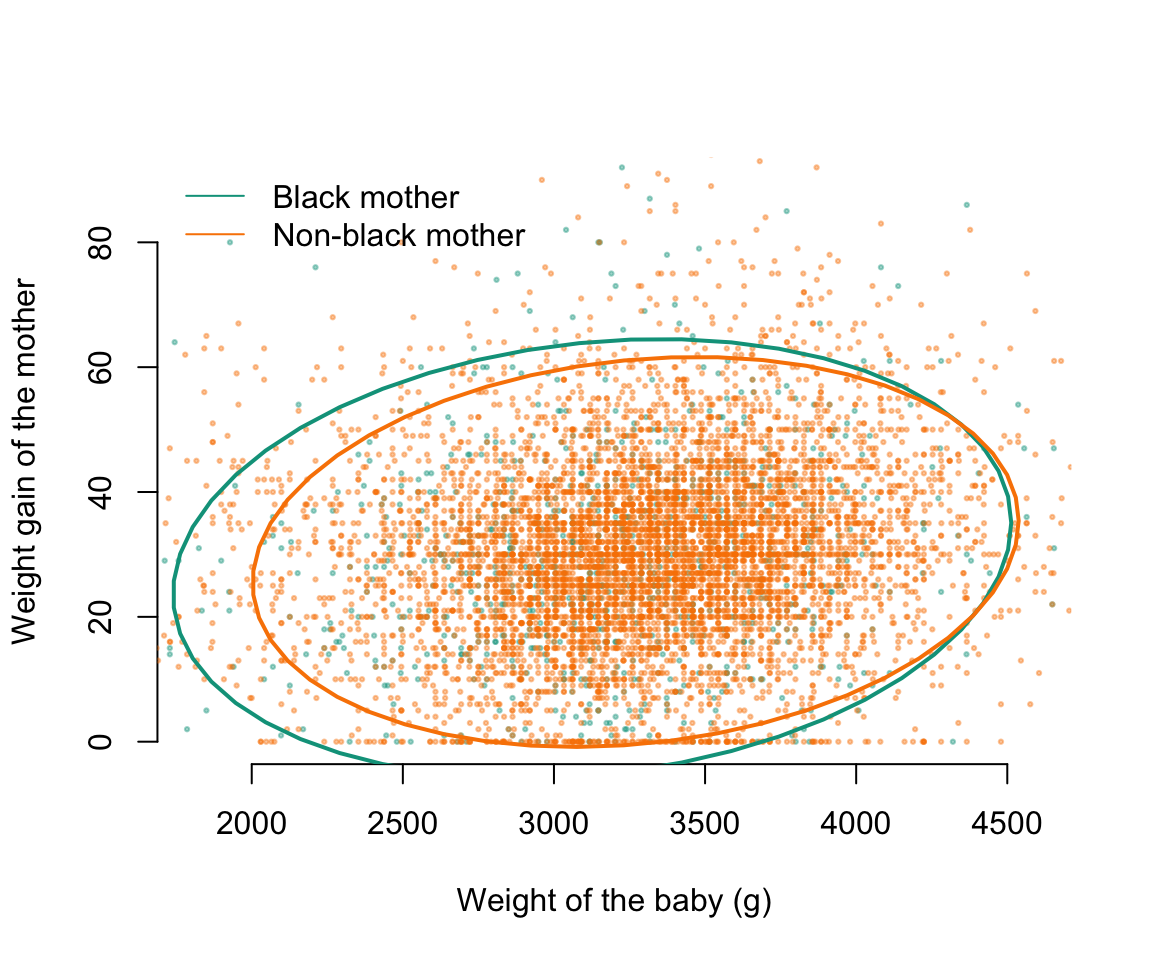}
     \includegraphics[width=.49\textwidth]{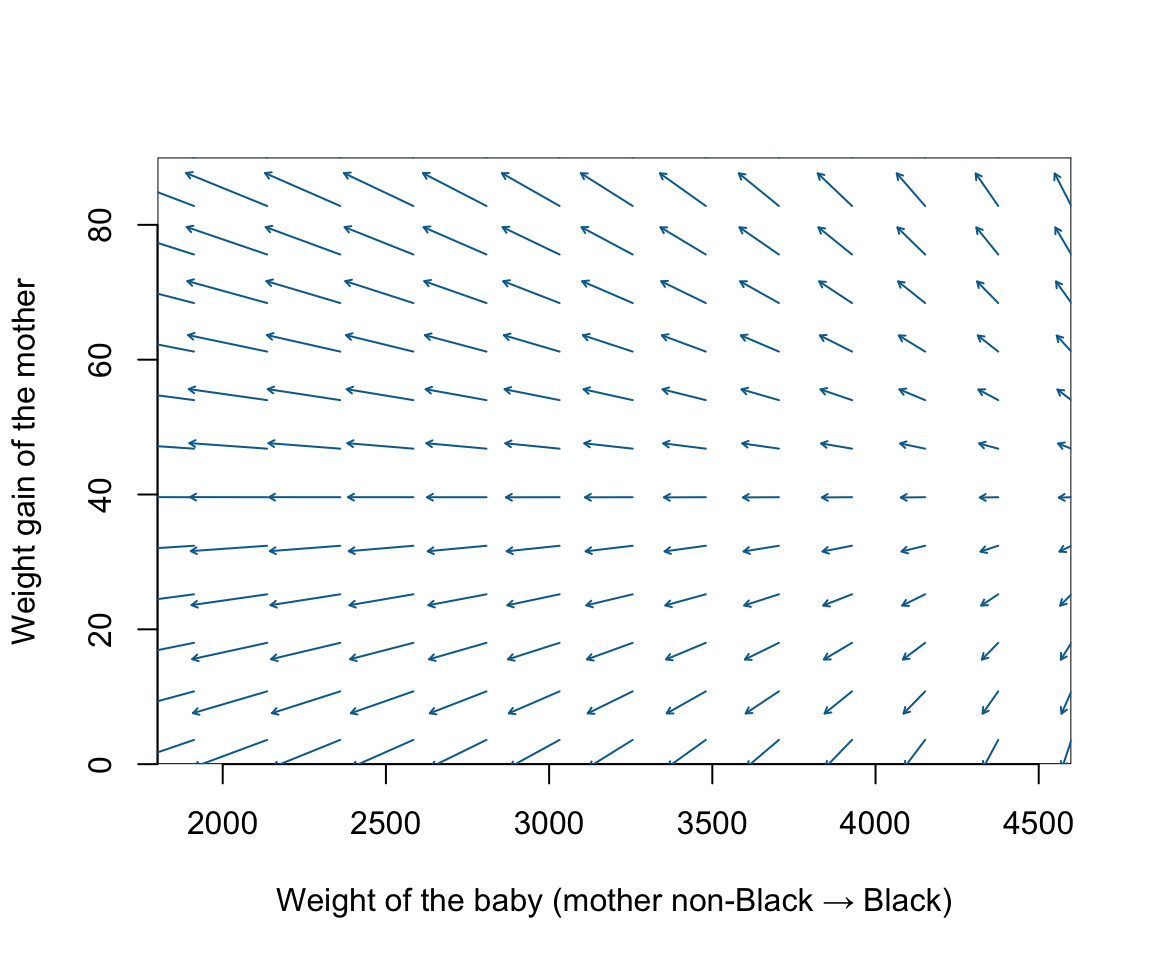}

    \caption{Joint distributions of $\boldsymbol{X}$ (weight of the newborn infant and weight gain of the mother), conditional on the treatment $T$, when $T$ indicates whether the mother is Black or not on the left. On the right, vector field associated with optimal Gaussian transport, in dimension two (weight of the newborn infant and weight gain of the mother). Some numerical values are given in Table~\ref{tab:transport:appendix}. On the right, the origin of the arrow is $\boldsymbol{x}$ in the control group (non-Black pregnant mother) and the arrowhead is $\widehat{\mathcal{T}}_{\mathcal{N}}(\boldsymbol{x}$ in the treated group (Black pregnant mother). See Figure~\ref{fig:joint-ellipse-arrow-black-biv-1x2:appendix} in Appendix~\ref{app:3} for similar graphs when $T$ indicates whether the pregnant mother is a smoker or not.}
    \label{fig:joint-arrow-grid-biv-1x2}
\end{figure}

The left-hand side of Figure~\ref{fig:joint-arrow-grid-biv-1x2} displays a scatter plot of $\boldsymbol{x}=(x_1,x_2)$, where $x_1$ represents the weight of the newborn infant while $x_2$ shows the weight gain of the mother, conditional on the treatment $T$, when $T$ indicates whether the mother is Black or not (see Figure~\ref{fig:joint-ellipse-arrow-black-biv-1x2:appendix} in Appendix~\ref{app:3} for similar graphs when $T$ indicates whether the mother is a smoker or not). The ellipses are the iso-density curves under a Gaussian assumption, such that $95\%$ of the points lie in the ellipse. The right-hand side of Figure~\ref{fig:joint-arrow-grid-biv-1x2}, shows $\mathcal{T}_{\mathcal{N}}$ on the same frame, $\boldsymbol{x}=(x_1,x_2)$, with, respectively, the weight of the newborn infant on the $x$-axis and weight gain of the mother on the $y$-axis. The origin of an arrow corresponds to $\boldsymbol{x}=(x_1,x_2)$, while its end corresponds to $\widehat{\mathcal{T}}_{\mathcal{N}}(\boldsymbol{x})$. Note that all the arrows point to the left. Regardless of the weight of the mother, had the latter been Black, the weight of the newborn would have been lower. Nevertheless, the length of the arrows varies according to the weight of the newborn. For infants whose weight is relatively high, for example for $x_1$ close to 4500g, had the mother been Black, the newborn’s weight would have been almost the same. For newborns whose weight $x_1$ is much lower than 4500g, had the mother been Black, the baby's weight would have been much smaller. Some numerical values are given in Table~\ref{tab:transport:appendix} in Appendix~\ref{app:3}. For instance, if we consider a non-Black mother with a baby weighting 2584g, who gained 10.8lbs, the counterfactual is a Black mother with a baby weighting 2392g, who gained 7.6lbs.

\begin{figure}[!ht]
    \centering
     \includegraphics[width=.48\textwidth]{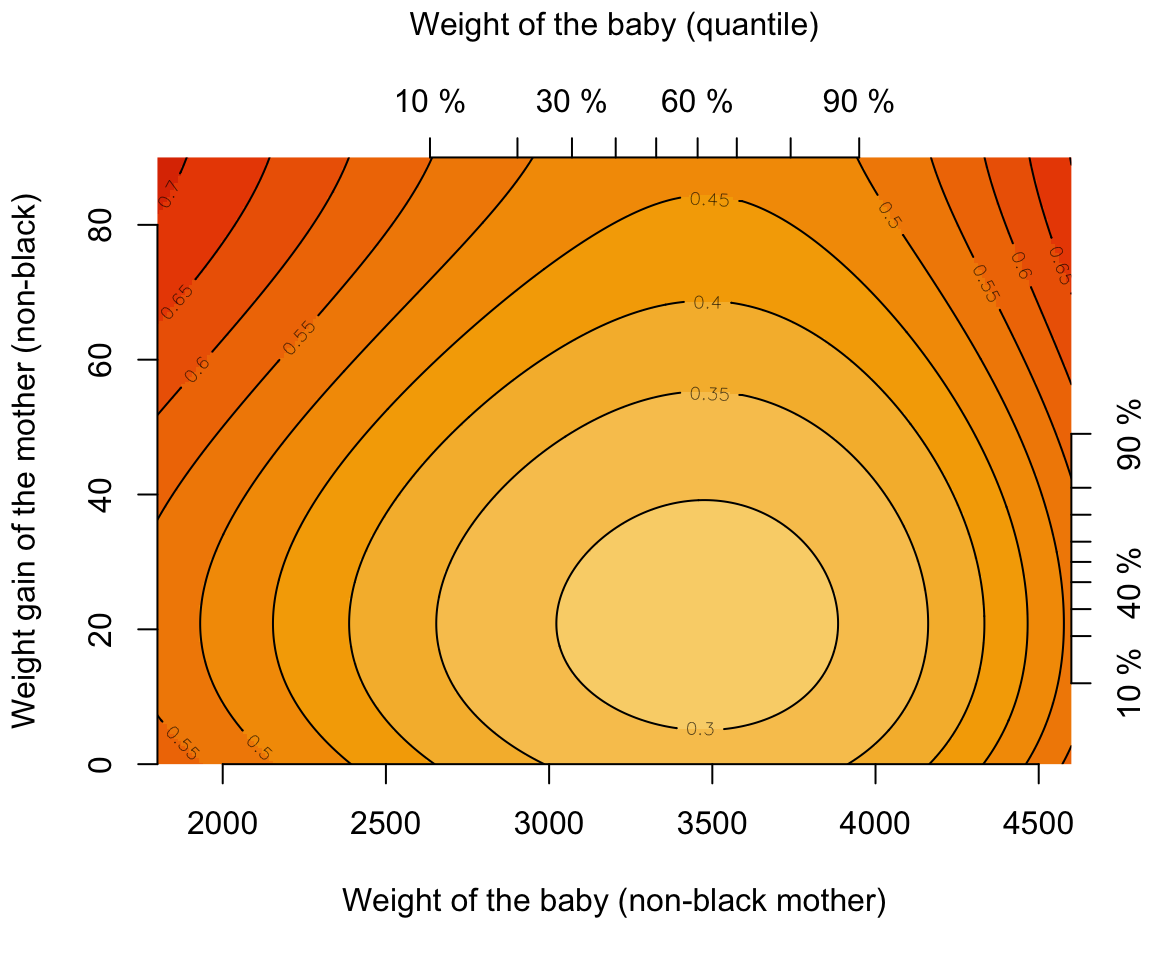}
     \includegraphics[width=.48\textwidth]{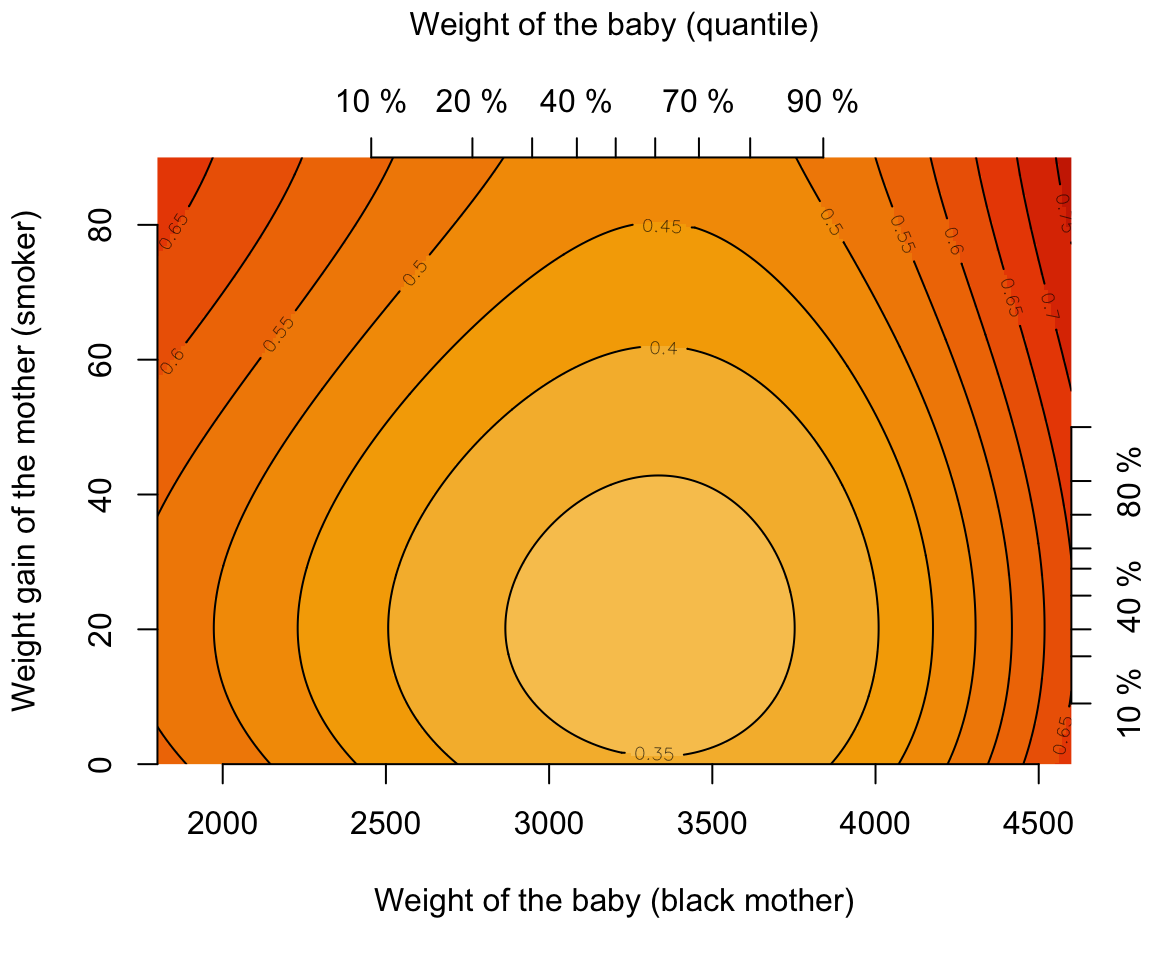}

    \centering
     \includegraphics[width=.48\textwidth]{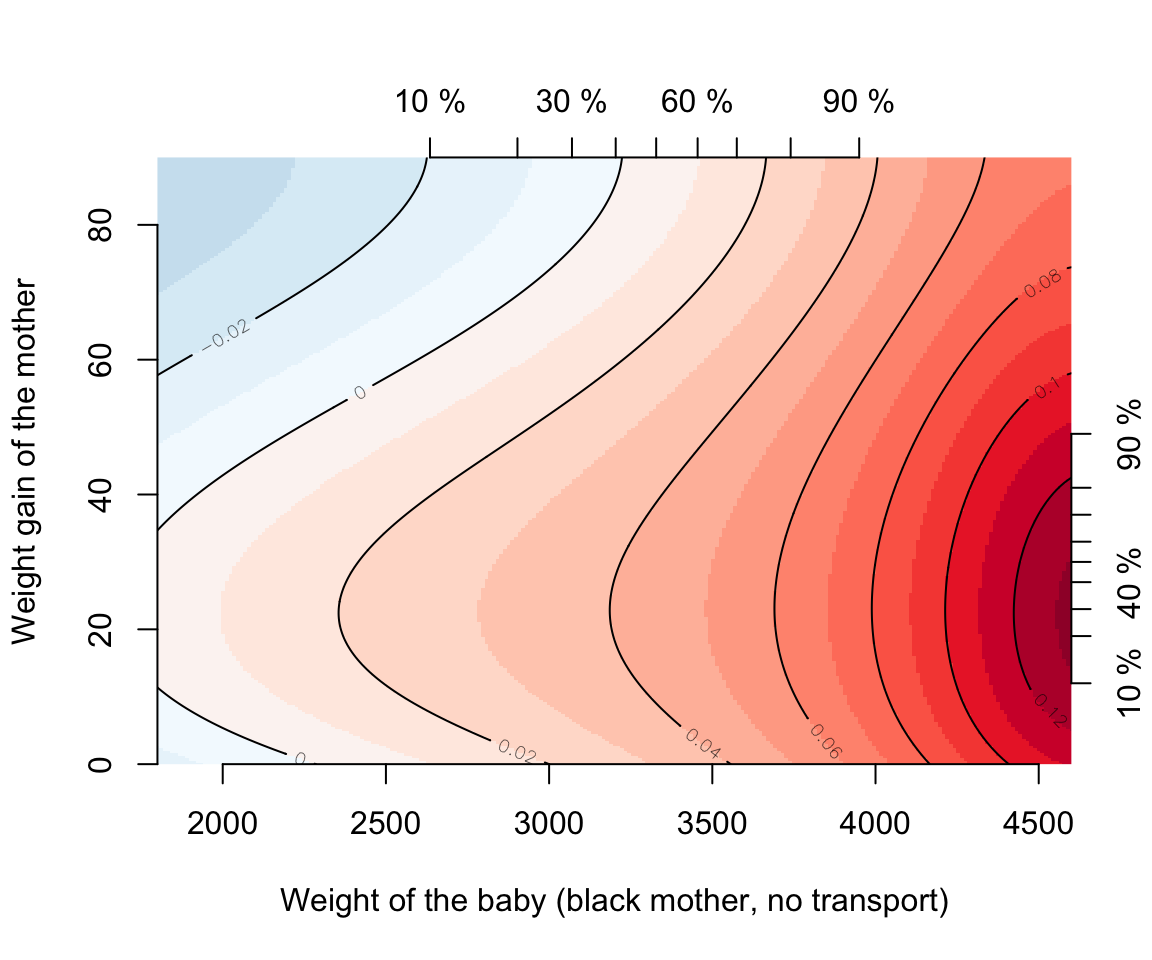}
     \includegraphics[width=.48\textwidth]{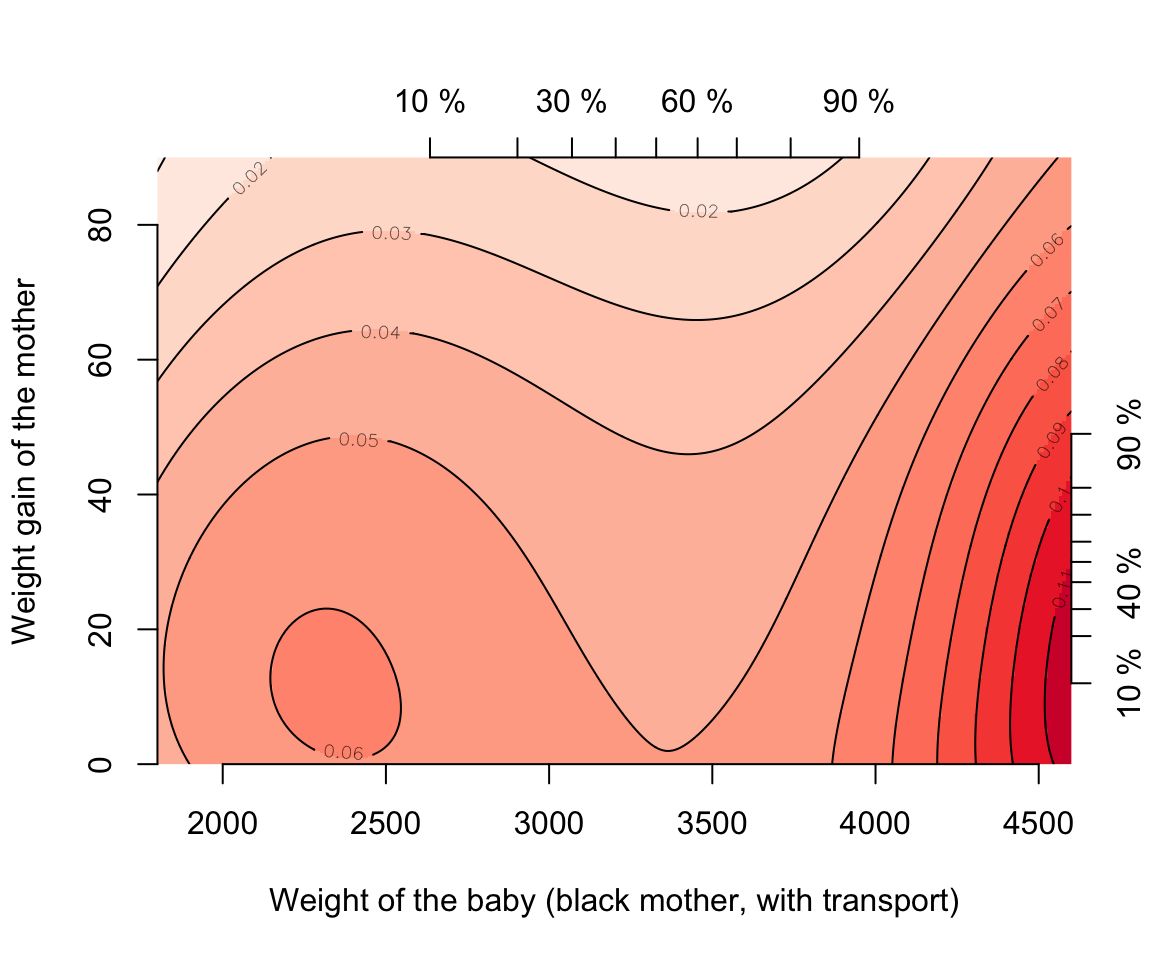}

    \centering
     \includegraphics[width=.48\textwidth]{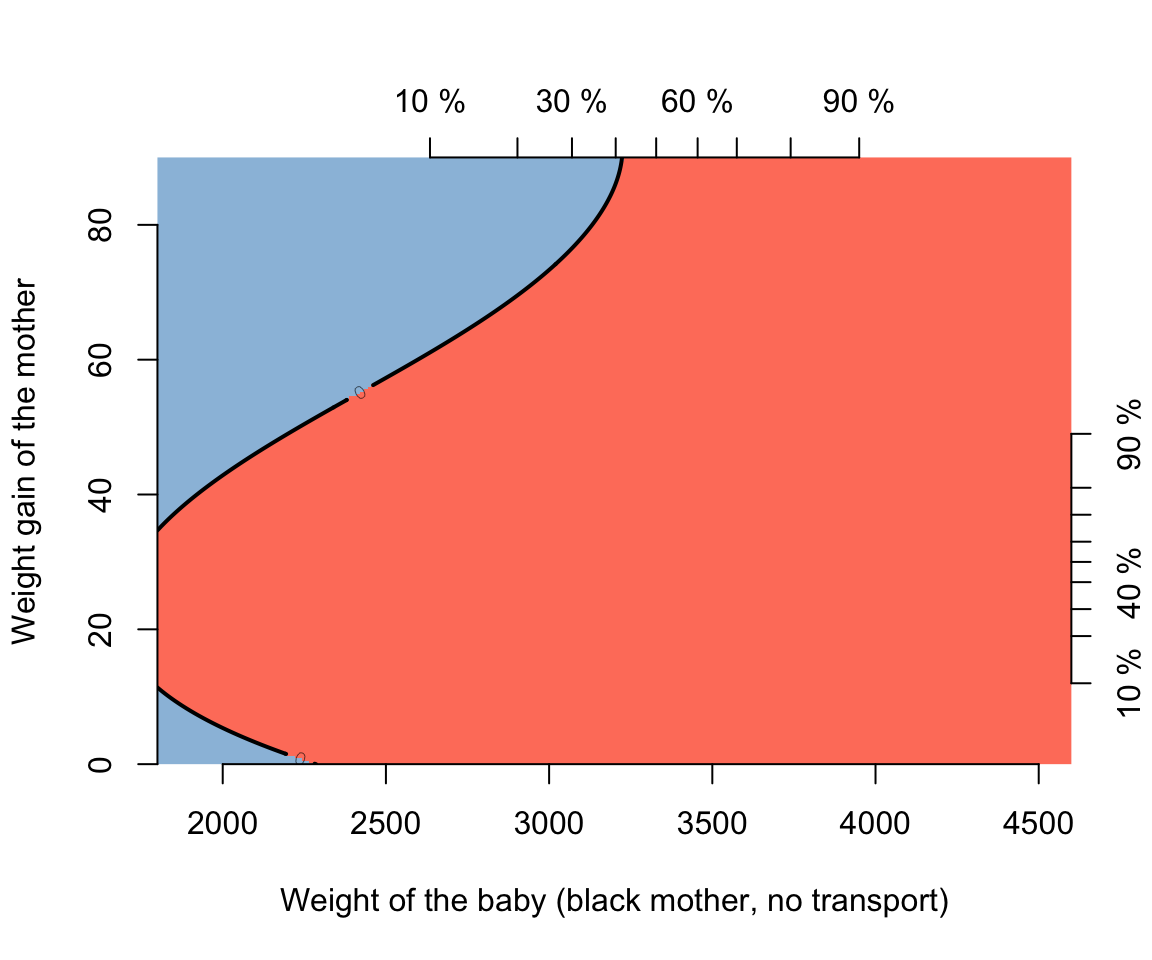}
     \includegraphics[width=.48\textwidth]{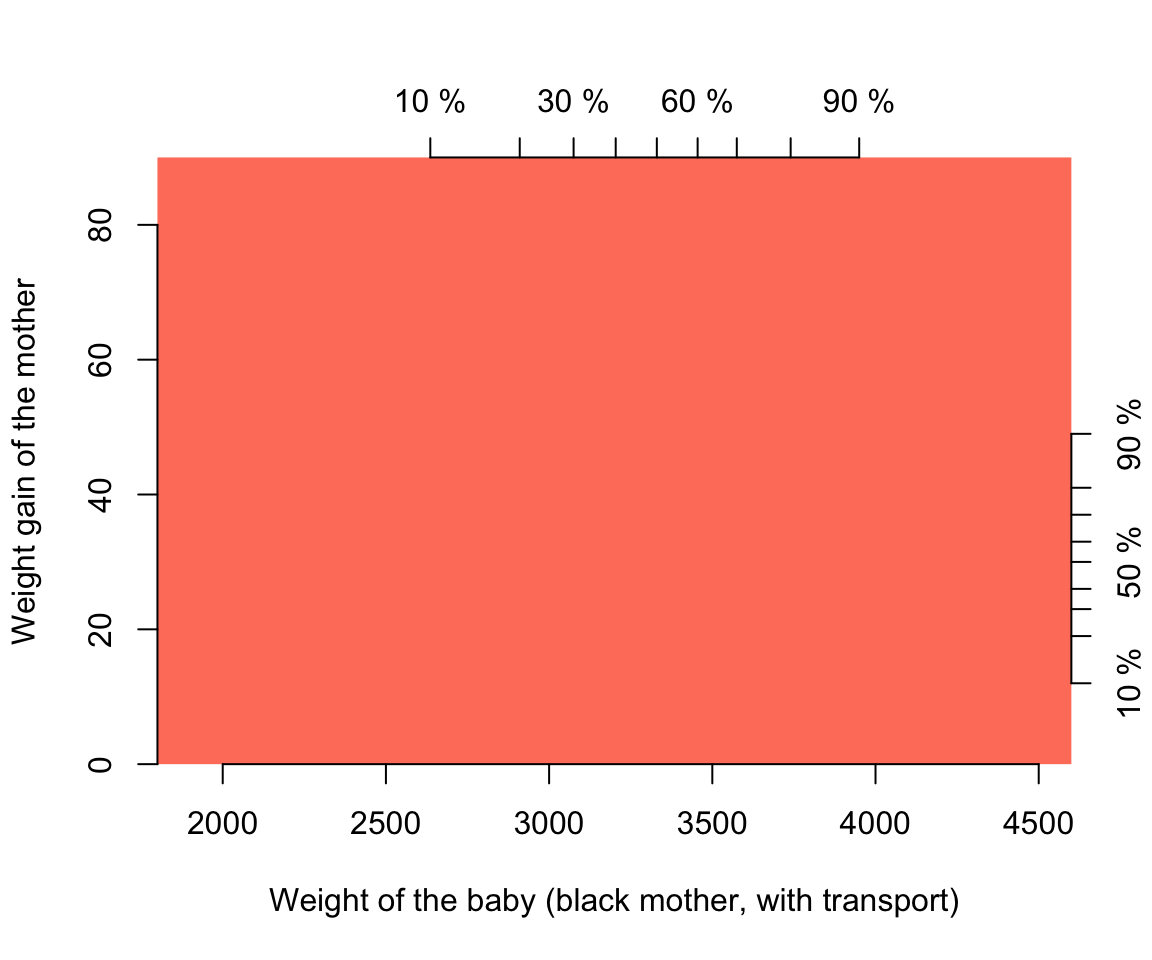}
    
   \caption{On top, contours of $\boldsymbol{x}\mapsto\mathbb{E}[Y|\boldsymbol{X}=\boldsymbol{x},T=0]$ and $\boldsymbol{x}\mapsto\mathbb{E}[Y|\boldsymbol{X}=\boldsymbol{x},T=1]$ when $T$ indicates whether a mother is Afro-American or not, estimated with logistic GAM models (cubic splines).
    In the middle, contours of the {\em ceteris paribus} $\boldsymbol{x}\mapsto\text{CATE}[\boldsymbol{x}]$ without any transport on the left, and $\boldsymbol{x}\mapsto\text{SCATE}[\boldsymbol{x}]$ {\em mutatis mutandis} on the right. At the bottom, positive/negative distinction for the conditional average treatment effect. See Figure~\ref{fig:CATE-biv-2x3-GAM-2-pred-C:appendix} in Appendix~\ref{app:3} for similar graphs when $T$ indicates whether the mother is a smoker or not.}
    \label{fig:CATE-biv-2x3-GAM-1-pred-B}
\end{figure}

\begin{figure}[!ht]
    \centering
     \includegraphics[width=.48\textwidth]{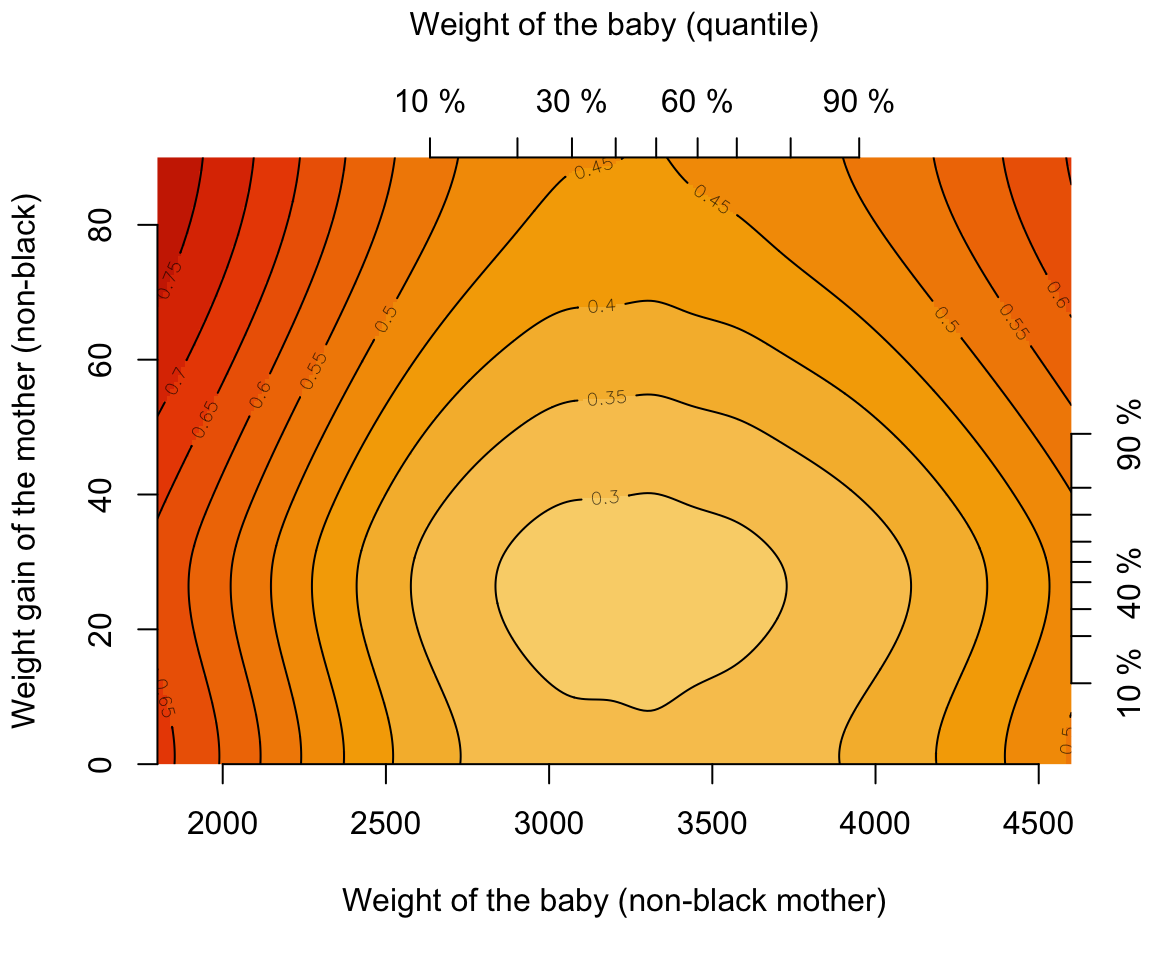}
     \includegraphics[width=.48\textwidth]{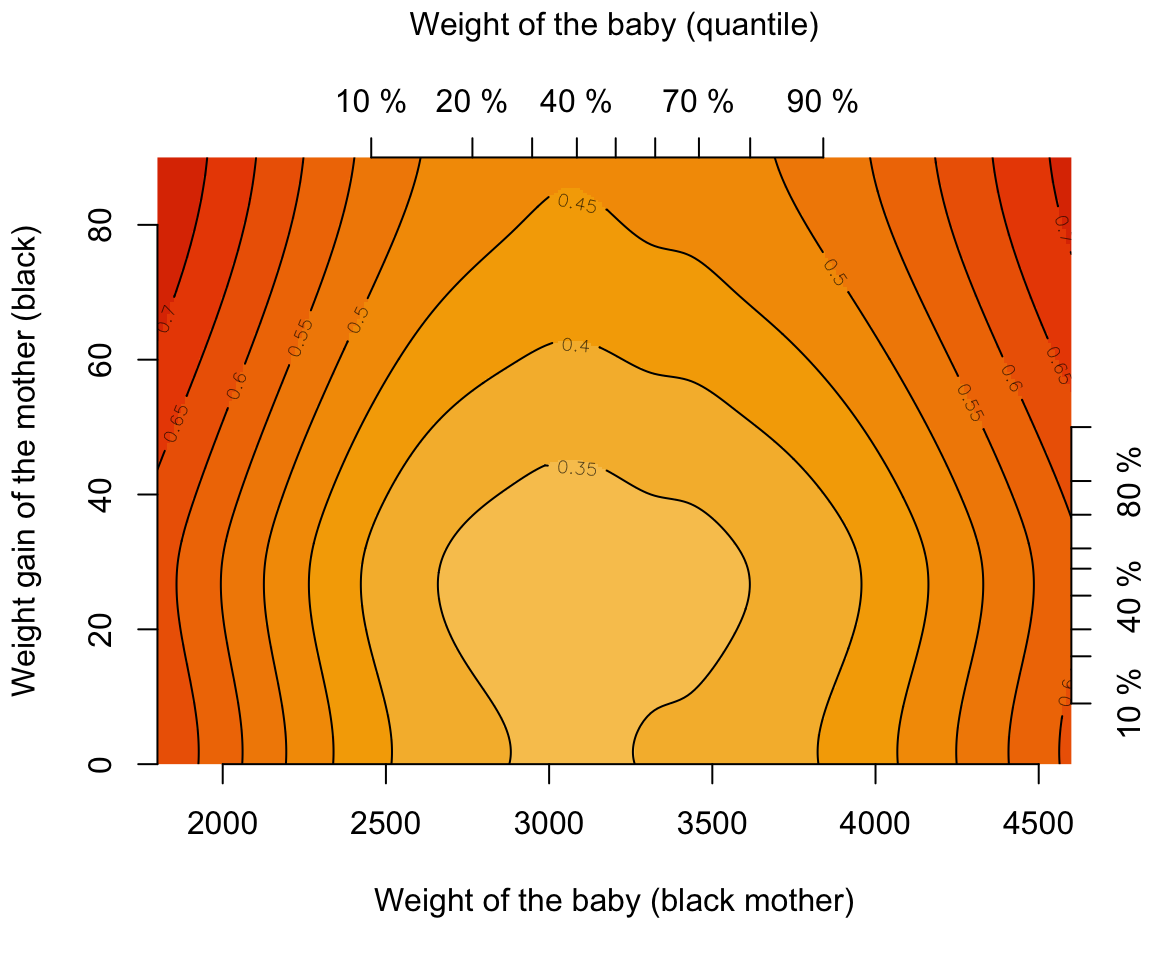}

    \centering
     \includegraphics[width=.48\textwidth]{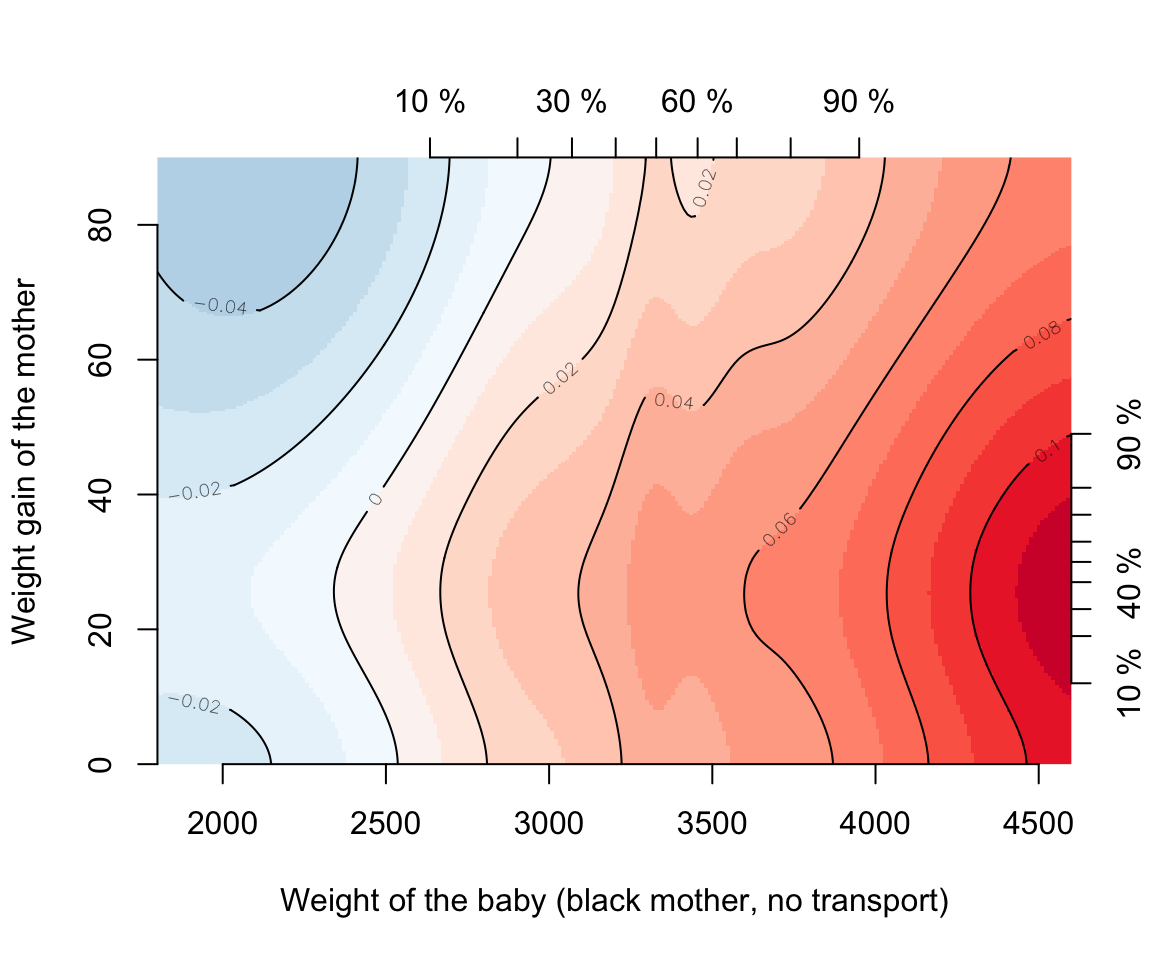}
     \includegraphics[width=.48\textwidth]{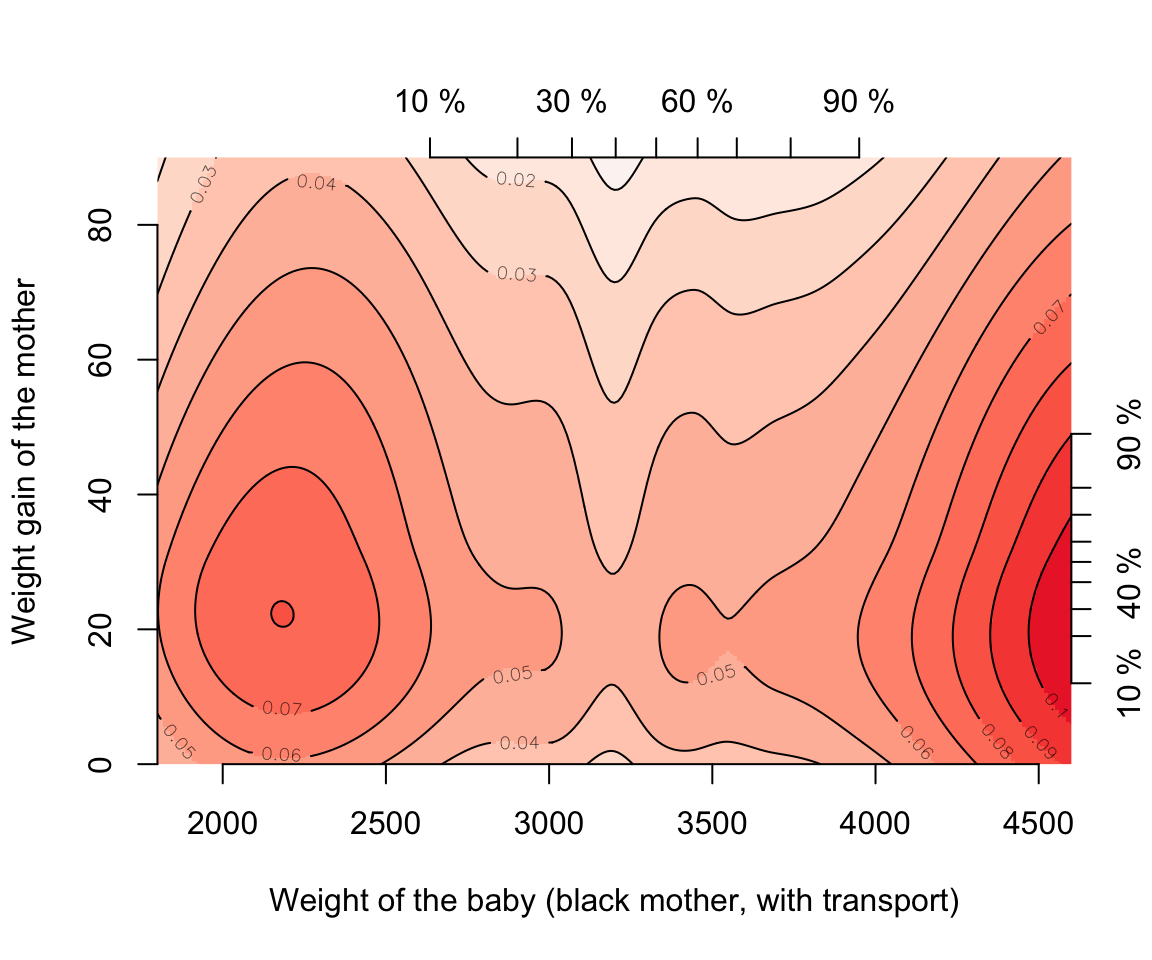}

    \centering
     \includegraphics[width=.48\textwidth]{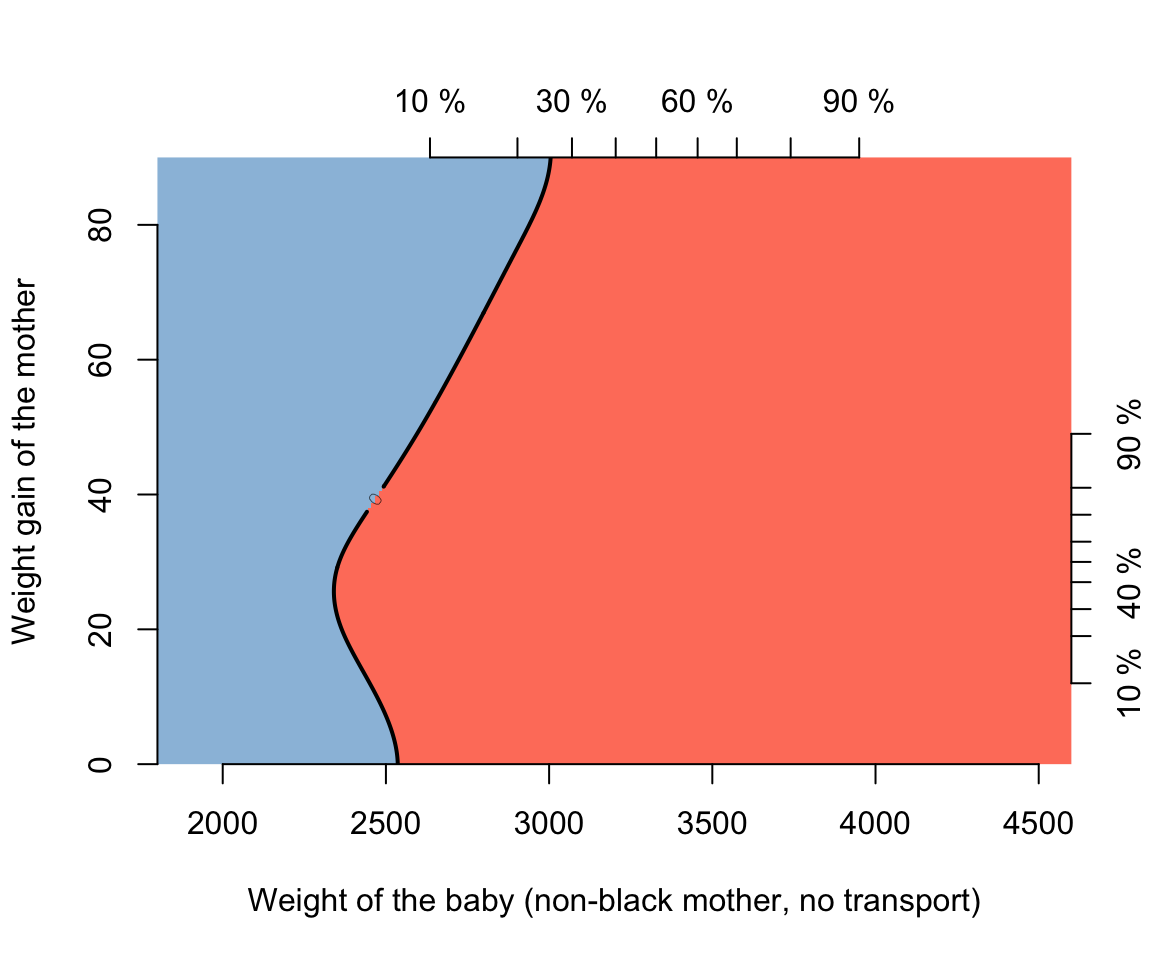}
     \includegraphics[width=.48\textwidth]{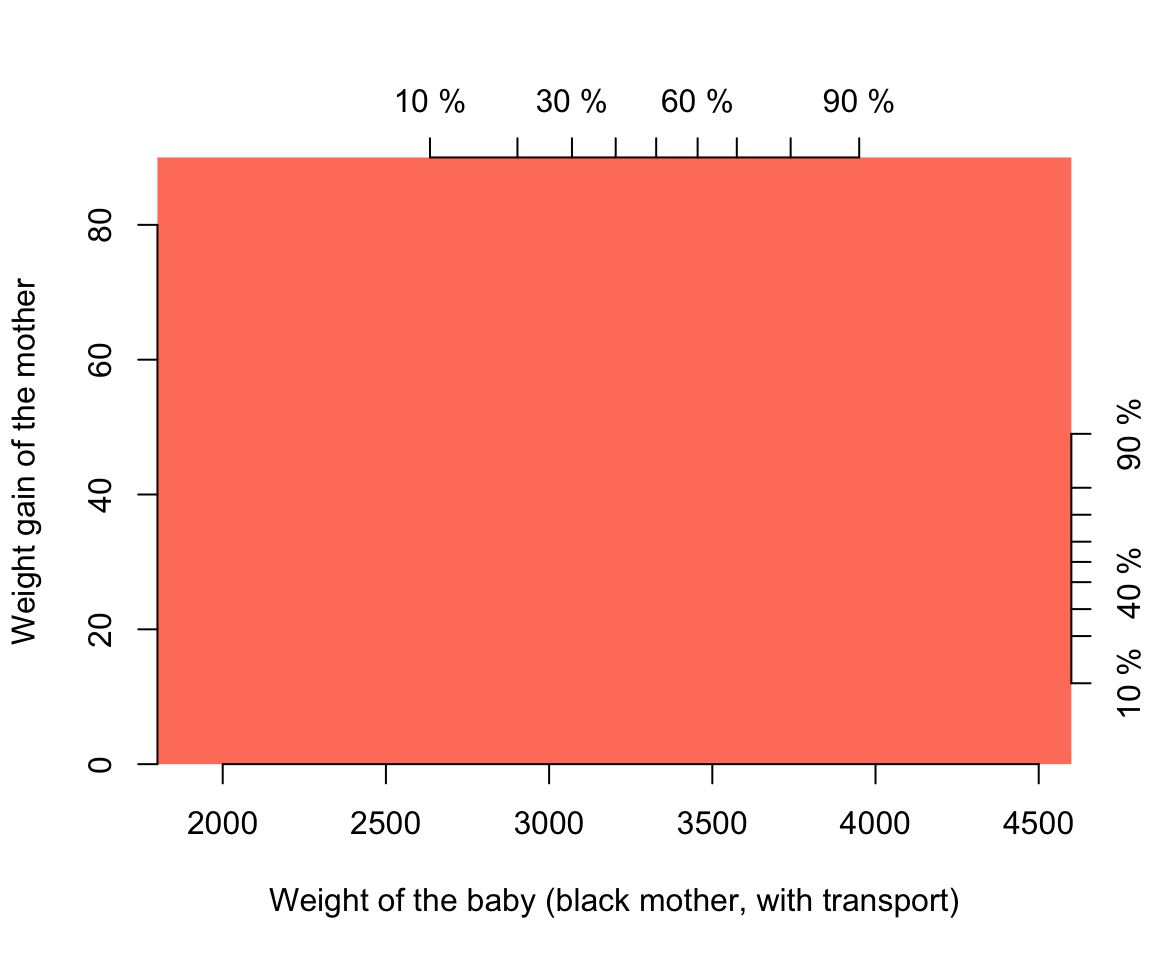}

   \caption{On top, contours of $\boldsymbol{x}\mapsto\mathbb{E}[Y|\boldsymbol{X}=\boldsymbol{x},T=0]$ and $\boldsymbol{x}\mapsto\mathbb{E}[Y|\boldsymbol{X}=\boldsymbol{x},T=1]$ when $T$ indicates whether a mother is Afro-American or not, estimated with logistic GAM models (cubic splines, {\bf with more knots and degrees of freedom}).
    In the middle, contours of the {\em ceteris paribus} $\boldsymbol{x}\mapsto\text{CATE}[\boldsymbol{x}]$ without any transport on the left, and $\boldsymbol{x}\mapsto\text{SCATE}[\boldsymbol{x}]$ {\em mutatis mutandis} on the right. At the bottom, positive/negative distinction for the conditional average treatment effect. See Figure~\ref{fig:CATE-biv-2x3-GAM-2-pred-C:appendix} in Appendix~\ref{app:3} for similar graphs when $T$ indicates whether the mother is a smoker or not.}
    \label{fig:CATE-biv-2x3-GAM-2-pred-B}
\end{figure}

The top panel of Figures~\ref{fig:CATE-biv-2x3-GAM-1-pred-B} shows the level curves of $\widehat{m}_0:\boldsymbol{x}\mapsto\mathbb{E}[Y|\boldsymbol{X}=\boldsymbol{x},T=0]$ (left-hand side) and $\widehat{m}_1:\boldsymbol{x}\mapsto\mathbb{E}[Y|\boldsymbol{X}=\boldsymbol{x},T=1]$ (right-hand side), when the treatment $T$ indicates whether a mother is Black or not, estimated with logistic GAM models (cubic splines). The middle-level panel displays curves of the {\em ceteris paribus} $\boldsymbol{x}\mapsto\text{CATE}[\boldsymbol{x}]$ without any transport (on the left), and $\boldsymbol{x}\mapsto\text{SCATE}[\boldsymbol{x}]$ {\em mutatis mutandis} (on the right). Lastly, the bottom panel shows a positive/negative distinction for the conditional average treatment effect (positive is red, negative is blue). Figure~\ref{fig:CATE-biv-2x3-GAM-2-pred-B} provides different results using more knots in the cubic splines.
% Without any transport, it looks like mothers with large baby (over 3.5kg) who did not gain too much weight (less than 40 lbs) would be more likely to get a non-natural delivery if they were smoking. If we take into account the fact that smoking mothers have smaller babies, it looks like mothers with large baby who did not gain too much weight would {\em not} be more likely to get a not natural delivery if they were smoking, while it would be the case for mothers who did not gain too much weight with {\em small} babies (less than 3.5kg).
% Figures~\ref{fig:CATE-biv-2x3-GAM-1-pred-B} and~\ref{fig:CATE-biv-2x3-GAM-2-pred-B}, present similar graphs to those in Figures~\ref{fig:CATE-biv-2x3-GAM-1-pred-C} and~\ref{fig:CATE-biv-2x3-GAM-2-pred-C}, except that here, $T$ is the indicator that the mother is Afro-American. Again, $\widehat{m}_0$ and $\widehat{m}_1$ are estimated using cubic splines, with some slight overfit in Figure~\ref{fig:CATE-biv-2x3-GAM-2-pred-B}. 
We can observe that all mothers are more likely to get a non-natural delivery would they be Black, whatever the weight of the baby (the {\em ceteris paribus} approach would suggest that mothers with small babies, below 2.5kg would be less likely to get a non-natural delivery if they were Black).

% \begin{figure}[!ht]
%     \centering
%      \includegraphics[width=.49\textwidth]{figures/rf-biv-black-image-0-1.png}
%      \includegraphics[width=.49\textwidth]{figures/rf-biv-black-image-1-1}

%     \caption{`Contours' of $\boldsymbol{x}\mapsto\mathbb{E}[Y|\boldsymbol{X}=\boldsymbol{x},T=0]$ and $\boldsymbol{x}\mapsto\mathbb{E}[Y|\boldsymbol{X}=\boldsymbol{x},T=1]$ when $T$ is the indicator of a mother is Black or not, estimated with a random forest (and a classification trees).}
%     \label{fig:CATE-biv-2x1-RF-1-pred-B}
% \end{figure}

As briefly discussed earlier, optimal matching or coupling in high dimension can be computationally intensive, since matrices $n_0\times n_1$ are involved. For instance, when $t$ is the sex of the newborn, the cost matrix is a matrix with 3,000 billion entries. Thus, it is quite natural to consider sub-sampling techniques (since our dataset is quite large). For convenience, we can use optimal matching on groups of size $n$, and study the robustness of estimated, as a function of $n$. Some simulations are mentioned in the Appendix.

\appendix
\section{Appendix}\label{app:1}

% \subsection{The dataset}

\subsection{Estimation of $\text{CATE}$ in a Gaussian framework}\label{app:2}

With a correlation $r$ (in the simulations, we considered $r=0.4$), consider the following SEM, 
$$\displaystyle{\begin{cases}
T = \boldsymbol{1}(\epsilon_t<0),\\
{X}^m_1 ={\epsilon}_{1m}+(T=1)\cdot\big(2+0.2{\epsilon}_{1m}) \\
{X}^m_2 =r{\epsilon}_{1m}+\sqrt{1-r^2}\epsilon_{2m}-0.2(T=1)\cdot\big(r{\epsilon}_{1m}+\sqrt{1-r^2}\epsilon_{2m}\big),\\
{X}^c = {\epsilon}_c  \\
Y = 2+T +{X}^m_1-{X}^m_2+{X}^c+\epsilon_y \\
\end{cases}}$$
where $\epsilon$'s are independent $\mathcal{N}(0,1) $ variables.
The two interventions yield
\begin{center}
  \begin{tabular}{cc}
  $do(T=0)$ & $do(T=1)$\\
$\displaystyle{\begin{cases}
T \leftarrow 0 \\
{X}^m_1 =\epsilon_{1m} \\
{X}^m_2 =r{\epsilon}_{1m}+\sqrt{1-r^2}\epsilon_{2m} \\
{X}^c = {\epsilon}_c  \\
Y_{T\leftarrow 0} = 2+{X}^m_1-{X}^m_2+{X}^c+\epsilon_y \\
\end{cases}}$& $\displaystyle{\begin{cases}
T \leftarrow 1 \\
{X}^m_1 =2+1.2\epsilon'_{1m} \\
{X}^m_2 =0.8 r\epsilon'_{1m}+\sqrt{1-r^2}\epsilon'_{2m} \\
{X}^c = {\epsilon}'_c  \\
Y_{T\leftarrow 1} = 3+{X}^m_1-{X}^m_2+{X}^c+\epsilon'_y \\
\end{cases}}$
  \end{tabular}
\end{center}
and if we consider $do(T=0)$, when $X_1^m=x_1$, we have
$$\displaystyle{\begin{cases}T \leftarrow 0 \\
{X}^m_1 =x_{1}^m \\
{X}^m_2 = r x_{1}^m+ \sqrt{1-r^2}{\epsilon}_{2m} \\
{X}^c = {\epsilon}_c  \\
Y_{T\leftarrow 0} = 2+{x}^m_1-(r x_{1}^m+ \sqrt{1-r^2}{\epsilon}_{2m} )+{\epsilon}_c+\epsilon_y \\
\end{cases}}$$
while if we consider $do(T=1)$, when $X_1^m=x_1'$ 
$$\displaystyle{\begin{cases}
T \leftarrow 1 \\
{X}^m_1 =x_1' \\
{X}^m_2 = 0.8 r (x_1'-2)/1.2+\sqrt{1-r^2}\epsilon'_{2m}  \\
{X}^c = {\epsilon}'_c  \\
Y_{T\leftarrow 1} = 3+x_1^m -\big(0.8 r (x_1'-2)/1.2+\sqrt{1-r^2}\epsilon'_{2m}\big)+{\epsilon}'_c +\epsilon'_y \\
\end{cases}}$$
so that
$$\displaystyle{\begin{cases}
Y_{T\leftarrow 0} = 2+(1-r){x}'_1 + \sqrt{1-r^2}{\epsilon}_{2m} +{\epsilon}_c+\epsilon_y \\
Y_{T\leftarrow 1} = 3+1.6r/1.2 +(1-0.8r/1.2)x_1' -\sqrt{1-r^2}\epsilon'_{2m}+{\epsilon}'_c +\epsilon'_y 
\end{cases}}$$
Since ${X}^m_1 =\epsilon'_{1m}$ when $T\leftarrow 0$, while ${X}^m_1 =2+1.2\epsilon'_{1m}$ when $T\leftarrow 1$, it is legitimate to assume that if $x_{1,T\leftarrow 0}^m=x_1$, then $x_{1,T\leftarrow 1}^m=2+1.2x_{1,T\leftarrow 0}^m$, in a {\em mutatis mutandis} approach,
$$\displaystyle{\begin{cases}
Y_{T\leftarrow 0} = 2+(1-r){x}_1 + \sqrt{1-r^2}{\epsilon}_{2m} +{\epsilon}_c+\epsilon_y \\
Y_{T\leftarrow 1} = 3+1.6r/1.2 +(1-0.8r/1.2)(2+1.2x_1) -\sqrt{1-r^2}\epsilon'_{2m}+{\epsilon}'_c +\epsilon'_y 
\end{cases}}$$
Thus,
$$
ATE = \mathbb{E}[Y_{T\leftarrow 1}-Y_{T\leftarrow 0}]=3
$$
while the {\em mutatis mutandis} CATE is
$$
CATE(x_1) = \mathbb{E}[Y_{T\leftarrow 1}|x_{1,T\leftarrow 1}^m]-\mathbb{E}[Y_{T\leftarrow 0}|x_{1,T\leftarrow 0}^m=x_1]
$$
that is
$$
CATE(x_1) = \big[3+1.6r/1.2 +(1-0.8r/1.2)(2+1.2x_1)\big] -  \big[2+(1-r){x}_1 \big] %3+0.2\cdot x_1^m
$$
i.e.,
$$
CATE(x_1) = 3 + 0.2 (1+r) {x}_1%3+0.2\cdot x_1^m
$$
that is linear in $x_1^m$, with slope $0.2 (1+r)$ in this {\em mutatis mutandis} case. 

\subsection{Additional applications (smoker and sex of newborn)}\label{app:3}

In this section, similar graphs to the one presented earlier are produced.

\begin{figure}[!ht]
    \centering
    \includegraphics[width=.49\textwidth]{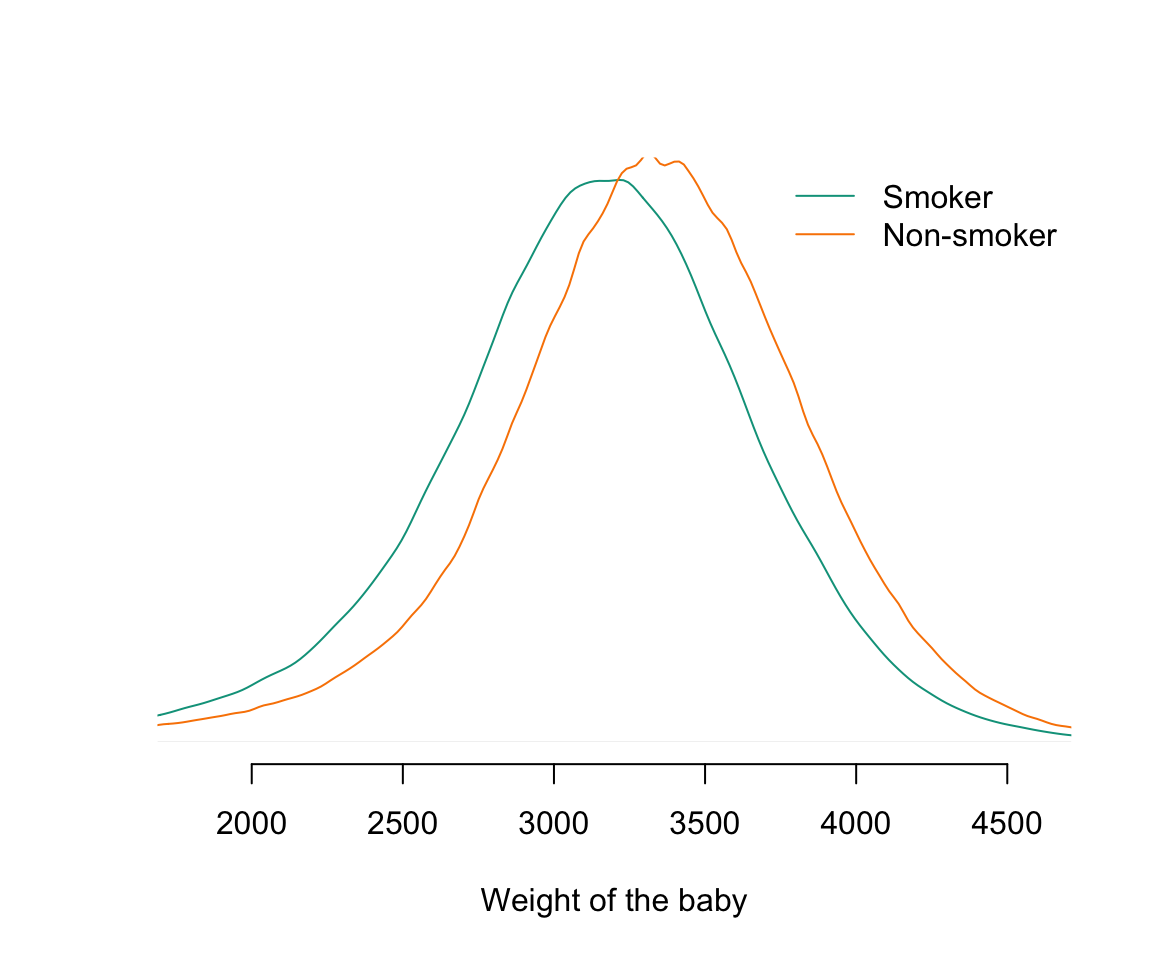}
     \includegraphics[width=.49\textwidth]{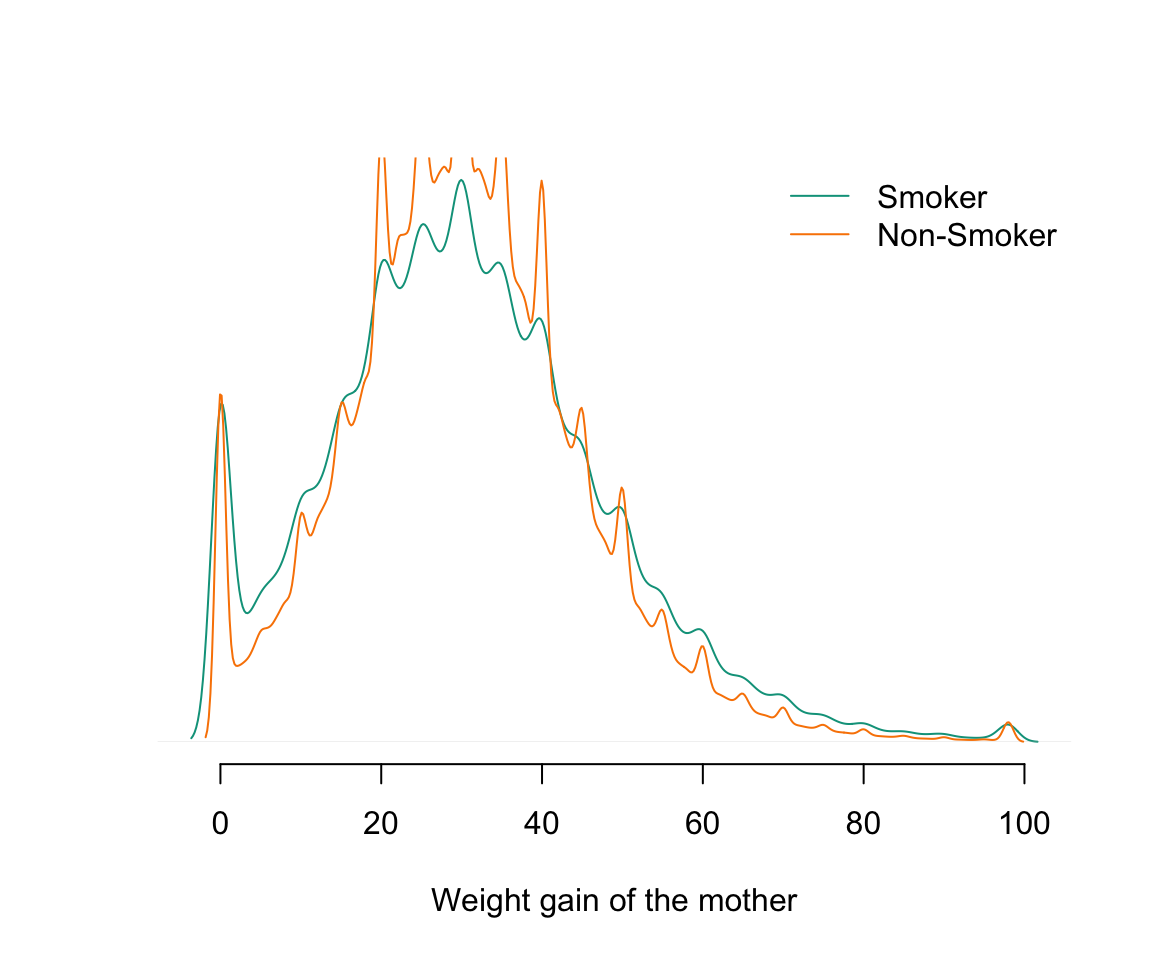}

    % \centering
    % \includegraphics[width=.49\textwidth]{figures/density-weight-black-1.png}
    % \includegraphics[width=.49\textwidth]{figures/density-weight-gain-black-1.png}

    \centering
    \includegraphics[width=.49\textwidth]{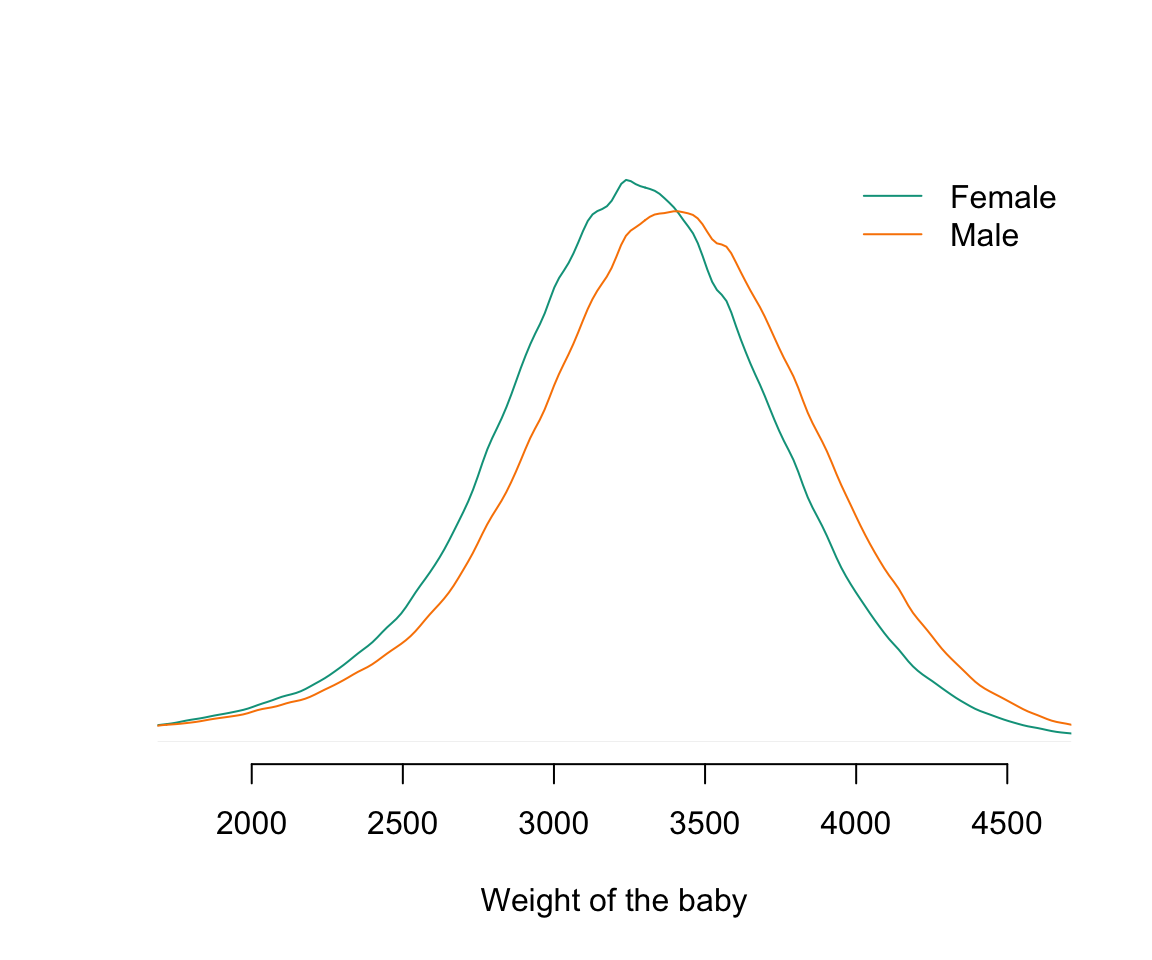}
    \includegraphics[width=.49\textwidth]{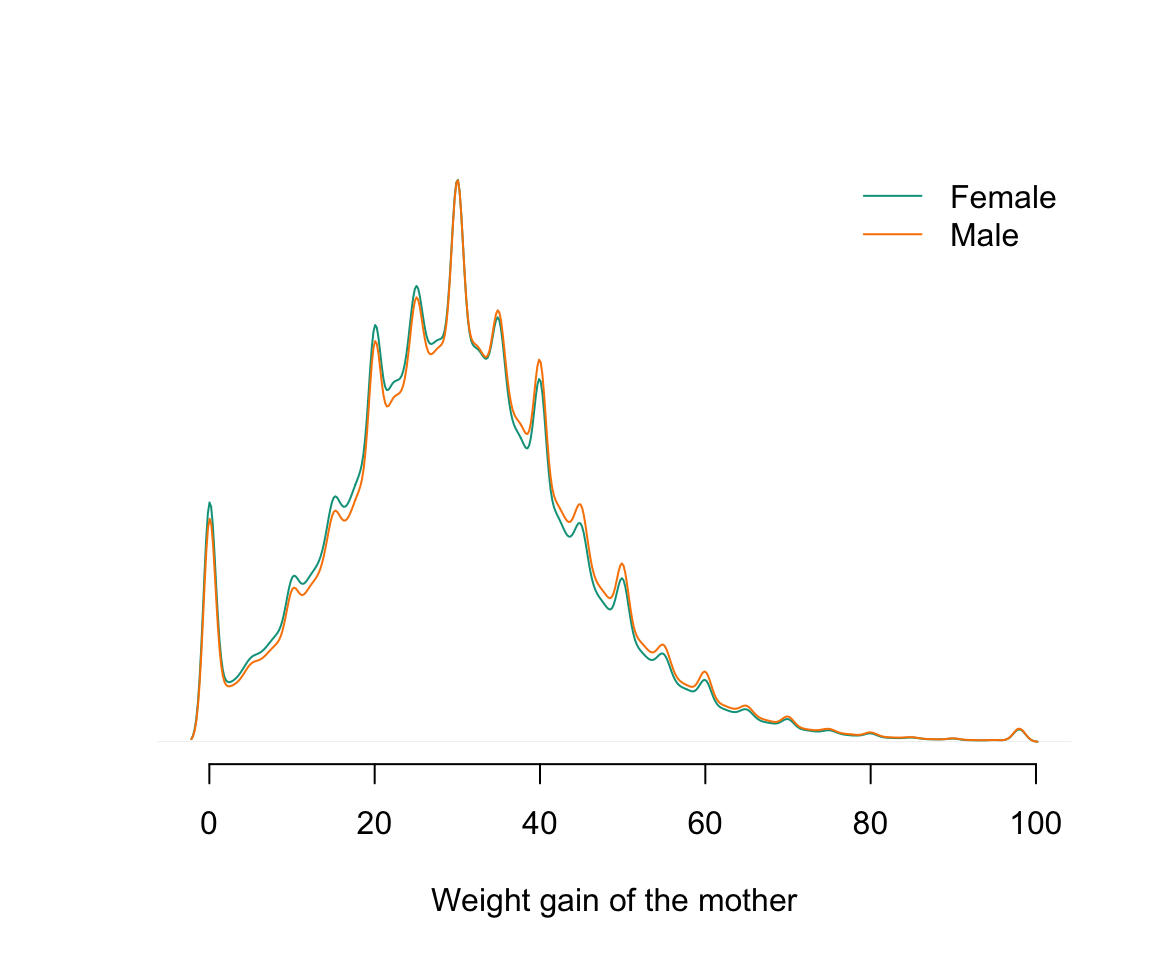}
    
    \caption{Distribution of the weight of the newborn infant (in grams) on the left and distribution of the weight gain of the mother on the right, when the mother is a smoker or not on top, and depending on the sex of the newborn infant at the bottom.}
    \label{fig:densite-2x3-conditional-t:appendix}
\end{figure}

\begin{table}[!ht]
    \centering
    \begin{tabular}{c|cccccc}\hline\hline
  \multicolumn{7}{c}{$t$: mother smoker}\\
  \hline
  $x$ (newborn's weight) & 2000 & 2500 & 3000 & 3500 & 4000 & 4500 \\\hline
  $u$ & 2.75\% &  7.44\% & 24.99\% & 64.14\% & 91.75\% & 98.86\% \\
 $\text{CATE}_0(x)$ (GAM) & -4.41\% & -2.55\% & -0.69\% &  0.79\% &  1.97\% &  2.50\%  \\
 % & $\text{CATE}(x)$ (propensity) & -5.69\% &  -2.33\% &  0.35\% &   1.94\% &   2.43\% &   0.93\%   \\
 $\widehat{\mathcal{T}}(x)$ & 1775   &   2280   &   2802   &   3317 &     3830  &    4337\\
 $\text{SCATE}(x)$ (GAM) & -0.08\% &      1.15\% &      1.15\% &      0.50\% &     -1.07\% &     -4.39\%  \\
 $\widehat{\mathcal{T}}_{\mathcal{N}}(x)$ &  1786& 2295& 2805& 3314& 3824& 4333\\
$\text{SCATE}_{\mathcal{N}}(x)$ (GAM) &  -0.28\% &  0.88\% &  1.12\% &  0.50\% & -1.15\% & -4.53\%  \\
$\text{SCATE}_{\mathcal{N}}(x)$ (kernel) &   -0.80\% &  0.24\% &  1.72\% &  0.15\% & -1.75\% & -2.78\%     \\\hline
 %  Black & $u$ & 2.67\% &  7.46\% & 25.13\% & 64.13\% & 91.73\% & 98.87\% \\
 % & $\text{CATE}_0(x)$ (GAM) & 0.58\% &  1.99\% &  3.24\% &  4.86\% &  7.78\% & 11.70\% \\
 % & $\text{CATE}(x)$ (propensity) & -6.64\% &  -3.67\% &  -0.56\% &   1.85\% &   3.39\% &   2.98\%     \\
 %  &  $\widehat{\mathcal{T}}(x)$ &1595   &   2301  &    2863   &   3375  &    3890  &    4415\\
 % & $\text{SCATE}(x)$ (GAM) &   7.94\% &    5.53\% &     4.53\% &      4.42\% &      5.16\% &      7.46\%  \\
 %  &  $\widehat{\mathcal{T}}_{\mathcal{N}}(x)$  &1758 &2297 &2836& 3376& 3915& 4455\\
 % & $\text{SCATE}_{\mathcal{N}}(x)$ (GAM) &  5.15\%& 5.60\%& 4.82\%& 4.42\%& 5.71\%& 9.41\% \\
 % & $\text{SCATE}_{\mathcal{N}}(x)$ (kernel) & 6.98\%& 6.64\%& 4.34\%& 4.53\%& 5.34\%& 7.13\%   \\\hline

 \multicolumn{7}{c}{$t$: sex of the newborn}\\
 \hline
 $x$ (newborn's weight) & 2000 & 2500 & 3000 & 3500 & 4000 & 4500 \\\hline
  $u$ & 2.79\% &  7.26\% & 23.00\% & 60.32\% & 90.04\% & 98.55\% \\
 $\text{CATE}_0(x)$ (GAM)  &-0.24\% & -1.66\% & -2.24\% & -2.00\% & -0.77\% &  2.38\%   \\
 % & $\text{CATE}(x)$ (propensity) &  -4.22\% &  -2.51\% &  -0.12\% &   1.24\% &   0.28\% &  -5.39\%    \\
  $\widehat{\mathcal{T}}(x)$& 1960 &     2438   &   2892   &   3374    &  3856   &   4338  \\
 $\text{SCATE}(x)$ (GAM) &  0.71\% &      -0.38\% &      -0.96\% &      -2.04\% &      -3.38\% &      -5.14\%  \\
  $\widehat{\mathcal{T}}_{\mathcal{N}}(x)$  &1947& 2424& 2901& 3377 &3854& 4331\\
 $\text{SCATE}_{\mathcal{N}}(x)$ (GAM) &  1.02\% & -0.08\% & -1.07\% & -2.05\% & -3.41\% & -5.43\% \\
 $\text{SCATE}_{\mathcal{N}}(x)$ (kernel) &  2.06\%& -0.27\%& -1.02\%& -2.30\%& -3.53\% &-5.10\%    \\\hline\hline
    \end{tabular}
    \caption{Estimation of the conditional average treatment (CATE), on the probability to have a non-natural birth ($y$), as a function of the weight of the baby ($x$, in g.), for different ``treatments'' ($t$): when the mother is a smoker, and when the baby is a boy. Several weights $x$ are considered, from $2$ to $4.5$ kg. $u$ is the probability associated with $x$, in the baseline population ($t=0$). $\text{CATE}_0$ is simply the difference  $\widehat{m}_1(x)-\widehat{m}_0(x)$, where both $\widehat{m}_0$ and $\widehat{m}_1$ are GAMs. $\widehat{\mathcal{T}}(x)$ is the quantile based transport function ($\widehat{\mathcal{T}}(x)= \widehat{F}_1^{-1}\circ \widehat{F}_0(x)$), while $\widehat{\mathcal{T}}_{\mathcal{N}}(x)$ is the Gaussian one. Thus, $\text{SCATE}(x)$ is the {\em mutatis mutandis CATE} $
\text{SCATE}(x)=\widehat{m}_1\big(\widehat{\mathcal{T}}(x)\big) - \widehat{m}_0\big(x\big)
$, while $
\text{SCATE}_{\mathcal{N}}(x)=\widehat{m}_1\big(\widehat{\mathcal{T}_{\mathcal{N}}}(x)\big) - \widehat{m}_0\big(x\big)
$, where both $\widehat{m}_0$ and $\widehat{m}_1$ are GAMs. Finally, the last estimate is obtained when $\widehat{m}_0$ and $\widehat{m}_1$ are simple local averages, using kernels.}
    \label{tab:num:CATE:w:appendix}
\end{table}

% & $\text{CATE}_0(x)$ (GAM) & 0.06\% &  0.84\% &  0.69\% & -0.08\% & -1.26\% & -2.59\%   \\
 
\begin{table}[!ht]
    \centering
    \begin{tabular}{c|cccccc}\hline\hline
    \multicolumn{7}{c}{$t$: mother smoker}\\
    \hline
  $x$ (weight gain of the mother) &  5& 15& 25& 35& 45& 55 \\\hline
  $u$ & 4.61\% & 14.50\% & 37.49\% & 67.12\% & 86.54\% & 95.05\% \\
  $\text{CATE}_0(x)$ (GAM) & 0.06\% &  0.84\% &  0.69\% & -0.08\% & -1.26\% & -2.59\%   \\
 % & $\text{CATE}(x)$ (propensity) & 1.36\% &   2.02\% &   1.72\% &   0.81\% &  -0.45\% &  -1.77\%      \\
   $\widehat{\mathcal{T}}(x)$ &1     &   13 &       25  &      37      &  49   &     60 \\
  $\text{CATE}(x)$ (GAM) &  1.60\% &      1.21\% &      0.69\% &      0.14\% &     -0.38\% &    -1.08\%   \\
  $\widehat{\mathcal{T}}_{\mathcal{N}}(x)$ &1 &13 &24 &36 &48 &59 \\
  $\text{CATE}_{\mathcal{N}}(x)$ (GAM) &  1.63\% &  1.29\% &   0.71\% &   0.01\% &  -0.71\% &  -1.33\% \\
  $\text{CATE}_{\mathcal{N}}(x)$ (kernel) &   0.31\% &  1.03\% &  0.98\% &  0.29\% & -1.30\% & -1.08\%     \\\hline
 %  Black & $u$   & 4.57\% & 14.34\% & 37.15\% & 66.81\% & 86.34\% & 94.94 \\
 % & $\text{CATE}_0(x)$  (GAM) & 3.79\% & 4.79\% & 5.06\% & 4.82\% & 4.18\% & 3.26\%  \\
 % & $\text{CATE}(x)$ (propensity) &  -0.10\% &   0.95\% &   1.19\% &   0.86\% &   0.11\% &  -0.92\%     \\
 %  & $\widehat{\mathcal{T}}(x)$  & 1  &      12  &      24  &      35    &    47   &     58 \\
 % & $\text{CATE}(x)$  (GAM) &  5.25\% &      5.25\% &      5.04\% &      4.82\% &      4.69\% &      4.19\%    \\
 % &  $\widehat{\mathcal{T}}_{\mathcal{N}}(x)$  &1& 12& 23& 34& 46& 57\\
 % & $\text{CATE}_{\mathcal{N}}(x)$ (GAM) & 5.22\% &  5.21\% &  5.03\% &  4.74\% &  4.33\% &  3.78\%    \\
 % & $\text{CATE}_{\mathcal{N}}(x)$ (kernel) &  3.78\% & 5.49\% & 5.31\% & 4.49\% & 4.12\% & 3.61\%   \\\hline
 \multicolumn{7}{c}{$t$: sex of the newborn}\\\hline
 $x$ (weight gain of the mother) &  5& 15& 25& 35& 45& 55 \\\hline
    $u$   & 4.68\% & 14.40\% & 36.73\% & 65.81\% & 85.58\% & 94.55\%  \\
  $\text{CATE}_0(x)$ (GAM) &  -1.79\% &     -1.60\% &     -1.60\% &     -1.73\% &     -1.90\% &     -2.02\%   \\
 % & $\text{CATE}(x)$ (propensity) & 0.19\% &   0.24\% &   0.14\% &  -0.04\% &  -0.22\% &  -0.32\%     \\
   $\widehat{\mathcal{T}}(x)$ & 5& 15& 25& 35& 45& 55\\
  $\text{CATE}(x)$  (GAM)&  -1.79\% &     -1.60\% &     -1.60\% &     -1.73\% &     -1.90\% &     -2.02\%   \\
   $\widehat{\mathcal{T}}_{\mathcal{N}}(x)$ & 4 &14 &24 &34 &44 &54\\
  $\text{CATE}_{\mathcal{N}}(x)$ (GAM) &   -1.52\% &  -1.47\% &  -1.61\% &  -1.87\% &  -2.17\% &  -2.41\%  \\
  $\text{CATE}_{\mathcal{N}}(x)$ (kernel) &  -1.58\% & -1.37\% & -1.66\% & -1.79\% & -2.22\% & -2.88\%    \\\hline\hline
    \end{tabular}
\caption{Estimation of the conditional average treatment (CATE), on the probability to have a non-natural birth ($y$), as a function of the weight gain of the mother ($x$, in lbs), for different ``treatments'' ($t$): when the mother is a smoker, and when the baby is a boy. Several weight gains $x$ are considered, from $5$ to $55$ lbs. $u$ is the probability associated with $x$, in the baseline population ($t=0$). $\text{CATE}_0$ is simply the difference  $\widehat{m}_1(x)-\widehat{m}_0(x)$, where both $\widehat{m}_0$ and $\widehat{m}_1$ are GAMs.  $\widehat{\mathcal{T}}(x)$ is the quantile based transport function ($\widehat{\mathcal{T}}(x)= \widehat{F}_1^{-1}\circ \widehat{F}_0(x)$), while $\widehat{\mathcal{T}}_{\mathcal{N}}(x)$ is the Gaussian one. Thus, $\text{SCATE}(x)$ is the {\em mutatis mutandis} CATE $
\text{SCATE}(x)=\widehat{m}_1\big(\widehat{\mathcal{T}}(x)\big) - \widehat{m}_0\big(x\big)
$, while $
\text{SCATE}_{\mathcal{N}}(x)=\widehat{m}_1\big(\widehat{\mathcal{T}_{\mathcal{N}}}(x)\big) - \widehat{m}_0\big(x\big)
$, where both $\widehat{m}_0$ and $\widehat{m}_1$ are GAMs. Finally, the last estimate is obtained when $\widehat{m}_0$ and $\widehat{m}_1$ are simple local averages, using kernels.}
    \label{tab:num:CATE:g:appendix}
\end{table}

\begin{figure}[!ht]
    \centering
     \includegraphics[width=.49\textwidth]{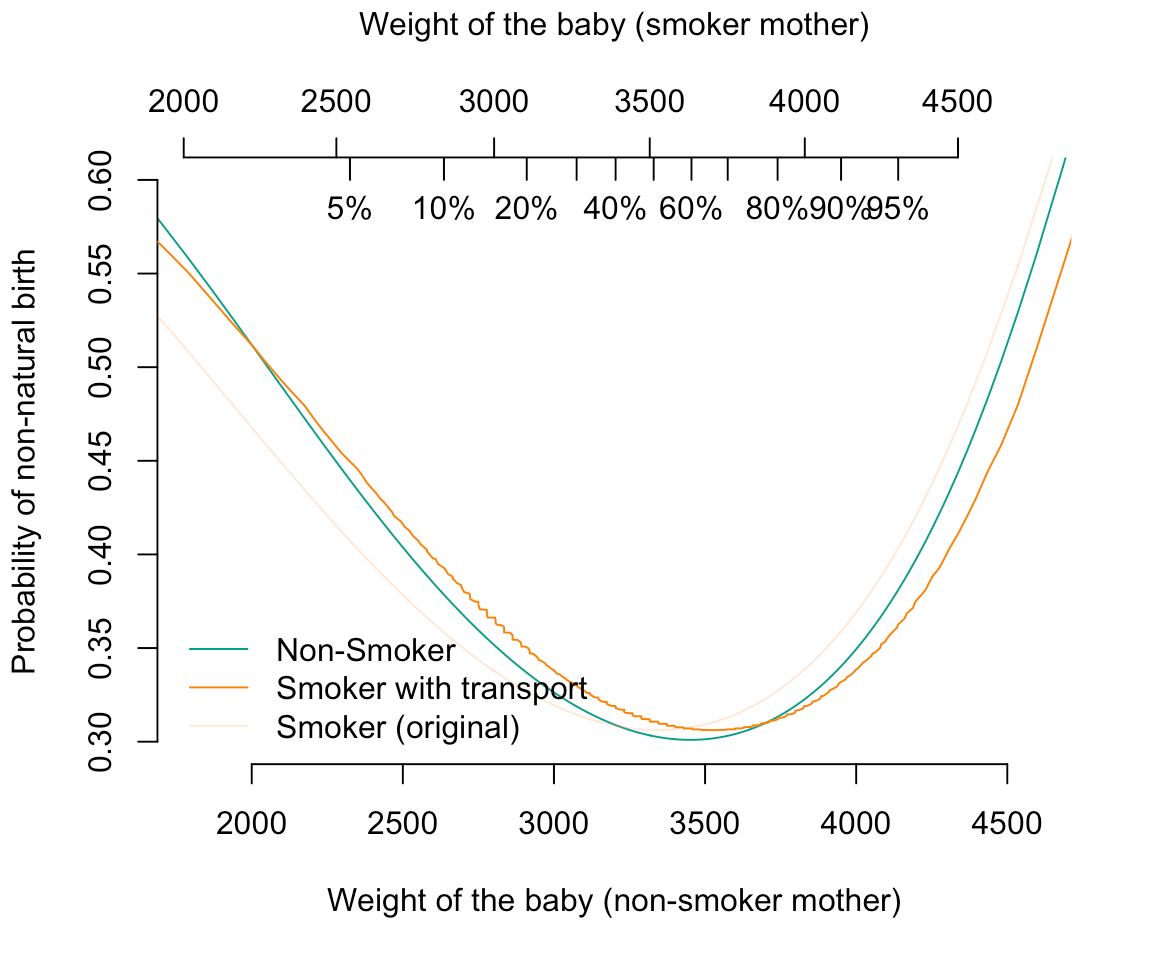}
     \includegraphics[width=.49\textwidth]{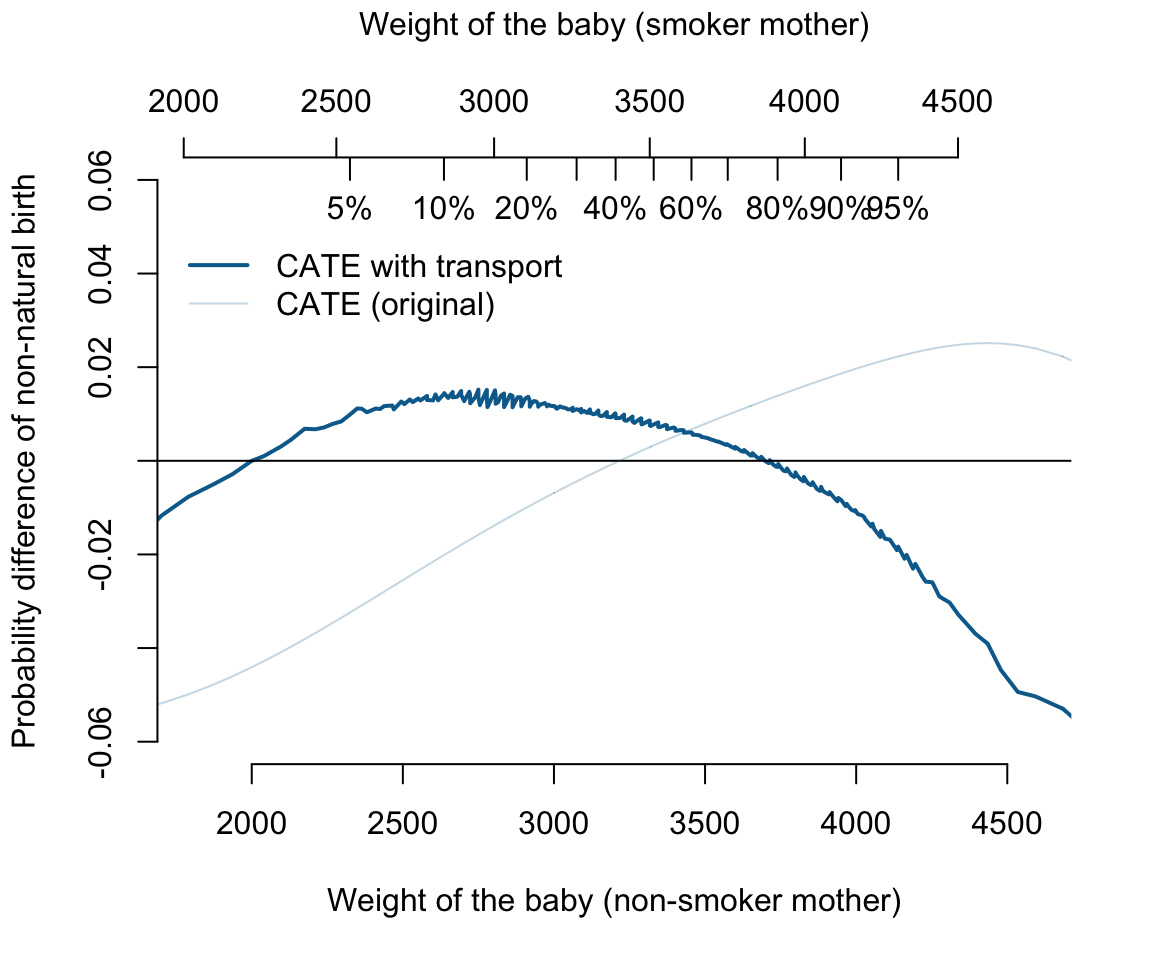}

    % \centering        
    % \includegraphics[width=.49\textwidth]{figures/gam-univariate-black-weight-compare-transport-1}
    %  \includegraphics[width=.49\textwidth]{figures/gam-univariate-diff-CATE-black-weight-compare-transport-1}

    \centering
     \includegraphics[width=.49\textwidth]{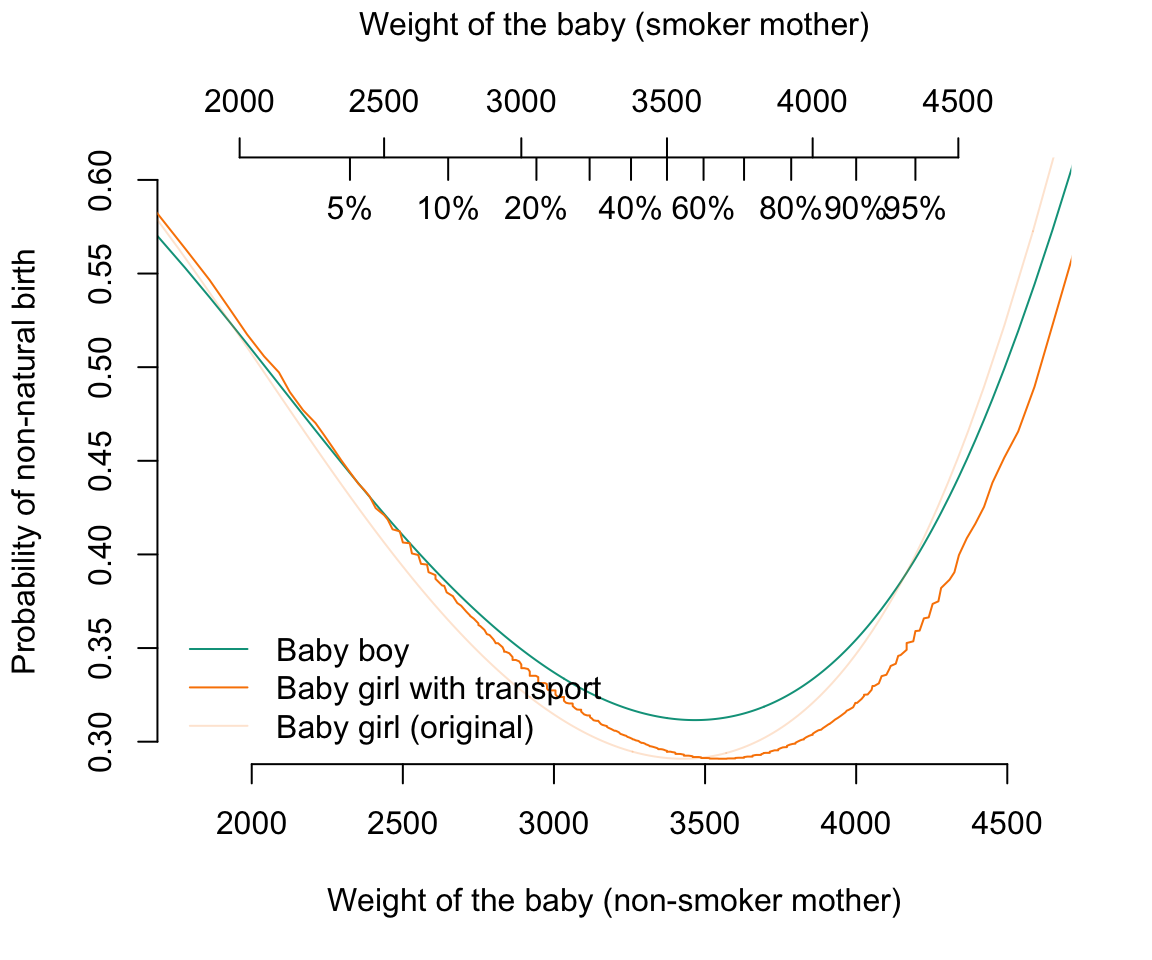}
     \includegraphics[width=.49\textwidth]{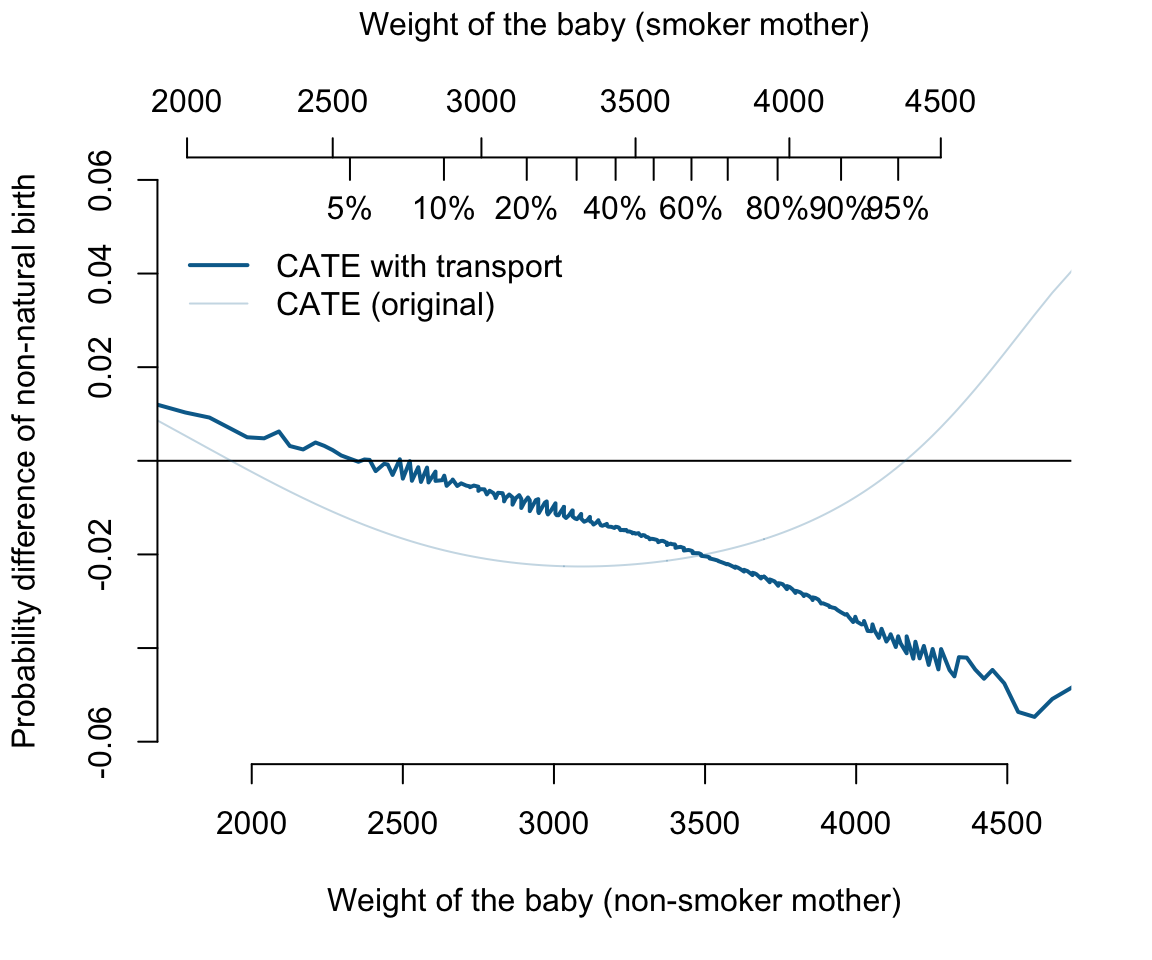}

    \caption{On the left, evolution of $x\mapsto\mathbb{E}[Y|X_{T\leftarrow t}=x,T=t]$, estimated using a logistic GAM model, when $Y=\boldsymbol{1}(\text{non-natural delivery})$, and $X$ is the weight of the newborn infant, respectively when $T$ indicates whether the mother is a smoker or not (on top), or whether the the newborn infant is a boy at the bottom. On the right, evolution of  $x\mapsto\text{SCATE}[Y|X=x]$.}
    \label{fig:CATE-compare-quantiles-2x3-weight:appendix}
\end{figure}

\begin{figure}[!ht]
    \centering
     \includegraphics[width=.49\textwidth]{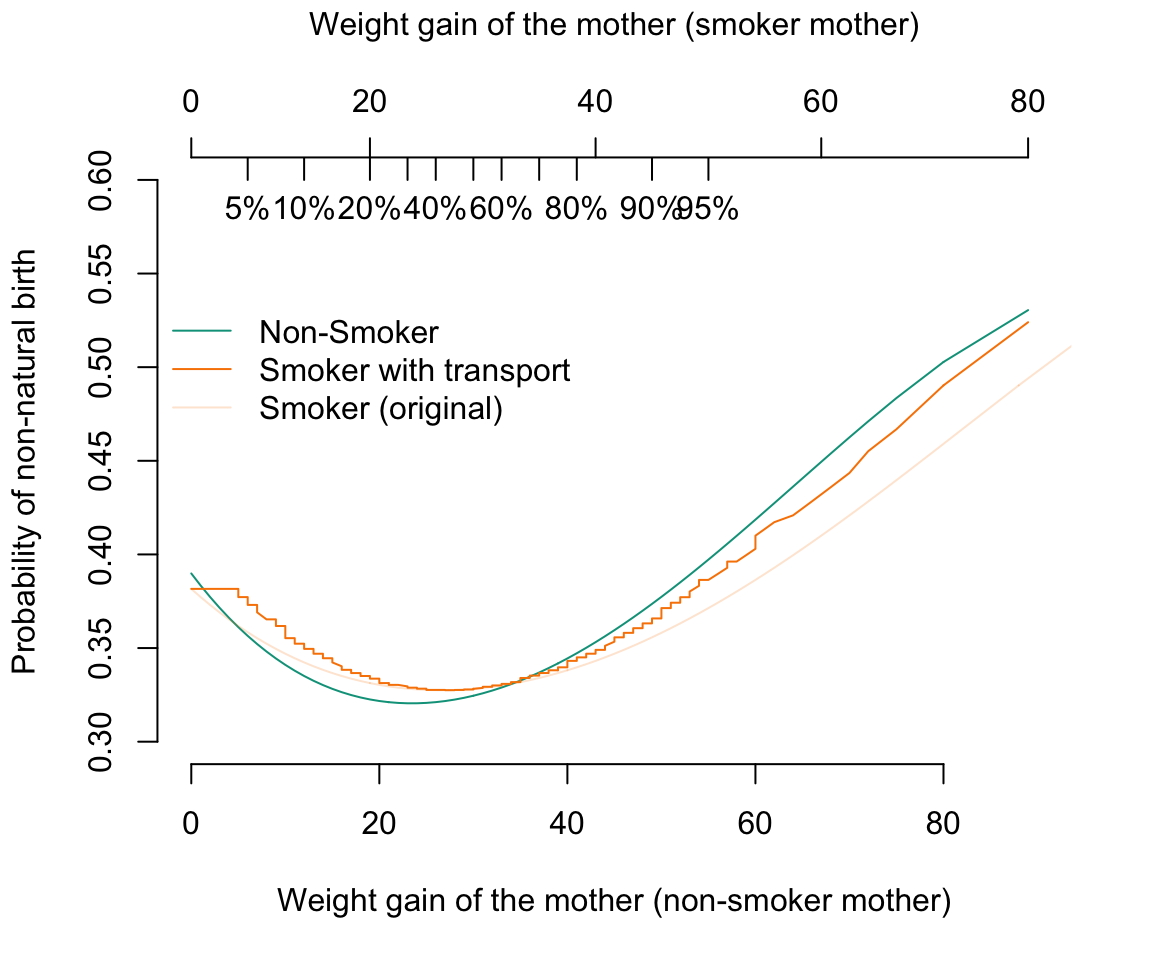}
     \includegraphics[width=.49\textwidth]{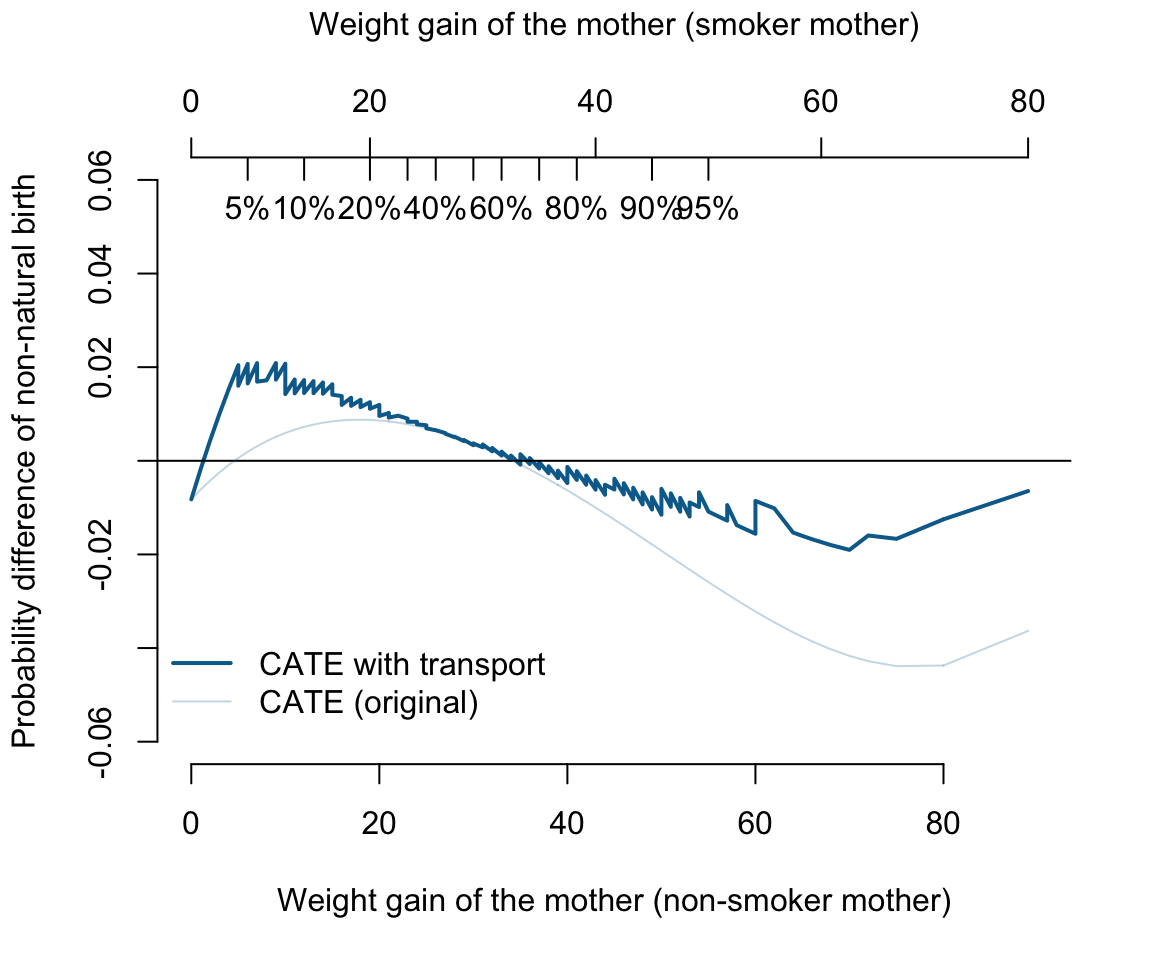}

    % \centering        
    % \includegraphics[width=.49\textwidth]{figures/gam-univariate-black-weight-gain-compare-transport-1}
    %  \includegraphics[width=.49\textwidth]{figures/gam-univariate-diff-CATE-black-weight-gain-compare-transport-1}

    \centering
     \includegraphics[width=.49\textwidth]{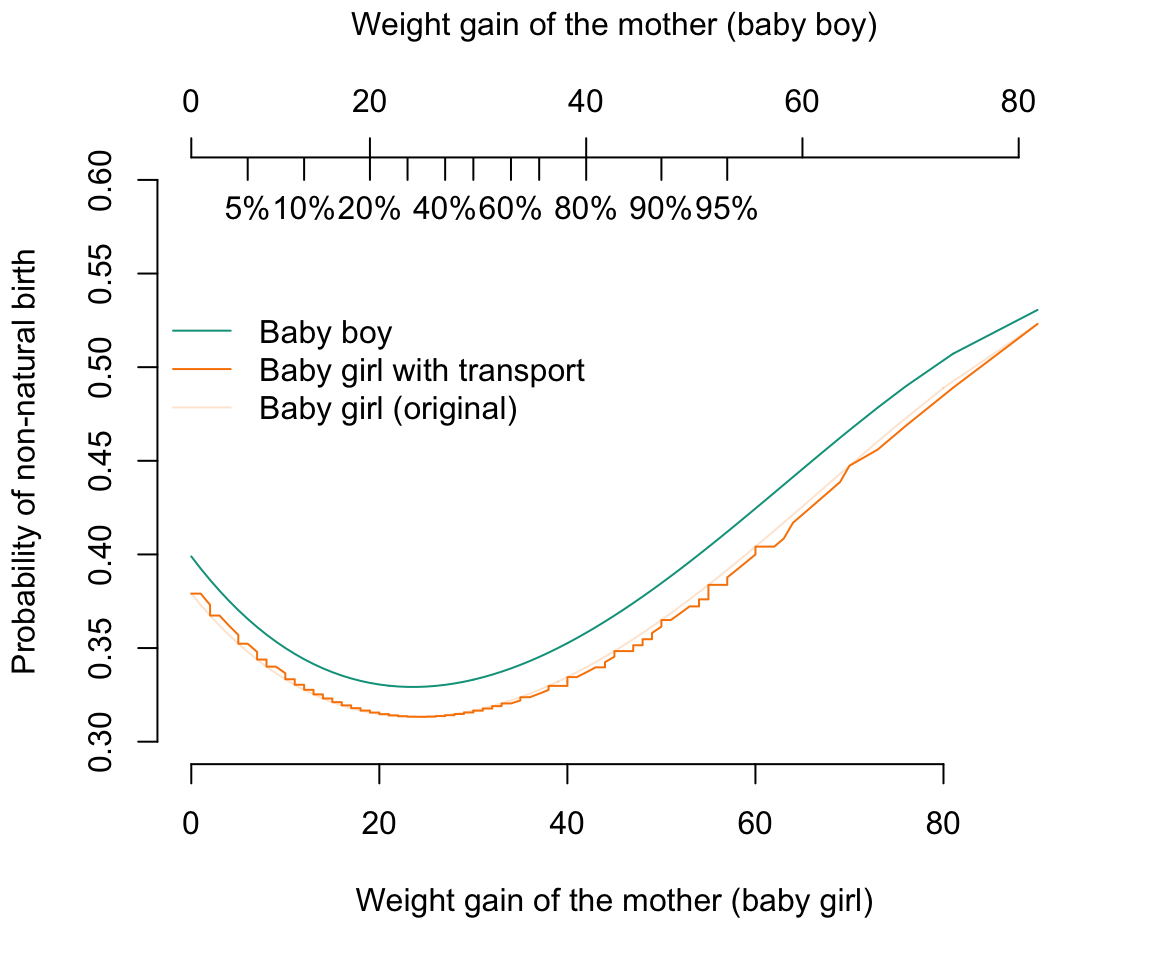}
     \includegraphics[width=.49\textwidth]{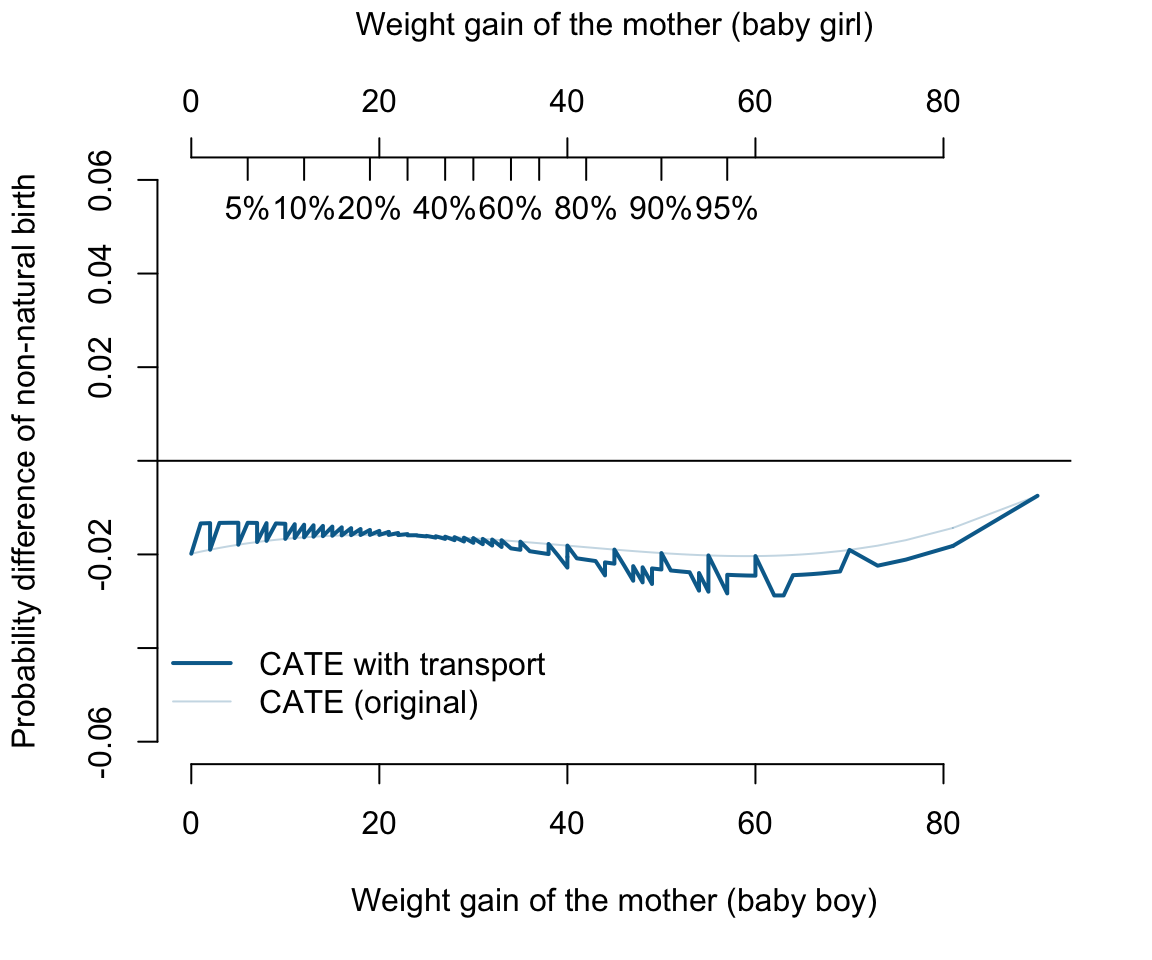}

    \caption{On the left, evolution of $x\mapsto\mathbb{E}[Y|X_{T\leftarrow t}=x,T=t]$, estimated using a logistic GAM model, when $Y=\boldsymbol{1}(\text{non-natural delivery})$, and $X$ is the weight gain of the mother, respectively when $T$ indicates whether the mother is a smoker or not on top, or weather the newborn infant is a girl or not, below. On the right, evolution of  $x\mapsto\text{SCATE}[Y|X=x]$.}
    \label{fig:CATE-compare-quantiles-2x3-weight-gain:appendix}
\end{figure}

\begin{figure}[!ht]
    \centering
     \includegraphics[width=.49\textwidth]{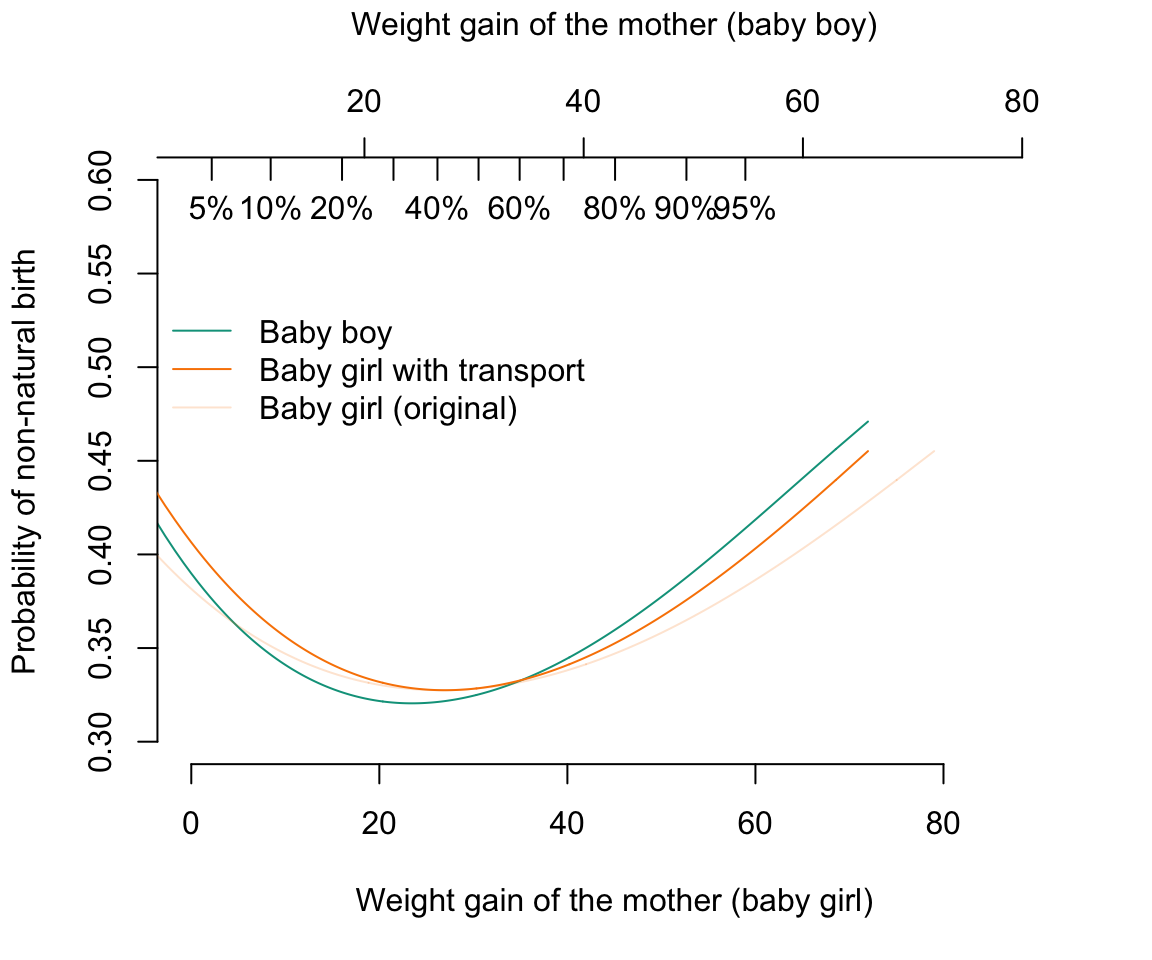}
     \includegraphics[width=.49\textwidth]{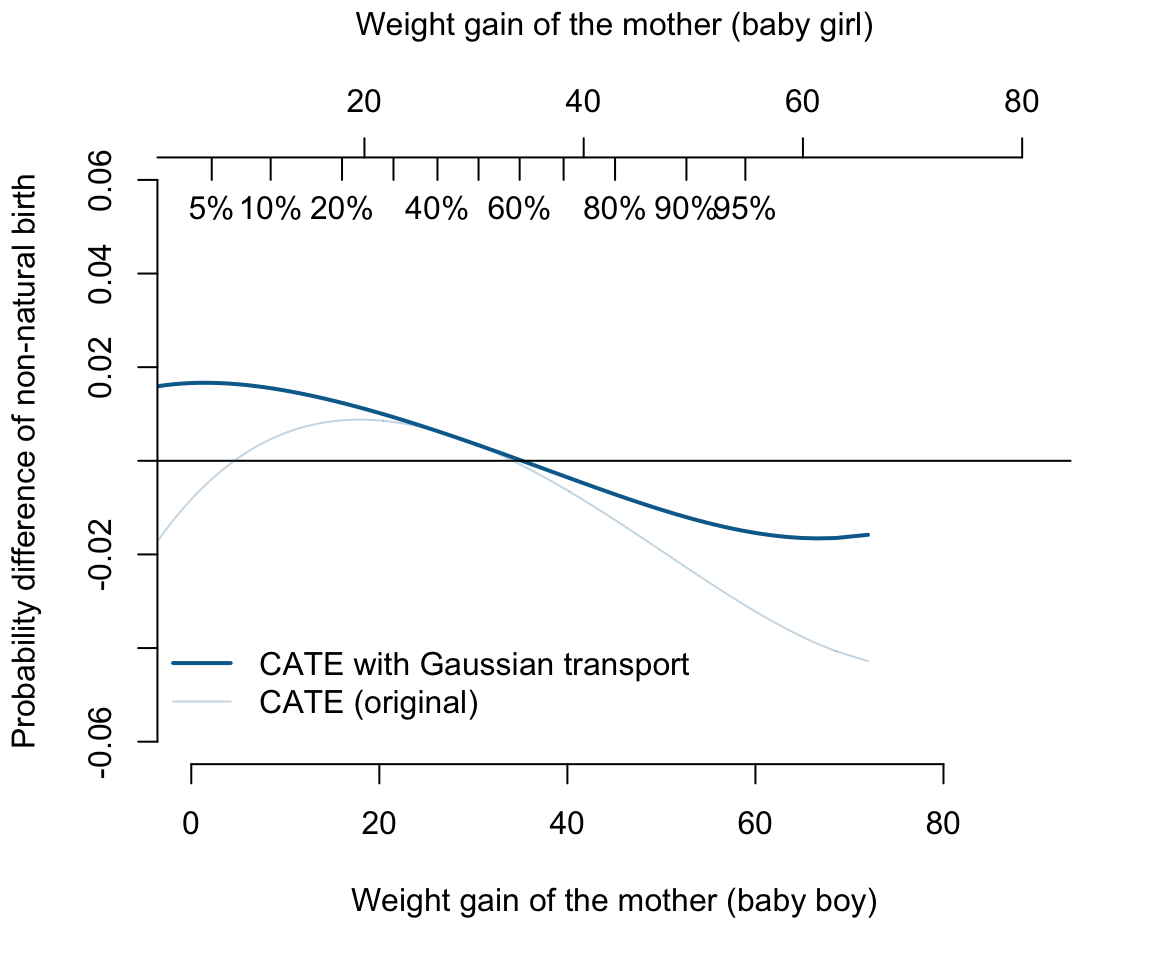}

    \caption{On the left, evolution of $x\mapsto\mathbb{E}[Y|X_{T\leftarrow t}=x,T=t]$, estimated using a logistic GAM model, when $Y=\boldsymbol{1}(\text{non-natural delivery})$, and $X$ is the weight gain of the mother, respectively when $T$ indicates whether the mother is a smoker or not. On the right, evolution of  $x\mapsto\text{SCATE}_{\mathcal{N}}[Y|X=x]$.}
    \label{fig:CATE-compare-gaussian-2x2-weight-gain:appendix}
\end{figure}

\begin{figure}[!ht]
    \centering
     \includegraphics[width=.49\textwidth]{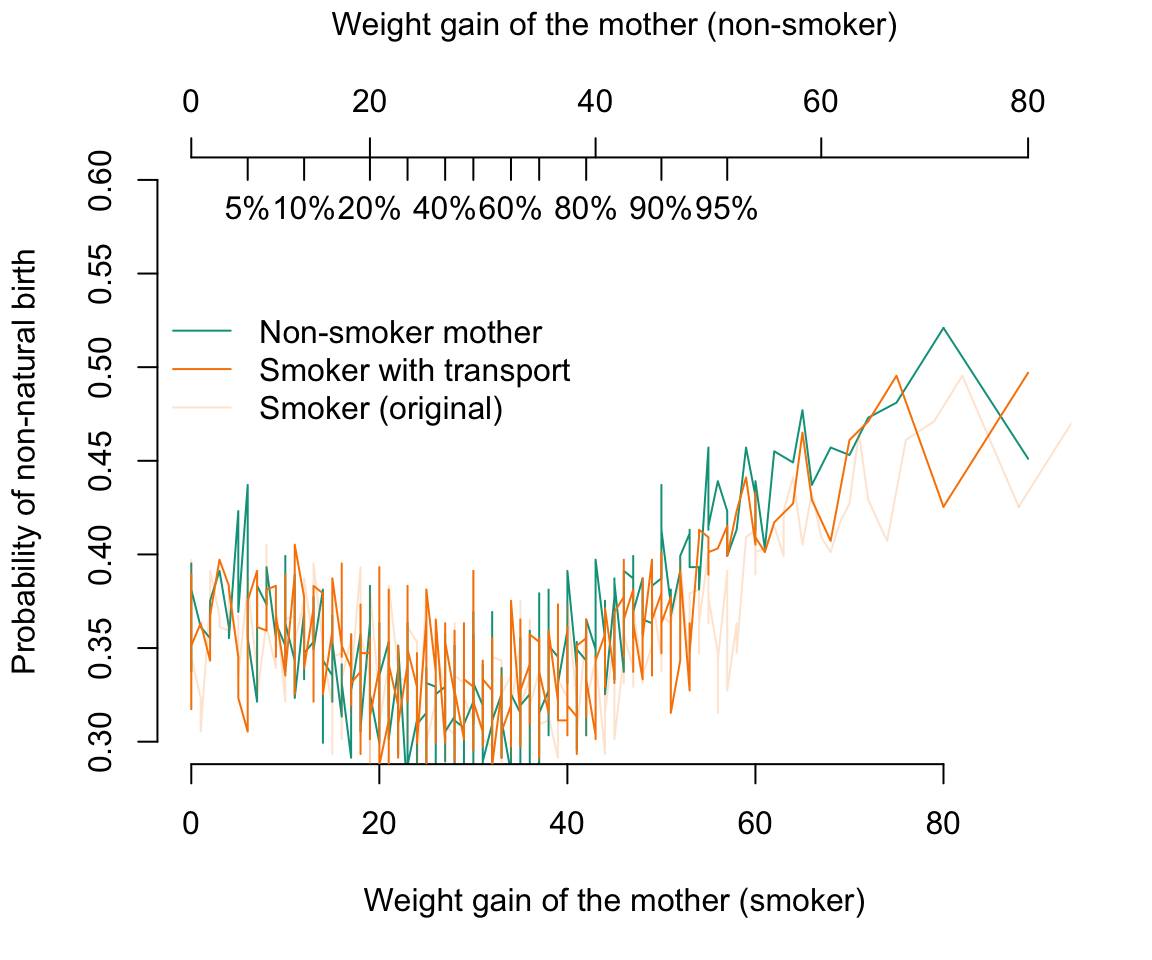}
     \includegraphics[width=.49\textwidth]{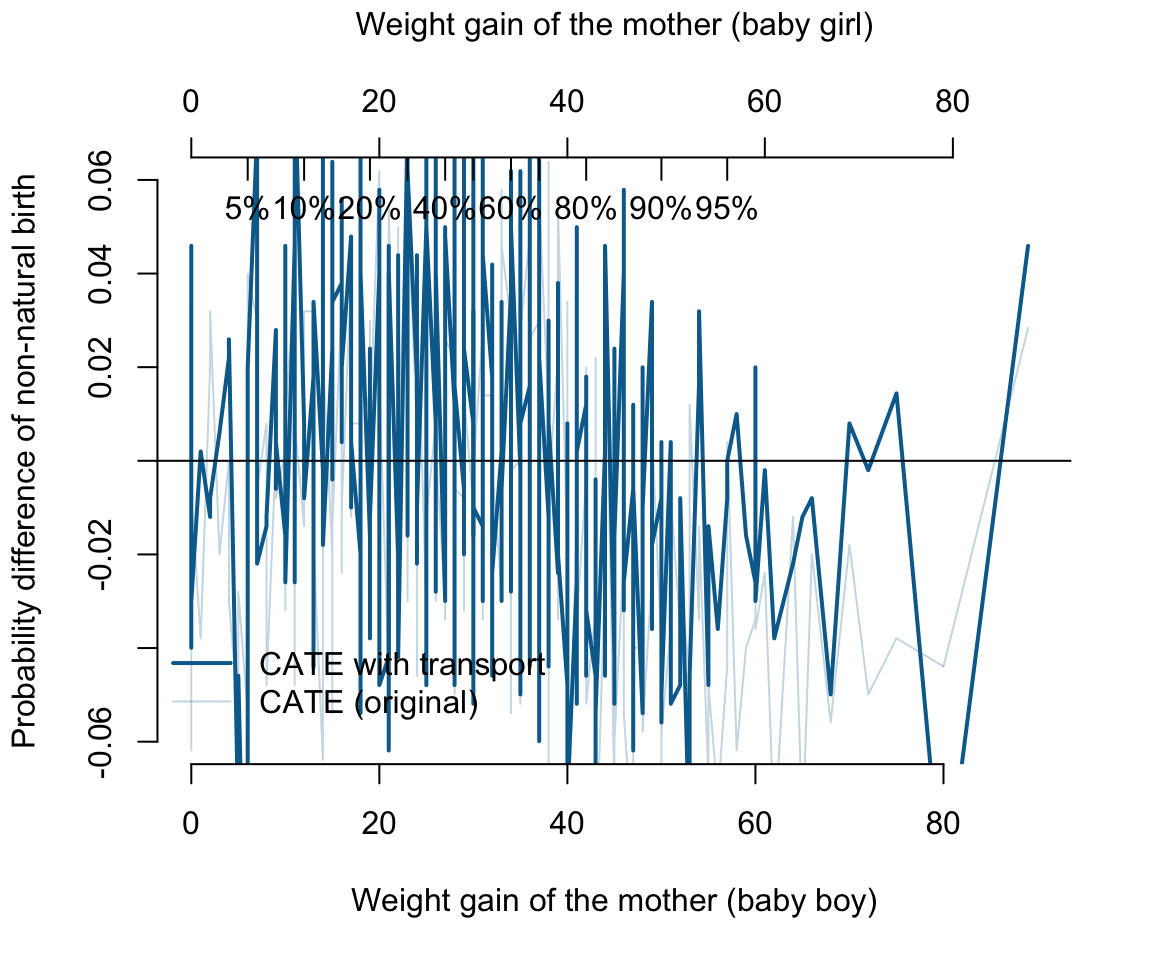}

    \caption{On the left, evolution of $x\mapsto\mathbb{E}[Y|X_{T\leftarrow t}=x,T=t]$, estimated using $k$-nearest neighbors, when $Y=\boldsymbol{1}(\text{non-natural delivery})$, and $X$ is the weight gain of the mother, when $T$ indicates whether the mother is a smoker or not. On the right, evolution of  $x\mapsto\text{SCATE}_{\mathcal{N}}[Y|X=x]$ with and without transport.}
    \label{fig:CATE-compare-knn-2x1-weight-gain:appendix}
\end{figure}

\begin{figure}[!ht]
    \centering
     \includegraphics[width=.49\textwidth]{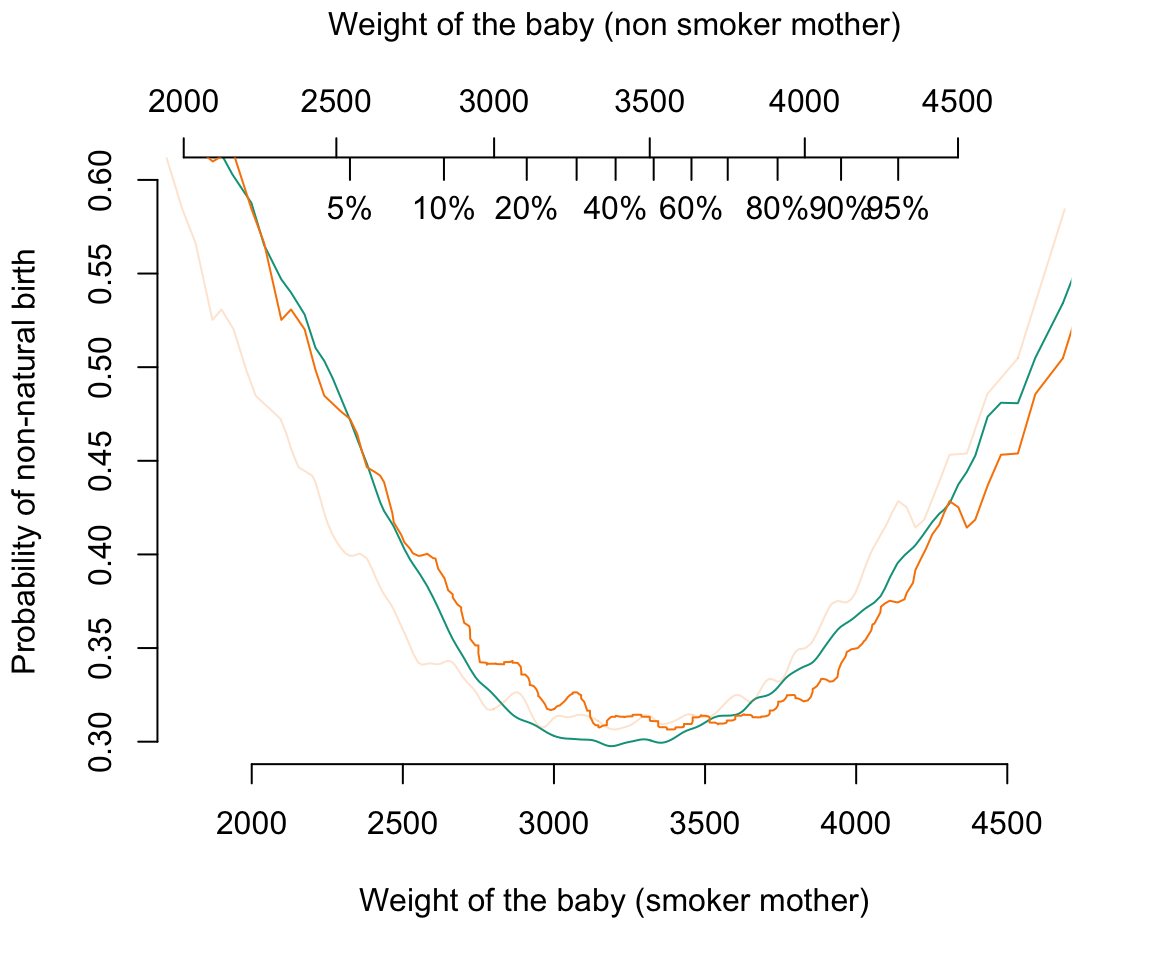}
     \includegraphics[width=.49\textwidth]{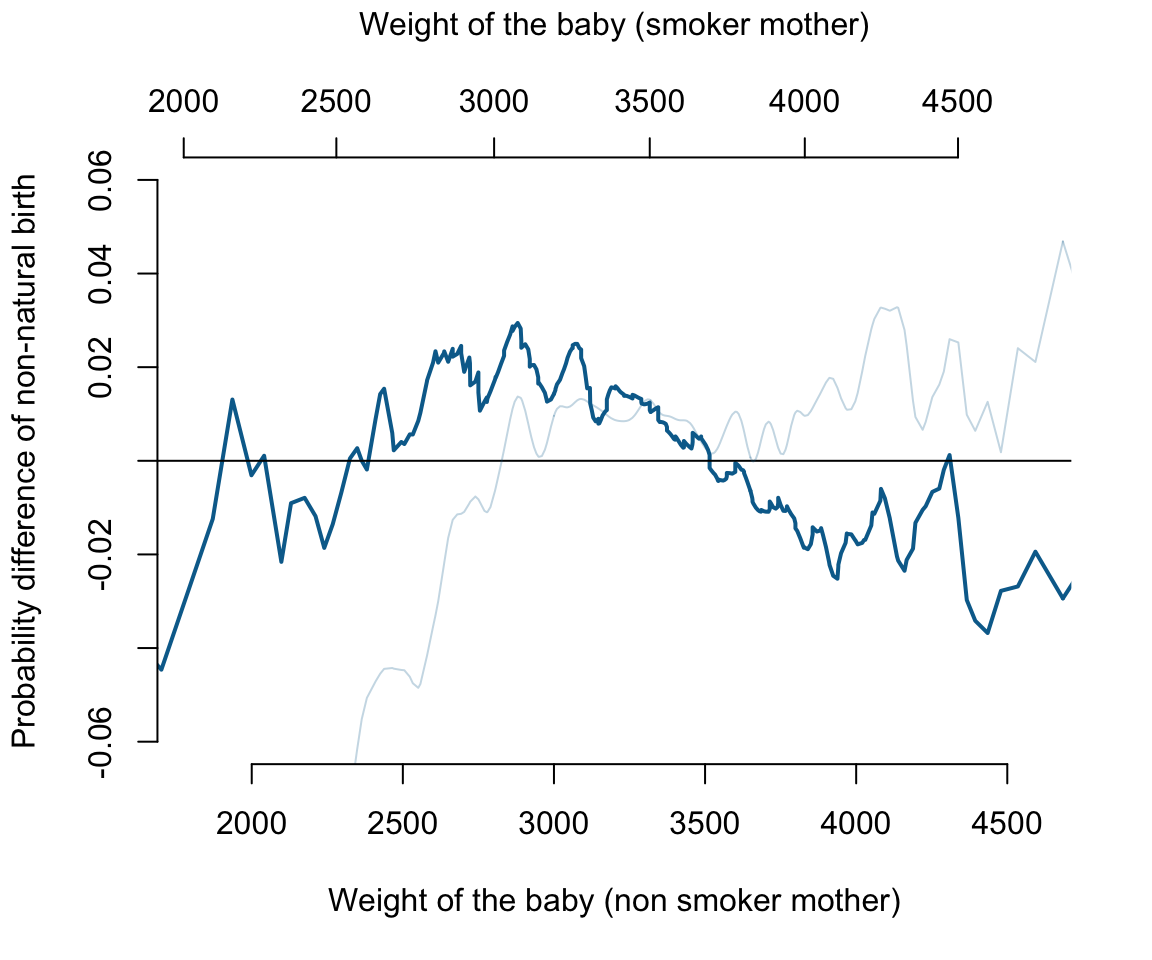}

    \centering        
    \includegraphics[width=.49\textwidth]{figures/kernel-univariate-black-weight-compare-gaussian-transport-1}
     \includegraphics[width=.49\textwidth]{figures/kernel-univariate-CATE-black-weight-compare-gaussian-transport-1.png}

    \centering
     \includegraphics[width=.49\textwidth]{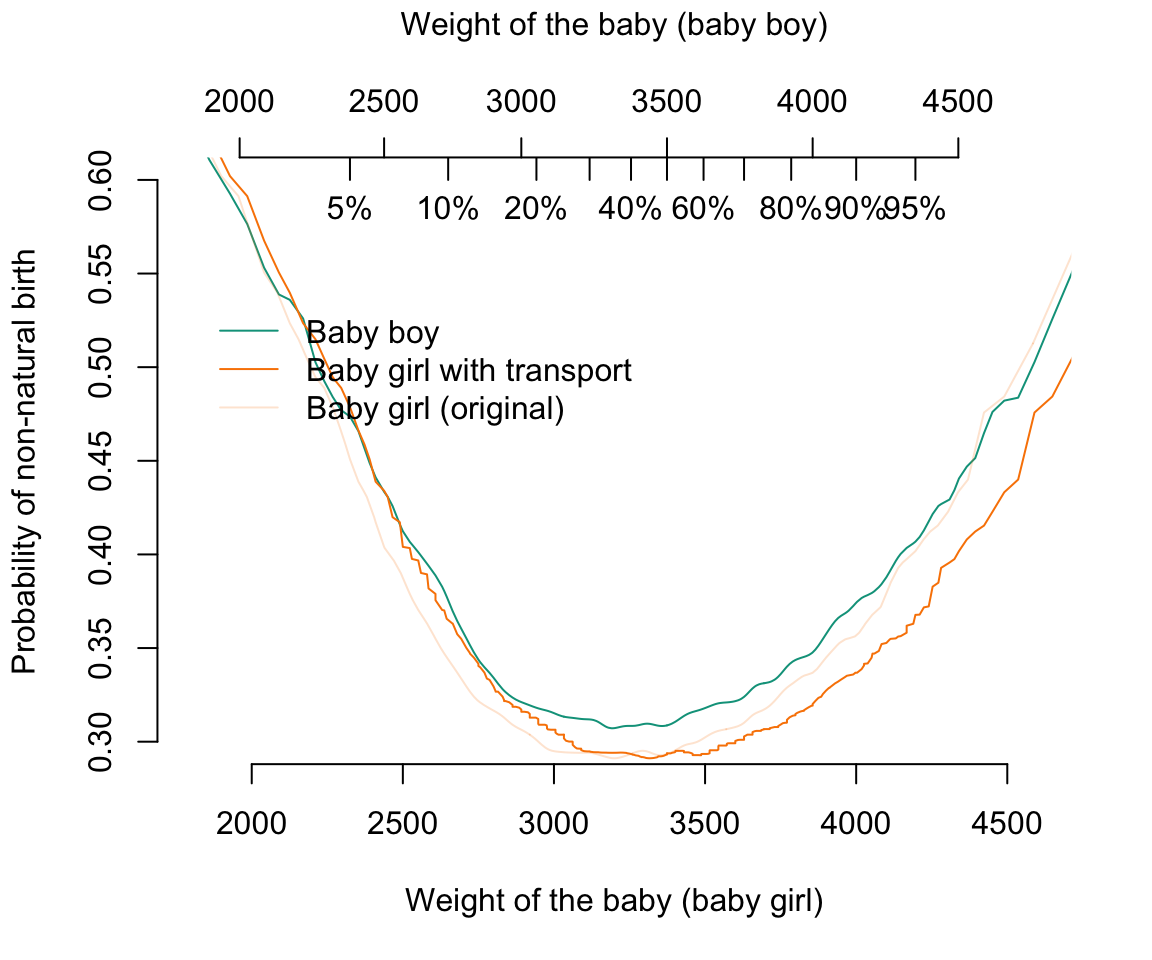}
     \includegraphics[width=.49\textwidth]{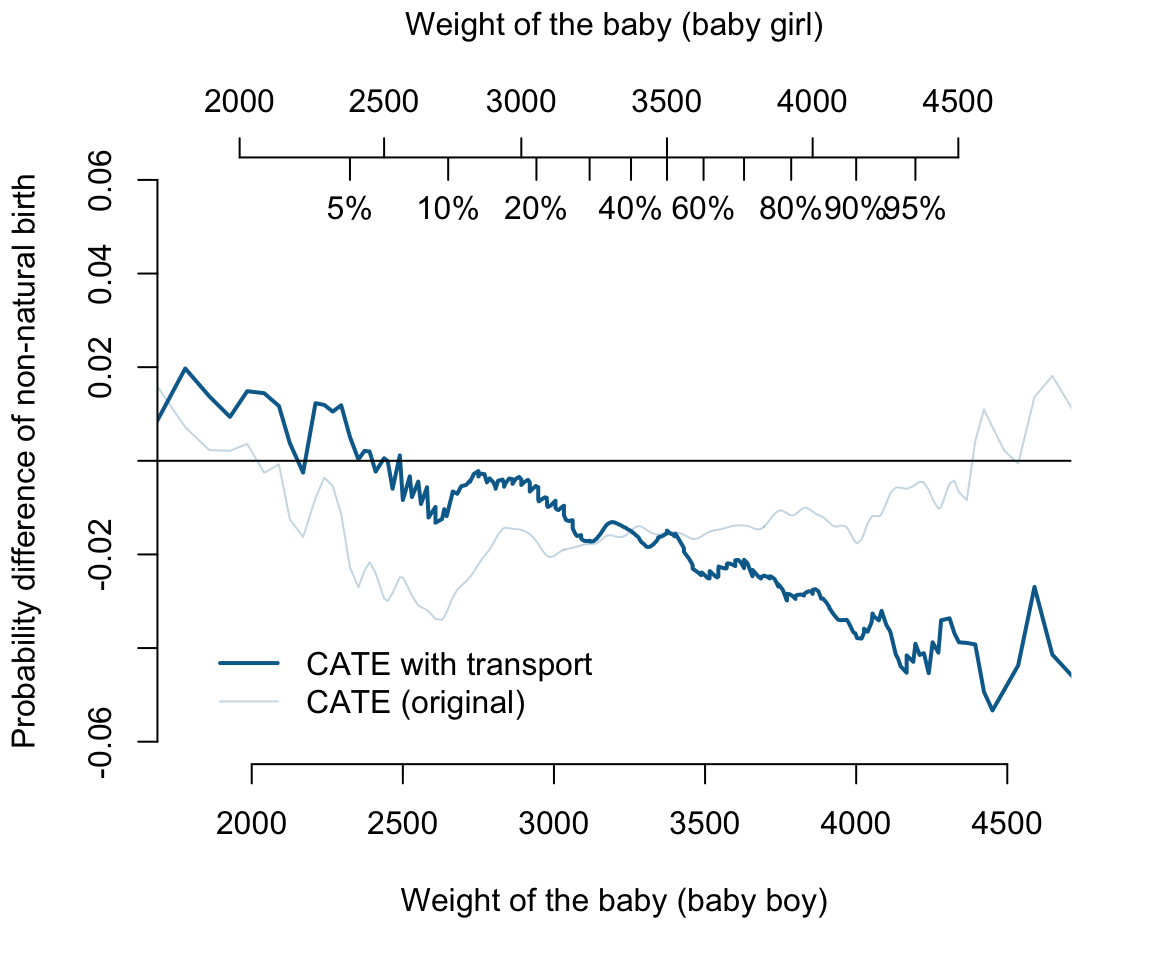}

    \caption{On the left, evolution of $x\mapsto\mathbb{E}[Y|X_{T\leftarrow t}=x,T=t]$, estimated using a kernel based local average, when $Y=\boldsymbol{1}(\text{non-natural delivery})$, and $X$ is the weight of the newborn infant, respectively when $T$ indicates whether the mother is a smoker or not (on top), whether the mother is Black or not in the middle, and whether the newborn infant is a boy at the bottom. On the right, evolution of  $x\mapsto\text{SCATE}_{\mathcal{N}}[Y|X=x]$ with an without transport, based on a Gaussian transport.}
    \label{fig:CATE-compare-kernel-2x3-weight:appendix}
\end{figure}

\begin{figure}[!ht]
    \centering
     \includegraphics[width=.49\textwidth]{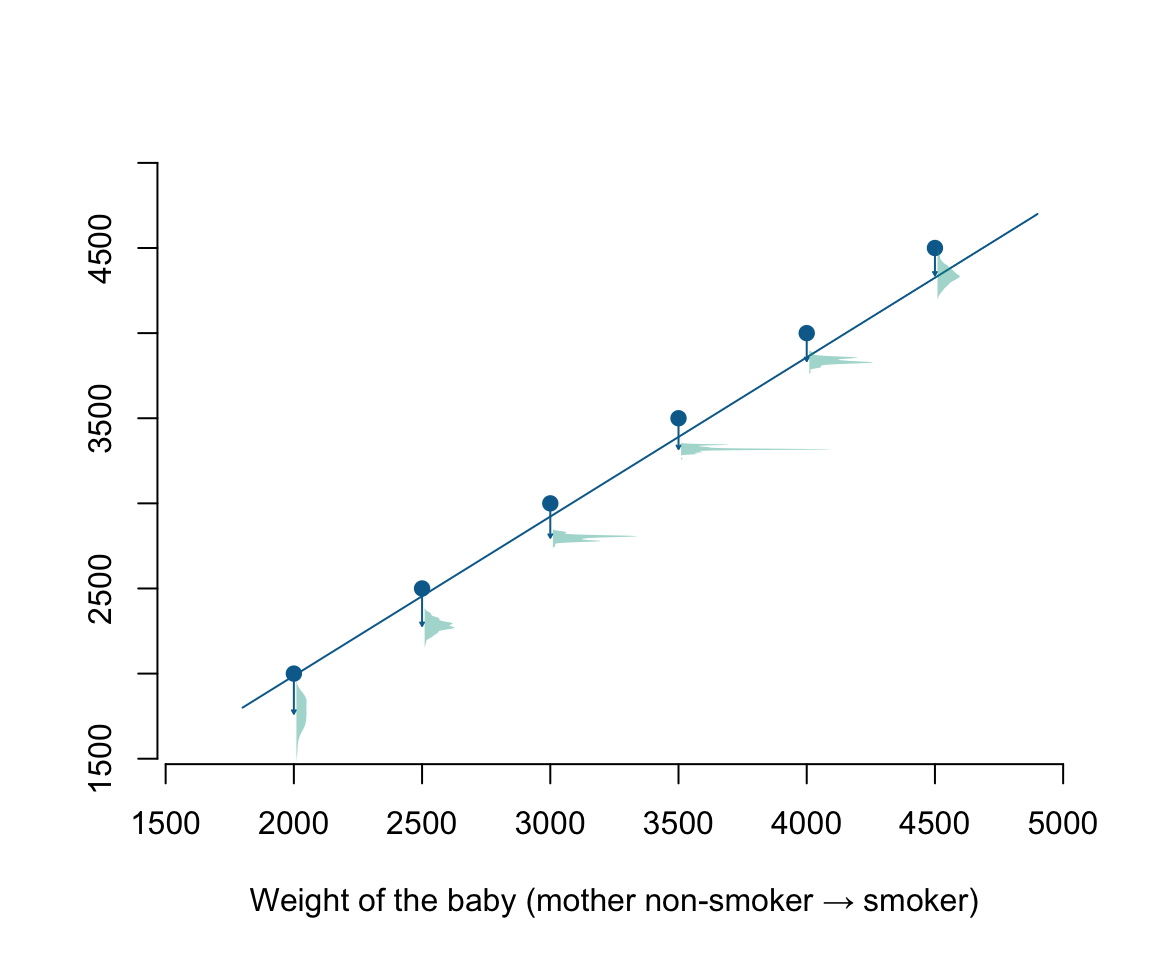}
     \includegraphics[width=.49\textwidth]{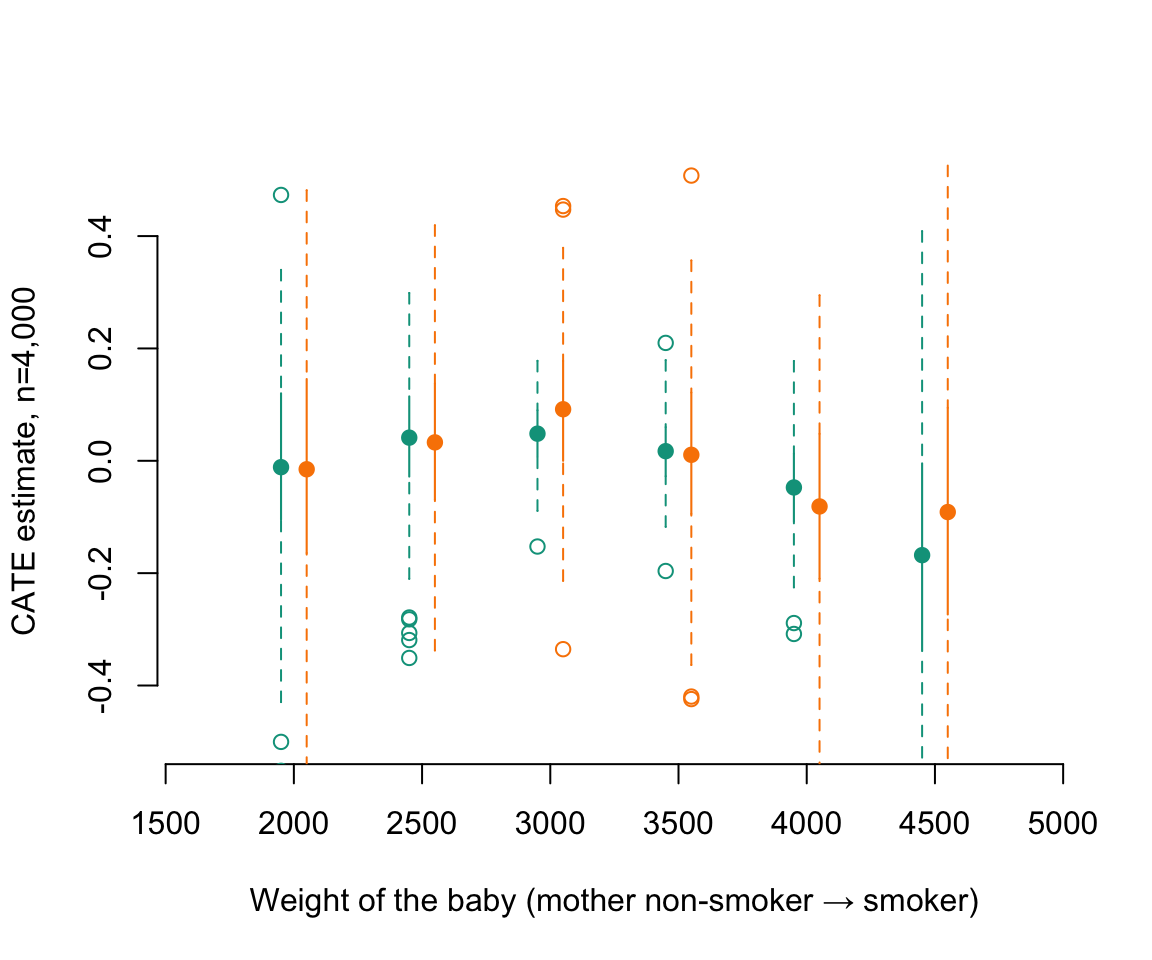}

    \centering        
    \includegraphics[width=.49\textwidth]{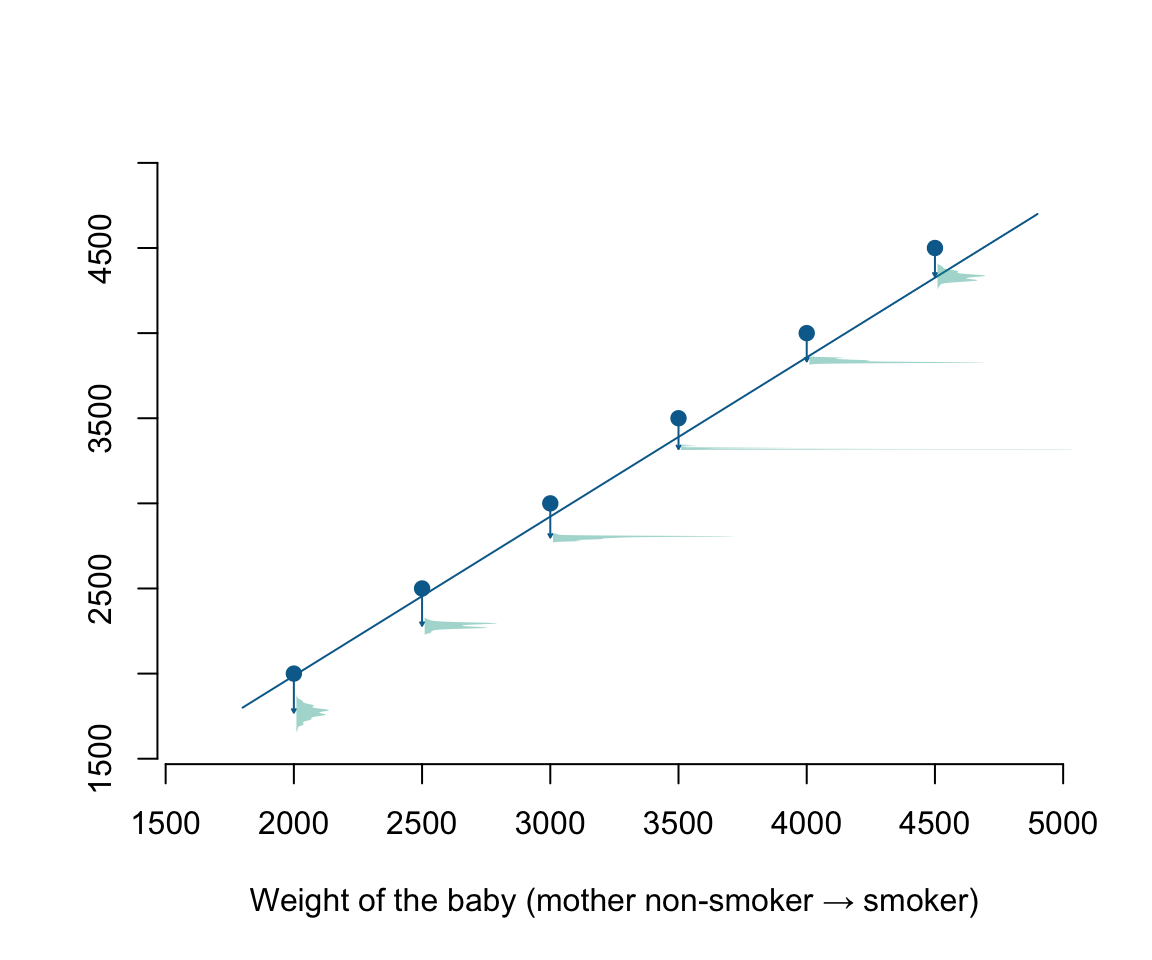}
     \includegraphics[width=.49\textwidth]{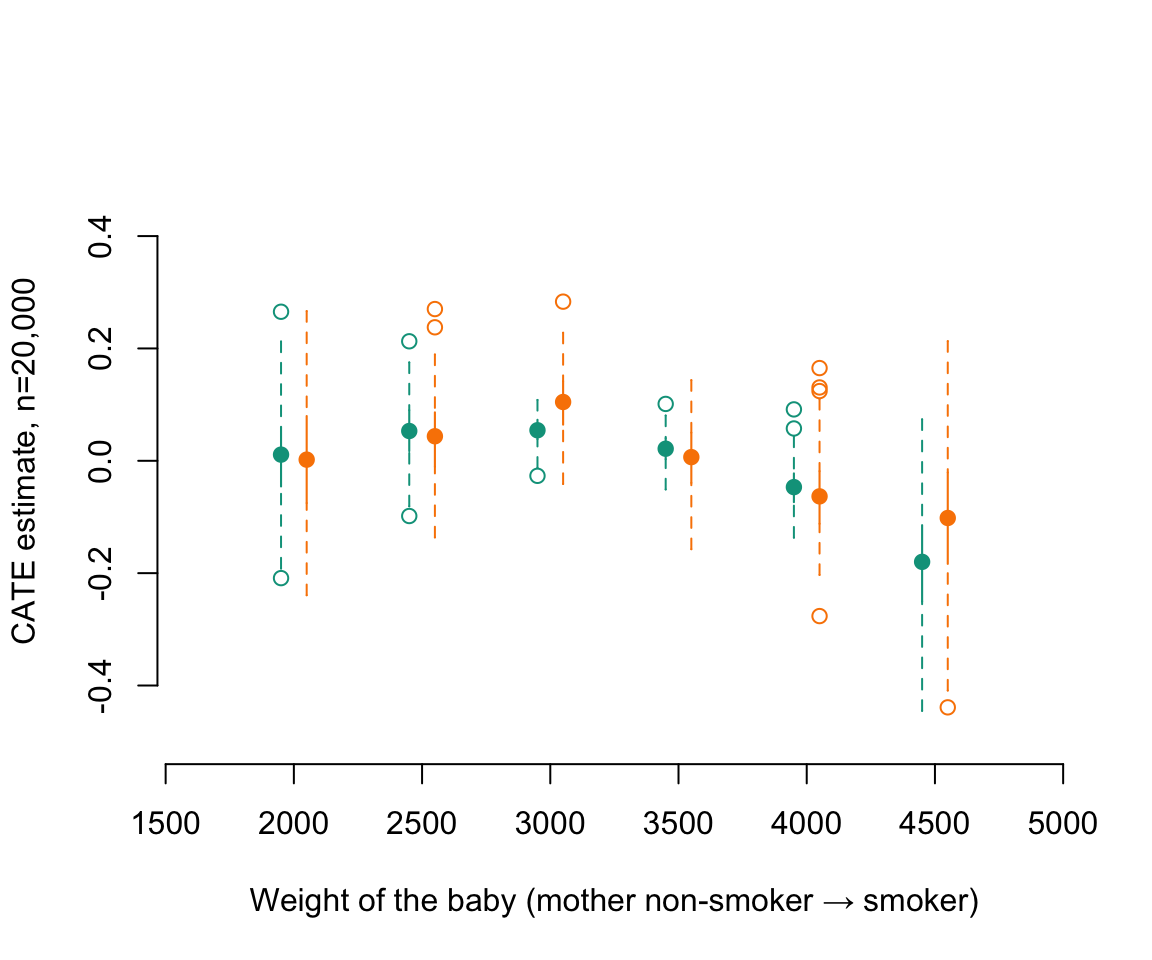}

    \centering
     \includegraphics[width=.49\textwidth]{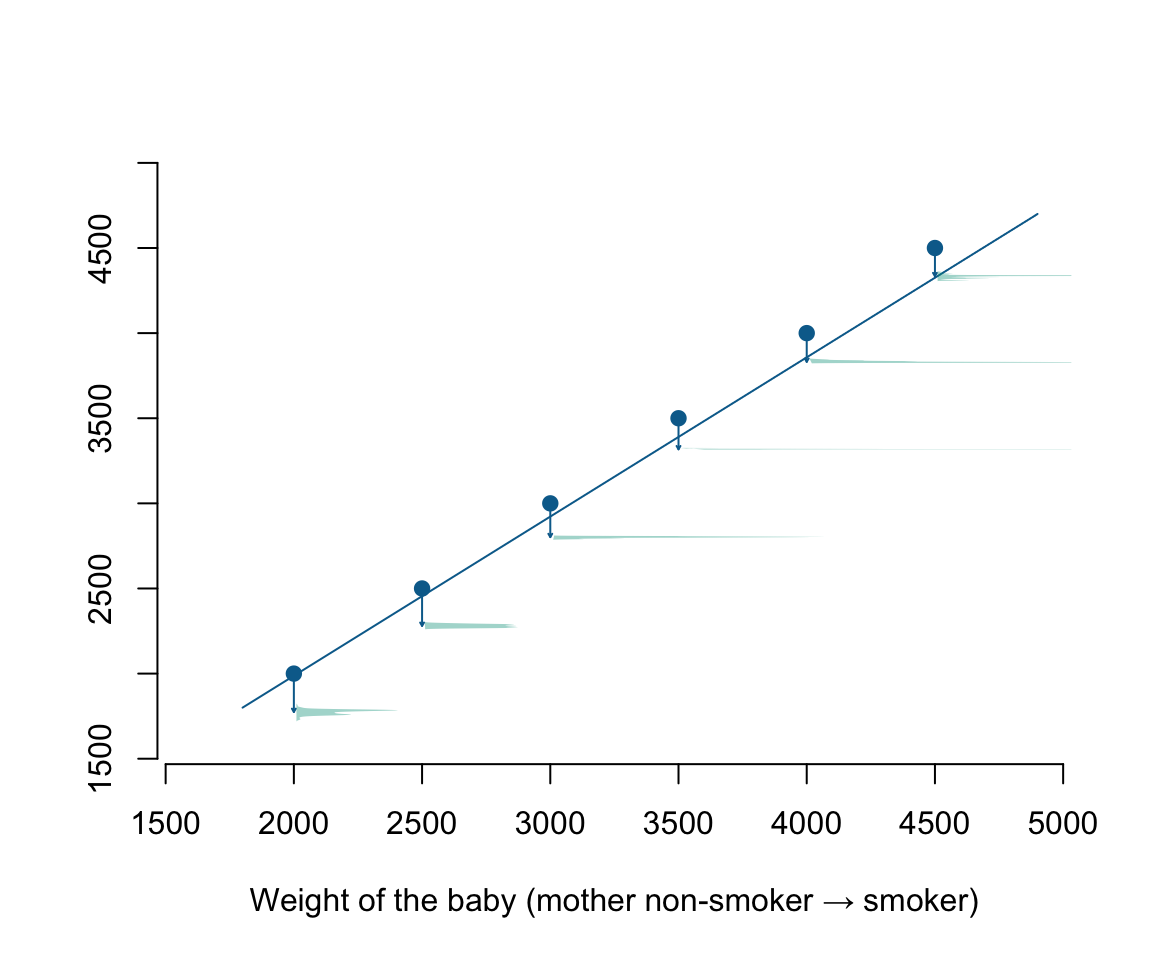}
     \includegraphics[width=.49\textwidth]{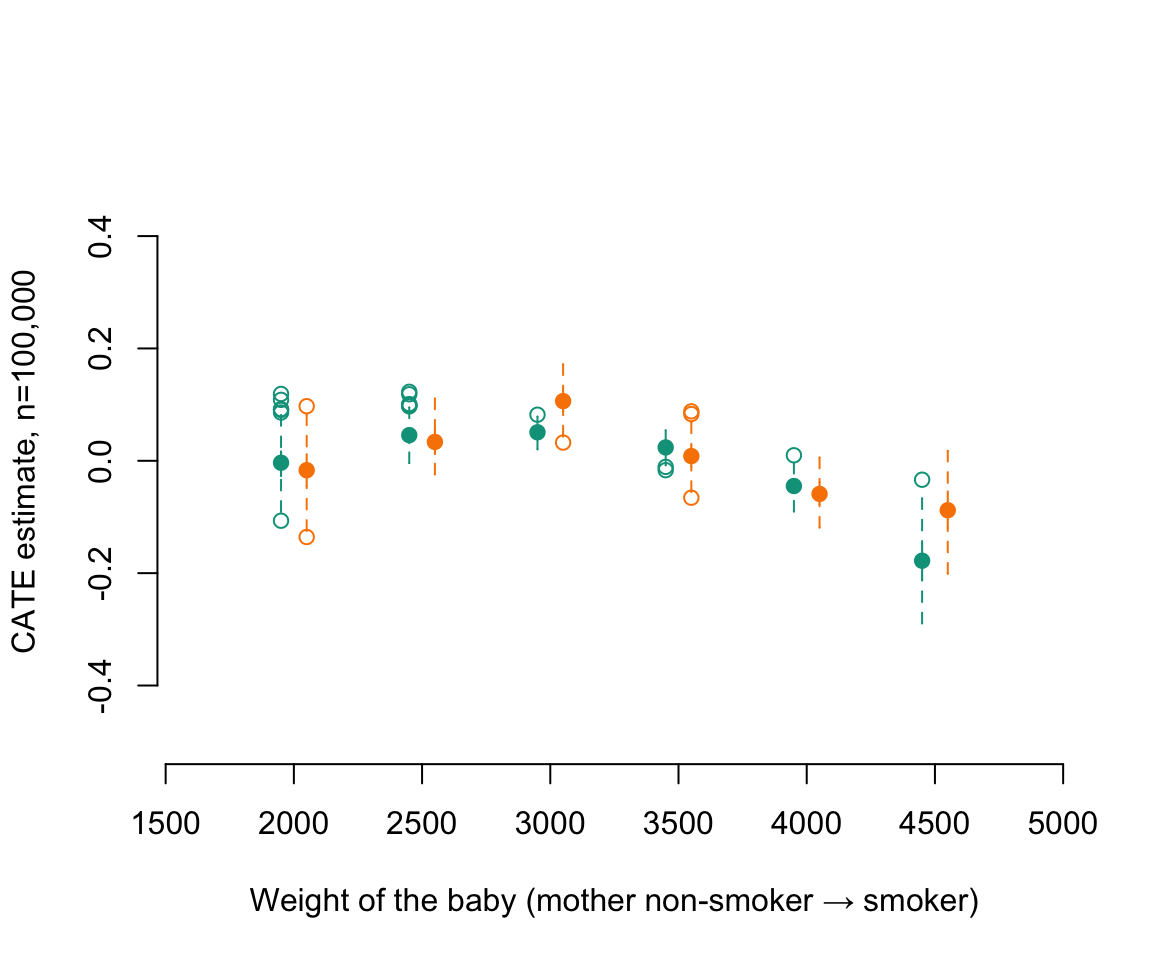}

    \caption{On the left, distribution of $\widehat{\mathcal{T}}(x)$, when $x$ is the weight of a newborn infant, when $n$ goes to $4,000$ (on top) to $20,000$ (in the middle) and $100,000$ (at the bottom), and when $T$ indicates whether the mother is a smoker or not. On the right, boxplots of the estimation of $\text{SCATE}(x)$, with two GAM models, when $x\in\{2000,2500,\cdots,4000,4500\}$.}
    \label{fig:CATE-sample-size-ranks-weight-smoker:appendix}
\end{figure}

\begin{figure}[!ht]
    \centering
     \includegraphics[width=.49\textwidth]{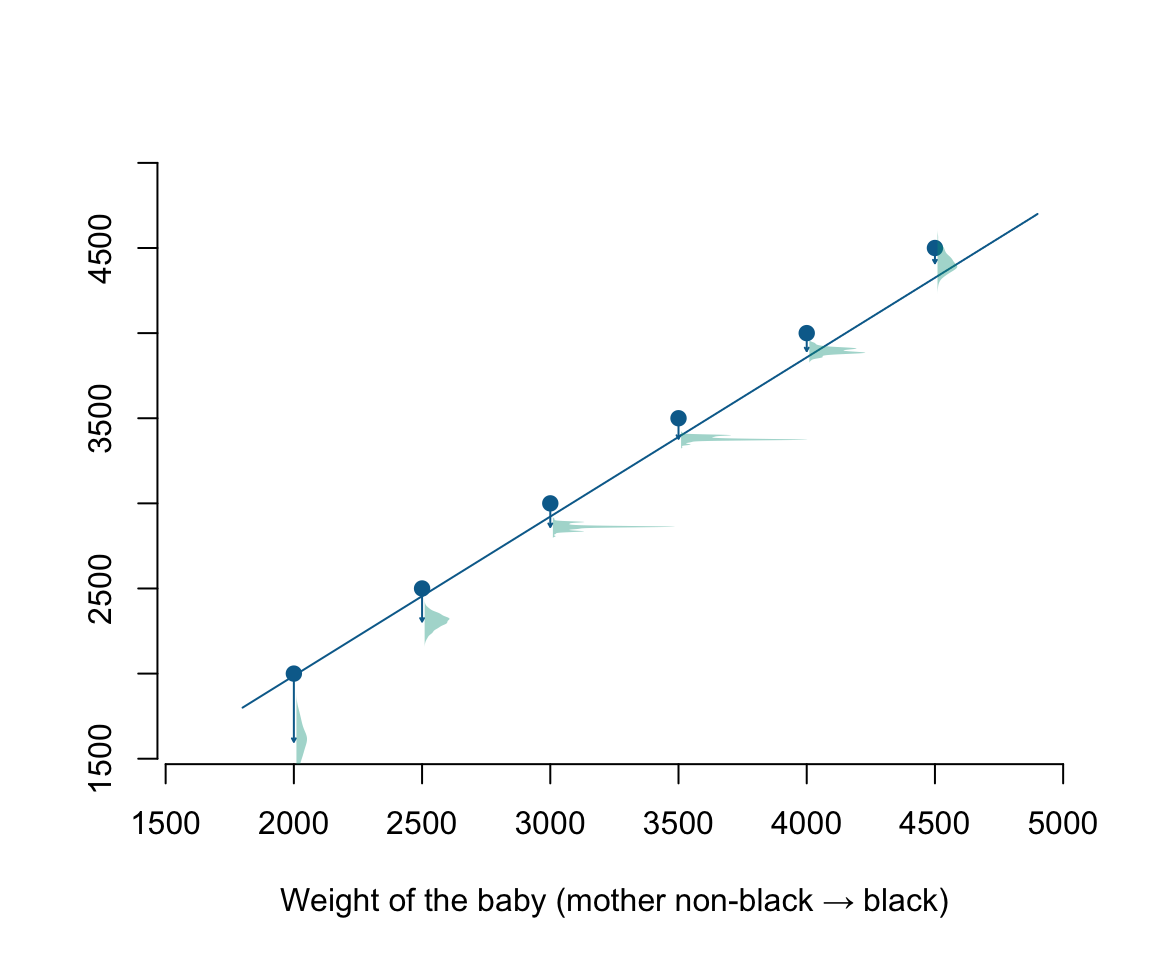}
     \includegraphics[width=.49\textwidth]{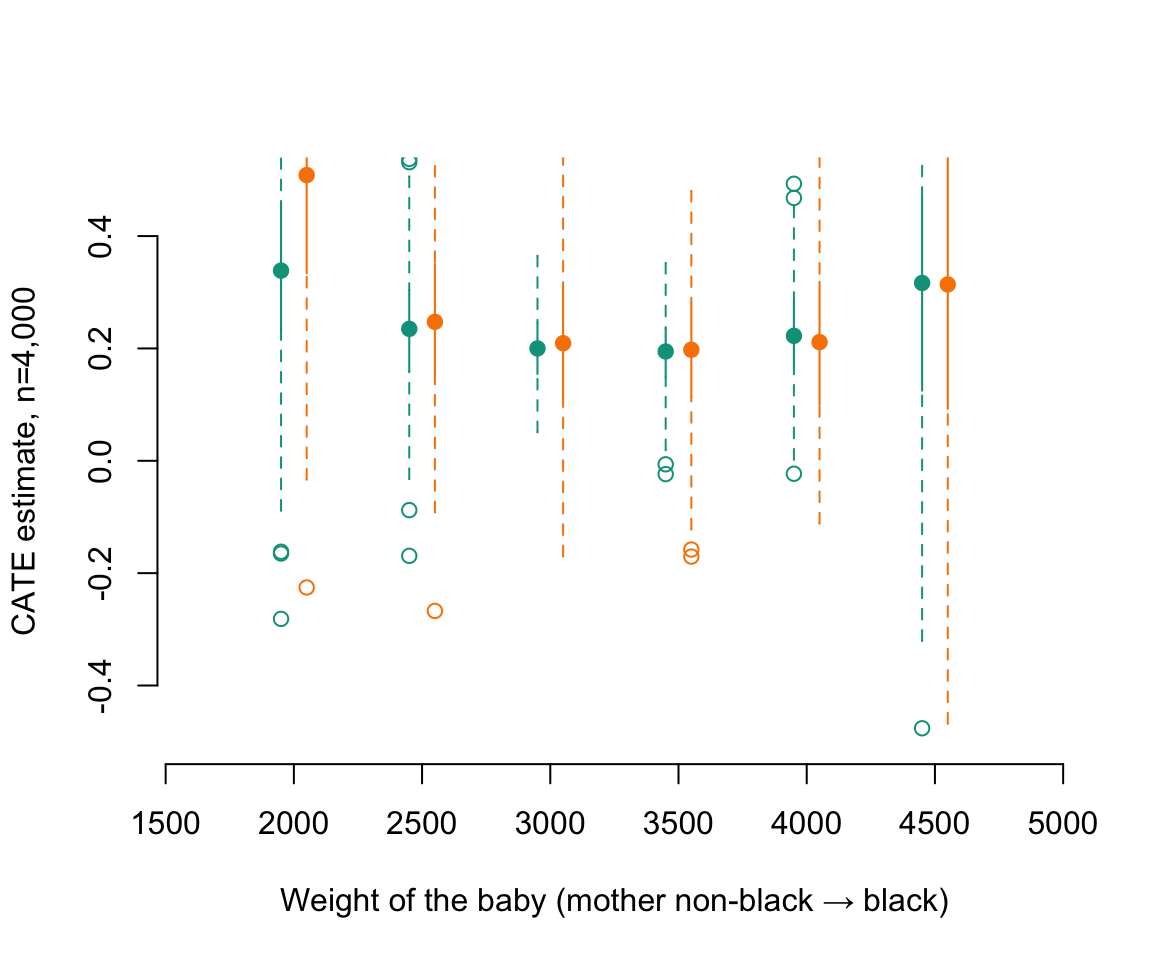}

    \centering        
    \includegraphics[width=.49\textwidth]{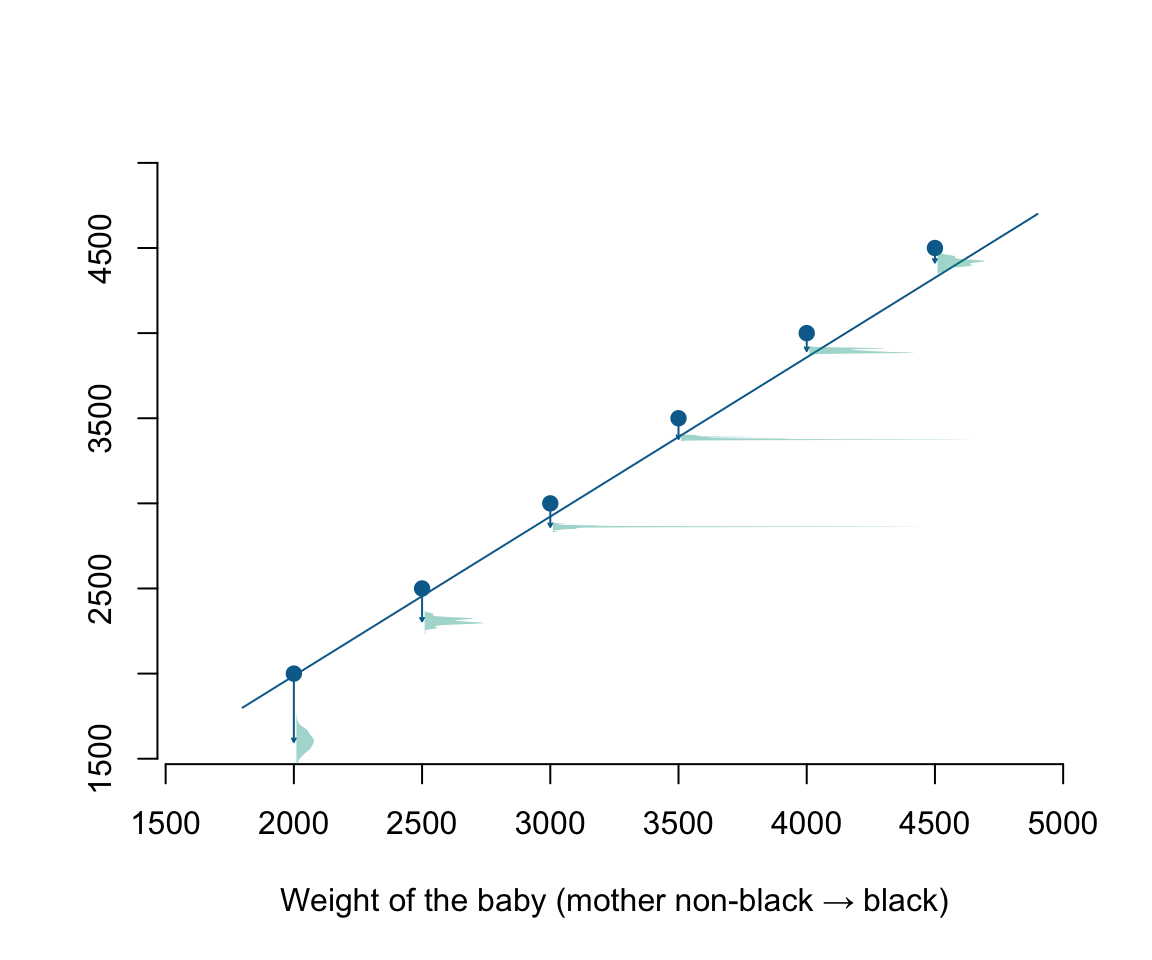}
     \includegraphics[width=.49\textwidth]{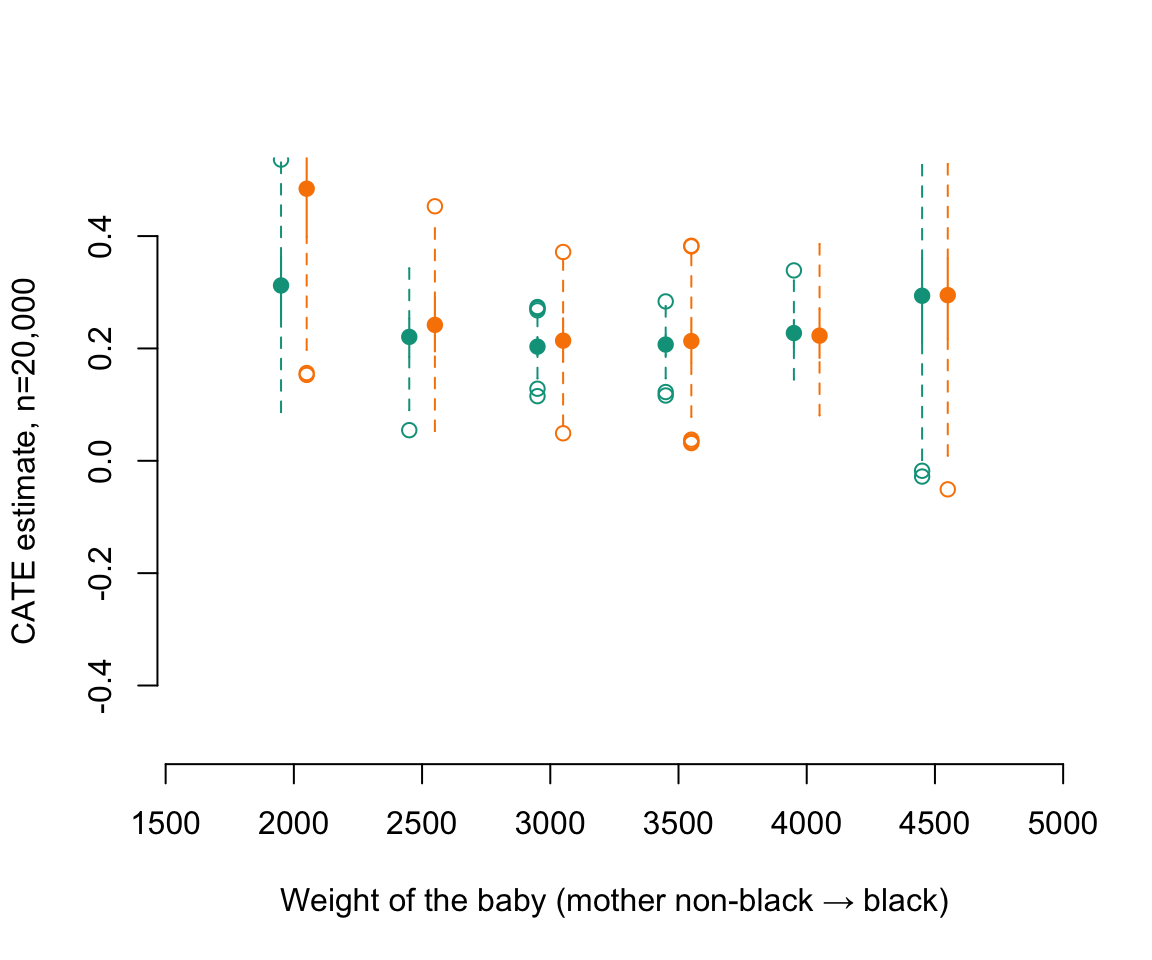}

    \centering
     \includegraphics[width=.49\textwidth]{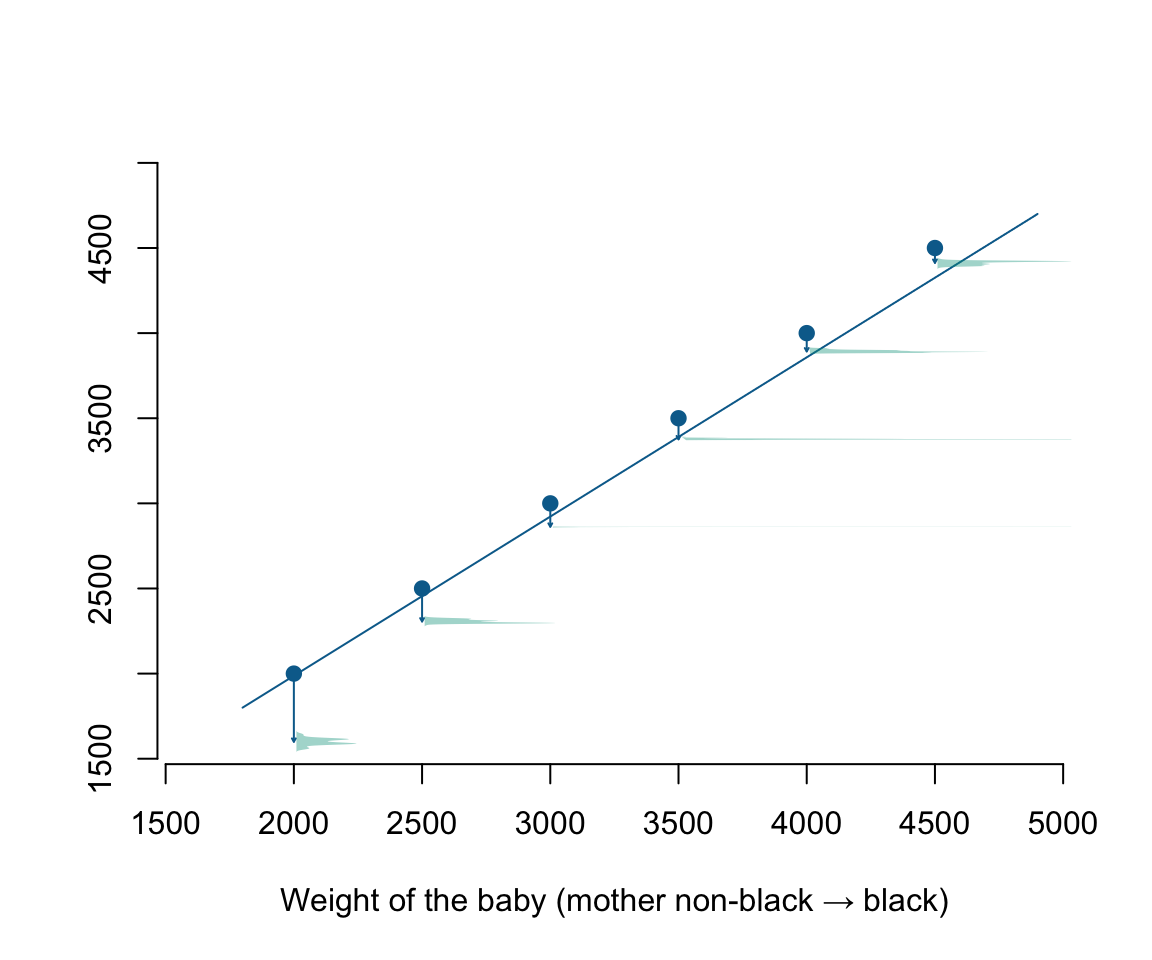}
     \includegraphics[width=.49\textwidth]{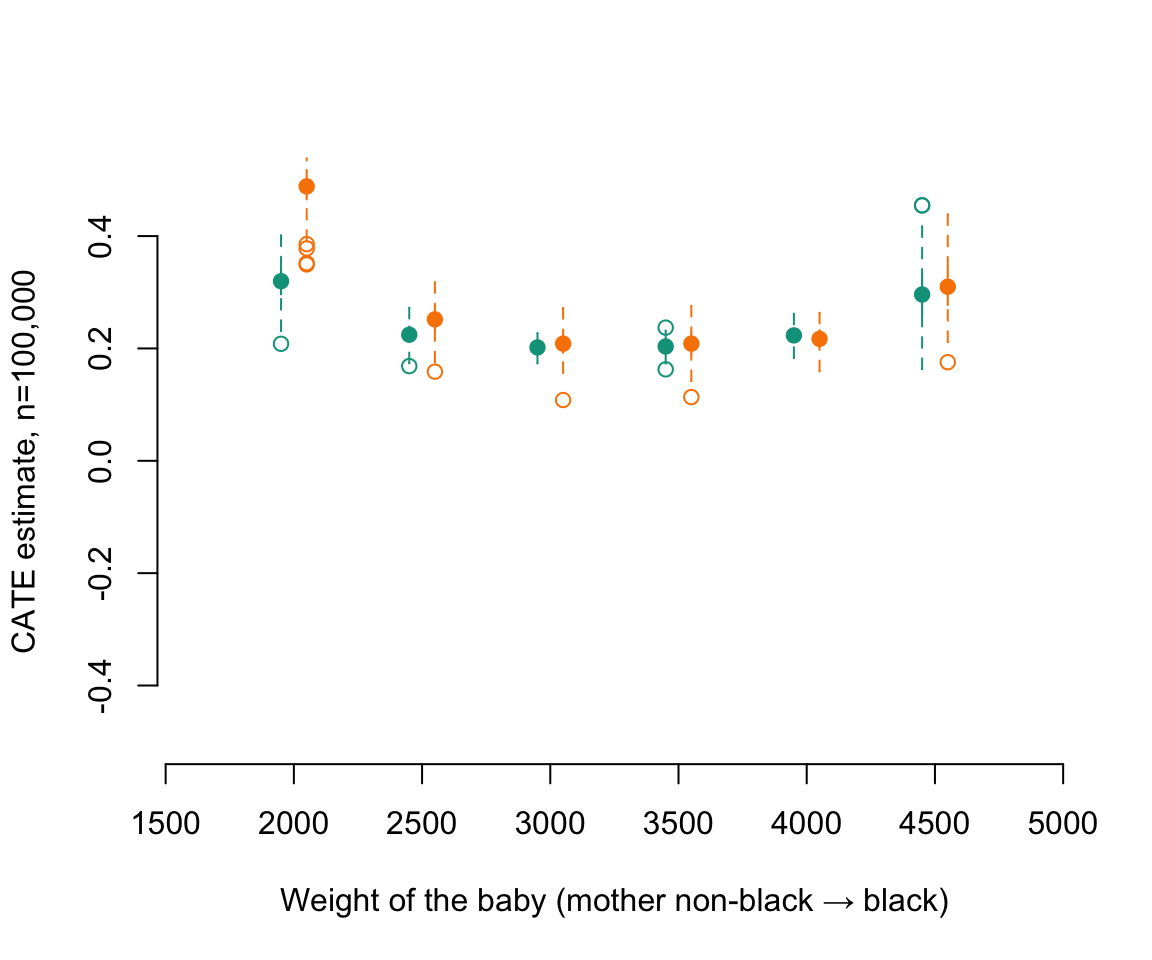}

    \caption{On the left, distribution of $\widehat{\mathcal{T}}(x)$, when $x$ is the weight of a newborn infant, when $n$ goes to $4,000$ (on top) to $20,000$ (in the middle) and $100,000$ (at the bottom), and when $T$ indicates whether the mother is Afro-American or not. On the right, boxplots of the estimation of $\text{SCATE}(x)$, with two GAM models, when $x\in\{2000,2500,\cdots,4000,4500\}$.}
    \label{fig:CATE-sample-size-ranks-weight-black:appendix}
\end{figure}

\begin{table}[!ht]
    \centering
    \begin{tabular}{cc|cc}\hline\hline
    \multicolumn{4}{c}{$t$: mother is smoker} \\
    \multicolumn{2}{c}{$t=0$} & \multicolumn{2}{c}{$t=1$}\\
    \multicolumn{2}{c}{non-smoker} & \multicolumn{2}{c}{smoker}\\
    \hline
\multicolumn{2}{c|}{$\boldsymbol{x}=(x_1,x_2)$}&
\multicolumn{2}{c}{$\mathcal{T}_{\mathcal{N}}(\boldsymbol{x})$}\\
weight & gain & weight & gain \\
    \hline
      2584 &10.8 & 2353.1 &7.6\\
2584 &46.8 & 2414.8 &49.5\\
4152 &10.8 & 3938.1 &7.9\\
4152 &46.8 & 3999.8 &49.8 \\\hline\hline
    \end{tabular}~~    \begin{tabular}{cc|cc}\hline\hline
    \multicolumn{4}{c}{$t$: mother is Afro-American} \\
    \multicolumn{2}{c}{$t=0$} & \multicolumn{2}{c}{$t=1$}\\
    \multicolumn{2}{c}{non-Black} & \multicolumn{2}{c}{Black}\\
    \hline
\multicolumn{2}{c|}{$\boldsymbol{x}=(x_1,x_2)$}&
\multicolumn{2}{c}{$\mathcal{T}_{\mathcal{N}}(\boldsymbol{x})$}\\
weight & gain & weight & gain \\
    \hline
2584& 10.8&  2392.5 & 7.6 \\
2584& 46.8&  2382.0 & 47.6 \\ 
4152& 10.8&  4086.1 & 7.5 \\ 
4152& 46.8&  4075.6 & 47.6 \\\hline\hline
    \end{tabular}~~    \begin{tabular}{cc|cc}\hline\hline
    \multicolumn{4}{c}{$t$: Sex of the newborn} \\
    \multicolumn{2}{c}{$t=0$} & \multicolumn{2}{c}{$t=1$}\\
    \multicolumn{2}{c}{boy} & \multicolumn{2}{c}{girl}
    \\
    \hline
\multicolumn{2}{c|}{$\boldsymbol{x}=(x_1,x_2)$}&
\multicolumn{2}{c}{$\mathcal{T}_{\mathcal{N}}(\boldsymbol{x})$}\\
weight & gain & weight & gain \\
    \hline
2584& 10.8&  2513.4 & 10.2 \\
2584& 46.8&  2493.0 & 45.9 \\ 
4152& 10.8&  4012.4 & 10.1 \\ 
4152& 46.8&  3992.0 & 45.8 \\\hline\hline
    \end{tabular}
    \caption{Bivariate optimal transport, $\boldsymbol{x}\mapsto \mathcal{T}_{\mathcal{N}}(\boldsymbol{x})$, for  the three treatments, for four different individuals $\boldsymbol{x}$ in the control group.}
    \label{tab:transport:appendix}
\end{table}

\begin{figure}[!ht]
    \centering
     \includegraphics[width=.49\textwidth]{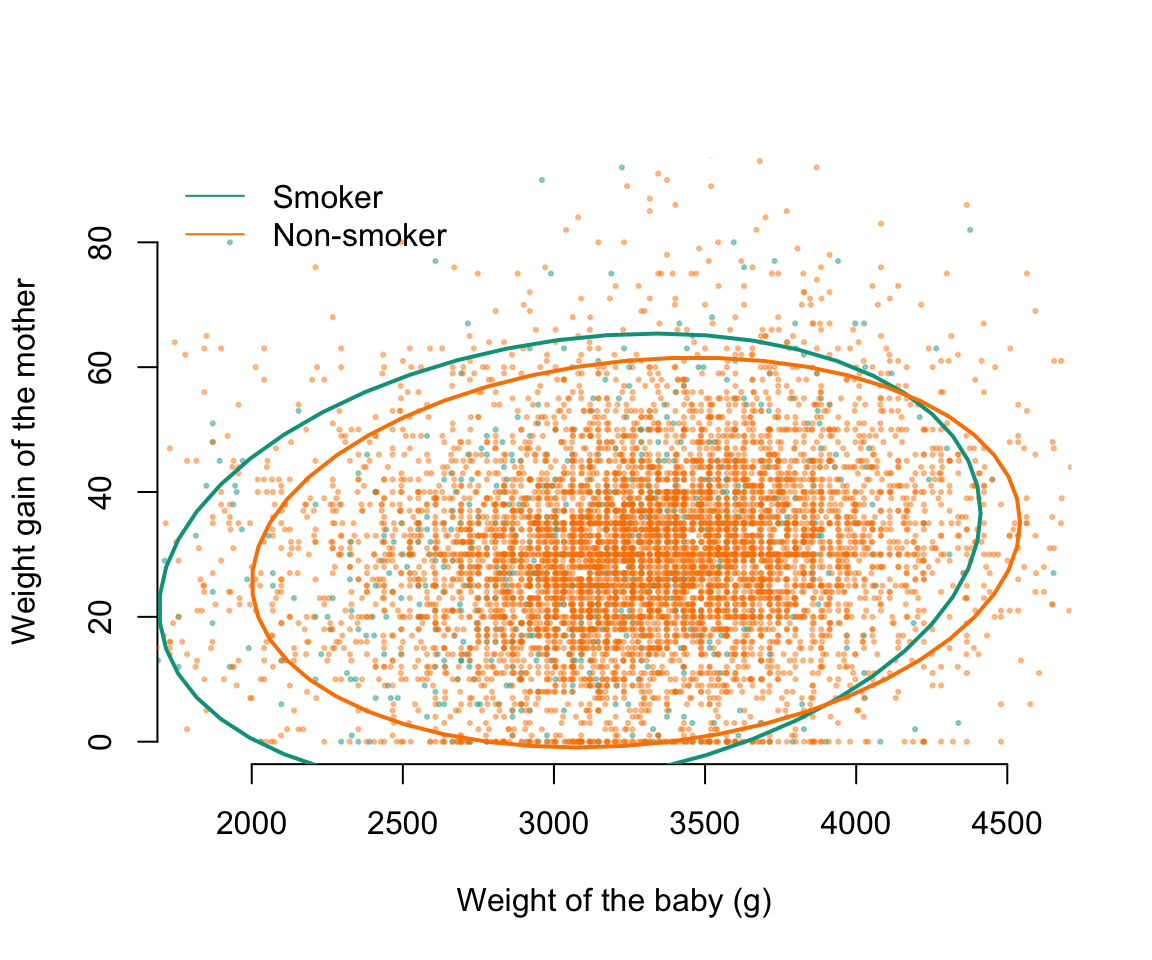}
     \includegraphics[width=.49\textwidth]{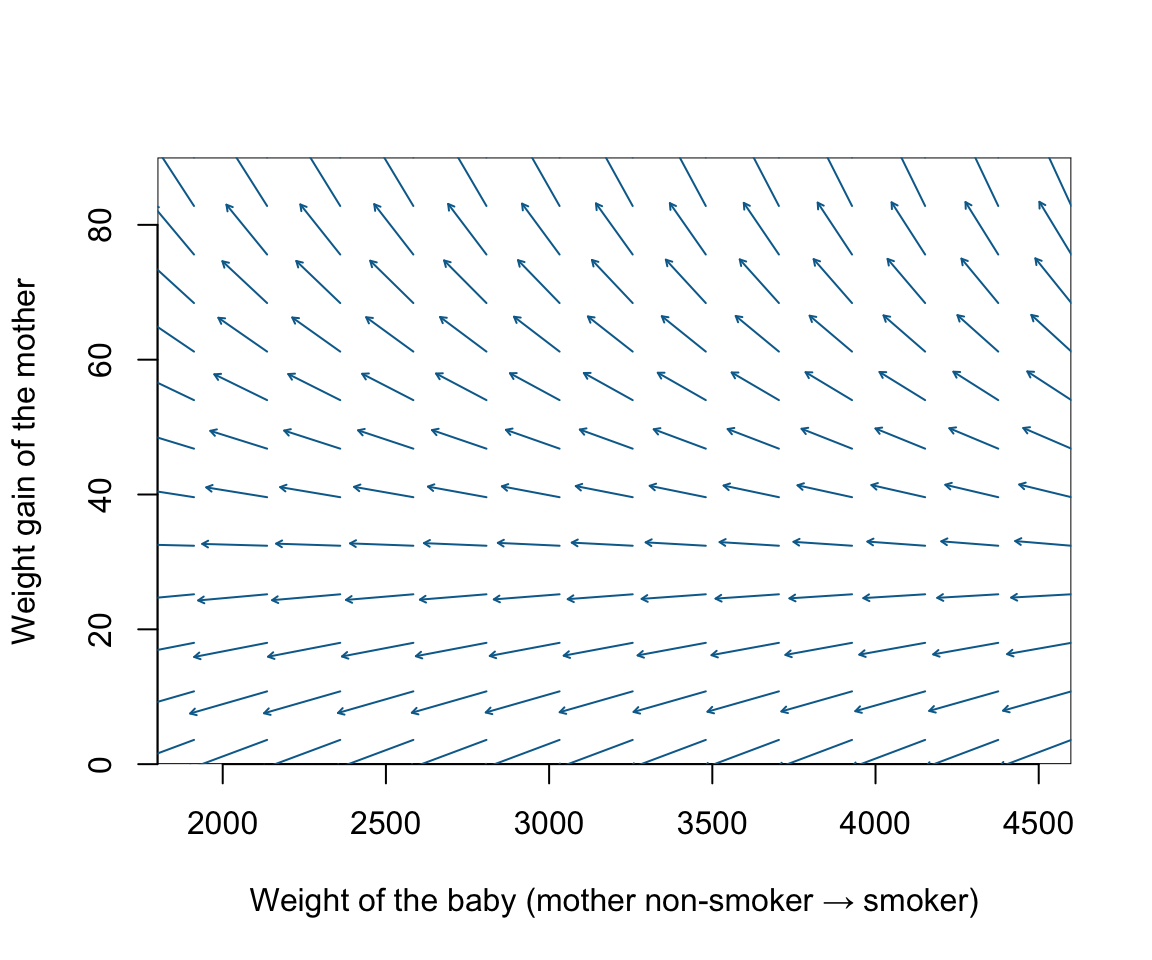}

    \caption{Joint distributions of $\boldsymbol{X}$ (weight of the newborn infant and weight gain of the mother), conditional on the treatment $T$, when $T$ indicates whether the mother is a smoker or not on the left. On the right, vector field associated with optimal Gaussian transport, in dimension two (weight of the newborn infant and weight gain of the mother), when the treatment $T$ indicates whether the mother is a smoker or not. Some numerical values are given in Table~\ref{tab:transport:appendix}. On the right, the origin of the arrow is $\boldsymbol{x}$ in the control group (non-smoker) and the arrowhead is $\widehat{\mathcal{T}}_{\mathcal{N}}(\boldsymbol{x}$ in the treated group (smoker)).}
    \label{fig:joint-ellipse-arrow-black-biv-1x2:appendix}
\end{figure}

\begin{figure}[!ht]
    \centering
    \includegraphics[width=.49\textwidth]{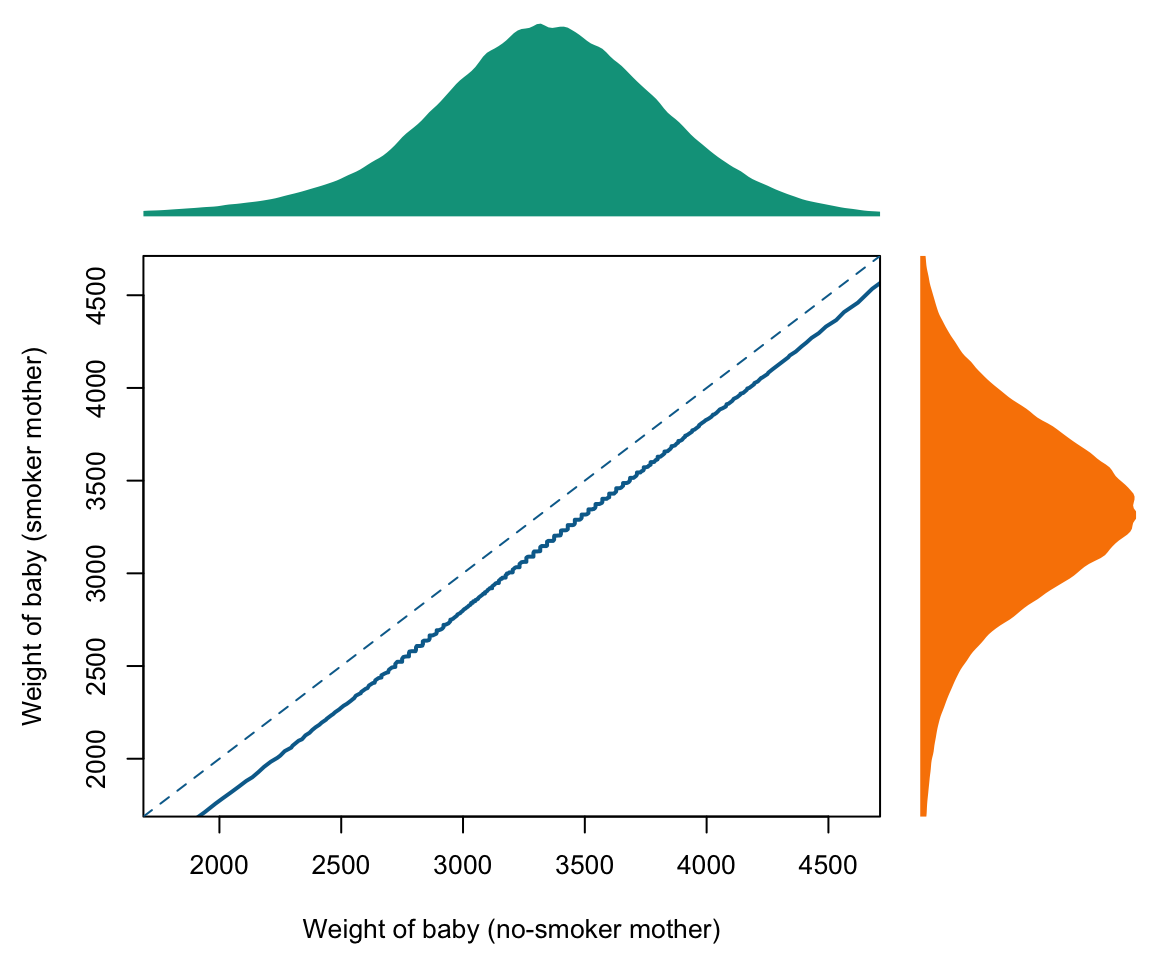}
    \includegraphics[width=.49\textwidth]{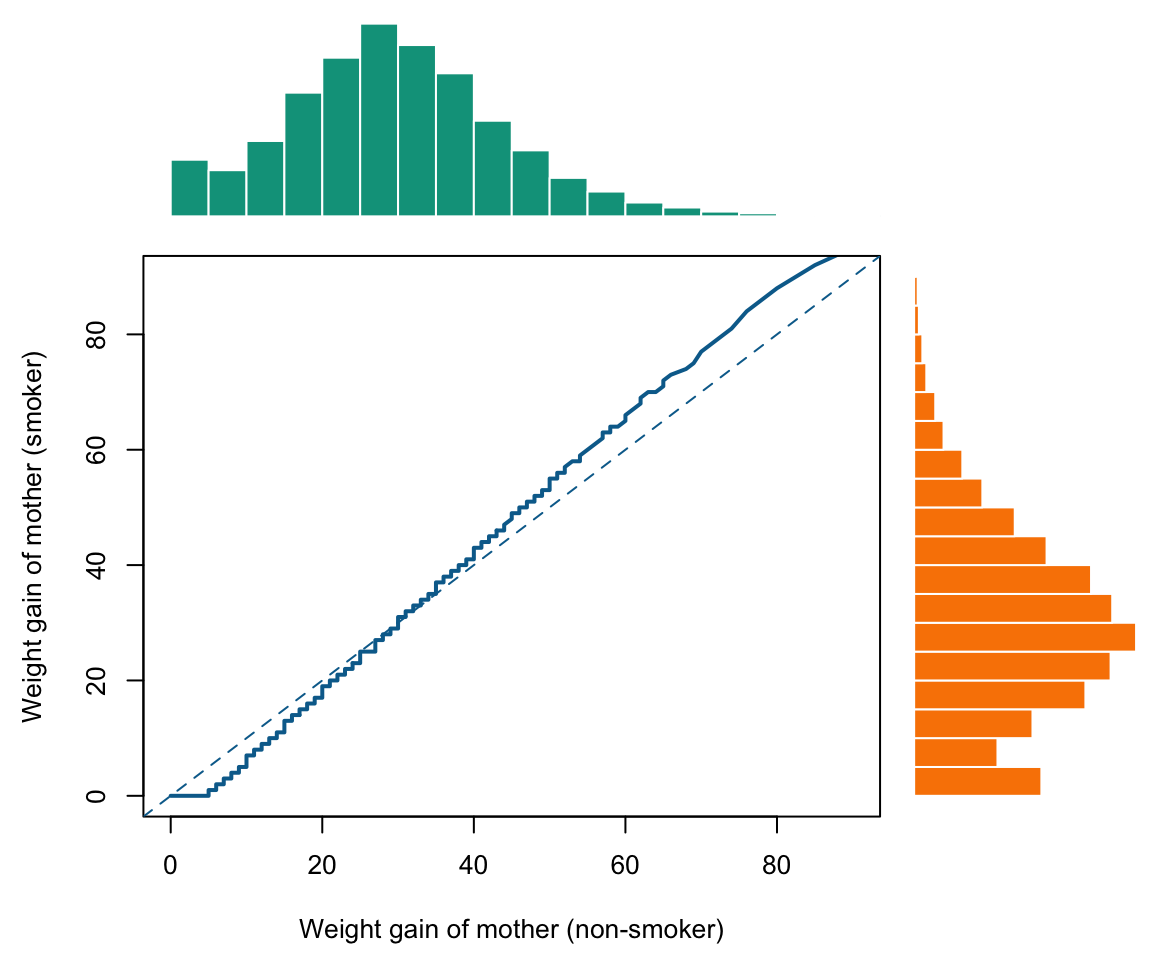}

    % \centering    
    % \includegraphics[width=.49\textwidth]{figures/joint-transport-black-weight-1}
    % \includegraphics[width=.49\textwidth]{figures/joint-transport-black-weight-gain-1}

    \centering
    \includegraphics[width=.49\textwidth]{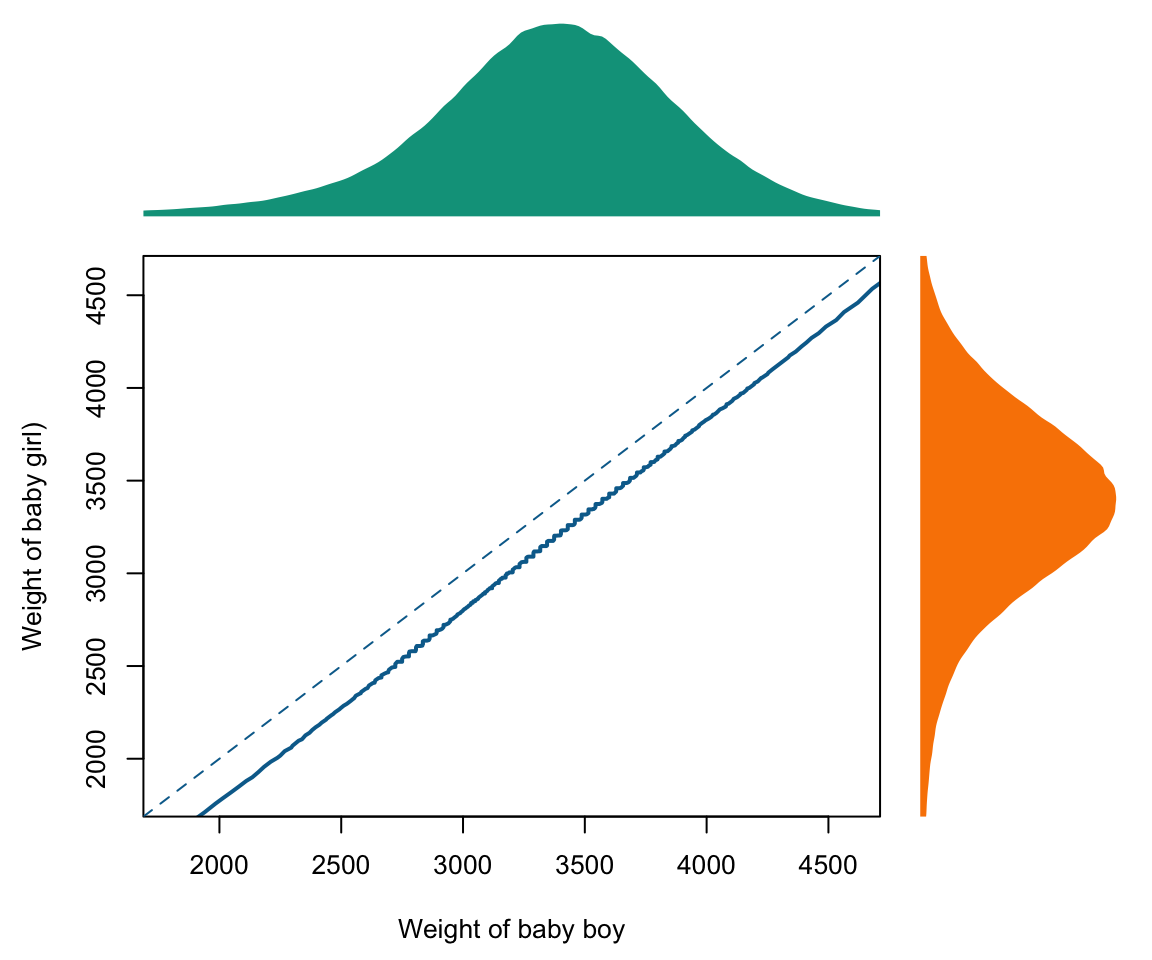}
    \includegraphics[width=.49\textwidth]{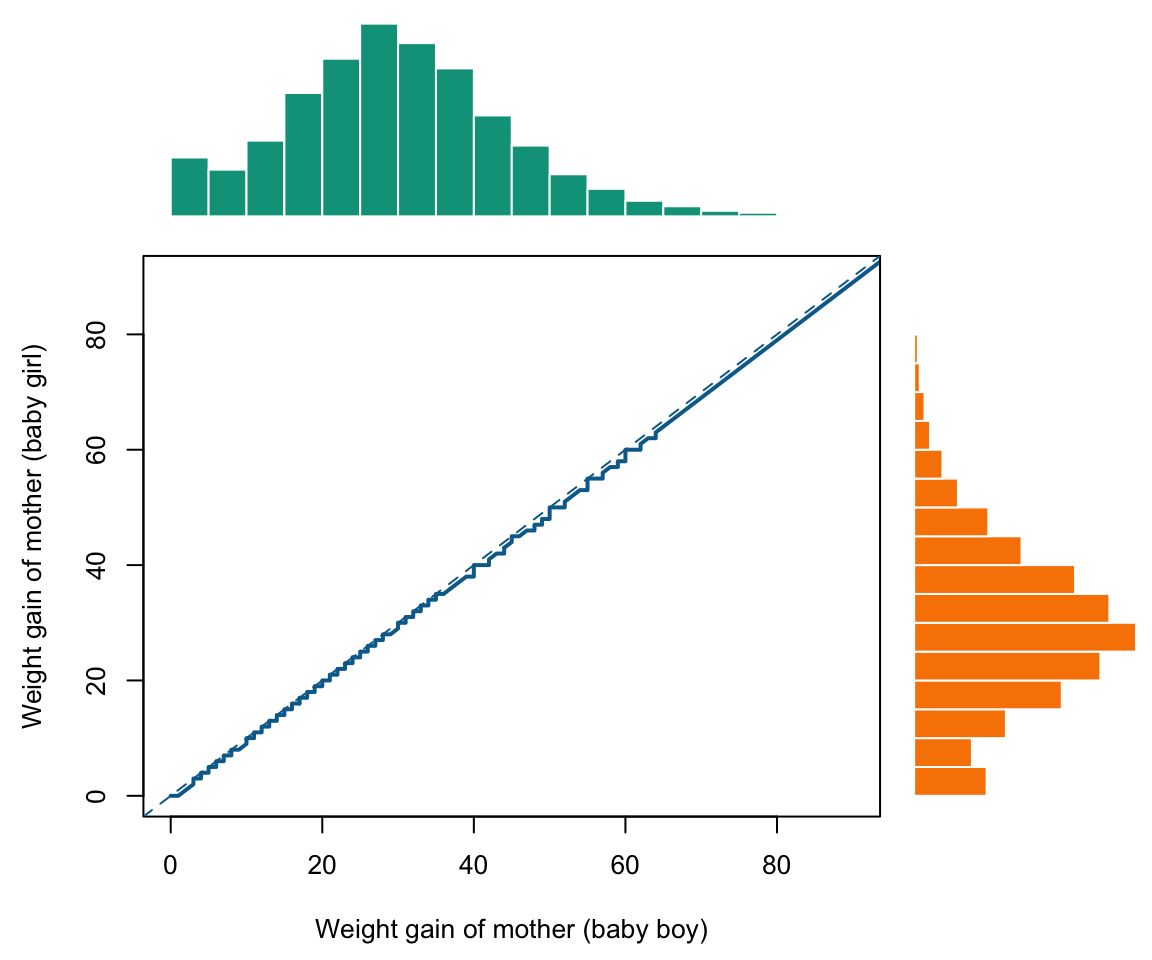}
    
    \caption{Optimal transport (quantile based) when $X$ is the weight of the newborn infant on the left, and the weight gain of the mother on the right, when $T$ indicates whether the mother is a smoker on top, and whether the newborn infant is a boy at the bottom.}
    \label{fig:opt-transport-quantiles-2x3:appendix}
\end{figure}

% \begin{figure}[!ht]
%     \centering
%      \includegraphics[width=.49\textwidth]{figures/distrib-joint-weight-gain-ellipse-smoker-1}
%      \includegraphics[width=.49\textwidth]{figures/distrib-joint-weight-gain-ellipse-black-1}

%     \caption{Joint distributions of $\boldsymbol{X}$ (weight of the newborn infant and weight gain of the mother), conditional on the treatment $T$, when $T$ is the indicator that the mother is smoking or not on the left, and the indicator that the mother is Black or not on the right.}
%     \label{fig:joint-ellipse-biv-1x2:appendix}
% \end{figure}

% \begin{figure}[!ht]
%     \centering
%      \includegraphics[width=.49\textwidth]{figures/transport-smoker-arrows-0-1-1}
%      \includegraphics[width=.49\textwidth]{figures/transport-black-arrows-v2-0-1-1}

%     \caption{Vector field associated with optimal gaussian transport, in dimension two (weight of the newborn infant and weight gain of the mother), when the treatment $T$ is the indicator that the mother is smoking or not on the left, and the indicator that the mother is Black or not, on the right. Some numerical values are given in Table~\ref{tab:transport}. On the left, the origin of the arrow is $\boldsymbol{x}$ in the control group (non-smoker) and the arrowhead is $\widehat{\mathcal{T}}_{\mathcal{N}}(\boldsymbol{x}$ in the treated group (smoker).}
%     \label{fig:joint-arrow-grid-biv-1x2:appendix}
% \end{figure}

\begin{figure}[!ht]
    \centering
     \includegraphics[width=.49\textwidth]{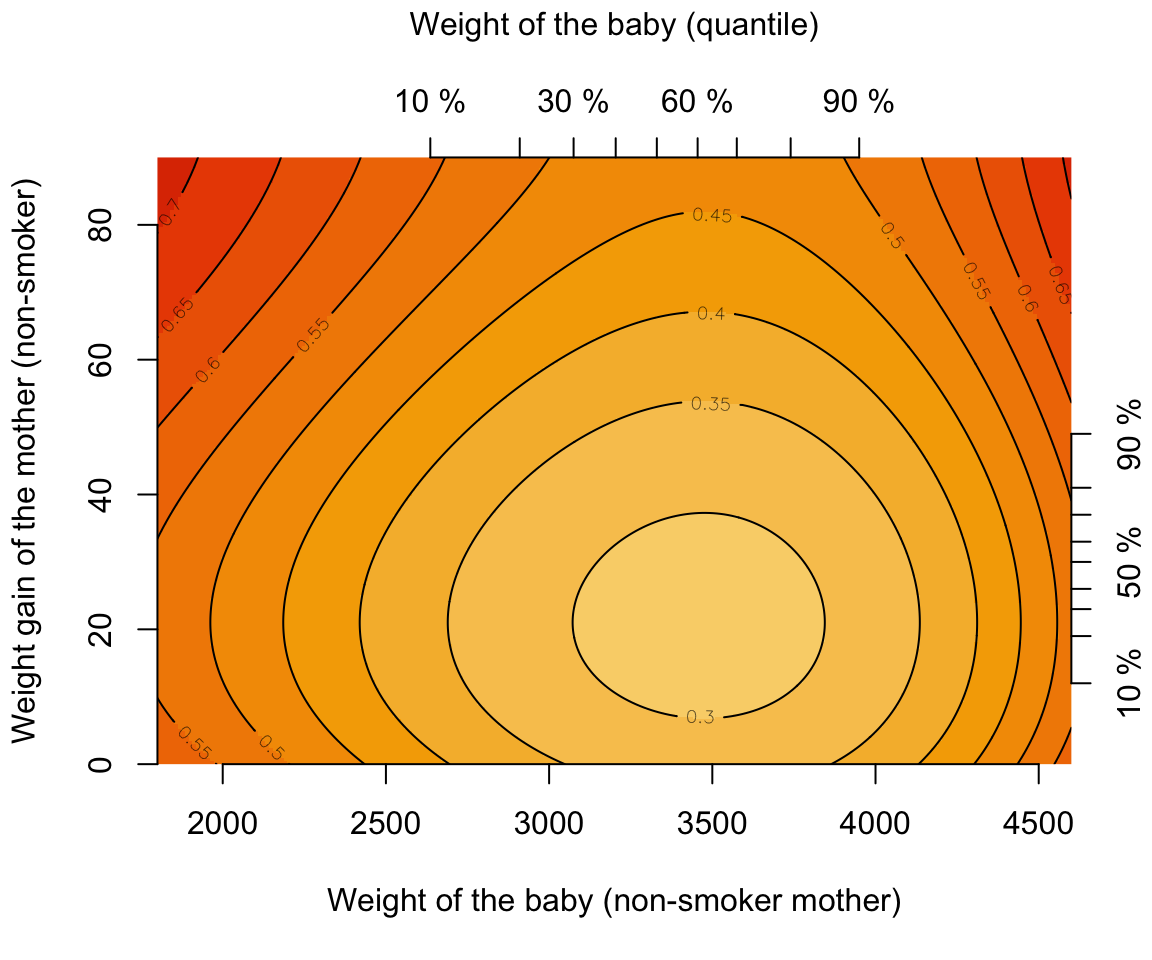}
     \includegraphics[width=.49\textwidth]{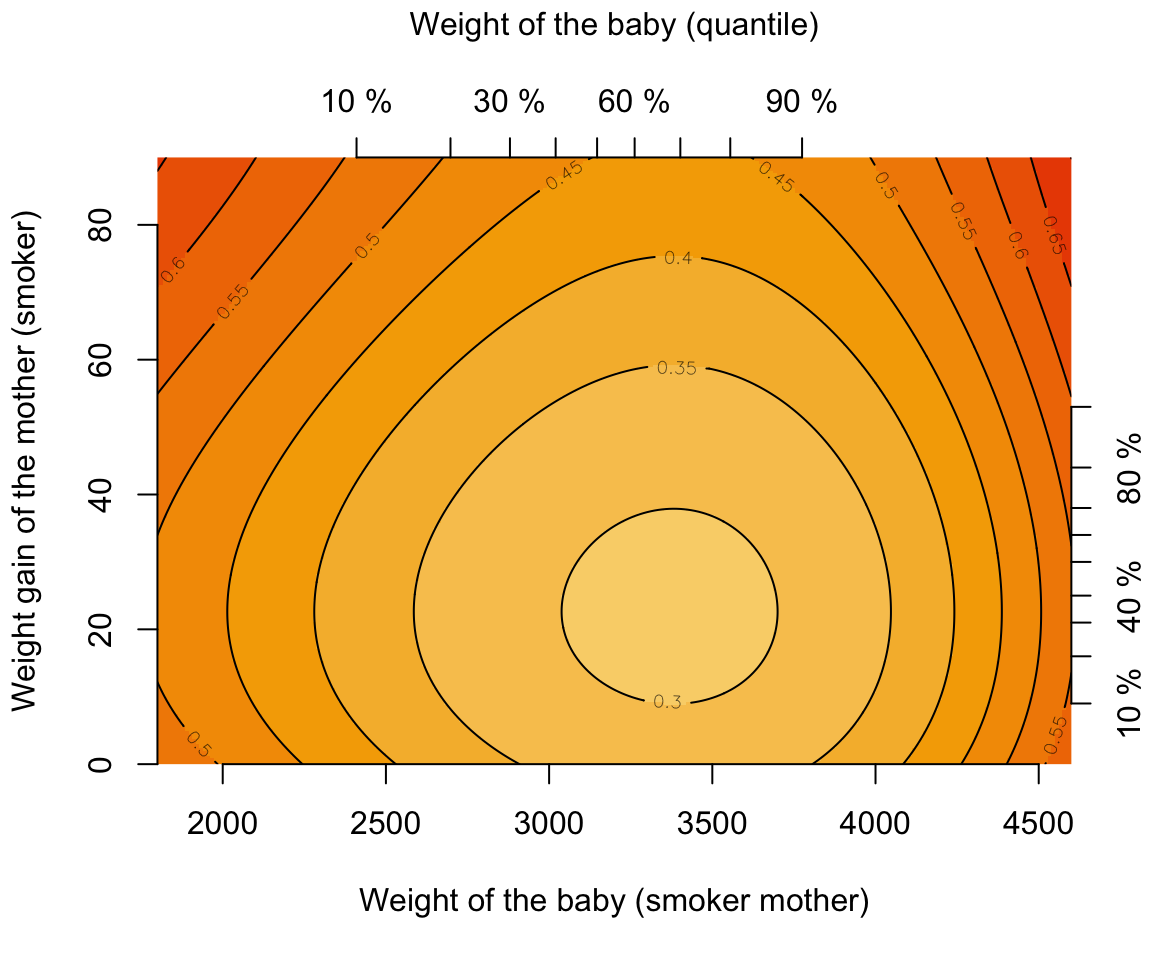}

    \centering
     \includegraphics[width=.49\textwidth]{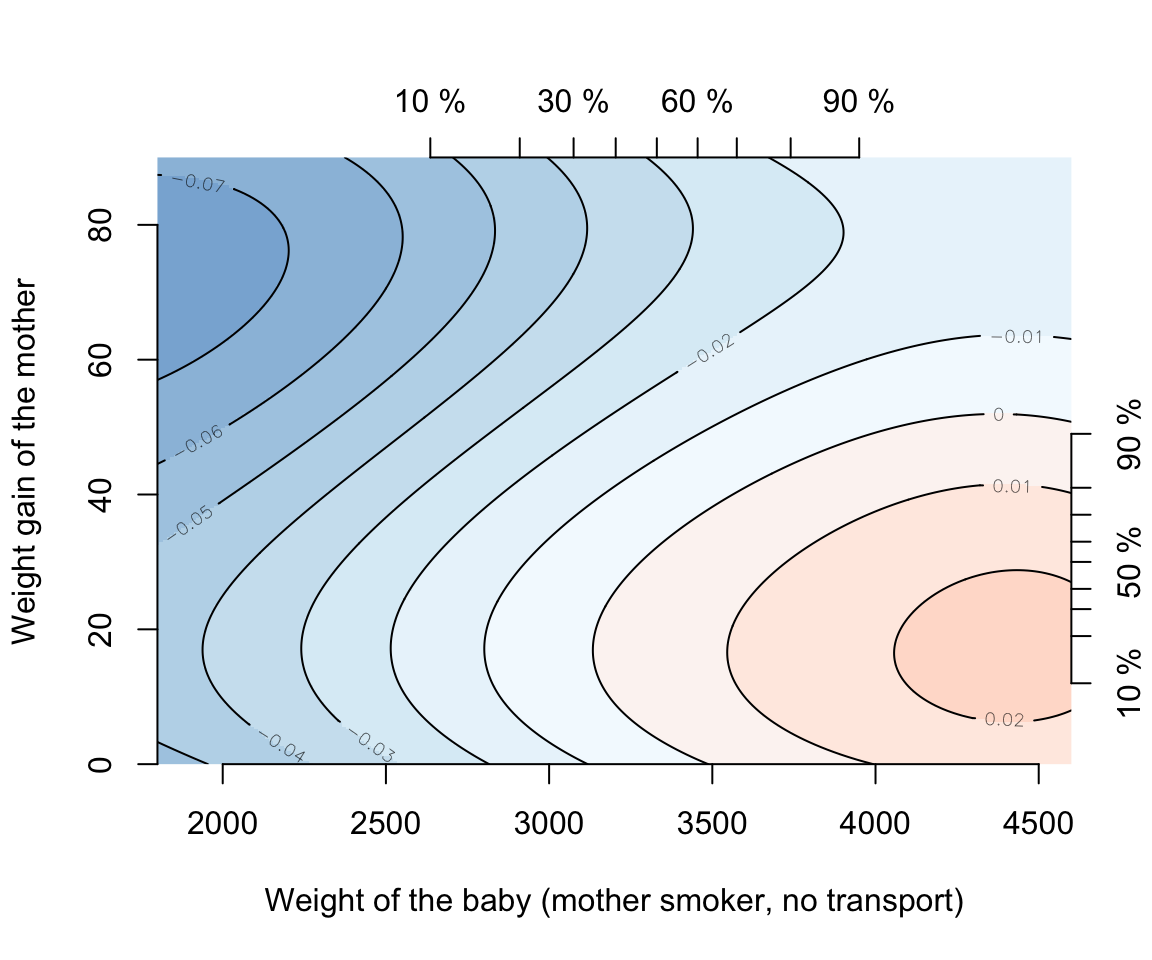}
     \includegraphics[width=.49\textwidth]{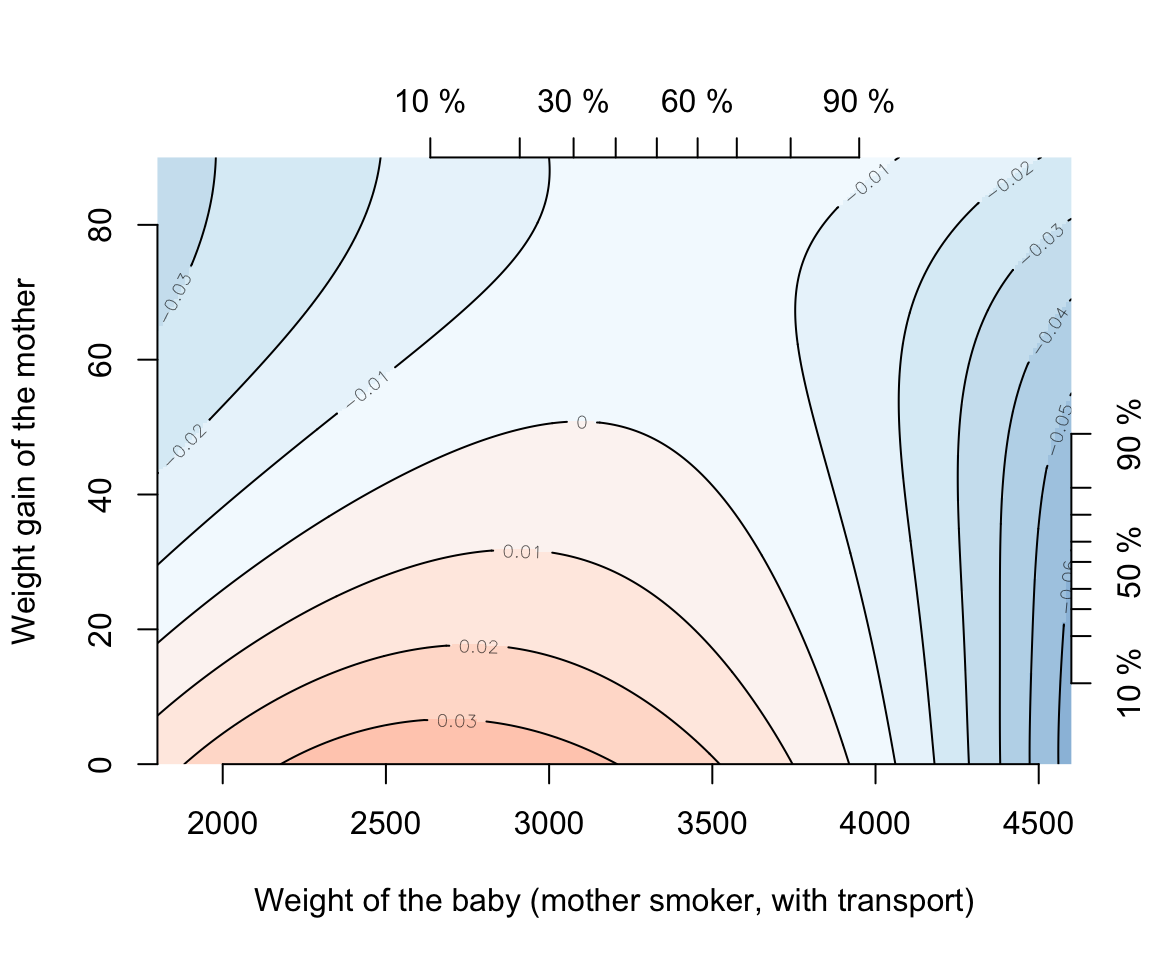}

    \centering
     \includegraphics[width=.49\textwidth]{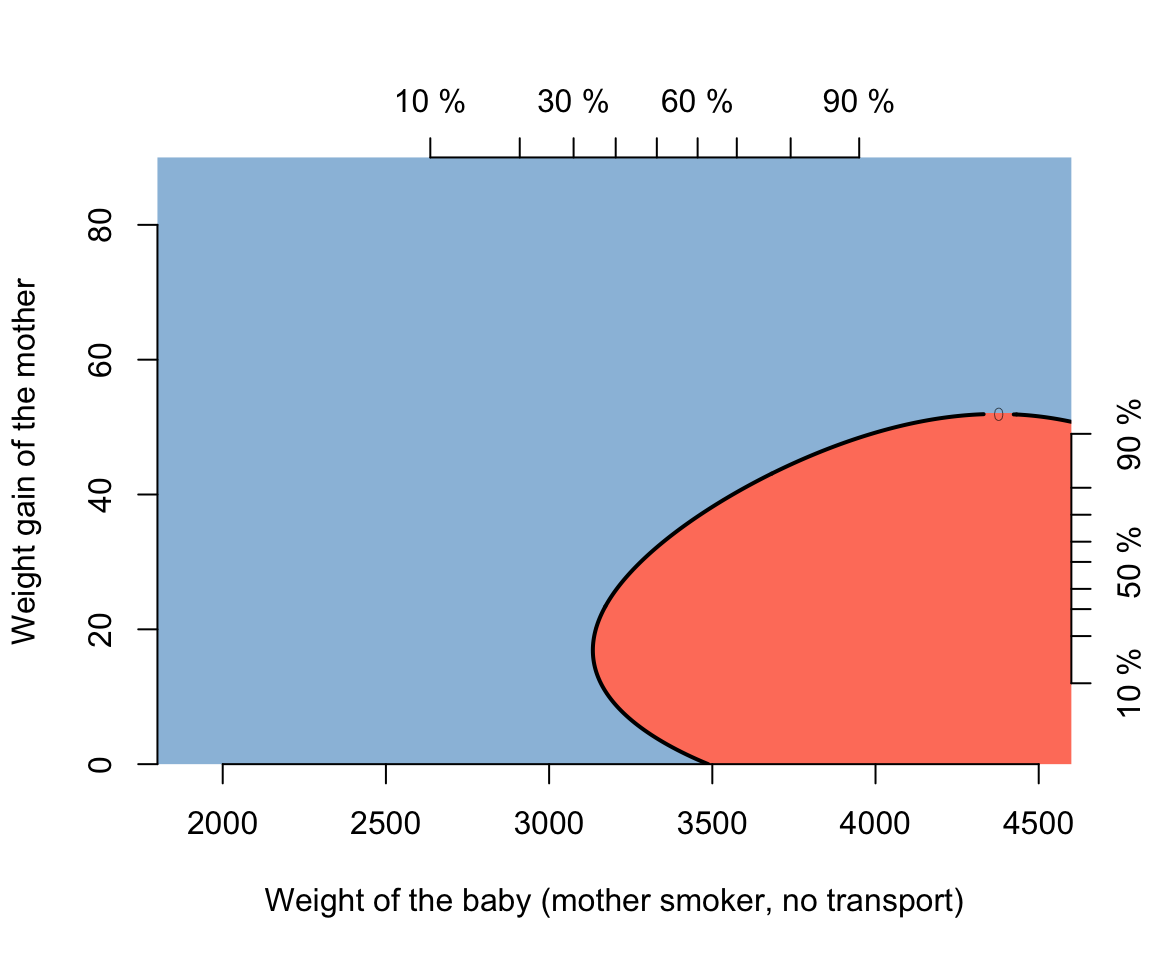}
     \includegraphics[width=.49\textwidth]{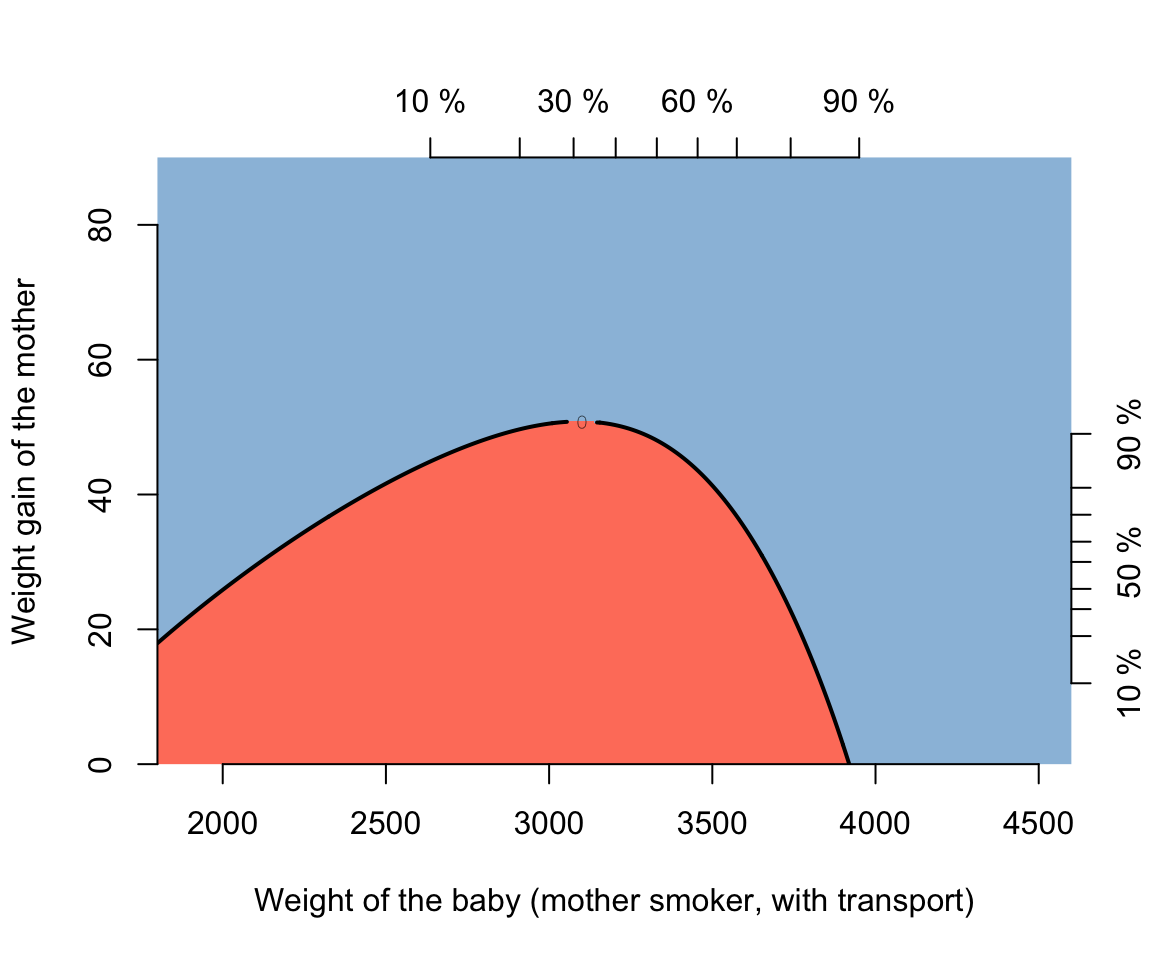}
    
    \caption{On top, contours of $\boldsymbol{x}\mapsto\mathbb{E}[Y|\boldsymbol{X}=\boldsymbol{x},T=0]$ and $\boldsymbol{x}\mapsto\mathbb{E}[Y|\boldsymbol{X}=\boldsymbol{x},T=1]$ when $T$ indicates whether a mother is a smoker or not, estimated with logistic GAM models (cubic splines).
    In the middle contours of the {\em ceteris paribus} $\boldsymbol{x}\mapsto\text{CATE}[\boldsymbol{x}]$ without any transport on the left, and $\boldsymbol{x}\mapsto\text{SCATE}[\boldsymbol{x}]$ {\em mutatis mutandis} on the right. At the bottom, positive/negative distinction for the conditional average treatment effect.}
    \label{fig:CATE-biv-2x3-GAM-1-pred-C:appendix}
\end{figure}

\begin{figure}[!ht]
    \centering
     \includegraphics[width=.49\textwidth]{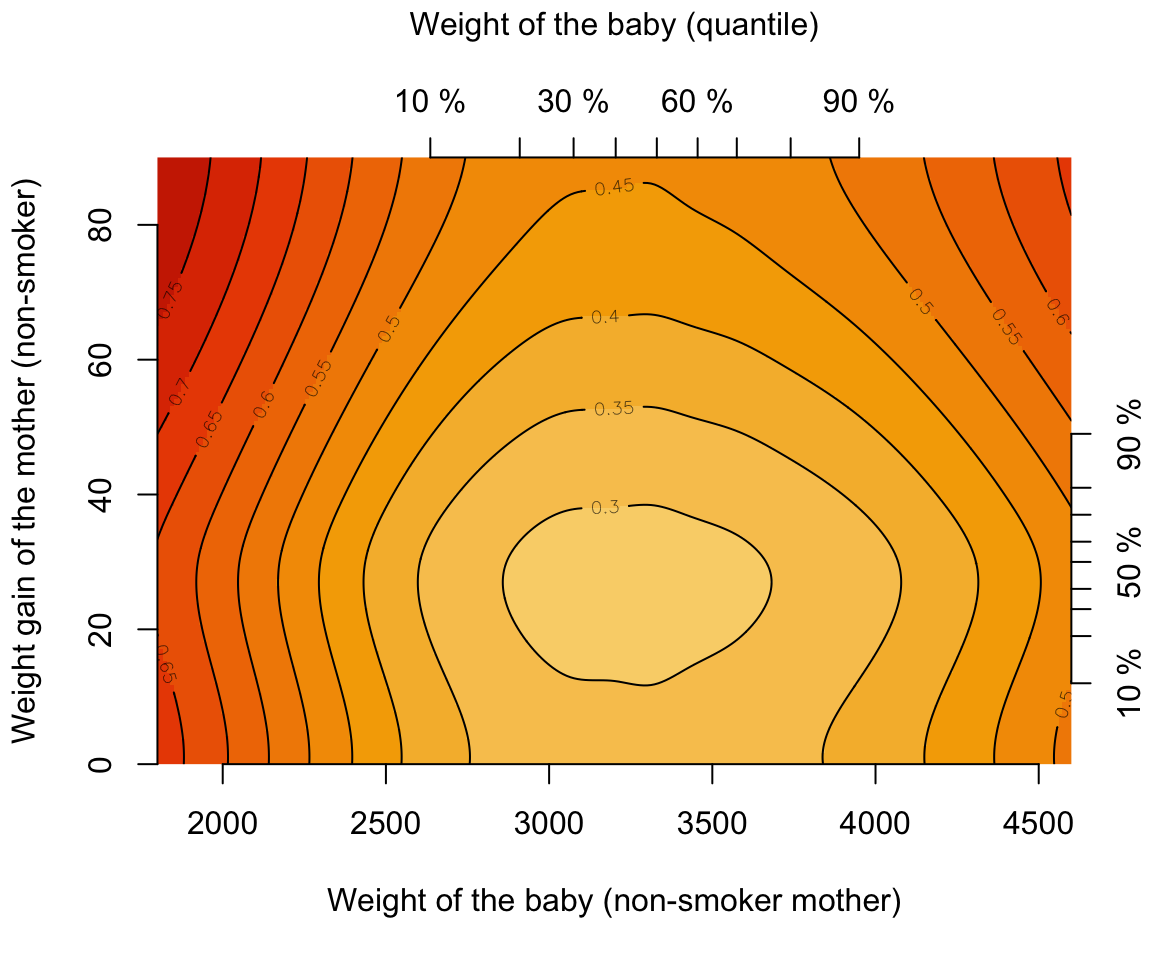}
     \includegraphics[width=.49\textwidth]{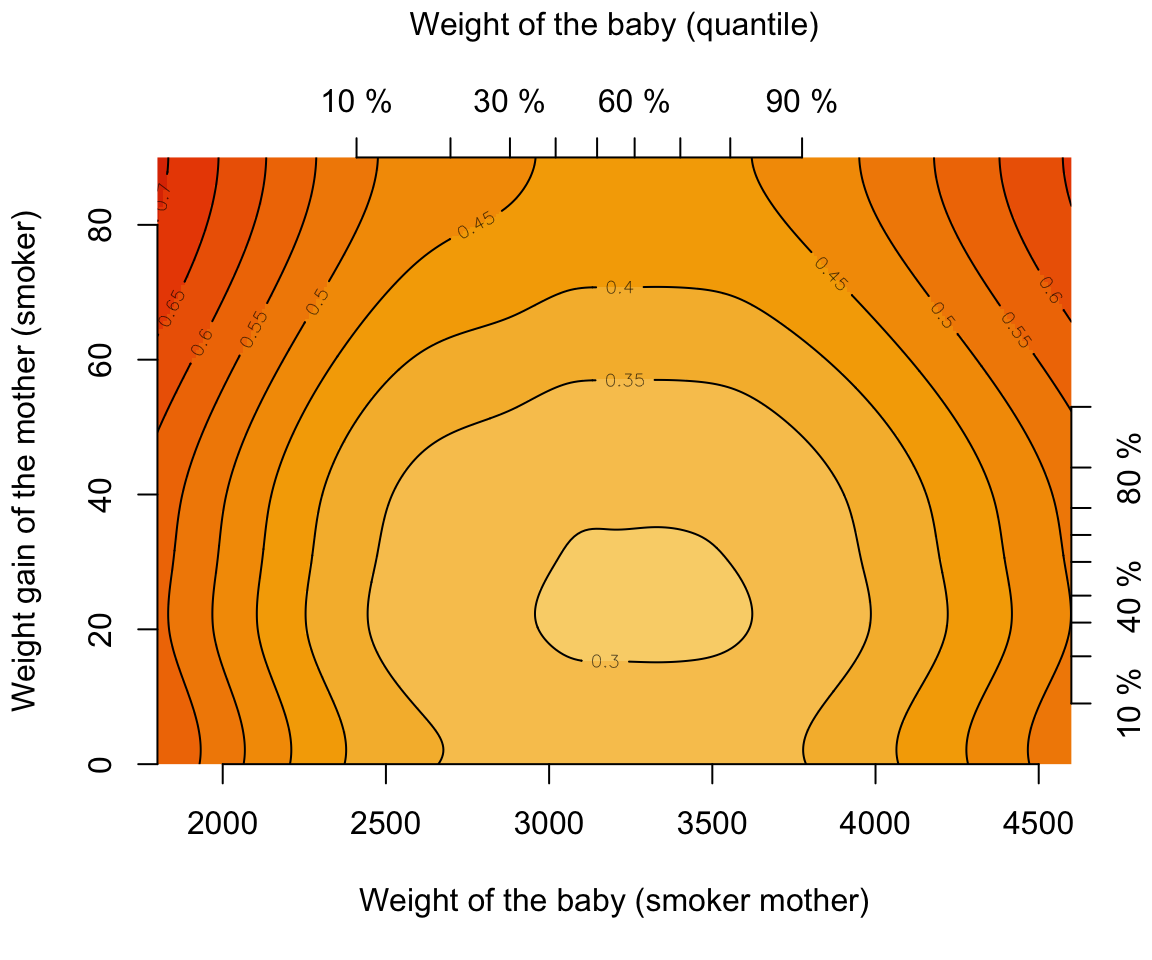}

    \centering
     \includegraphics[width=.49\textwidth]{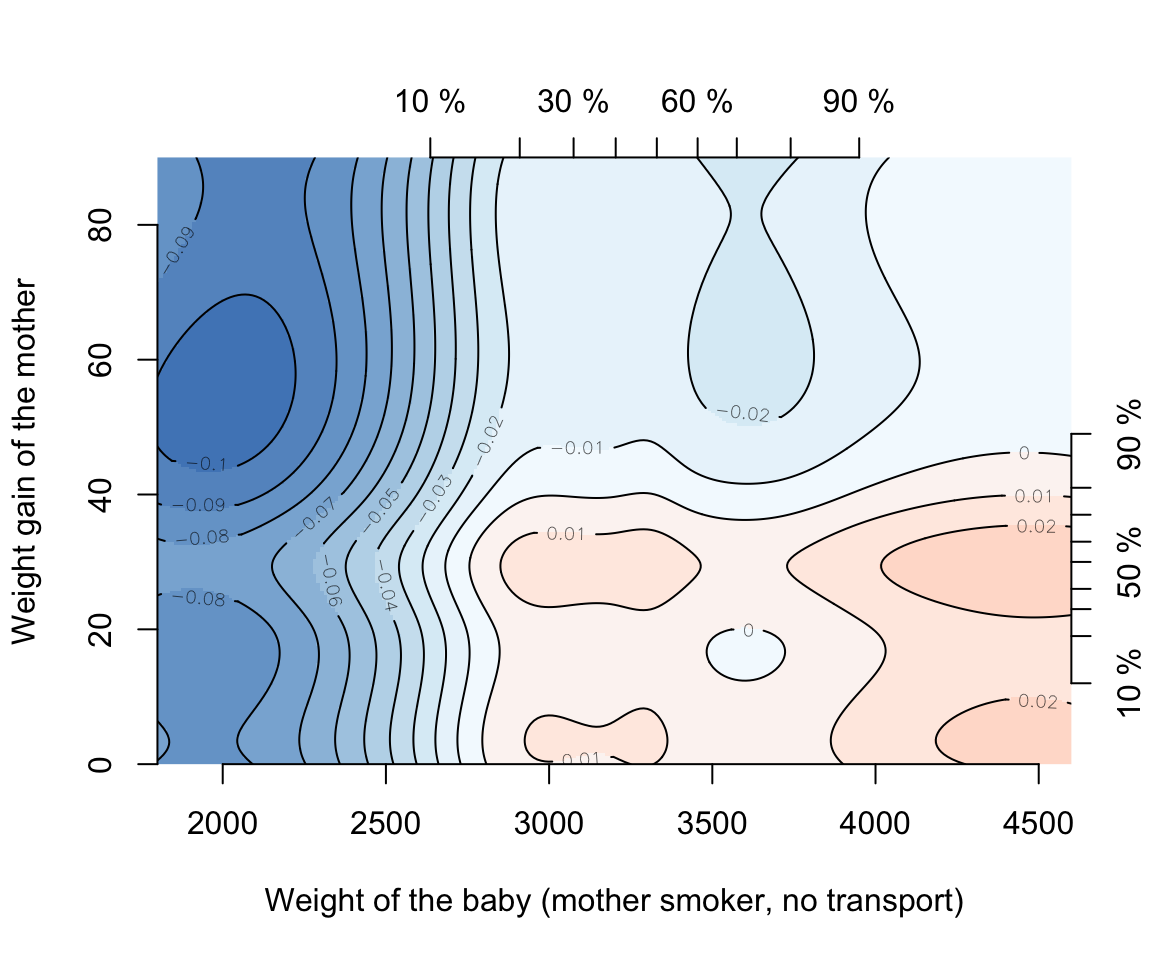}
     \includegraphics[width=.49\textwidth]{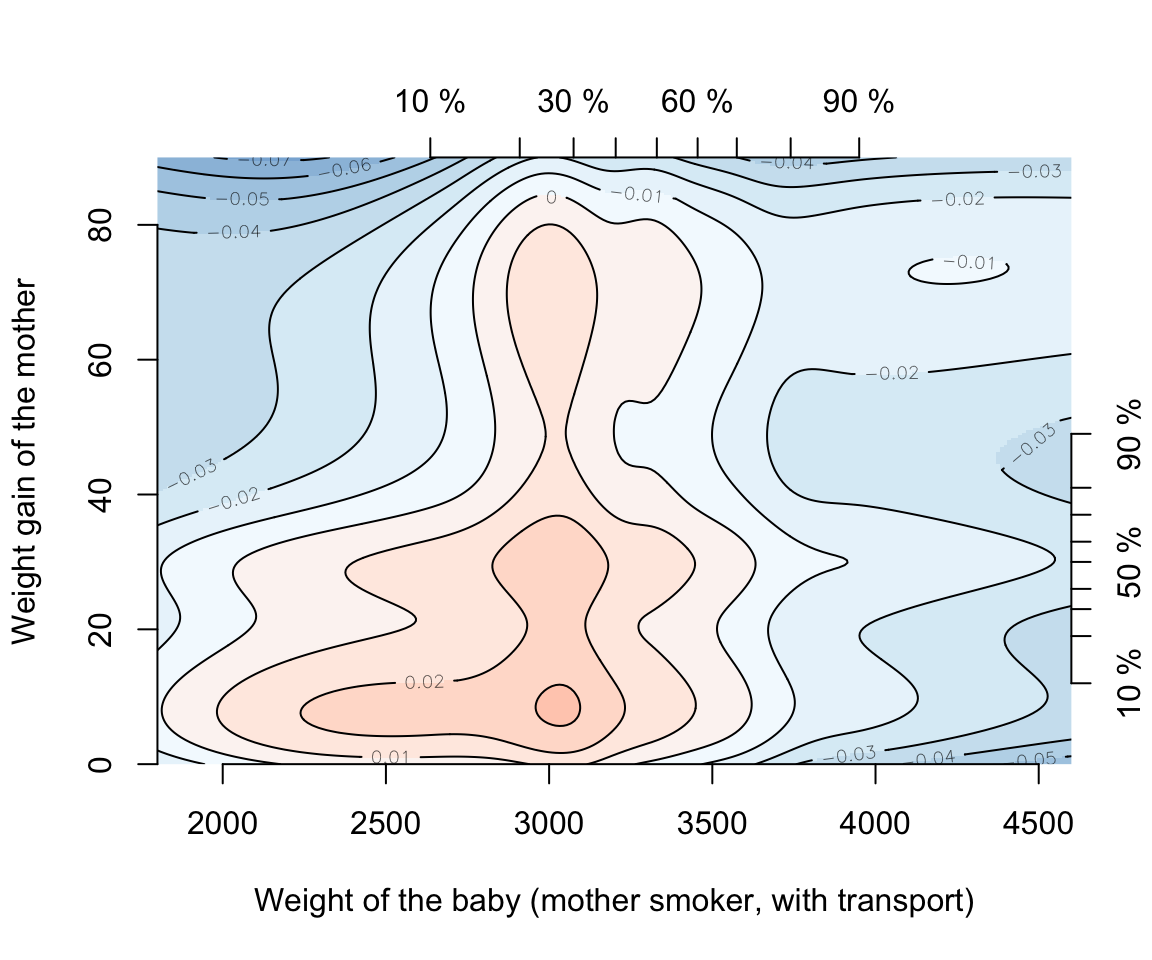}

    \centering
     \includegraphics[width=.49\textwidth]{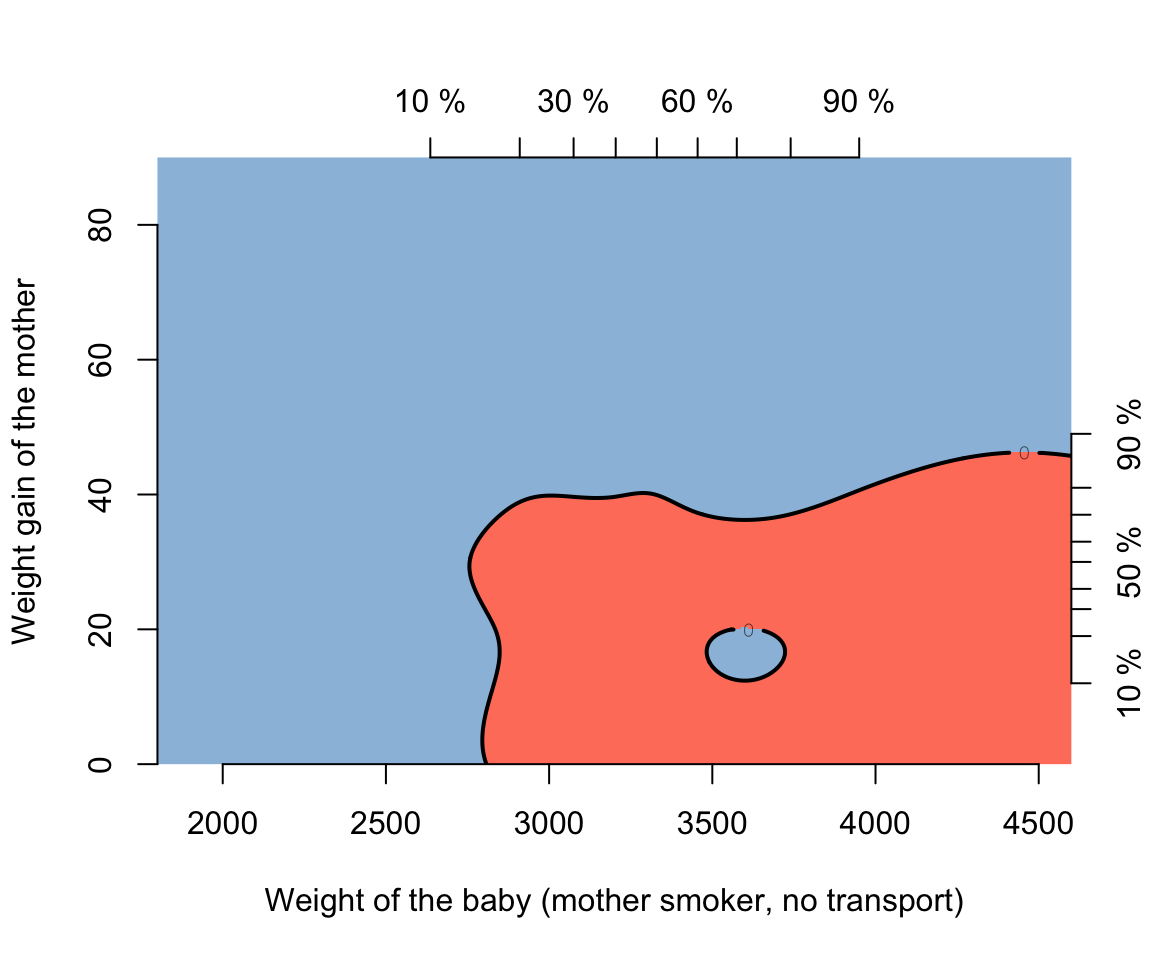}
     \includegraphics[width=.49\textwidth]{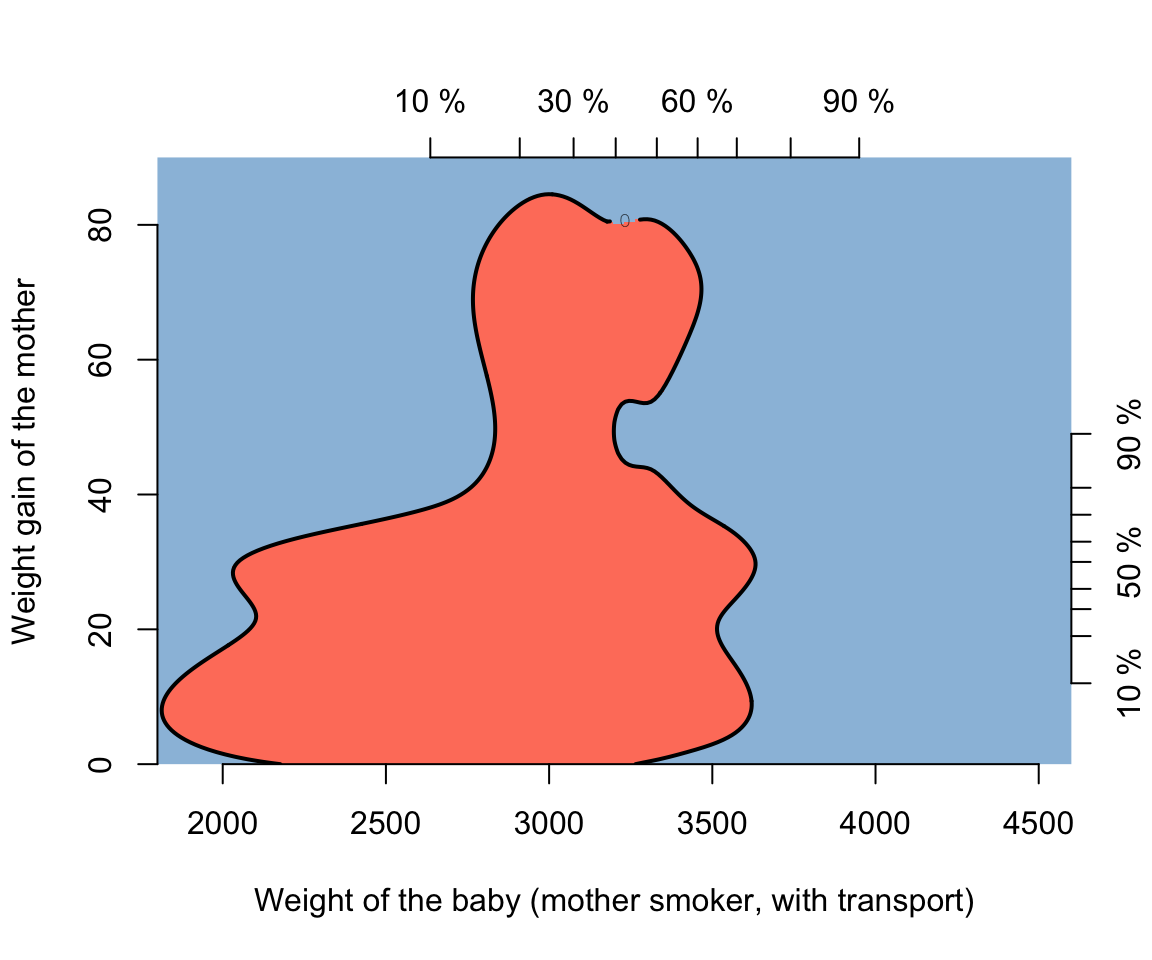}
    
     \caption{On top, contours of $\boldsymbol{x}\mapsto\mathbb{E}[Y|\boldsymbol{X}=\boldsymbol{x},T=0]$ and $\boldsymbol{x}\mapsto\mathbb{E}[Y|\boldsymbol{X}=\boldsymbol{x},T=1]$ when $T$ indicates whether a mother is a smoker or not, estimated with logistic GAM models (cubic splines, with more knots).
    In the middle contours of the {\em ceteris paribus} $\boldsymbol{x}\mapsto\text{CATE}[\boldsymbol{x}]$ without any transport on the left, and $\boldsymbol{x}\mapsto\text{SCATE}[\boldsymbol{x}]$ {\em mutatis mutandis} on the right. At the bottom, positive/negative distinction for the conditional average treatment effect.}
    \label{fig:CATE-biv-2x3-GAM-2-pred-C:appendix}
\end{figure}

\newpage

\begin{figure}[!ht]
    \centering
     \includegraphics[width=.49\textwidth]{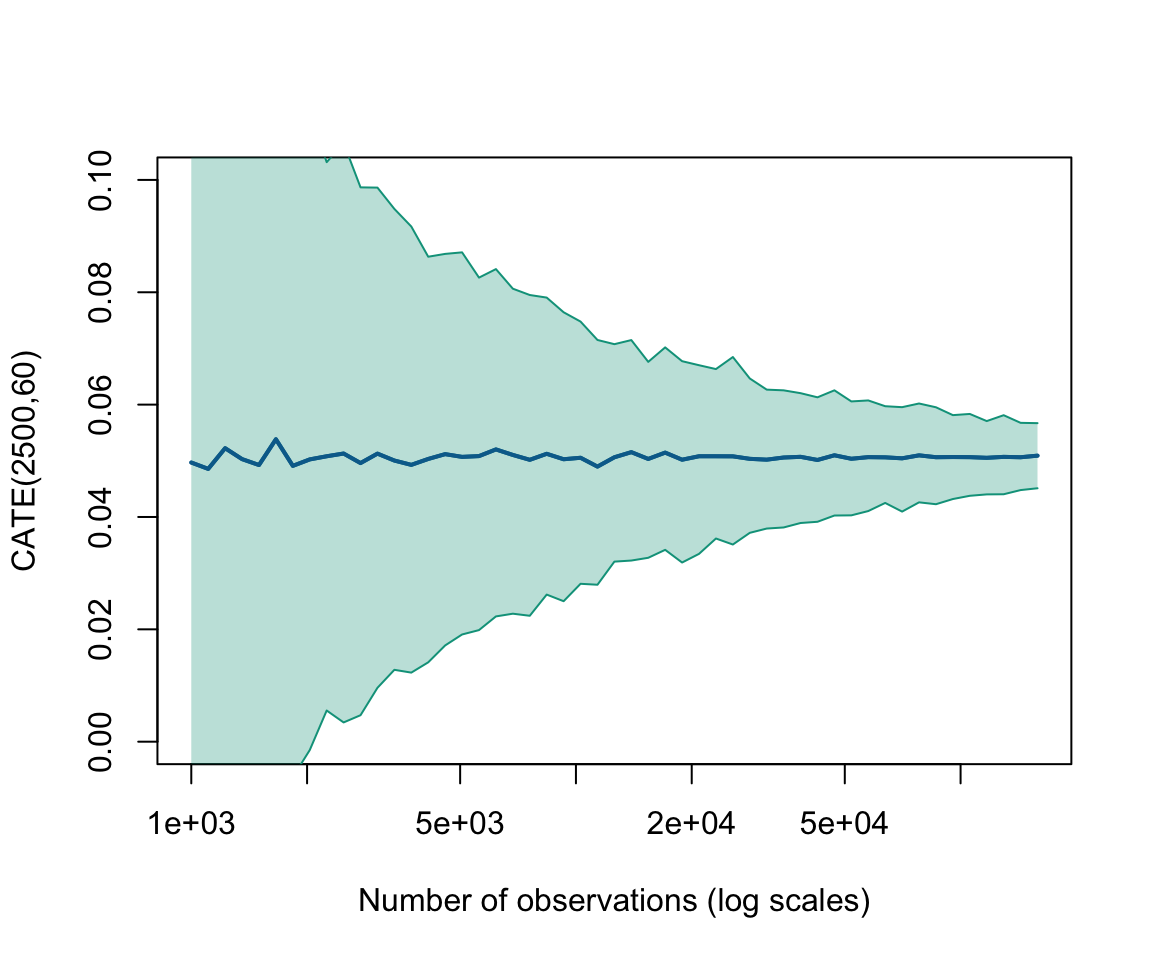}
     \includegraphics[width=.49\textwidth]{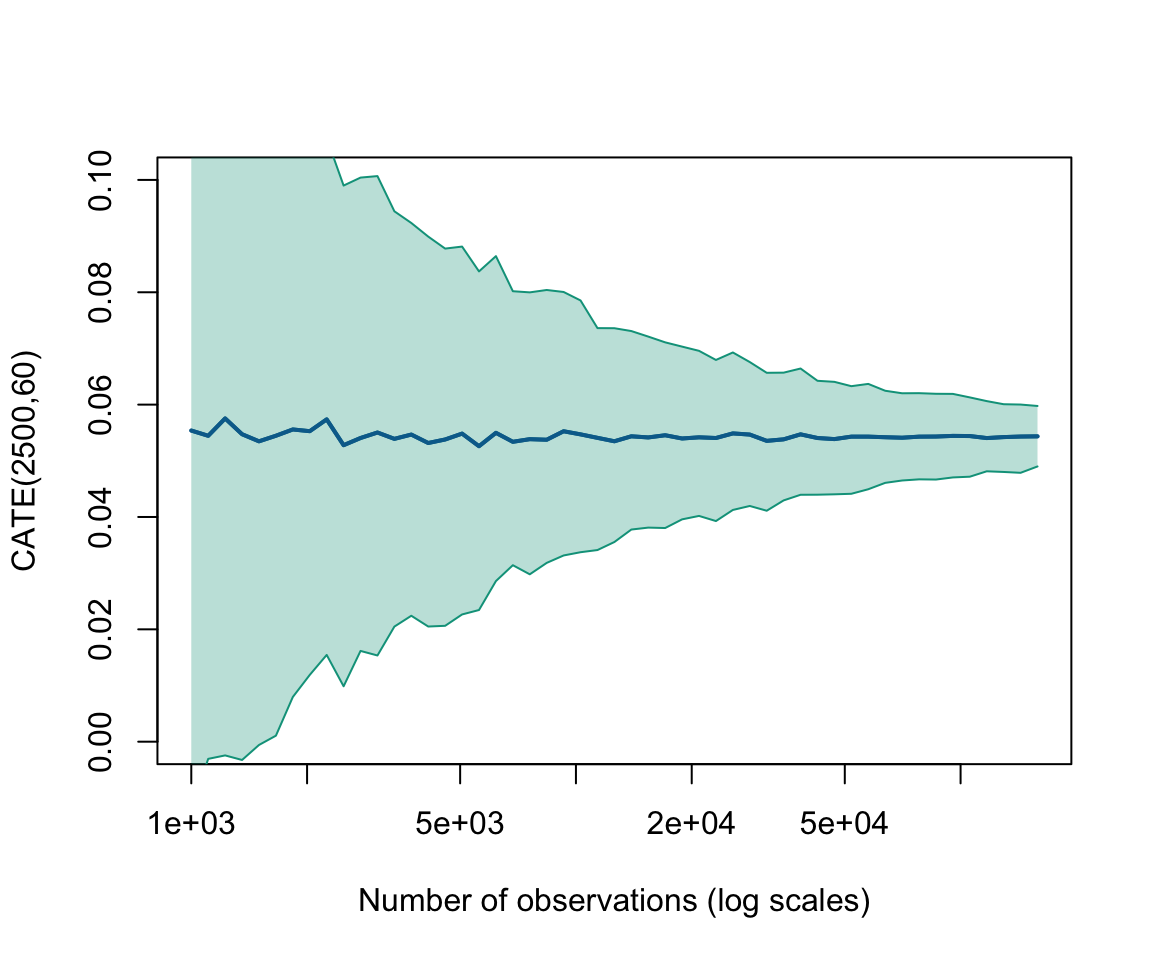}
     
    \centering
     \includegraphics[width=.49\textwidth]{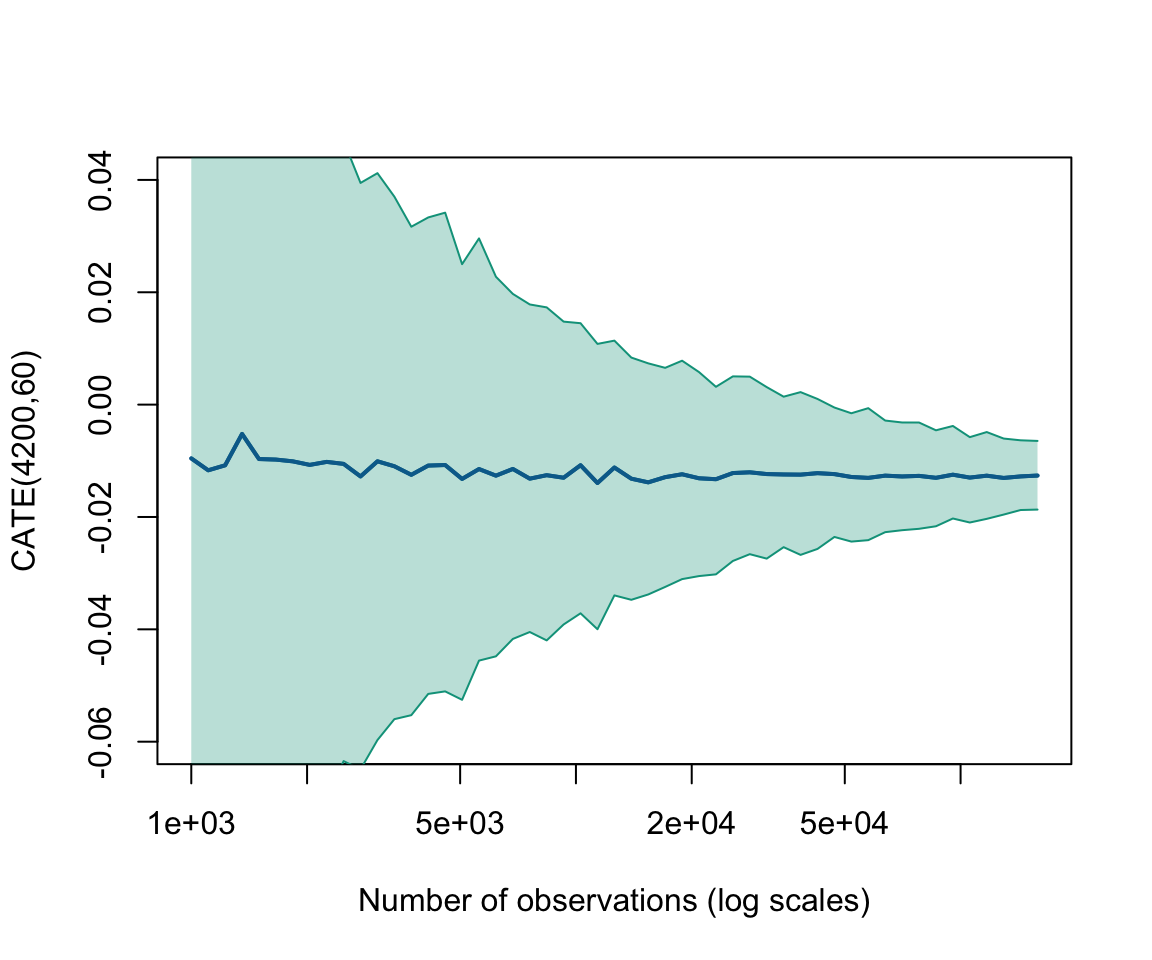}
     \includegraphics[width=.49\textwidth]{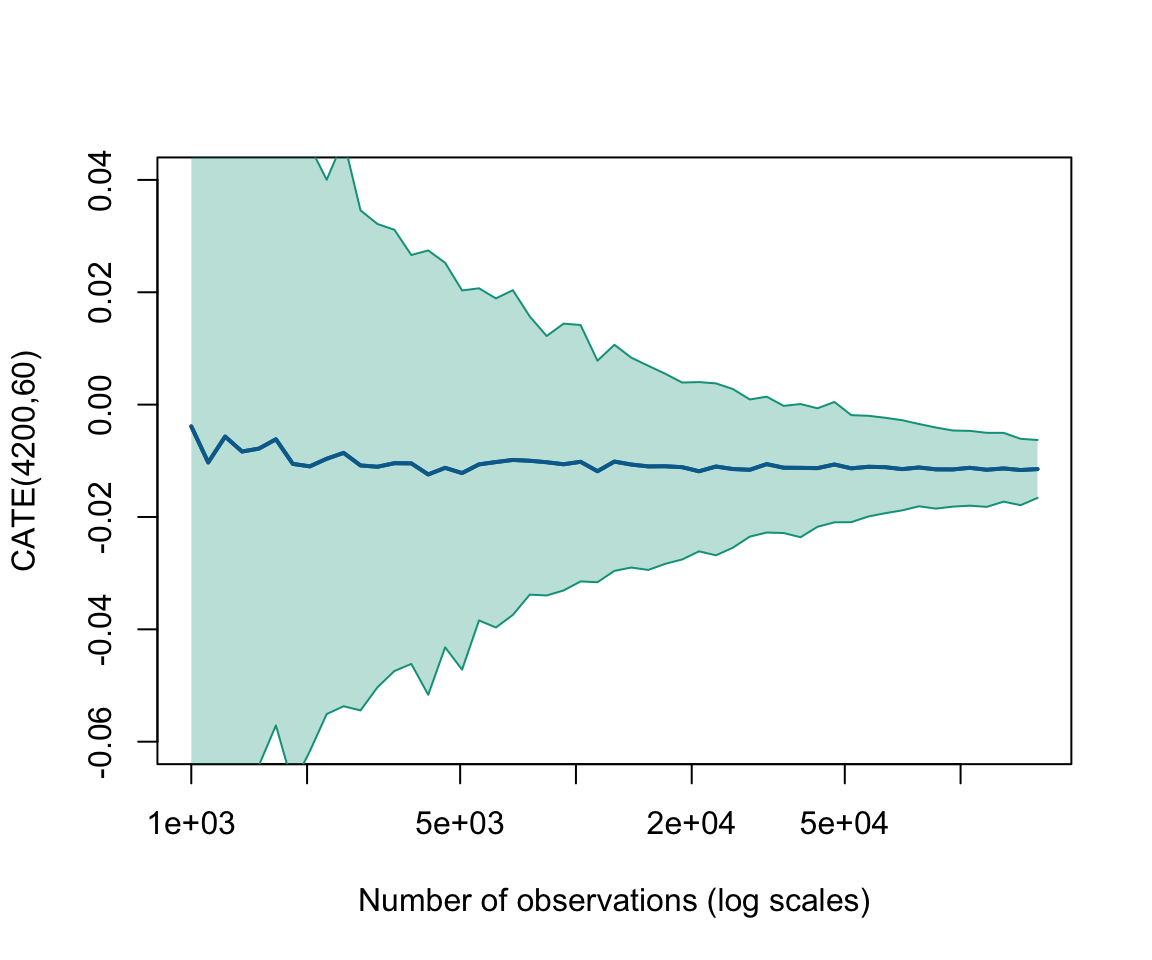}
     
    \centering
     \includegraphics[width=.49\textwidth]{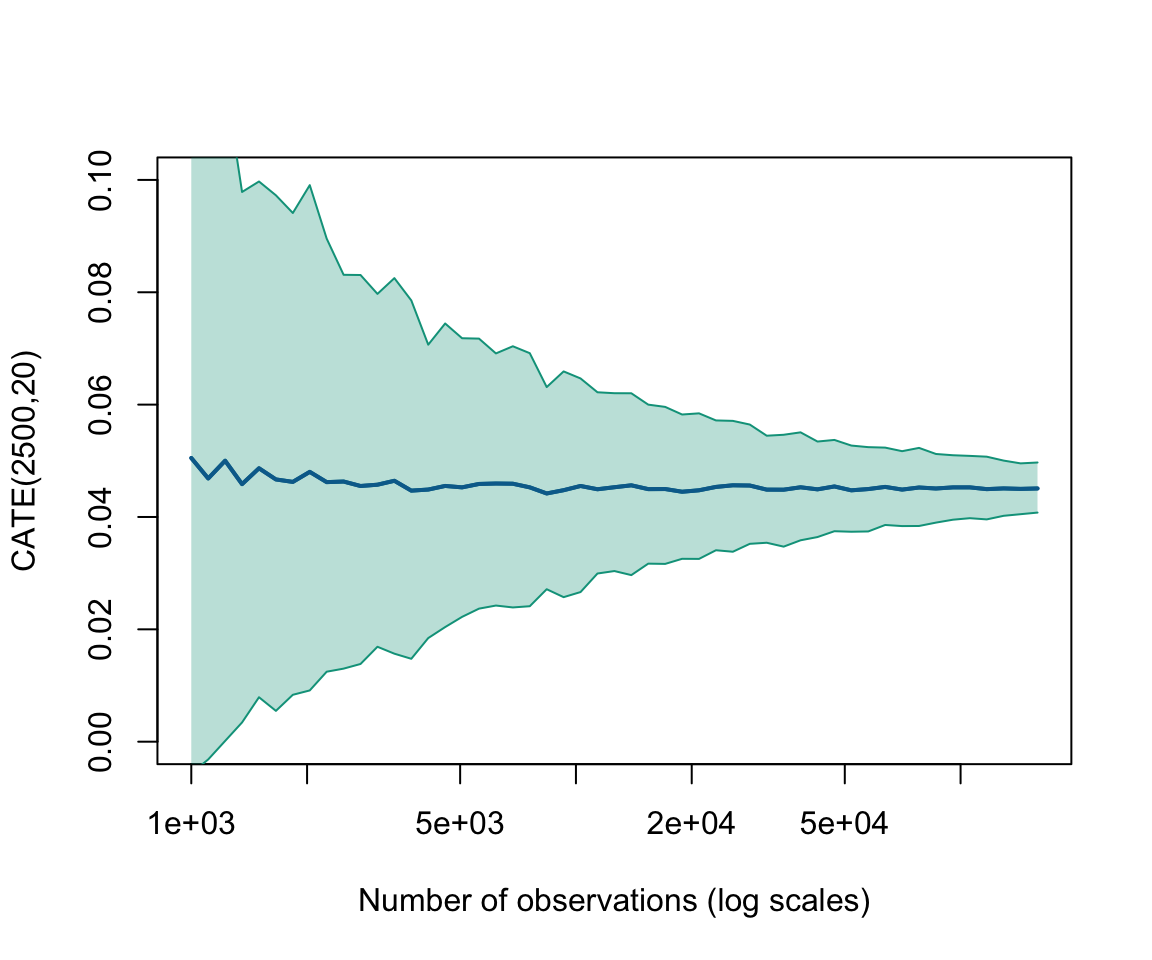}
     \includegraphics[width=.49\textwidth]{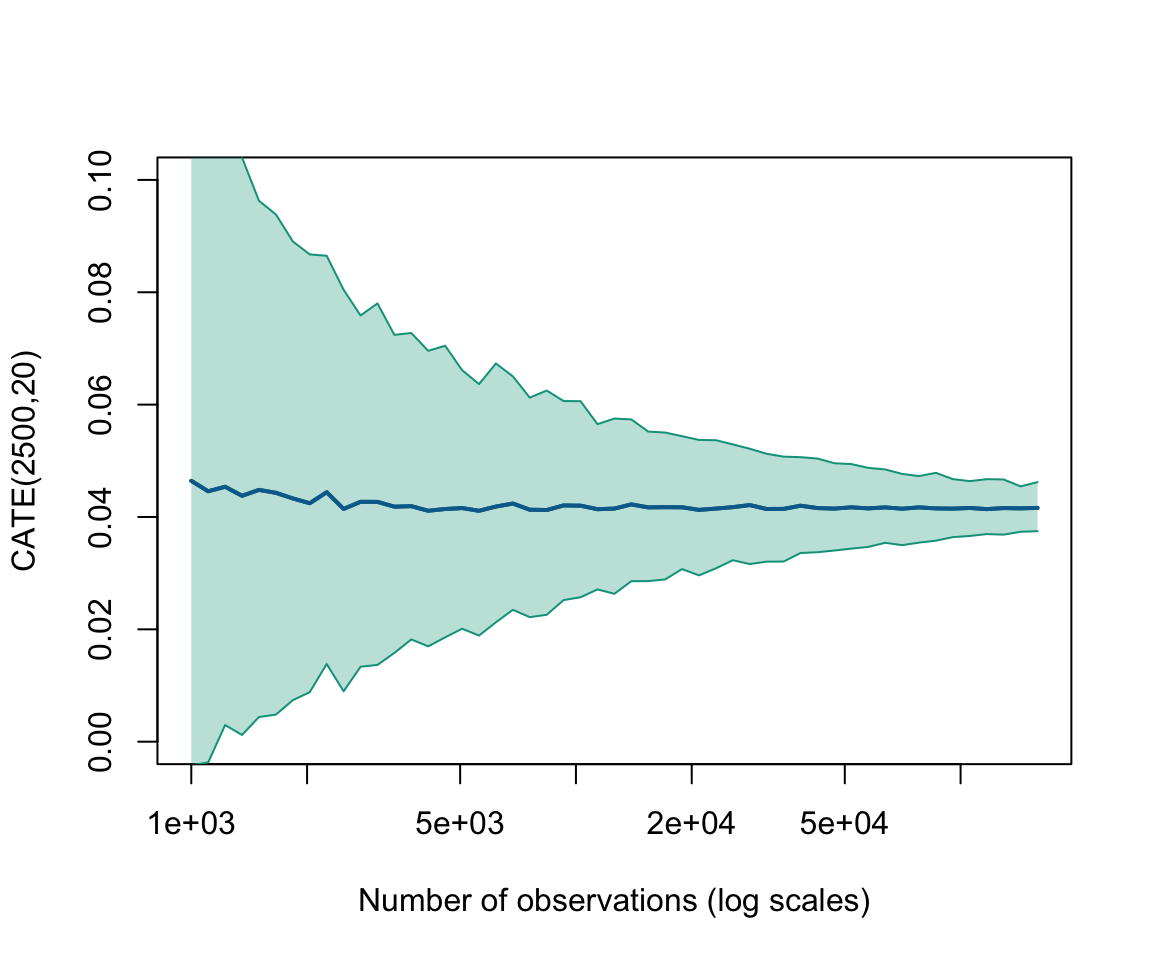}

    \caption{On the left, estimation of $\boldsymbol{x}\mapsto\text{SCATE}_{\mathcal{N}}[Y\boldsymbol{x}]$, estimated using a logistic GAM model with Gaussian transport estimated on $n$ observations (with $n$ increasing from $1,000$ to $150,000$), when $Y=\boldsymbol{1}(\text{non-natural delivery})$, and $\boldsymbol{X}$ is the weight of the newborn infant and the weight gain of the mother, respectively when $T$ indicates whether the mother is a smoker or not (on the left), whether the mother is Black or not (on the right). On top $\boldsymbol{x}=(2500,60)$, in the middle $\boldsymbol{x}=(4200,60)$ and at the bottom, $\boldsymbol{x}=(2500,20)$.}
    \label{fig:CATE-2x3-gaussian-transport-conf-int:appendix}
\end{figure}

\begin{figure}
    \centering
    \includegraphics[width=.49\textwidth]{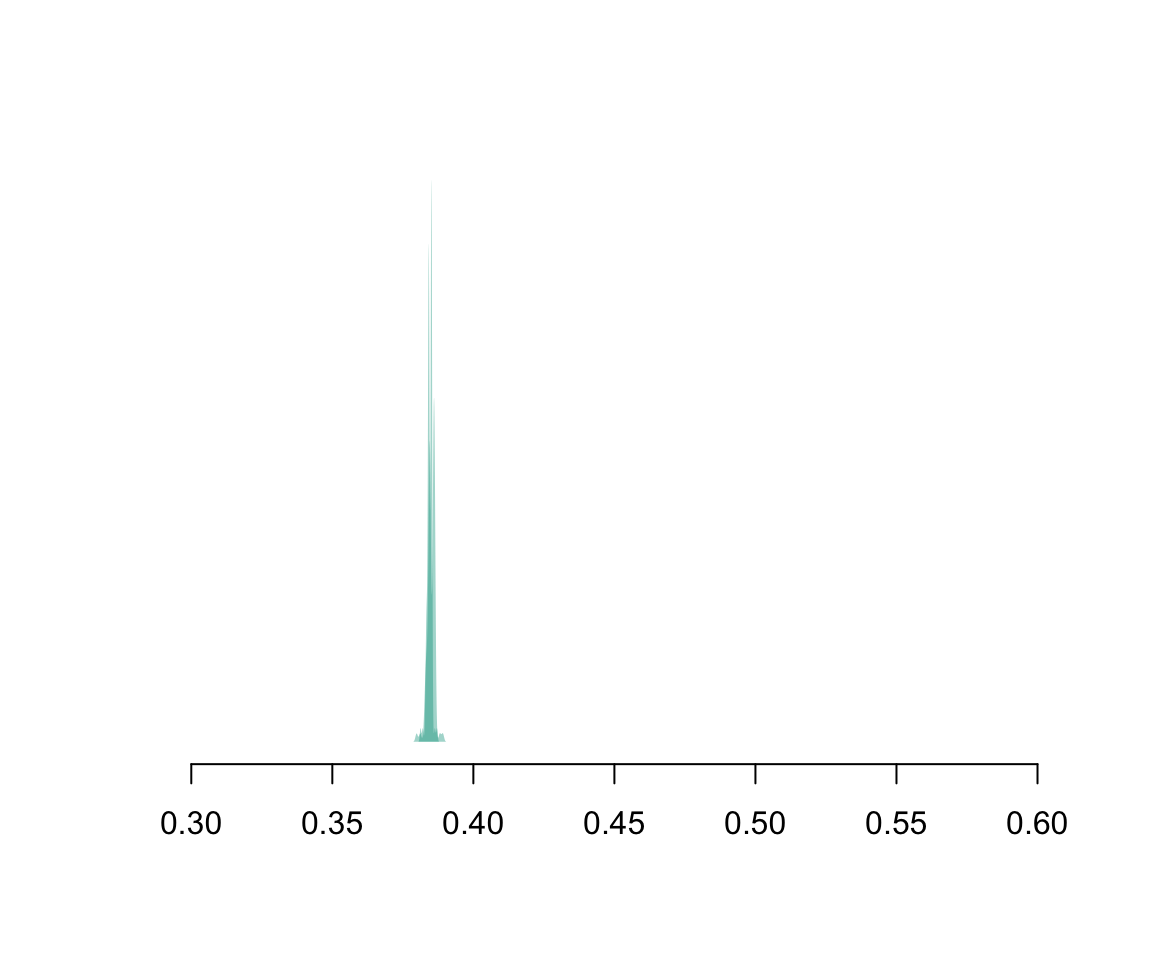}
     \includegraphics[width=.49\textwidth]{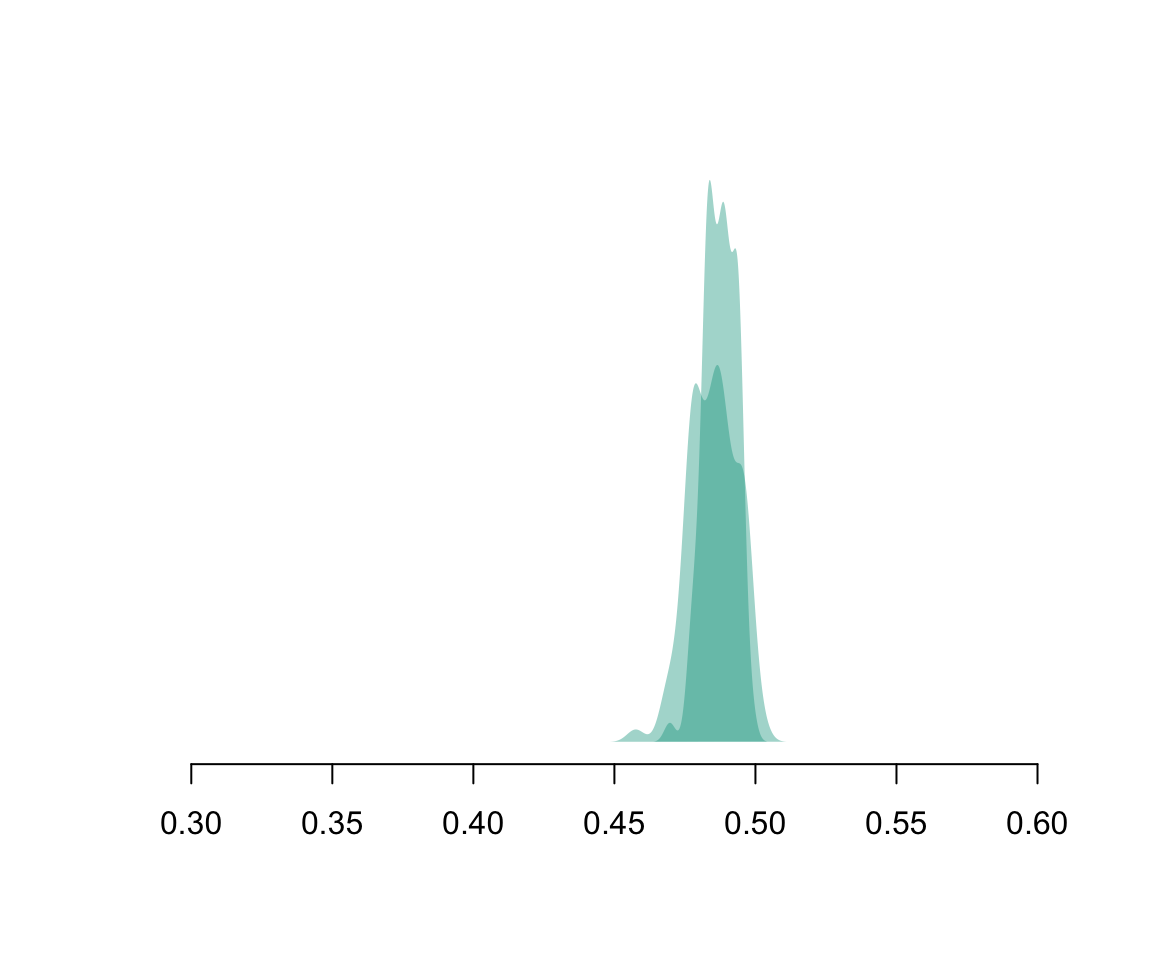}
     
    \centering
    \includegraphics[width=.49\textwidth]{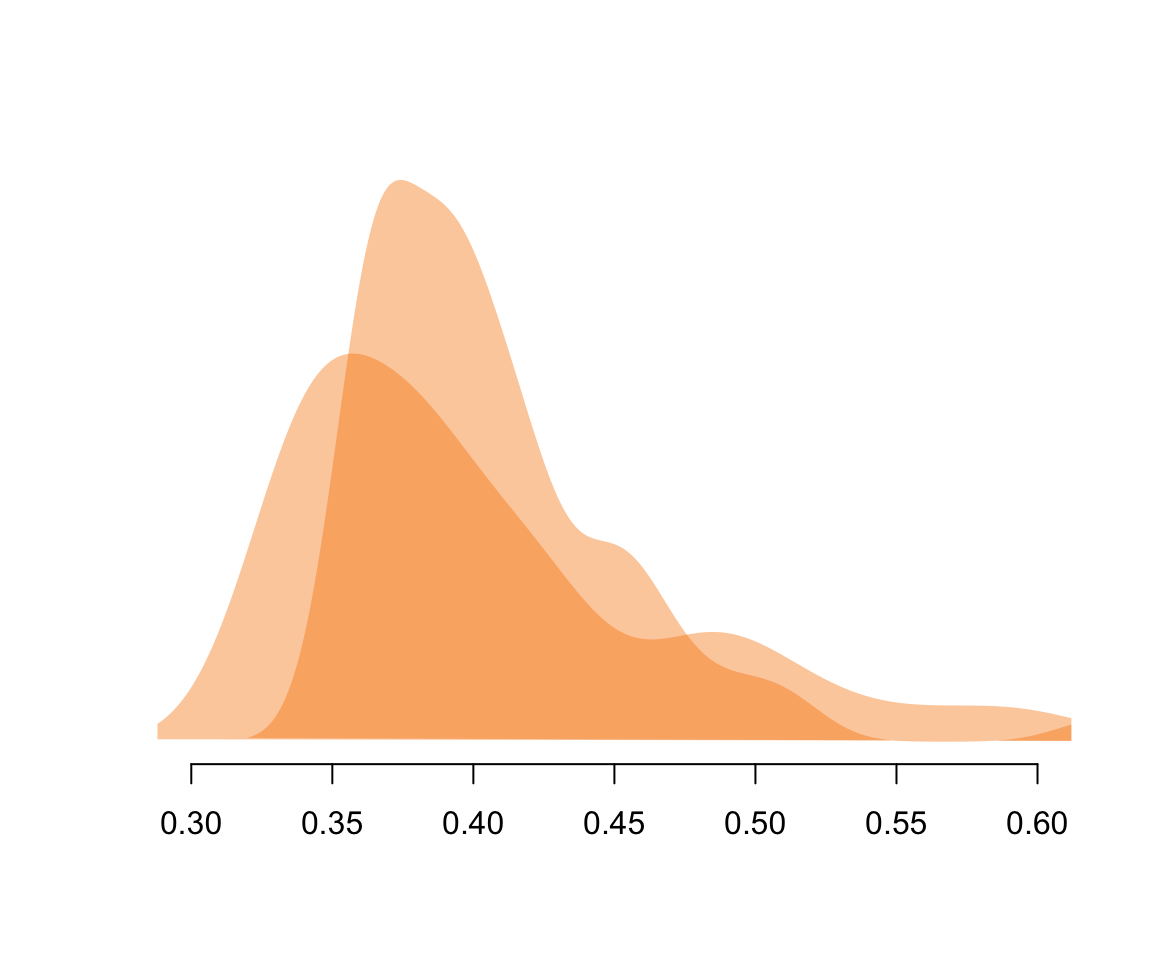}
     \includegraphics[width=.49\textwidth]{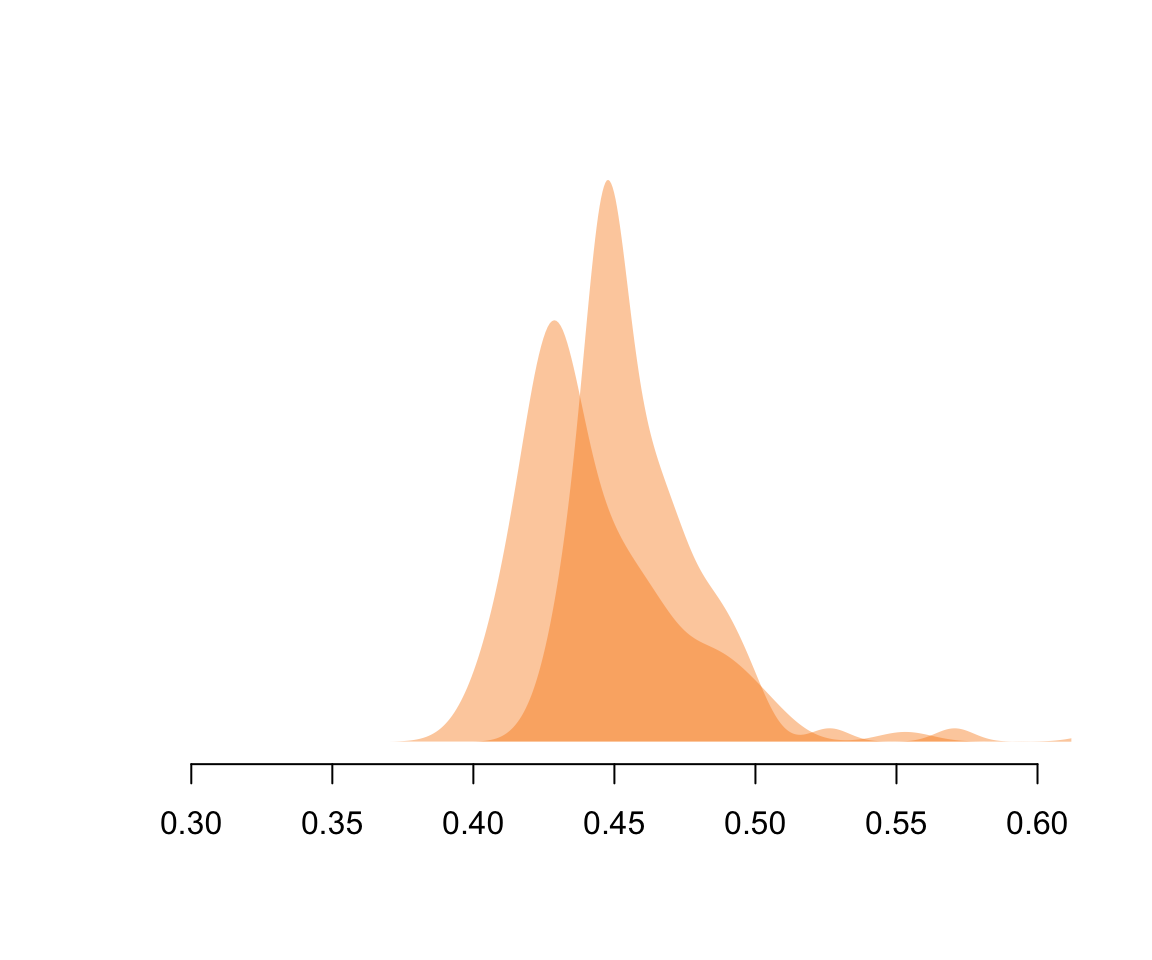}
     
    \centering
    \includegraphics[width=.49\textwidth]{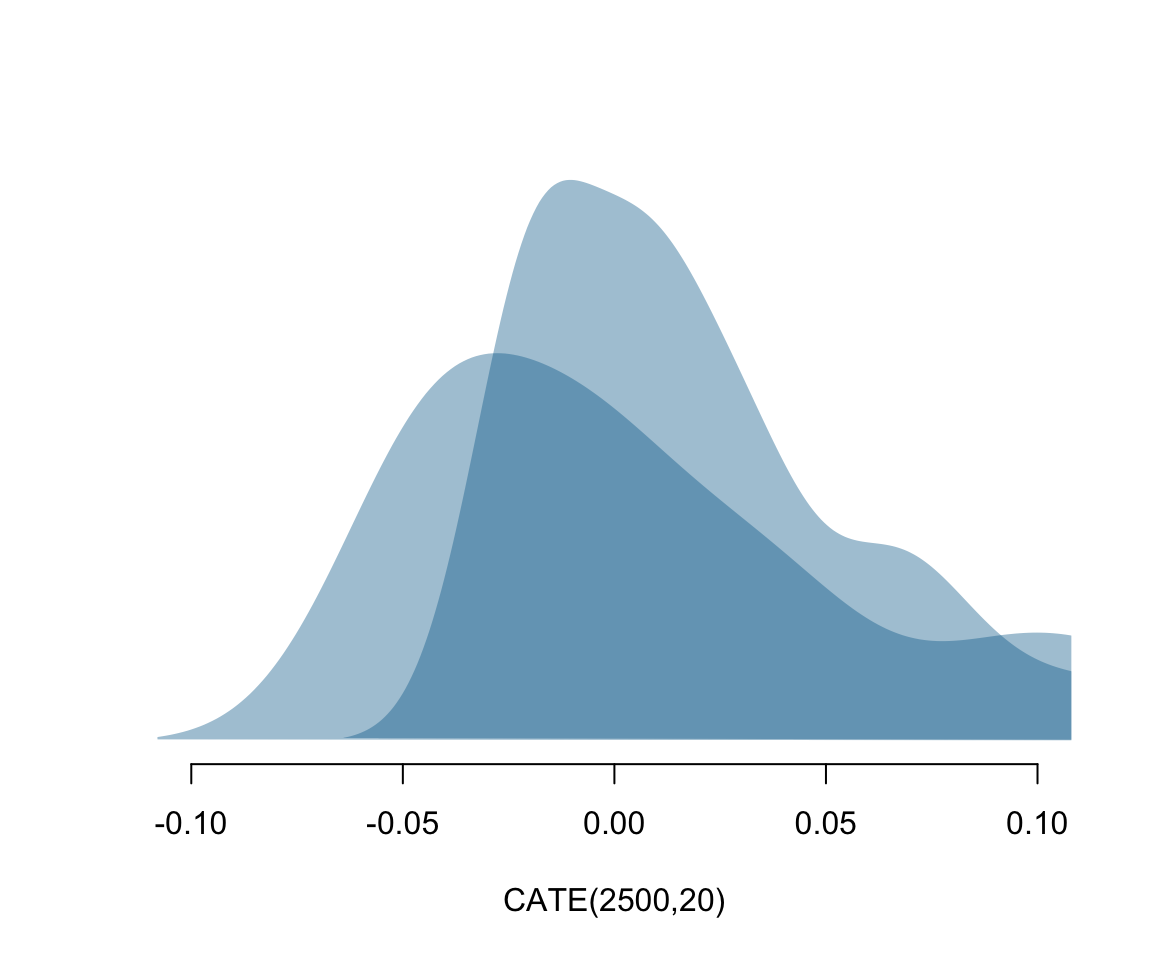}
     \includegraphics[width=.49\textwidth]{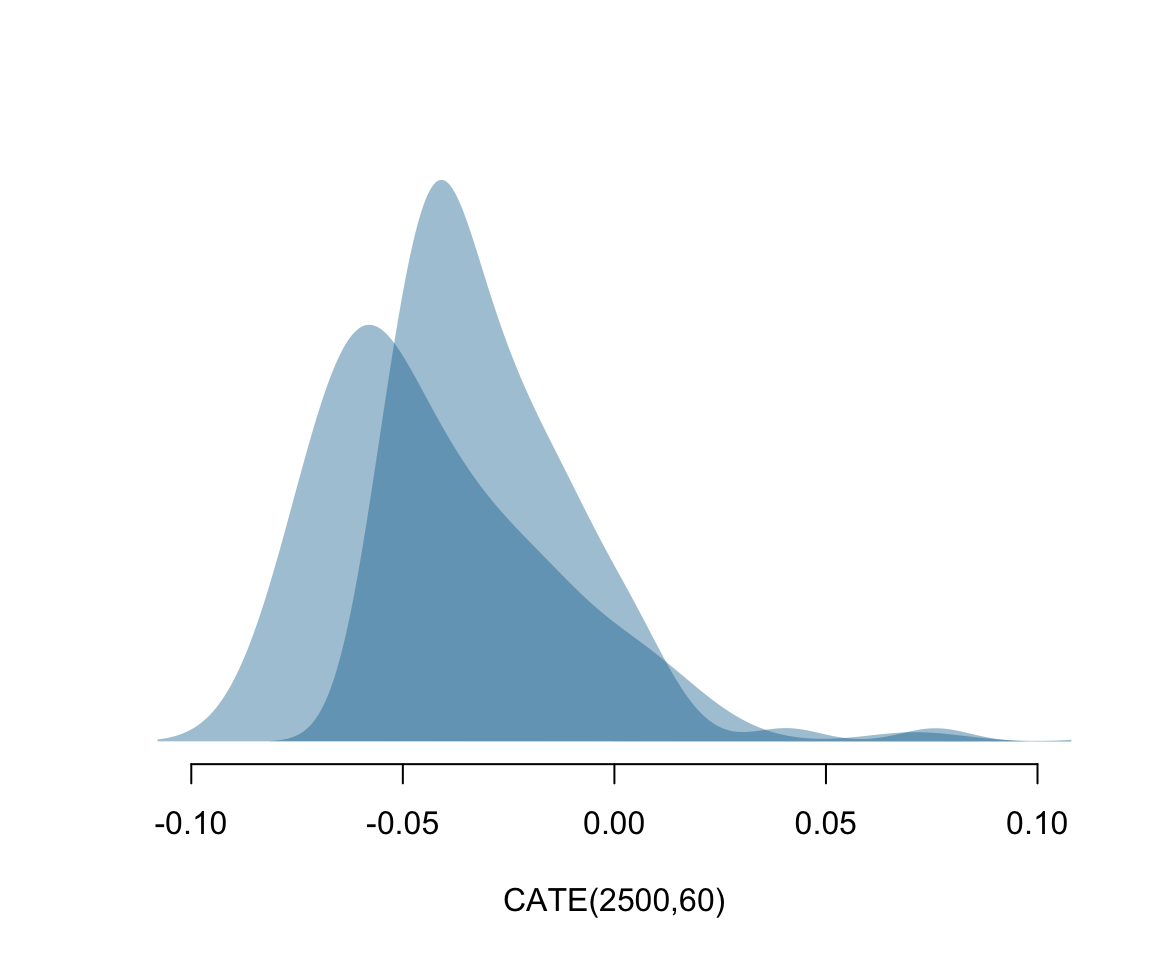}
    
    \caption{Distribution of $\widehat{m}_0(\boldsymbol{x})$ on top, $\widehat{m}_1(\widehat{\mathcal{T}}(\boldsymbol{x}))$ in the middle, and $\text{SCATE}(\widehat{\mathcal{T}}(\boldsymbol{x}))=\widehat{m}_1(\widehat{\mathcal{T}}(\boldsymbol{x}))-\widehat{m}_0(\boldsymbol{x})$ at the bottom, estimated on boostrapped samples of size $n=20,000$. On the left, $\boldsymbol{x}=(2500,20)$ and on the right $\boldsymbol{x}=(2500,60)$. The two densities are based on the fact that two GAM models are considered.}  
    \label{fig:densite-2x3-conditional-transport-simul:appendix}
\end{figure}

In Figure~\ref{fig:CATE-2x3-gaussian-transport-conf-int:appendix}, $\text{SCATE}(\boldsymbol{x})$ is estimated for various values of $\boldsymbol{x}$ ($\boldsymbol{x}=(2500,60)$ on top, $\boldsymbol{x} = 4200,60$ in the middle and $\boldsymbol{x}=(2500,20)$ at the bottom), depending on sample size $n$. The solid line is the average value, that is quite stable, but, as expected, the confidence interval is quite large when $n$ is small. In Figure~\ref{fig:densite-2x3-conditional-transport-simul:appendix}, we can visualize the distribution of $\widehat{m}_0(\boldsymbol{x})$ on top, $\widehat{m}_1(\widehat{\mathcal{T}}(\boldsymbol{x}))$ in the middle, and $\text{SCATE}(\widehat{\mathcal{T}}(\boldsymbol{x}))=\widehat{m}_1(\widehat{\mathcal{T}}(\boldsymbol{x}))-\widehat{m}_0(\boldsymbol{x})$ at the bottom, estimated on boostrapped samples of size $n=20,000$. On the left, $\boldsymbol{x}=(2500,20)$ and on the right $\boldsymbol{x}=(2500,60)$. The two densities are based on the fact that two GAM models are considered, with more or less knots. Confidence intervals are obtained by bootstrap.

\section*{Acknowledgments}
Arthur Charpentier acknowledges the financial support of the AXA Research Fund through the joint research initiative {\em use and value of unusual data in actuarial science}, as well as NSERC grant 2019-07077.

Emmanuel Flachaire and Ewen Gallic acknowledge the financial support of the French National Research Agency Grant ANR-17-EURE-0020, the Excellence Initiative of Aix Marseille University -- A*MIDEX
\clearpage

\bibliographystyle{erae}
\bibliography{biblio.bib}

%\begin{thebibliography}{9}
%\end{thebibliography}
\end{document}